\pdfoutput=1
\documentclass[smallcondensed, nospthms]{svjour3}

\usepackage{amsmath,amsfonts,amssymb,amscd,amsthm,amsbsy}
\usepackage{bm}
\newcommand\mc{\mathcal}

\allowdisplaybreaks

\usepackage[nointegrals]{wasysym}
\newcommand{\EarthSym}{\oplus}
\newcommand{\MoonSym}{\text{\leftmoon}}
\newcommand{\SunSym}{\odot}
\def\E{\EarthSym}
\def\M{\MoonSym}
\def\S{\SunSym}

\usepackage{amsthm}

\usepackage{upgreek}
\usepackage[cal=dutchcal]{mathalfa} 

\newcommand{\upw}[1]{\upomega^{#1}}
\newcommand{\upwb}{\upw{\mathsf b}}
\newcommand{\upwM}{\upw{\M}}
\newcommand{\upwS}{\upw{\S}}
\newcommand{\upwE}{\upw{\E}}

\DeclareRobustCommand{\rb}{%
  \mathord{%
    \mathchoice
      {\pmb{\displaystyle\mathcal{r}}}
      {\pmb{\textstyle\mathcal{r}}}
      {\pmb{\scriptstyle\mathcal{r}}}
      {\pmb{\scriptscriptstyle\mathcal{r}}}%
  }%
}

\newcommand{\vb}{\bm{v}}

\usepackage{graphicx}

\usepackage{epstopdf}
\usepackage{subfigure}

\usepackage{natbib}
\bibpunct{(}{)}{;}{a}{}{,}

\usepackage{appendix}
\usepackage{xr}
\usepackage{siunitx}           
\usepackage{booktabs}
\usepackage{makecell}

\usepackage{pdflscape}
\usepackage{longtable}
\usepackage{tabularx}

\usepackage[table]{xcolor}  
\definecolor{rowgray}{gray}{0.97} 

\newcommand{\zonebreak}{\multicolumn{4}{@{}l}{\hspace{0.8em}\rule{0.95\textwidth}{0.3pt}}\\[1ex]}

\usepackage[table]{xcolor}
\usepackage{booktabs}
\usepackage{colortbl}

\definecolor{tblBase}{HTML}{FBF7EF}   
\definecolor{tblZebra}{HTML}{F2EDE3}  
\definecolor{tblHeader}{HTML}{2D2620} 
\definecolor{tblAccent}{HTML}{E4EDF0} 
\definecolor{tblRule}{HTML}{5A4A3F}   
\definecolor{tblText}{HTML}{201A15}   
\definecolor{tblTerm}{HTML}{1F2E3D}   
\definecolor{tblHeadText}{HTML}{FBF7EF}

\usepackage{caption}
\usepackage{mathtools} 

\newcommand{\res}[3]{\(#1\!:\!#2#3\)}
\newcommand{\Res}[3]{%
  \quad $\mathmakebox[1.6em][r]{#1}\!:\!\mathmakebox[0.5em][l]{#2}\,#3$%
}

\makeatletter
\renewcommand{\makeheadbox}{}

\date{}
\makeatother
    
\begin{document}	

\title{The Astrodynamics Primer on Cislunar and Translunar Space}

\titlerunning{The Astrodynamics Primer on Cislunar and Translunar Space}

\author{
Aaron J. Rosengren \and 
Anjali Rawat \and \\
Bhanu Kumar \and
Shane D. Ross
}

\authorrunning{A. J. Rosengren et al.}

\institute{Aaron J. Rosengren \at Department of Mechanical and Aerospace Engineering, UC San Diego, La Jolla, CA 92037, USA \\ 
\email{arosengren@ucsd.edu}
\and Anjali Rawat \at Department of Aerospace and Ocean Engineering, Virginia Tech, Blacksburg, VA 24061, USA
\and Bhanu Kumar \at Department of Mathematics, University of Michigan, Ann Arbor, MI 48109, USA
\and Shane D. Ross \at Department of Aerospace and Ocean Engineering, Virginia Tech, Blacksburg, VA 24061, USA
\and Some aspects of this paper were presented at the {\itshape Advanced Maui Optical and Space Surveillance Technologies Conference}, 
Maui, Hawaii, 2021, 2024, and at the  {\itshape 76th International Astronautical Congress}, Sydney, Australia, 2025.}

\maketitle

\begin{abstract}
The dynamical structure of the Earth--Moon environment is multiscale and strongly nonuniform, yet it is often discussed as though it were a single cislunar regime. In reality, the region between Earth and the outer limit of meaningful Earth-bound motion is partitioned by changes in the perturbation hierarchy, by gateway topology, and by families of secular and resonant structures that differ qualitatively across the circumterrestrial domain. This paper develops a unified spatiographic description of that structure, combining perturbative theory, semi-analytical resonance cartography, restricted multi-body dynamics, and direct numerical mapping.

The terrestrial--cislunar transition is sharpened through the Laplace radius, beyond which lunisolar torques overtake the classical oblateness-dominated terrestrial picture. Earthward of the Moon, cislunar space divides into a secularly dominated inner zone and an outer zone structured by interior lunar mean-motion resonances. Near the Moon, circumlunar space emerges as a distinct dynamical enclave organized by the Earth--Moon gateway geometry and the lunar sphere of influence. Beyond the Moon, translunar space forms an outer circumterrestrial province in which the Moon acts as an interior perturber, the Sun remains an exterior perturber, and the resulting dynamics acquire a mixed lunisolar secular and resonant architecture before weakening outward toward heliocentric behavior.

These analytical and restricted-problem results are synthesized through MEGNO and fate-class cartographies across six numerical map domains derived from the spatiographic partition, using two complementary numerical models: an elliptic Earth--Moon restricted model that isolates the lunar resonant and gateway architecture, and an Earth--Moon--Sun ephemeris model that restores solar forcing and tests the persistence, displacement, or reorganization of that architecture. The maps reveal where quasiperiodic confinement survives, where resonance overlap and gateway transport produce sticky residence and organized escape, and where solar forcing becomes a qualitative architectural ingredient rather than a perturbative correction. Placed alongside a curated catalog of historic and current cislunar and translunar satellites and debris, the resulting framework provides a dynamical geography of the Earth--Moon environment that clarifies the transitions among cislunar, circumlunar, and translunar motion and offers a more precise basis for interpreting stability, transport, and long-term Earth-bound behavior.
\end{abstract}


\bigskip 
\begin{quote}
\itshape
If God wanted man to become a spacefaring species, He would have given man a moon.

\hfill{\normalfont\small---Krafft A.~Ehricke (1985), \emph{Lunar Industrialization and Settlement}}
\end{quote}

\clearpage 

\section{Introduction}

Between Earth and the onset of heliocentric escape lies a circumterrestrial region whose internal dynamics are far from uniform. The space interior to the Moon, the lunar-vicinity gateway, and the weakly bound domain beyond lunar distance are governed by different perturbation hierarchies, different resonant structures, and different transport geometries \citep{kE62a, kE62b}. Yet in contemporary policy, legal, security, and technical discourse these distinct provinces are routinely gathered under a single broad operational banner, \emph{cislunar} \citep{mH21, APL22, NCSTS22, bB24, aW25, dK25, tP26}. The consequence is not merely terminological looseness. It is the loss of a dynamical taxonomy precisely where one is most needed.

This paper develops a dynamical geography of the Earth--Moon environment. Its central claim is that the coupled Earth--Moon--Sun problem is best understood not as a single homogeneous beyond-GEO shell, but as a circumterrestrial hierarchy of distinct provinces: terrestrial, cislunar, circumlunar, translunar, and finally heliocentric. Those provinces are not introduced as matters of vocabulary alone. They correspond to genuine dynamical handoffs: from oblateness-dominated terrestrial motion, to lunisolar secular cislunar dynamics beyond the Laplace radius, to the outer cislunar region structured by interior lunar mean-motion resonances, to the circumlunar gateway organized by the Earth--Moon Lagrange-point passages, and finally to the translunar realm beyond the Moon, where the Moon acts as an interior perturber, the Sun remains an exterior perturber, and the resulting outer circumterrestrial dynamics acquire a mixed secular and resonant architecture before fading into heliocentric behavior. As a result, stability, predictability, observability, and even the most natural coordinate description of motion are regime dependent.

This framing also places the familiar libration-point architecture in its proper context \citep{rB56a, wKeB58, kE59, vE59, bT59, gC60, gC61, gEaD61, gN66, rFaK73}. The Earth--Moon collinear points and their associated Lyapunov, halo, Lissajous, and related orbit families are not treated here as the definition of cislunar space, but as gateway structures embedded within the larger spatiography: $L_1$-associated families mediate the cislunar--circumlunar transition, while $L_2$-associated families connect the circumlunar gateway to the exterior translunar realm. The same regime dependence applies to representation. Classical geocentric orbital elements do not simply fail beyond geosynchronous orbit; over most of the cislunar domain and much of the translunar domain they remain meaningful instantaneous circumterrestrial descriptors, even though they cease to behave as slowly varying Keplerian labels and must eventually give way to selenocentric, rotating-frame, or manifold-based descriptions near the lunar gateway and within the circumlunar enclave.

This corrective is not merely semantic. It is also historical. Much of what is now presented as a newly emerging cislunar problem was already examined, often in remarkable depth, during the first decades of the Space Age. Early astronautical authors did not lack either a taxonomy or a dynamical intuition for the Earth--Moon environment. On the contrary, Ehricke and his contemporaries articulated a coherent circumterrestrial picture in which \emph{terrestrial}, \emph{cislunar}, \emph{circumlunar}, and \emph{translunar} denoted distinct regions and modes of motion rather than interchangeable labels for everything beyond GEO \citep{kE55, kE56, kE57, kE58a, kE58b, kE59, kE62a, kE62b, hS58, hS59, WEC59}. In parallel, the mathematical restricted-problem literature was already constructing much of the dynamical backbone on which modern transport theory---the geometric description of phase-space structures that organize motion---now rests: Broucke's catalogs and bifurcation diagrams furnished a topological atlas of periodic-orbit families in the Earth--Moon circular restricted three-body problem (CR3BP) and elliptic extension \citep{rB68, rB69}, while Egorov's three-dimensional classifications provided an early inventory of cislunar and circumlunar multi-revolution, resonant, and quasi-periodic trajectories in higher-fidelity settings \citep{vE58, vE69}. 

Around these foundational works lay a broader and now often neglected wave of studies on translunar flight, lunar-vicinity operations, repeated-passage trajectories, and restricted multi-body structure, including early trajectory and navigation analyses \citep{pP61, aP61, jM63, rT63, gH64}, symmetric free-return and circumlunar/cislunar flyby studies \citep{aS63}, repeated close-passage and cycler-like orbit families \citep{rN59, sH62, rA63, sHcW63, rHbW68, jKjL68}, restricted four-body formulations \citep{sH60, sH61, jdV64, jCetal64, jD65, lW66, rKlC67, ySmE67, lMjK67, hS68, rKlC68, jCetal68, bTbS68, aKjB70, pB77}, and circumlunar dynamics \citep{yK63, aR68a, aR68b, gG70, wB71}. The Earth--Moon environment is thus not a newly discovered dynamical province; what is new is the present need to synthesize that older taxonomic, operational, and restricted-problem literature into a single framework that is dynamically explicit and usable across the full circumterrestrial problem.

The present manuscript offers that description by combining four complementary viewpoints. First, orbit-averaged perturbation theory identifies the principal secular transitions and clarifies how the roles of the Moon and Sun change across circumterrestrial space. Second, semi-analytical resonance atlases furnish a local cartography of the dominant cislunar and translunar mean-motion commensurabilities in orbital-element space. Third, restricted multi-body phase-space structure---especially through periodic orbits, bifurcations, separatrices, and Poincar\'e sections---reveals the global resonant skeleton that local width estimates alone cannot capture. Finally, direct MEGNO and fate-class cartographies show where those structures survive, where they are eroded, and where they give way to broader transport, capture, collision, or escape. For this final numerical layer, we deliberately separate two complementary dynamical models. The first is an elliptic Earth--Moon restricted model initialized from JPL Horizons at the chosen epoch, in which the Earth and Moon subsequently evolve as an isolated point-mass pair; this baseline isolates the lunar resonant, gateway, and outer Earth--Moon transport architecture. The second is a JPL-Horizons-initialized Earth--Moon--Sun point-mass ephemeris model, in which solar forcing is restored and the persistence, displacement, or destruction of the Earth--Moon structures can be assessed. The result is not a single-model picture, but a layered one: perturbative, semi-analytical, restricted, and numerical, each contributing a different scale of dynamical legibility.

The paper is therefore cast as a \emph{Primer}: not simply to compile mission classes or isolated resonances, but to provide a coherent language and architecture for motion in the Earth–Moon environment. That framework is intended to be useful both scientifically and operationally: scientifically, because it reconnects perturbation theory, resonance structure, and restricted-problem dynamics into a unified picture; operationally, because the demands of navigation, custody, disposal, and infrastructure are all sensitive to the regime in which an object actually resides.

The paper is organized as follows. Section~\ref{sec:history} revisits the historical lexicon and the modern umbrella use of \emph{cislunar}, motivating the need for a more precise dynamical partition. Section~\ref{sec:catalog} then introduces the curated catalog of historic and current cislunar and translunar satellites and debris, and uses it to provide a first empirical view of how the circumterrestrial population beyond GEO is distributed in element space. Section~\ref{sec:perturbed} develops the perturbative treatment of distant circumterrestrial motion, distinguishing secular architecture from mean-motion commensurabilities and clarifying the different roles of the Moon and Sun inside and beyond lunar distance. Section~\ref{sec:boundaries} constructs the spatiographic partition itself, introducing the principal physical scales and dynamical boundaries that organize the Earth--Moon environment, including the Laplace radius, lunisolar tidal handoff criteria, and sphere-of-influence and Hill-type demarcations, and then using them to define the secularly dominated cislunar interior, the resonant cislunar zone, the circumlunar gateway, and the translunar provinces beyond. Section~\ref{sec:restricted} turns to restricted multi-body dynamics, showing how periodic orbits, bifurcations, separatrices, and transport channels organize the local and global phase-space structure in the Earth--Moon and patched Earth--Moon--Sun problems. Finally, Section~\ref{sec:cartographies} presents numerical astro-cartographies---MEGNO and fate-class maps organized into six cartographic domains derived from the cislunar--translunar spatiographic partition---that place the analytical, semi-analytical, and restricted-problem results into a single global portrait, and then relate that portrait directly to the curated population introduced earlier.

The theme unifying all that follows is simple. The Earth--Moon environment is not one vast regime. It is a stratified dynamical country, whose borders, passages, and provinces are distinguished by different organizing mechanisms and whose internal transitions matter both mathematically and operationally. The task of this paper is to describe that geography and to show how cislunar, circumlunar, and translunar space fit together as distinct parts of one circumterrestrial whole.

\section{Historical Context and Persistent Misconceptions}
\label{sec:history}

As activity beyond geosynchronous orbit accelerates---navigation, rendezvous, resource prospecting, security, and infrastructure---the language used to describe the Earth--Moon environment has become operationally consequential. One term in particular, \emph{cislunar}, has drifted from its historically narrower sense into a modern umbrella label that now anchors policy documents, technical visions, and even engineering primers \citep{NASAUSSF20, mH21, NCSTS22, APL22, bB24, dK25, aW25, tP26, yHsCyW26}. This drift is sociolinguistically understandable, and often operationally convenient, but it can obscure distinctions that matter dynamically. This section argues that early astronautical literature already supplied a clearer taxonomy---one that modern usage has largely forgotten, but that remains naturally aligned with the underlying dynamics of the Earth--Moon system \citep{kE55, kE56, kE57, kE58a, hS58, hS59, kE62a, kE62b, mV84}.

\subsection{From Eclipse Discourse to Astronautics: A Lunar-Boundary Lexicon}

As summarized in the lexicographic timeline of Table~\ref{tab:ems_lexicon} in Appendix~\ref{app:glossary}, \emph{circumlunar} (1871) and \emph{cislunar} (1872) enter English in the eclipse--corona discourse of the early 1870s as observationally grounded spatial qualifiers: \emph{circumlunar} for phenomena ``round the Moon,'' and \emph{cislunar} for effects inferred to lie Earthward of the Moon (i.e., between Earth and lunar distance), often framed explicitly against atmospheric explanations. In doing so, \emph{cislunar} supplies a physical/astronomical alternative to the older metaphysical cosmography of \emph{sublunar} (1598) and \emph{sublunary} (1601), whose dominant historical sense marks the mutable ``earthly realm'' beneath the lunar sphere. \emph{Translunar} is older (1791), but it is the member of this broader ``lunar-boundary'' family that most fully survives into modern technical usage, acquiring an explicit astronautical sense during the Space Age alongside revived and standardized uses of \emph{cislunar} and \emph{circumlunar}.

\subsection{The Positional Meaning of \emph{Cislunar} and the Limits of Umbrella Usage}

Lexicographically, \emph{cislunar} is unambiguous: \emph{cis-} denotes ``on this side of,'' a geometric relation defined with respect to a reference boundary---here, the Moon or the Moon's orbit. Historically, the term denotes the Earthward region within the Moon's orbit, i.e., ``this side of the Moon,'' not ``everything in the Earth--Moon system.'' The term is therefore \emph{positional} by construction: it partitions space by \emph{where} an object lies with respect to the Moon, not by \emph{which dynamical regime} organizes its motion. That positional content matters. A large class of trajectories beyond GEO never approach the Moon at all \citep{bS26}. Some Earth--Moon transfers do not remain cleanly ``cis'' or ``trans'' in any enduring geometric sense (e.g., long weak-stability-boundary transfers, multi-loop phasing arcs, resonant tours). The Earth--Moon $L_3/L_4/L_5$ neighborhoods are neither ``this side'' nor ``beyond'' the Moon in any natural geometric sense; in the rotating frame they are equilibrium neighborhoods tied to the \emph{system} \citep{sS60, eRaS62, hSwH64, lSdV66}. Nor does \emph{translunar} repair the ambiguity when pressed into service as a substitute umbrella term: in its plain and historical sense \emph{trans-} marks passage \emph{across} the lunar-distance boundary, so the term naturally names either (i) transfer arcs that reach or cross that boundary (e.g., translunar injection) or (ii) the circumterrestrial region \emph{beyond} the Moon's orbit (i.e., the \emph{complement} of \emph{cislunar})---not the coupled Earth--Moon environment as a whole.

If the goal is to denote a gravitationally and dynamically coupled Earth--Moon environment, then ``cislunar'' is too narrow and too positional for that role. A term for that environment must be \emph{relational, not positional; framework-based, not monolithic; dynamically grounded, yet topologically faithful}. In other words, it should name the \emph{coupled system} and its organizing structures, rather than borrow a geometric prefix whose meaning is fixed by the Moon as a boundary.

\subsection{The Sociolinguistic Drift and its Operational Cost.}

Despite its narrow meaning, \emph{cislunar} has been widely adopted to denote a broad operational domain including Earth--Moon transfers, libration-point neighborhoods, lunar-adjacent orbits, and sometimes even volumes well beyond the Moon \citep{mH21, NCSTS22, APL22, tP26}. Representative definitions now describe cislunar space as (i) the volume ``influenced by the Earth and/or Moon,'' (ii) the region beyond GEO ``within the gravitational influence of the Earth and/or the Moon,'' (iii) a sphere enclosing the Earth, Moon, and Lagrange points, and (iv) a catch-all aligned with ``xGEO'' usage. The pattern is strikingly consistent across communities. The common features are that the term is already English, rhetorically symmetric with \emph{translunar}, short and policy-friendly, and avoids technical dynamical language (e.g., ``restricted three-body regime''). In short: the modern adoption reads as sociolinguistic convergence rather than scientific---an expedient label that grew by usage, not by definition.

But the drift has costs. When ``cislunar'' is used to denote a single homogeneous region, it obscures the fact that Earth--Moon space is partitioned by qualitatively different dynamical regimes---oblateness-dominated terrestrial space, lunisolar-torque-dominated cislunar space, resonance-structured corridors, and Sun--Earth organized behavior beyond lunar distance (see, e.g., Table~\ref{tab:spatio}, described in \textsection\ref{sec:boundaries}). The boundaries are not merely semantic: they determine precession structure, stability timescales, observability, station-keeping cost, disposal pathways, and even which coordinate frames and element sets remain well conditioned.

\subsection{A Forgotten Space-Age Taxonomy: \emph{Circumterrestrial} as the Umbrella}

The early Space Age did not lack a clean lexicon. It had one---explicitly systematized by Krafft Ehricke and others---built around the broader term \emph{circumterrestrial} \citep{kE55, kE56, kE62a}. In this view, circumterrestrial is the umbrella domain of Earth-bound motion, within which distinct subdomains are recognized: \emph{terrestrial} (near-Earth, where oblateness, atmosphere, and surface irregularities are prominent), \emph{cislunar} (predominantly Earthward of the Moon, where lunar and solar perturbations are no longer negligible), and \emph{translunar} (beyond the lunar orbit but still within Earth's ability to retain a satellite, hence still circumterrestrial in the broad sense). \emph{Circumlunar} space, however, is conceptually different: it denotes the Moon's immediate vicinity, where lunar gravity dominates---roughly, within the Moon's sphere of influence---and capture, escape, and selenocentric motion are organized by the $L_1/L_2$ gateway geometry. It is therefore best understood as a local lunar enclave embedded within the broader Earth--Moon system, furnishing the transition between cislunar and translunar motion.

This framework is both lexically faithful and dynamically useful: it preserves \emph{cis/trans} as geometric qualifiers while letting \emph{circumterrestrial} carry the burden of naming the system domain. It also clarifies how libration points should be spoken about: $L_1$ (between Earth and Moon) and $L_2$ (beyond the Moon) are naturally \emph{cislunar} and \emph{translunar}, respectively, in the Earth--Moon problem, whereas $L_3/L_4/L_5$ lie on or near the lunar distance and are more naturally treated as \emph{system} equilibria than as ``cis'' or ``trans'' artifacts. These historical terms are not merely period language: they align naturally with the dynamical partitions revealed by the modern analysis developed in Sections~\ref{sec:perturbed}--\ref{sec:cartographies}, including resonance structure, secular-torque regimes, and the gateway geometry of Hill's regions and invariant-manifold transport, and therefore remain operationally meaningful.

\begin{figure}[htp!]
	\begin{center}
	\includegraphics[width=0.875\textwidth]{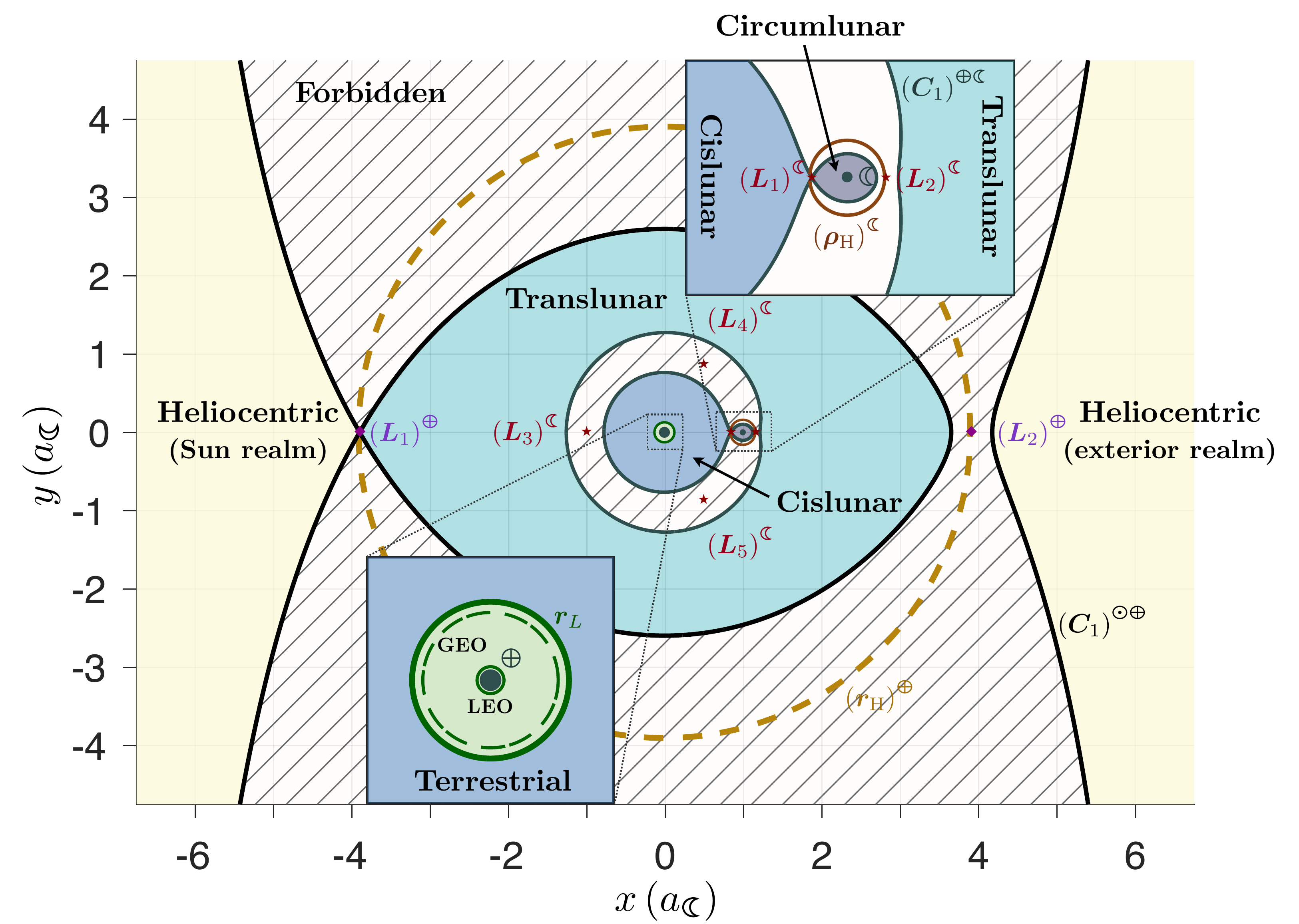}
	\includegraphics[width=0.875\textwidth]{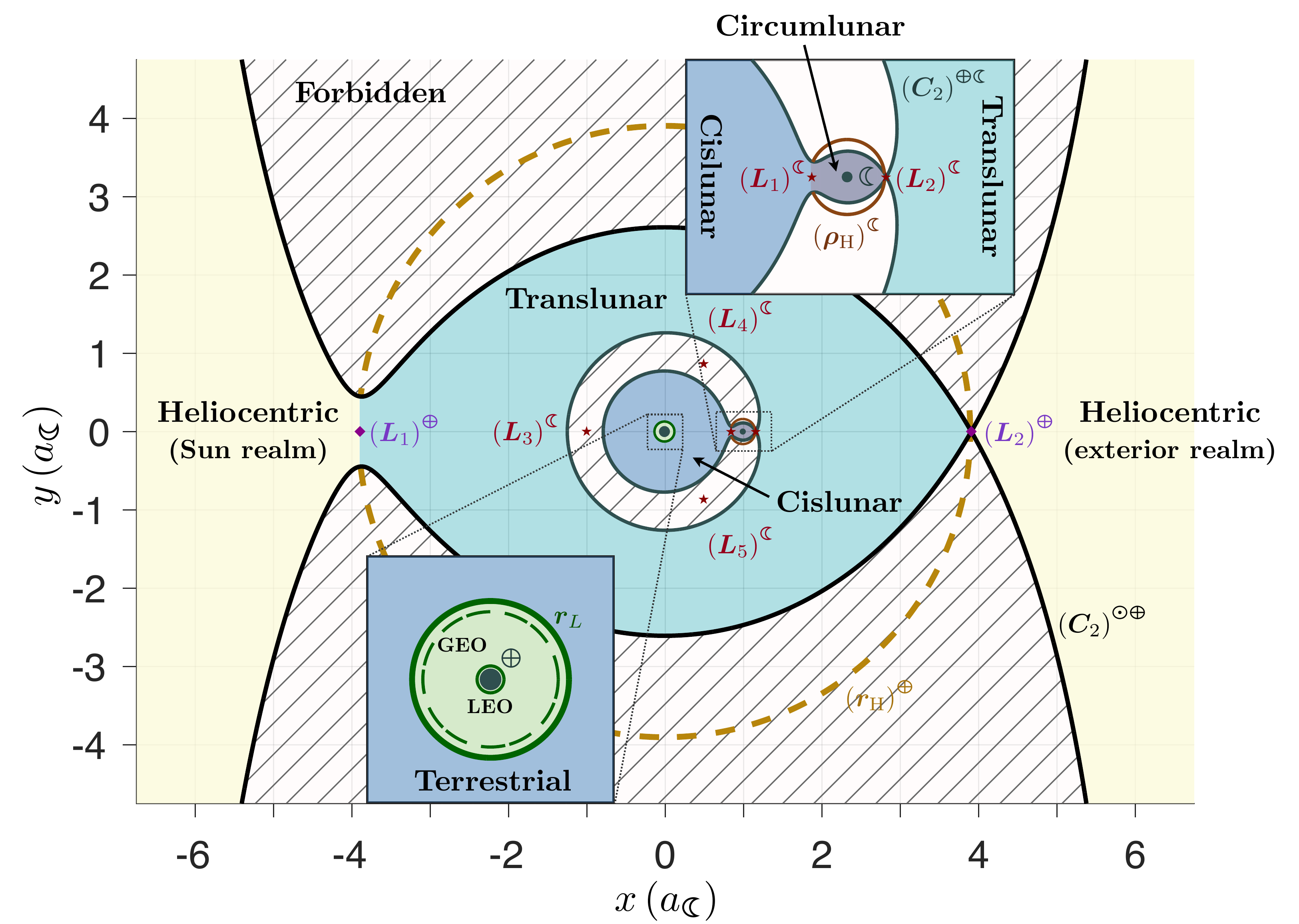}	
	\vskip -0.075in
	\caption{\small 
A dynamical ``spatiography'' of the Earth--Moon environment constructed by patching two complementary restricted problems at different scales (schematic; adapted from \citet{kE62b} and \citet{sH60, sH61, sH69}). The {\it top panel} patches the Earth--Moon and Sun--(Earth\texttt{+}Moon) restricted problems at the first critical Jacobi level of each, with the corresponding $C_1$ zero-velocity boundaries shown; the \emph{bottom panel} repeats the construction at the second critical level, with the corresponding $C_2$ boundaries shown. In each panel, the {\it dark curves} are the zero-velocity boundaries of the corresponding restricted problem, drawn at lunar scale, and the {\it hatched regions} denote forbidden motion. The inner patch identifies the cislunar (Earth), circumlunar (Moon), and translunar (exterior) realms associated with the Earth--Moon CR3BP. The outer Sun--(Earth\texttt{+}Moon) curves indicate the larger circumterrestrial lobe (Earth\texttt{+}Moon realm) separating Earth-associated translunar motion from the heliocentric Sun and exterior realms. The terrestrial {\it inset} marks the near-Earth region---from low-Earth orbit to just beyond GEO---bounded by the Laplace radius $r_L$, while the circumlunar {\it inset} shows the $L_1/L_2$ gateway geometry and the lunar Hill sphere $\rho_H$ separating local selenocentric motion from adjacent cislunar and translunar domains. The {\it dashed gold curve} marks the Earth Hill scale $r_H$ in the Sun--(Earth\texttt{+}Moon) problem, and {\it shading} indicates the distinct terrestrial, cislunar, circumlunar, translunar, and heliocentric provinces.
        }
	\label{fig:patched}
	\end{center}
	\vskip -0.075in
	\hspace{2cm}\rule{12.5cm}{0.5pt}
\end{figure}

\subsection{Spatiography and the Patched-CR3BP Viewpoint}
\label{sec:spatio}

In the early Space Age, Hubertus Strughold proposed \emph{spatiography} as a literal ``geography of space'': a topographical description of the region \emph{between} celestial bodies, intended to support orientation, navigation, and the classification of space operations. The Oxford English Dictionary defines \emph{spatiography} as ``the scientific description of the features of (outer) space; the topography of space,'' and records first use (1958) in Strughold's formulation that ``we need a topographical description of space \dots\ in analogy to geography \dots\ `spatiography'\,''~\citep{hS58, hS59}. In Strughold's scheme, orbital paths serve as topographical demarcation lines, and Ehricke's cis-/trans- nomenclature supplies corresponding regional names---provided those names are used with their intended boundary meanings.

Ehricke's mature treatment then supplies the dynamical backbone for this spatiography by effectively \emph{patching} two restricted problems at different scales (qv.~Fig.~\ref{fig:patched}): the Earth--Moon system organizes the stability and gateways of cislunar and circumlunar motion, while the Sun--(Earth\texttt{+}Moon) system organizes the stability structure of translunar motion \citep{kE62b}.\footnote{Here $\left(C_i\right)^{\S\E}$ and $\left(C_i\right)^{\E\M}$ denote the critical Jacobi values associated with the collinear equilibrium points of the Sun--(Earth\texttt{+}Moon) and Earth--Moon circular restricted three-body problems, respectively. The superscript indicates the relevant primary system, while the index $i$ identifies the corresponding libration point. For notational economy, the full system designation is omitted from the symbols $L_i$ themselves, and the smaller primary is used to indicate which restricted problem is meant. When the context is unambiguous, we further shorten $\left(C_i\right)^{\E\M}$ to $C_i$.} In this view, the collinear Earth--Moon gateways and their associated boundary lobes delimit both the outer extent of practically Earth-bound cislunar transport and the capture geometry of circumlunar motion. Beyond these, the pertinent outer demarcation is set by the $C_1$ boundary of the Sun--(Earth\texttt{+}Moon) problem, which Ehricke plots in scale against the Earth--Moon boundary curves, interpreting the intervening region as a limited-stability translunar ``hang-on'' domain whose inner edge is strongly Moon-perturbed and whose outer edge becomes exquisitely sensitive to weak-gradient perturbations near the $C_1$ curve. Su-Shu Huang independently formalizes the same two-scale reduction (``broken down \dots\ into two restricted three-body problems'') and shows how the innermost/outermost contact surfaces of the small-scale Earth--Moon problem sit and rotate within the large lobe of the Sun--(Earth\texttt{+}Moon) problem, furnishing an operational criterion for classifying Earth-bound, lunar, and escaping trajectories \citep{sH61, sH69}.

Within this patched viewpoint, the organizing structures are simultaneously \emph{dynamical} and \emph{topological}: the $L_1/L_2$ gateway surfaces provide faithful demarcations for cislunar extent and circumlunar capture \citep{rB56a, vE59, bT59}, while the Sun--(Earth\texttt{+}Moon) $C_1$ gateway provides the natural outer boundary beyond which trajectories are no longer meaningfully Earth-bound and instead become organized by heliocentric phase-space structure. While Fig.~\ref{fig:patched} is drawn in the planar problem for clarity, these partitions lift to three dimensions as families of surfaces, preserving the same gateway logic for practical transport, capture, and escape.

\subsection{This Primer's Usage}

This Primer therefore adopts the older, more precise taxonomy and uses \emph{cislunar} in its historically narrow, lexicographically literal sense. When we must speak about the coupled Earth--Moon--Sun environment as a whole---transfer corridors, libration-point neighborhoods, circumlunar operating regions, and the broader system regime---we will avoid using ``cislunar'' as a catch-all. Instead we will use explicitly system-based terms such as \emph{Earth--Moon system space} or \emph{Earth--Moon dynamical domain}. As a compact relational alternative, we will also use \emph{geolunar} space \citep{bT59, kE81, kE85}.

The point is not to police language for its own sake; it is to prevent a policy convenience from hardening into a technical misconception. The Earth--Moon environment is not one regime. It is a structured, multiscale dynamical geography---a \emph{spatiography}---and the terminology should reveal, not obscure, that structure. A timeline of the relevant Earth--Moon--Sun spatial terms (with earliest attestations and historical senses) is given in Table~\ref{tab:ems_lexicon} in Appendix~\ref{app:glossary}.

\section{The Curated Cislunar and Translunar Catalog}
\label{sec:catalog}

\subsection{Perturbation Hierarchy and the Persistence of Circumterrestrial Elements}

The modern literature often begins from the premise that the Earth--Moon environment is not simply an enlarged extension of traditional circumterrestrial space, but a distinct dynamical regime in which three-body trajectories often evade the simple geometric intuitions inherited from the Kepler problem, and that two-line elements (TLEs) are a poor medium for cislunar catalog maintenance and trajectory exchange once lunar forcing becomes substantial \citep{mH21}. But those operational cautions are too strong if read as a blanket dismissal of classical orbital elements themselves. Already in the early astronautical literature, \citet{kE55} drew a sharper distinction: the cislunar satellite is not under the Earth's exclusive gravitational control, yet its orbit can still be defined ``in the conventional manner'' by six elements, and both terrestrial and cislunar satellites remain members of the broader circumterrestrial class so long as the Earth is still the domineering body. In his later taxonomy, terrestrial, cislunar, and translunar motion are likewise all treated as subdivisions of the wider circumterrestrial domain \citep{kE56, kE58a, kE62a}, rather than as regions where Earth-centered description somehow ceases to exist altogether.

\begin{figure}[t!]	
	\begin{center}
	\includegraphics[width=0.95\textwidth]{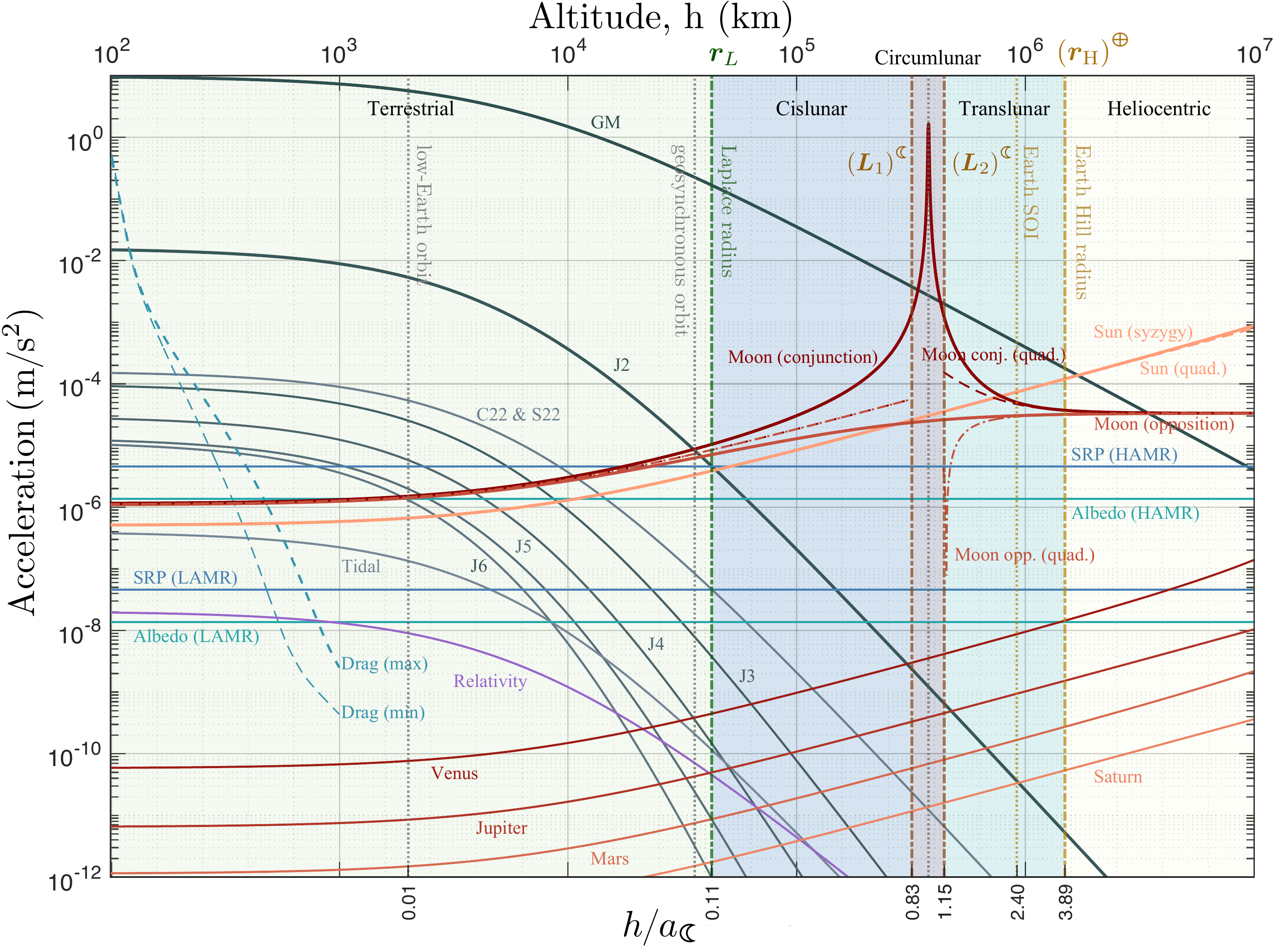}	
	\vskip -0.075in
	\caption{\small 
Order-of-magnitude circumterrestrial perturbation accelerations as functions of geocentric altitude, recast on the same radial partition used in the Earth--Moon--Sun spatiography (cf.~Fig.~\ref{fig:patched}). The {\it shaded bands} identify the terrestrial, cislunar, circumlunar, translunar, and heliocentric provinces, while the {\it vertical markers} indicate the Laplace radius $r_L$, the Earth--Moon $L_1$ and $L_2$ gateway distances, and the Earth sphere of influence and Hill radius. Within the terrestrial domain, the acceleration hierarchy is dominated by Earth's central field, low-order geopotential harmonics, drag, radiation pressure, tides, and relativistic corrections; beyond GEO, third-body forcing by the Moon and Sun progressively overtakes the zonal structure and reorganizes the dynamics.
        }
	\label{fig:accels}
	\end{center}
	\vskip -0.075in
	\hspace{2cm}\rule{12.5cm}{0.5pt}
\end{figure}

Figure~\ref{fig:accels} makes that point dynamically. Cast on the same radial partition as the patched-CR3BP spatiography of Fig.~\ref{fig:patched}, it shows not the disappearance of circumterrestrial orbital structure beyond GEO, but the orderly replacement of one perturbation hierarchy by another. In the terrestrial regime, the motion is organized primarily by Earth's central field and low-order geopotential, with drag, radiation pressure, tides, and relativistic corrections layered beneath \citep{kYeM04}. Moving outward, the Laplace radius marks the familiar transition at which lunisolar torque becomes comparable to the secular effect of Earth's oblateness \citep{aRdS14_ASR}; beyond that distance, the natural reference geometry shifts away from the equator toward the lunar-orbital and ecliptic geometry, exactly as Ehricke noted for cislunar orbits more strongly affected by the Moon than by $J_2$. This is a change in dominant perturbation and preferred reference plane, not a loss of orbital describability.

The most striking feature of Fig.~\ref{fig:accels} is the clear dynamical separation between the cislunar and translunar regimes. Inside the lunar distance, the lunar third-body term grows rapidly and the conjunction branch rises sharply as the direct lunar attraction on the satellite increasingly outpaces the indirect acceleration of the Earth. Near the Moon itself, that differential term becomes singular in the one-dimensional line-of-centers idealization, which is why the conjunction branch spikes at lunar distance. Beyond the Moon, however, the Earth-centered lunar perturbation no longer remains purely tidal. The direct lunar attraction on the satellite decays with the satellite--Moon separation, whereas the indirect term associated with the Moon's pull on the Earth remains at the fixed geocentric scale $\mu_\M/r_\M^2$. As a result, the two syzygy branches separate around a common asymptotic level: on the Moon-facing side the direct and indirect contributions reinforce one another, so the conjunction branch decays toward that indirect floor from above, while on the anti-lunar side they partially oppose one another, so the opposition branch approaches the same level from below. In other words, the figure makes visually explicit that ``beyond the Moon'' is not simply ``more of the same cislunar problem,'' but a distinct perturbative regime in which the lunar contribution ceases to be predominantly tidal and instead becomes organized by the interior-perturber structure of the Earth-centered formulation. This is precisely the kind of cislunar/translunar distinction Ehricke insisted on from the 1950s onward.

At still greater distance the solar third-body term overtakes the lunar term, regardless of syzygy configuration, and the translunar realm becomes progressively organized by the Sun--Earth problem with the Moon relegated to an internal perturbation. That two-scale handoff is exactly the one already built into Fig.~\ref{fig:patched}'s patched spatiography and stated explicitly by Ehricke in his 1962 space flight dynamics volume: the Earth--Moon system determines the principal stability structure of cislunar motion, while the Sun--(Earth\texttt{+}Moon) system determines the principal stability structure of translunar motion; the stable circumlunar capture region is confined to the small lobe between $L_1$ and $L_2$, whereas a much larger but less robust translunar ``hang-on'' region persists farther out for temporarily Earth-bound motion \citep{kE62b}. Read together, the two figures therefore recover a forgotten Space-Age insight: cislunar and translunar are not rhetorical variants for one undifferentiated beyond-GEO volume, but distinct dynamical provinces governed by different restricted-problem backbones \citep{sH61, sH69}.

This matters for orbital representation. \citet{mH21} are entirely right that TLEs are not an adequate language for cislunar (and translunar) operations: their mean-element compression, model dependence, and limited numerical precision make them unsuitable for sharing or propagating trajectories subject to strong lunar and solar perturbations over useful timescales \citep{cLwM11, fH21}. But that is an argument against TLEs as an operational exchange format, not against osculating circumterrestrial elements as local descriptors of an instantaneous Earth-referenced state. In Ehricke's own treatment, cislunar trajectories remained conventionally elementized even when their precession had to be referred to the plane of the Moon's orbit rather than the terrestrial equator, and even when their equations were most naturally developed in barycentric coordinates and only later transformed to geocentric form. The right conclusion, then, is narrower and more precise: in cislunar and much of translunar space, classical elements are no longer slowly varying orbital parameters and are no longer suitable as a compact long-arc catalog product, but they remain well-defined as instantaneous circumterrestrial elements.

Only upon entering the circumlunar domain do circumterrestrial elements cease to be the natural primary orbital language. In the patched-CR3BP spatiography of Fig.~\ref{fig:patched}, that threshold is identified geometrically by the Earth--Moon $L_1$ and  $L_2$ gateway surfaces: outside them, even strongly Moon-perturbed motion remains usefully describable as Earth-bound and hence circumterrestrial; inside them, the local organizing center is the Moon, not the Earth, and capture, escape, and orbit geometry are more naturally expressed selenocentrically. The transition is therefore generally not from `elements'' to `no elements,'' but from one osculating two-body surrogate to another. Circumterrestrial elements remain meaningful through the terrestrial, cislunar, and much of the translunar domain; once a trajectory crosses the gateway into the Moon-centered lobe, selenocentric elements become the natural osculating two-body descriptor.

There is, nevertheless, an important exception in the immediate neighborhoods of the libration-point orbit families themselves. For Lyapunov, halo, Lissajous, and related gateway orbits---and for the invariant manifolds attached to them---neither geocentric nor selenocentric Keplerian elements are especially illuminating as primary coordinates. In those regions, the natural description is the rotating Earth--Moon phase space itself, or coordinates adapted to the relevant invariant object, such as manifold coordinates or local normal-form/action--angle parametrizations \citep[e.g.,][]{aJjV98,aJ99,dSrE24,lPetal24}. This exception does not undermine the catalog-level use of osculating geocentric or selenocentric elements; it simply marks the gateway as a region where the organizing object is no longer a Keplerian surrogate about either primary, but the libration-point phase-space structure of the Earth--Moon problem.

\subsection{Hill-Region Topology and Orbital-Element Projections}

The change in organizing center can be made more precise through the Hill regions of the Earth--Moon circular restricted three-body problem. In that setting, the Jacobi constant $C_J = 2\mathcal{U}_{\E\M}(\rb) - v^2$---where $\mathcal{U}_{\E\M}$ is the CR3BP effective potential and $v$ the speed in the rotating frame---is conserved along every trajectory, partitioning configuration space into energetically allowed and forbidden regions of motion. The zero-velocity surfaces, defined by the condition $v = 0$ and hence $C_J = 2\mathcal{U}_{\E\M}$, bound these regions in three dimensions; their planar intersections are the familiar zero-velocity curves. (The nomenclature, rotating-frame equations of motion, the effective potential $\mathcal{U}_{\E\M}$, and the formal definition of $C_J$ are given in \textsection\ref{sec:phenomenology}.) As $C_J$ decreases, the topology of the allowed region changes in the familiar sequence of neck openings at the collinear libration points. In the coplanar limit, this structure admits an associated projection in the semi-major axis--eccentricity plane through contours of the lunar Tisserand parameter, $T_\M$, introduced formally in Eq.~\eqref{eq:tisserand_moon} as an approximate orbital-element analogue of the Jacobi constant. These contours should therefore be read as accessibility guides rather than exact zero-velocity boundaries, but they provide a useful element-space view of the underlying three-body constraints.

\begin{figure}[t!]	
	\begin{center}
	\includegraphics[width=0.925\textwidth]{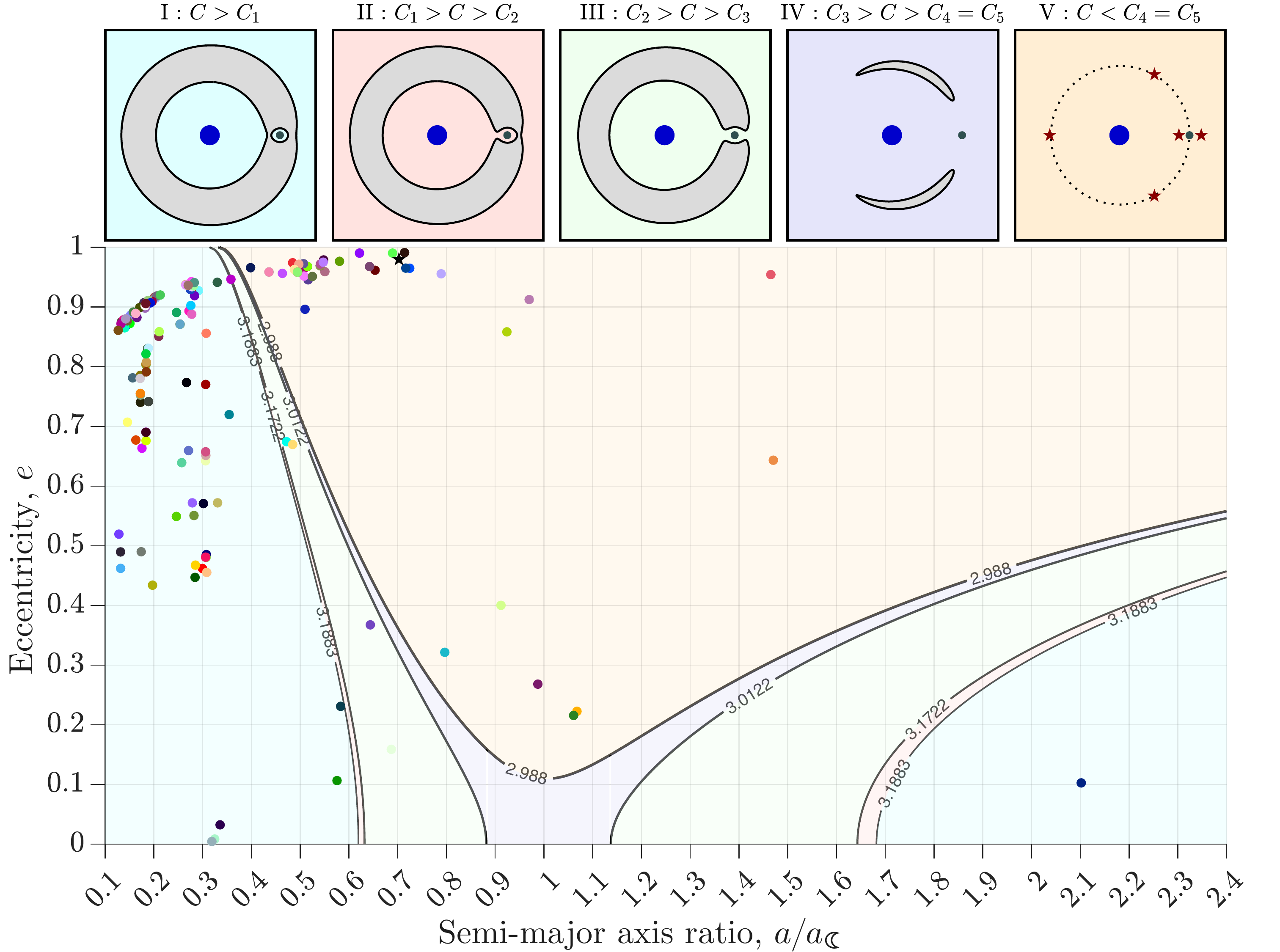}	
	\vskip -0.075in
	\caption{\small 
Historic and contemporary cataloged cislunar and translunar objects projected into the geocentric semi-major axis--eccentricity ($a$--$e$) plane. The {\it colored markers} denote the osculating geocentric elements inferred from the latest available TLE for each object, while the {\it black curves} show coplanar contours of constant Tisserand parameter with respect to the Moon (see Eq.~\eqref{eq:tisserand_moon} in \textsection\ref{sec:phenomenology}). The {\it banner panels} above summarize the five classical zero-velocity (planar) topologies of the Earth--Moon circular restricted three-body problem, corresponding to the successive critical Jacobi levels associated with the libration points. Read together, the Tisserand contours and the zero-velocity cases indicate where Earth-centered orbital elements remain physically informative, where transfer between circumterrestrial and circumlunar motion becomes possible, and where entry into the lunar lobe or into libration-point neighborhoods calls for a transition from purely geocentric elements to either a local Earth--Moon-system parameterization or, where appropriate, a selenocentric description (www.space-track.org. Assessed 1 May 2025.)
        }
	\label{fig:hills}
	\end{center}
	\vskip -0.075in
	\hspace{2cm}\rule{12.5cm}{0.5pt}
\end{figure}

Figure~\ref{fig:hills} shows the curated cislunar and translunar catalog in this projected $a$--$e$ space, where the colored markers denote osculating geocentric elements obtained from the most recent available state representations for the objects in the catalog, and the dark curves indicate the corresponding coplanar Tisserand contours. The banner panels summarize the five classical (planar) zero-velocity topologies associated with the critical values $C_1, C_2, C_3$, and $C_4 = C_5$. Read together, the panels and the $a$--$e$ projection supply a compact map of where circumterrestrial orbital elements remain physically informative, where passage between Earth-centered and Moon-centered motion becomes possible, and where the geocentric description begins to lose its local dynamical privilege.

In Case~I, $C_J > C_1$, the Earth-centered, Moon-centered, and external Hill regions are mutually disconnected. Motion is confined to whichever component it occupies, so geocentric elements are appropriate for the Earth-centered and exterior circumterrestrial branches, while selenocentric elements are appropriate for the Moon-centered circumlunar branch. In Case~II, $C_1 > C_J > C_2$, the $L_1$ neck opens and low-energy passage between the Earth-centered and Moon-centered regions becomes possible, while the exterior region remains disconnected. In Case~III, $C_2 > C_J > C_3$, both the $L_1$ and $L_2$ necks are open, so transit between the Earth and Moon is possible and escape through the exterior lunar gateway can occur. In Cases~IV and V, $C_3 > C_J > C_4 = C_5$ and $C_J < C_4 = C_5$, respectively, the allowable geometry expands further and high-energy lunar flybys and broad translunar access become permissible.

The essential point is that orbital elements do not abruptly fail once a trajectory becomes substantially non-Keplerian. Rather, they remain well defined and physically useful over most of circumterrestrial phase space, including strongly perturbed cislunar and much of translunar motion, provided they are interpreted as instantaneous osculating descriptors rather than nearly conserved integrals. What the Hill-region structure clarifies is where that description becomes locally inadequate as a \emph{primary} language: namely, in the immediate gateway neighborhoods and, more decisively, upon entry into the Moon-centered Hill lobe itself. There the geocentric element set becomes poorly conditioned as a descriptor of the local motion, and a switch to piecewise or fully selenocentric elements becomes the more faithful representation.

Figure~\ref{fig:hills} should therefore be read not as evidence against the use of orbital elements beyond GEO, but as a phase-space guide to where different element sets are most natural. Away from the Earth--Moon gateway region, the cataloged objects admit a perfectly intelligible circumterrestrial projection in $a$--$e$ space. Near and within the circumlunar lobe, however, the patched spatiography of Fig.~\ref{fig:patched} reasserts itself: the Moon becomes the relevant local center, and the appropriate orbital language changes accordingly.

\subsection{Mapping the Catalog onto the Earth--Moon Synodic Frame}

The orbital-element projection of Fig.~\ref{fig:hills} is useful precisely because it remains circumterrestrial: it shows that a large fraction of the curated catalog can still be represented meaningfully in geocentric element space, even when the underlying motion is strongly perturbed. But the same objects, when mapped into the Earth--Moon synodic frame, reveal something equally important: the catalog is not organized as a static cloud occupying a fixed geometric sector of space. Rather, it unfolds into a time-dependent distribution of arcs and ``comet-like'' tails, reflecting the continual precession, the controlling influence of mean-motion and secular resonances, and the gateway-mediated transport that accompany motion in the Earth--Moon environment.

\begin{figure}[t!]	
	\begin{center}
	\includegraphics[width=0.925\textwidth]{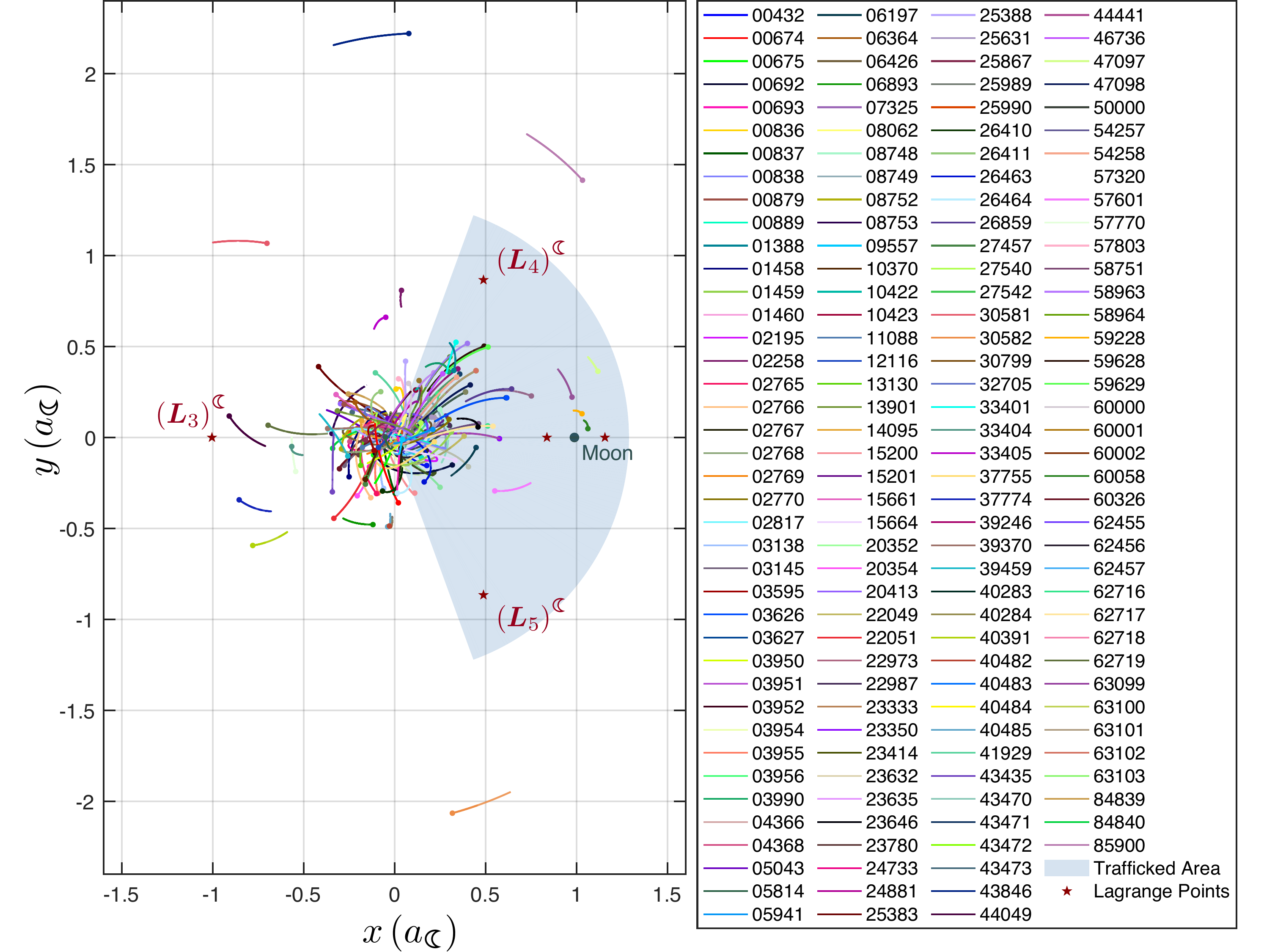}	
	\vskip -0.075in
	\caption{\small 
The curated catalog mapped into the Earth--Moon synodic frame. Each {\it colored arc} shows cataloged object propagated into the rotating frame from the last available osculating state in the 1 May 2025 set, over one Earth sidereal day, producing the short ``comet-like'' tails characteristic of strongly perturbed circumterrestrial motion. The {\it legend} identifies each trajectory by its Satellite Catalog Number, as given in the corresponding TLE. The {\it red stars} mark the Earth--Moon Lagrange points, the Moon is shown at its synodic location, and the {\it shaded sector} indicates the Moon-facing ``trafficked area'' proposed by \citet{bB24}. The figure shows that the catalog is not confined to a static wedge, but spreads across a broader and time-dependent synodic geography, including regions outside the nominal traffic sector.
        }
	\label{fig:synodic}
	\end{center}
	\vskip -0.075in
	\hspace{2cm}\rule{12.5cm}{0.5pt}
\end{figure}

Figure~\ref{fig:synodic} shows this complementary view. Each cataloged object is mapped into the rotating Earth--Moon frame and propagated for one sidereal day, so that its state appears not as a single point but as a short synodic tail. The result is therefore a dynamical portrait rather than a static occupancy map. Some objects remain clustered near the familiar Moon-facing corridor, but many do not: their traces spread across a much broader portion of synodic phase space, including regions exterior to the immediate Earth--Moon line and, in several cases, into the anti-lunar sector. What emerges is not a bounded ``traffic lane,'' but a set of trajectories continually sheared and redistributed by the multiscale Earth--Moon--Sun dynamics.

This is where the now-common ``cislunar traffic'' sketch must be read with care. \citet{bB24} depict the cislunar region as a synodic-plane disk centered on the Earth and extending slightly beyond $L_2$, within which a single Moon-facing wedge is shaded as the ``trafficked area''; their figure also places $L_4$ and $L_5$ within the outer disk while excluding the anti-Moon sector near $L_3$. The wedge is best interpreted as an operationally defined region of interest: a notional high-traffic volume intended to capture where many current and planned lunar missions concentrate, rather than a boundary derived from invariant manifolds, resonance structure, Hill-region topology, or other intrinsic features of the multi-body dynamics. As a rough visual heuristic, the sketch has descriptive value, especially when read as a clustered-use delineation rather than a dynamical boundary \citep{tP26}. But taken as a dynamical delimiter, it is far too restrictive: it suppresses the anti-lunar sector, de-emphasizes the spatiographic character of the wider Earth--Moon--Sun system (see Fig.~\ref{fig:patched}), and encourages the false impression that cislunar activity is confined to a fixed angular stencil in the rotating frame. Indeed, as the authors note, that construction is presented as a region of practical concentration, not a genuine dynamical boundary.

Figure~\ref{fig:synodic} shows why such a stencil is inadequate. Transport beyond GEO is mediated not by static sectors but by resonant corridors, secular drift, and the invariant-manifold geometry associated with the Earth--Moon gateways. Short arcs of cataloged motion routinely cross any fixed angular cut, and the anti-lunar side cannot be dismissed as dynamically irrelevant merely because fewer missions are intentionally targeted there. In particular, the vicinity of $L_3$ belongs to the same global synodic geography as the $L_1$ and $L_2$ necks: it is part of the broader circumterrestrial phase space through which objects may evolve under repeated lunar and solar perturbations. A sector-based picture therefore risks both false negatives, by ignoring dynamically accessible regions outside the wedge, and false positives, by treating simple angular proximity as a proxy for genuine dynamical or operational relevance.

The distinction developed in the preceding sections becomes especially useful here. In the cislunar regime, the dominant organizing problem is well approximated by the Earth--satellite--Moon restricted three-body problem, with the Sun acting as an external perturbation. In the translunar regime, by contrast, the Earth--satellite--Sun problem becomes the larger scaffold and the Moon is demoted to an interior perturbation. The synodic tails in Fig.~\ref{fig:synodic} are a kinematic trace of that transition: objects interior to lunar distance are shaped primarily by Earth--Moon geometry, whereas those farther out begin to reflect the broader translunar organization already identified in Fig.~\ref{fig:accels}. The catalog is therefore better understood as occupying a connected dynamical geography than as filling a prescribed ``traffic sector.''

For SDA purposes, this point is not merely semantic. If sensing architectures, surveillance priorities, or operational taxonomies are built around fixed wedges in the synodic plane, they will systematically miss the time-dependent transport structure sampled by objects in cislunar and translunar space. This matters directly for the construction of cislunar SSA and STM frameworks, which depend on architectures capable of searching, tracking, characterizing, and cataloging objects across the broader Earth--Moon system rather than only within a nominal Moon-facing corridor \citep{aW25, mGkH25, yHsCyW26}. A more faithful framework begins from the spatiographic partition developed in \textsection\ref{sec:boundaries}: terrestrial, cislunar, circumlunar, translunar, and heliocentric regions, connected by gateway geometry and populated through resonant, secular, and manifold-mediated motion. Figure~\ref{fig:synodic} should thus be read as the synodic companion to Figs.~\ref{fig:patched}--\ref{fig:hills}: where Fig.~\ref{fig:patched} supplies the patched dynamical geography, Fig.~\ref{fig:accels} the perturbation hierarchy, and Fig.~\ref{fig:hills} the orbital-element projection, Fig.~\ref{fig:synodic} shows the catalog as it actually stretches and migrates in the Earth--Moon rotating frame.

\section{Perturbative Treatment of Distant Geocentric Orbits}
\label{sec:perturbed}

\subsection{The Perturbed Hamiltonian Formulation}

Many dynamical systems admit a useful separation into an underlying Hamiltonian ``skeleton'' that is nearly integrable and a hierarchy of perturbations that distort it. In multidimensional settings of this sort, global transport and long-term structure are governed to a large extent by the locations and widths of the dominant resonances \citep{aLmL92, gH99}. The circumterrestrial and circumlunar environments are especially rich in this respect because they contain motions spanning a wide range of characteristic frequencies: the terrestrial day, the lunar month, the solar year, and several much longer precession periods arising from the Sun's gravitational action on the Earth--Moon system \citep{kE62a, jS94}. Chief among the latter are the regression of the lunar line of nodes (period $\sim$$18.61$~yr) and the rotation of the lunar apsidal line (period $\sim$$8.85$~yr). This constellation of frequencies gives rise to an extensive variety of commensurabilities and resonant interactions.

Resonances shaping satellites in the traditional geocentric domains from LEO to GEO have been studied so thoroughly that many remaining questions are now regarded as matters of refinement \citep{aR19}. The medium-Earth orbit (MEO) problem is emblematic: although Earth's oblateness largely controls the satellite's apsidal and nodal precession, the slow regression of the Moon's line of nodes supplies an additional frequency that organizes a vast family of commensurabilities \citep{jDaR16}. Secular resonances involving integer combinations of the satellite's apsidal and nodal precession rates ($\dot{\omega}$ and $\dot{\Omega}$, largely forced by Earth's oblateness) together with the lunar nodal rate ($\dot{\Omega}_\M$, driven by the Sun) occur in profusion, so that terrestrial phase space is densely threaded by an exceedingly complicated web-like structure \citep{tEkH97, jDaR16, aCaR17, aR19, jD22, eLcE23}.

\subsubsection{The Lunar and Solar Disturbing Functions}
\label{sec:disturbing}

Taking the Earth as the central body, the disturbing functions (negative disturbing potentials in the usual celestial-mechanics sign convention) associated with the Moon and the Sun may be written in the standard Earth-centered form,
\begin{align}
	\mathcal{R}^\M = 
		\mu_\M \left( \frac{1}{\lvert \bm r_\M - \bm r \rvert} - 
			\frac{\bm r \cdot \bm r_\M}{r_\M^3} \right), \qquad
    	\mathcal{R}^\S = 
		\mu_\S \left( \frac{1}{\lvert \bm r_\S - \bm r \rvert} - 
			\frac{\bm r \cdot \bm r_\S}{r_\S^3} \right),
\end{align}
where $\bm r$ is the satellite position relative to Earth, $\bm r_{\mathsf b}$ is the position of perturber $\mathsf b\in\{\M,\S\}$ relative to Earth, and $\mu_{\mathsf b}$ is its gravitational parameter. The second term in each expression is the familiar \emph{indirect} contribution that accounts for the fact that the Earth-centered frame is not inertial: it removes the acceleration of the Earth induced by the perturber.

While these expressions are exact, they are typically expanded in Legendre polynomials to expose a small parameter and enable analytical perturbation theory \citep{jLgB10, rM13}. Writing,
\begin{equation}
	\bm r \cdot \bm r_{\mathsf b} = r\,r_{\mathsf b}\cos\psi_{\mathsf b},
    	\qquad
	\cos\psi_{\mathsf b} = \hat{\bm r}\cdot \hat{\bm r}_{\mathsf b},
\end{equation}
the direct term admits the classical series
\begin{equation}
	\frac{1}{\lvert \bm r_{\mathsf b} - \bm r \rvert}
    	=
	\frac{1}{r_{\mathsf b}} \sum_{l=0}^\infty \left(\frac{r}{r_{\mathsf b}}\right)^l 
	P_{l,0}(\cos\psi_{\mathsf b}),~~~~\text{when}~~r < r_{\mathsf b},
\end{equation}
and an analogous series in $(r_{\mathsf b}/r)$ when the satellite lies outside the perturber \citep{fFjL10, tG12, dL14, tI16, gMpG20, mC22, mC24}.

For the solar case, the satellite remains an internal body throughout circumterrestrial space ($r<r_\S$). The constant ($l=0$) term is an arbitrary potential offset and the dipole ($l=1$) term is canceled by the indirect contribution, so the solar disturbing function begins at quadrupole order:
\begin{align}
	\mathcal{R}^\S =
		\frac{\mu_\S}{r_\S} \sum_{l = 2}^\infty \left( \frac{r}{r_\S} \right)^l 
		P_{l,0}\! \left( \hat{\bm r}\cdot\hat{\bm r}_\S \right).
\end{align}

The lunar problem, however, requires a fundamental distinction depending on whether the satellite lies inside or outside the lunar distance.

\paragraph{Lunar-external expansion ($r < r_\M$; cislunar regime).}
When the satellite is internal to the lunar orbit, the Moon acts as an external perturber and the expansion parameter is $(r/r_\M)$. As in the solar case, the $l=0$ term is a removable constant and the $l=1$ term cancels with the indirect contribution, so the series begins at $l=2$:
\begin{align}
	\mathcal{R}^\M_{\mathrm{ext}} =
		\frac{\mu_\M}{r_\M} \sum_{l = 2}^\infty \left( \frac{r}{r_\M} \right)^l 
		P_{l,0}\! \left( \hat{\bm r}\cdot\hat{\bm r}_\M \right).
\end{align}

\paragraph{Lunar-internal expansion ($r > r_\M$; translunar regime).}
In the translunar realm (and, more generally, whenever $r > r_\M$), the Moon becomes an internal perturber relative to the satellite's geocentric distance and convergence requires an expansion in powers of $(r_\M/r)$. In this case the monopole ($l=0$) and dipole ($l=1$) terms of the direct potential remain, and the indirect term no longer cancels the $l=1$ contribution because it has a different functional dependence on $r$. It is therefore convenient to separate the disturbing function into its monopole term, the combined direct-dipole and indirect contribution, and a residual direct-potential Legendre series beginning at $l=2$ \citep{mC22, mC24}:
\begin{align}
	\label{eq:legendre_int}
	\mathcal{R}^\M_{\mathrm{int}} =
		\underbrace{\frac{\mu_\M}{r}}_{\text{monopole}} +
		\underbrace{\mu_\M \,\frac{\bm r \cdot \bm r_\M}{r^3} -
		\mu_\M \,\frac{\bm r \cdot \bm r_\M}{r_\M^3}}_{\text{direct dipole $-$ indirect}} + \frac{\mu_\M}{r} \sum_{l = 2}^\infty \left( \frac{r_\M}{r} \right)^l 
		P_{l,0}\!\left( \hat{\bm r}\cdot\hat{\bm r}_\M \right).
\end{align}

This switch in series structure is visible in the acceleration-level comparison of Fig.~\ref{fig:accels}: once $r$ exceeds the lunar distance, the satellite's direct lunar attraction decays as $O(r^{-2})$, while the Earth-centered formulation retains an indirect contribution of characteristic magnitude $\mu_\M/r_\M^2$. Consequently, the \emph{relative} lunar perturbation (satellite minus Earth) no longer has the purely tidal character appropriate inside the lunar orbit. On the Moon-facing side, the surviving direct and indirect contributions reinforce one another, so the Earth-centered differential acceleration approaches the indirect scale from above; on the anti-lunar side, they partially cancel, so the corresponding branch approaches the same scale from below. Beyond lunar distance, then, the Earth-centered lunar forcing is controlled not by a single tidal term but by the combined monopole, uncancelled dipole, indirect, and higher-order interior-series contributions, with the indirect term increasingly setting the overall acceleration level.

\subsubsection{Kaula Expansion of the Third-Body Disturbing Function}
\label{sec:kaula}

To express the disturbing functions in orbital elements, introduce the Keplerian elements
$(a, e, I, \Omega, \omega, M)$ for the satellite and
$(a_{\mathsf b}, e_{\mathsf b}, I_{\mathsf b}, \Omega_{\mathsf b}, \omega_{\mathsf b}, M_{\mathsf b})$
for the perturber $\mathsf b \in \{\M,\S\}$, all referred to a common reference plane.
The Kaula expansion may then be written in two complementary forms depending on whether the perturber is exterior or interior to the satellite orbit.
In the exact Legendre form, this distinction is based on the instantaneous radial ordering $r \lessgtr r_{\mathsf b}$; in the Kaula expansion, it is expressed in the usual hierarchical sense through the semi-major-axis ratio.

\paragraph{Exterior perturber ($a_{\mathsf b} > a$).}
For a hierarchical exterior perturber, the natural small parameter is
$\alpha = a/a_{\mathsf b} < 1$, and the disturbing function admits the standard Kaula form \citep{bE99, aCaR17}
\begin{align}
    \label{eq:kaula_ext}
	\mathcal{R}^{\mathsf b}_{\mathrm{ext}} =
		\frac{\mu_{\mathsf b}}{a_{\mathsf b}}
		\sum_{l = 2}^{\infty}
		\left( \frac{a}{a_{\mathsf b}} \right)^l
		\sum_{m = 0}^{l}
		\sum_{p = 0}^{l}
		\sum_{p^\prime = 0}^{l}
		\sum_{q = -\infty}^{\infty}
		\sum_{q^\prime = -\infty}^{\infty}
		\mathcal{K}^{\mathsf b}_{lmpq\,p^\prime q^\prime}
		\cos\phi^{\mathsf b}_{lmpq\,p^\prime q^\prime},
\end{align}
with phase angle
\begin{align}
	\label{eq:phase}
	\phi^{\mathsf b}_{lmpq\,p^\prime q^\prime} 
		& = (l - 2p) \, \omega + (l - 2p + q) \, M + m \, \Omega
		\nonumber \\[3pt]
		&\qquad - 
		\Big[ 
			(l - 2p^\prime) \, \omega_{\mathsf b} + 
			(l - 2p^\prime + q^\prime) \, M_{\mathsf b} +
			m \, \Omega_{\mathsf b}
		\Big],
\end{align}
and coefficient kernel
\begin{align}
	\mathcal{K}^{\mathsf b}_{lmpq\,p^\prime q^\prime} =
		\epsilon_m\,
		\frac{(l-m)!}{(l+m)!}\,
		F_{lmp}(I)\,
		F_{lmp^\prime}(I_{\mathsf b})\,
		H_{lpq}(e)\,
		G_{lp^\prime q^\prime}(e_{\mathsf b}),
\end{align}
where $\epsilon_0 = 1$ and $\epsilon_m = 2$ for $m \ge 1$; the primed indices refer to the perturbing body's orbital harmonics, while $F_{lmp}$ denotes Kaula inclination functions and $H_{lpq}$ and $G_{lp^\prime q^\prime}$ denote the corresponding Hansen-coefficient combinations in the standard Kaula expansion \citep{wK62, bE99, aCaR17}:
\begin{align}
	H_{lpq}(e) 
		&= X^{\,l,\,l-2p}_{\,l-2p+q}(e), \\
	G_{lp^\prime q^\prime}(e_{\mathsf b}) 
		&= X^{-(l+1),\,l-2p^\prime}_{\,l-2p^\prime+q^\prime}(e_{\mathsf b}).
\end{align}
This form applies directly to the solar disturbing function, since the satellite remains interior to the Sun throughout circumterrestrial space, and to the lunar disturbing function in the cislunar regime where $r < r_\M$.

\paragraph{Interior perturber ($a_{\mathsf b} < a$).}
For a hierarchical interior perturber, the natural small parameter is
$\alpha = a_{\mathsf b}/a < 1$.
For the Earth-centered lunar-internal problem, this compact Kaula form applies most naturally to the residual direct-potential
$l \ge 2$ part of Eq.~\eqref{eq:legendre_int}, while the monopole, dipole, and indirect terms are retained explicitly prior to averaging.
The homogeneous interior series may be written as
\begin{align}
    \label{eq:kaula_int}
	\mathcal{R}^{\mathsf b}_{\mathrm{int}, \, l \ge 2} ={}&
		\frac{\mu_{\mathsf b}}{a}
		\sum_{l = 2}^{\infty}
		\left( \frac{a_{\mathsf b}}{a} \right)^l
		\sum_{m = 0}^{l}
		\sum_{p = 0}^{l}
		\sum_{p^\prime = 0}^{l}
		\sum_{q = -\infty}^{\infty}
		\sum_{q^\prime = -\infty}^{\infty}
		\widetilde{K}^{\mathsf b}_{lmpq\,p^\prime q^\prime}
		\cos\phi^{\mathsf b}_{lmpq\,p^\prime q^\prime},
\end{align}
with the same phase angle $\phi^{\mathsf b}_{lmpq\,p^\prime q^\prime}$ as above, but with coefficient kernel
\begin{align}
	\widetilde{K}^{\mathsf b}_{lmpq\,p^\prime q^\prime} ={}&
		\epsilon_m\,
		\frac{(l-m)!}{(l+m)!}\,
		F_{lmp}(I)\,
		F_{lmp^\prime}(I_{\mathsf b})\,
		\widetilde{H}_{lpq}(e)\,
		\widetilde{G}_{lp^\prime q^\prime}(e_{\mathsf b}),
\end{align}
where the Hansen coefficients now appear in the complementary arrangement
\begin{align}
	\widetilde{H}_{lpq}(e) &=
		X^{-(l+1),\,l-2p}_{\,l-2p+q}(e),
	\\
	\widetilde{G}_{lp^\prime q^\prime}(e_{\mathsf b}) &=
		X^{\,l,\,l-2p^\prime}_{\,l-2p^\prime+q^\prime}(e_{\mathsf b}).
\end{align}
Thus, for the lunar-internal expansion one may write
\begin{align}
	\mathcal{R}^\M_{\mathrm{int}} =
		\frac{\mu_\M}{r} +
		\mu_\M \,\frac{\bm r \cdot \bm r_\M}{r^3} -
		\mu_\M \,\frac{\bm r \cdot \bm r_\M}{r_\M^3} +
		\mathcal{R}^\M_{\mathrm{int}, \, l \ge 2},
\end{align}
with $\mathcal{R}^\M_{\mathrm{int}, \, l \ge 2}$ given by the compact interior-Kaula series above.

\subsubsection{Doubly Averaged Third-Body Disturbing Function}

The usefulness of the Kaula forms in Eqs.~\eqref{eq:kaula_ext} and~\eqref{eq:kaula_int} is precisely that they make the usual perturbation-theoretic bookkeeping explicit: harmonics involving the mean anomalies correspond to short-periodic variations, the remaining slowly varying angle combinations supply long-periodic and resonant terms, and the angle-independent part gives the purely secular contribution.

For the long-term problem considered here, we average the disturbing function over the mean anomalies of both the satellite and the perturber, thereby removing the short-periodic terms while retaining the long-periodic and secular contributions.\footnote{For notational economy, the averaging operator $(\,\overline{\phantom{x}}\,)$ will henceforth be retained on the Hamiltonian and disturbing functions, but dropped from the orbital elements themselves; accordingly, $(a,e,I,\omega,\Omega)$ are to be understood from this point onward as doubly averaged (mean) elements unless otherwise noted.} 
Define
\begin{align}
	\overline{\mathcal{R}}^{\mathsf b}
		=
		\frac{1}{4 \pi^2}
		\int_0^{2\pi}
		\int_0^{2\pi}
		\mathcal{R}^{\mathsf b}
		\, dM \, dM_{\mathsf b}.
\end{align}
In the lunar-internal expansion, it is useful to treat the low-order terms of Eq.~\eqref{eq:legendre_int} separately before applying the Kaula development to the residual $l \ge 2$ series. Using the standard orbit-averaging identities,
\begin{align}
	\left\langle \frac{1}{r} \right\rangle
		&= \frac{1}{a},
	&
	\left\langle \frac{\bm r}{r^3} \right\rangle
		&= \bm 0,
	&
	\left\langle \bm r \right\rangle
		&= -\frac{3}{2} a \, {\bm e},
\end{align}
together with the corresponding identity
\begin{align}
	\left\langle \frac{\bm r_{\mathsf b}}{r_{\mathsf b}^3} \right\rangle
		= \bm 0
\end{align}
for the perturber orbit, one finds that the monopole term survives as a constant while both the direct-dipole and indirect terms vanish upon double averaging:
\begin{align}
	\overline{\left( \frac{\mu_\M}{r} \right)}
		&= \frac{\mu_\M}{a},
	&
	\overline{\left( \mu_\M \,\frac{\bm r \cdot \bm r_\M}{r^3} \right)}
		&= 0,
	&
	\overline{\left( - \mu_\M \,\frac{\bm r \cdot \bm r_\M}{r_\M^3} \right)}
		&= 0.
\end{align}
Accordingly, the doubly averaged lunar-internal disturbing function becomes
\begin{align}
	\label{eq:avg_lunar_int}
	\overline{\mathcal{R}}^\M_{\mathrm{int}}
		=
		\frac{\mu_\M}{a}
		+
		\overline{\mathcal{R}}^\M_{\mathrm{int}, \, l \ge 2},
\end{align}
where the first term is a constant that may be absorbed into the integrable part of the Hamiltonian if desired.

In the Kaula expansions, double averaging retains only those harmonics that are independent of both mean anomalies. From the phase angle
\begin{align}
	\phi^{\mathsf b}_{lmpq\,p^\prime q^\prime}
		&= (l - 2p) \, \omega + (l - 2p + q) \, M + m \, \Omega
		\nonumber \\[3pt]
		&\qquad -
		\Big[
			(l - 2p^\prime) \, \omega_{\mathsf b} +
			(l - 2p^\prime + q^\prime) \, M_{\mathsf b} +
			m \, \Omega_{\mathsf b}
		\Big],
\end{align}
this requires
\begin{align}
	l - 2p + q = 0,
	\qquad
	l - 2p^\prime + q^\prime = 0,
\end{align}
or equivalently
\begin{align}
	q = 2p - l,
	\qquad
	q^\prime = 2p^\prime - l.
\end{align}

The surviving orbit-averaged phase angle is therefore
\begin{align}
	\label{eq:avg_phase}
	\overline{\phi}^{\mathsf b}_{lmp\,p^\prime}
		&= (l - 2p) \, \omega + m \, \Omega
		\nonumber \\[3pt]
		&\qquad -
		\Big[
			(l - 2p^\prime) \, \omega_{\mathsf b}
			+
			m \, \Omega_{\mathsf b}
		\Big].
\end{align}

\paragraph{Exterior perturber ($a_{\mathsf b} > a$).}
For an exterior perturber, the doubly averaged disturbing function reduces to
\begin{align}
	\overline{\mathcal{R}}^{\mathsf b}_{\mathrm{ext}}
		=
		\frac{\mu_{\mathsf b}}{a_{\mathsf b}}
		\sum_{l = 2}^{\infty}
		\biggl(  \frac{a}{a_{\mathsf b}}  \biggr)^l
		\sum_{m = 0}^{l}
		\sum_{p = 0}^{l}
		\sum_{p^\prime = 0}^{l}
		\overline{\mathcal{K}}^{\mathsf b}_{lmp\,p^\prime}
		\cos \overline{\phi}^{\mathsf b}_{lmp\,p^\prime},
\end{align}
with orbit-averaged coefficient kernel
\begin{align}
	\overline{\mathcal{K}}^{\mathsf b}_{lmp\,p^\prime}
		=
		\epsilon_m\,
		\frac{(l-m)!}{(l+m)!}\,
		F_{lmp}(I)\,
		F_{lmp^\prime}(I_{\mathsf b})\,
		\overline{H}_{lp}(e)\,
		\overline{G}_{lp^\prime}(e_{\mathsf b}),
\end{align}
where
\begin{align}
	\overline{H}_{lp}(e)
		&= X^{\,l,\,l-2p}_{\,0}(e),
	\\
	\overline{G}_{lp^\prime}(e_{\mathsf b})
		&= X^{-(l+1),\,l-2p^\prime}_{\,0}(e_{\mathsf b}).
\end{align}
This form gives the doubly averaged solar disturbing function throughout circumterrestrial space, and the doubly averaged lunar disturbing function in the cislunar regime where the satellite lies interior to the lunar orbit.

\paragraph{Interior perturber ($a_{\mathsf b} < a$).}
For an interior perturber, the doubly averaged homogeneous series becomes
\begin{align}
	\overline{\mathcal{R}}^{\mathsf b}_{\mathrm{int}, \, l \ge 2}
		=
		\frac{\mu_{\mathsf b}}{a}
		\sum_{l = 2}^{\infty}
		\biggl( \frac{a_{\mathsf b}}{a} \biggr)^l
		\sum_{m = 0}^{l}
		\sum_{p = 0}^{l}
		\sum_{p^\prime = 0}^{l}
		\overline{\widetilde{\mathcal{K}}}^{\mathsf b}_{lmp\,p^\prime}
		\cos \overline{\phi}^{\mathsf b}_{lmp\,p^\prime}
\end{align}
with
\begin{align}
	\overline{\widetilde{\mathcal{K}}}^{\mathsf b}_{lmp\,p^\prime}
		=
		\epsilon_m\,
		\frac{(l-m)!}{(l+m)!}\,
		F_{lmp}(I)\,
		F_{lmp^\prime}(I_{\mathsf b})\,
		\overline{\widetilde{H}}_{lp}(e)\,
		\overline{\widetilde{G}}_{lp^\prime}(e_{\mathsf b}),
\end{align}
where
\begin{align}
	\overline{\widetilde{H}}_{lp}(e)
		&= X^{-(l+1),\,l-2p}_{\,0}(e),
	\\
	\overline{\widetilde{G}}_{lp^\prime}(e_{\mathsf b})
		&= X^{\,l,\,l-2p^\prime}_{\,0}(e_{\mathsf b}).
\end{align}
Thus, for the lunar-internal problem one may write Eq.~\eqref{eq:avg_lunar_int}, where the first term is an additive constant and the nontrivial secular contribution begins at $l = 2$.

\subsubsection{The Orbit-Averaged Hamiltonian}

The doubly averaged disturbing functions derived above furnish the natural Hamiltonian description for the long-term dynamics of distant geocentric motion. In Delaunay variables,
\begin{align}
	L &= \sqrt{\mu_\E a}, 
	&
	G &= L \sqrt{1 - e^2}, 
	&
	H &= G \cos I,
	\nonumber\\[3pt]
	l &= M,
	&
	g &= \omega,
	&
	h &= \Omega ,
\end{align}
the mean anomaly $l$ is cyclic once the orbital averaging has been carried out, so that $L$ is constant and the mean semi-major axis acts as a parameter.

At this stage, however, it is useful to distinguish between the \emph{orbit-averaged} Hamiltonian and the \emph{purely secular} Hamiltonian. The former retains not only the phase-independent secular terms, but also those long-periodic harmonics that depend on the slow angular variables of the satellite and perturbers. In the present Earth-centered problem, the most important explicit time dependence enters through the slow precession of the lunar line of nodes and apsides. Introducing appended canonical pairs
\begin{align}
	\tau_\Omega &= \Omega_\M,
	&
	\tau_\omega &= \omega_\M,
	&
	\dot{\tau}_\Omega &= \dot{\Omega}_\M,
	&
	\dot{\tau}_\omega &= \dot{\omega}_\M,
\end{align}
with conjugate actions $\Gamma_\Omega$ and $\Gamma_\omega$, renders the system autonomous.

With the disturbing functions defined as negative potentials, the extended orbit-averaged Hamiltonian may therefore be written as
\begin{align}
	\overline{\mathcal{H}}
		=
		-\frac{\mu_\E^2}{2L^2}
		+
		\overline{\mathcal{H}}^{\M}
		+
		\overline{\mathcal{H}}^{\S}
		+
		\dot{\Omega}_\M \Gamma_\Omega
		+
		\dot{\omega}_\M \Gamma_\omega,
\end{align}
where
\begin{align}
	\overline{\mathcal{H}}^{\mathsf b} = -\overline{\mathcal{R}}^{\mathsf b}
\end{align}
denotes the appropriate orbit-averaged third-body contribution for perturber $\mathsf b \in \{\M,\S\}$, using the exterior or interior branch as required.

For an exterior perturber, this contribution is
\begin{align}
	\label{eq:H_avg_ext}
	\overline{\mathcal{H}}^{\mathsf b}_{\mathrm{ext}}
		=
		-
		\frac{\mu_{\mathsf b}}{a_{\mathsf b}}
		\sum_{l = 2}^{\infty}
		\biggl( \frac{a}{a_{\mathsf b}} \biggr)^l
		\sum_{m = 0}^{l}
		\sum_{p = 0}^{l}
		\sum_{p^\prime = 0}^{l}
		\overline{\mathcal{K}}^{\mathsf b}_{lmp\,p^\prime}
		\cos \overline{\phi}^{\mathsf b}_{lmp\,p^\prime},
\end{align}
whereas for an interior perturber the homogeneous contribution is
\begin{align}
	\overline{\mathcal{H}}^{\mathsf b}_{\mathrm{int}, \, l \ge 2}
		=
		-
		\frac{\mu_{\mathsf b}}{a}
		\sum_{l = 2}^{\infty}
		\biggl( \frac{a_{\mathsf b}}{a} \biggr)^l
		\sum_{m = 0}^{l}
		\sum_{p = 0}^{l}
		\sum_{p^\prime = 0}^{l}
		\overline{\widetilde{\mathcal{K}}}^{\mathsf b}_{lmp\,p^\prime}
		\cos \overline{\phi}^{\mathsf b}_{lmp\,p^\prime}.
\end{align}
In particular, for the lunar-internal problem one has
\begin{align}
	\overline{\mathcal{H}}^{\M}_{\mathrm{int}}
		=
		-\frac{\mu_\M}{a}
		+
		\overline{\mathcal{H}}^{\M}_{\mathrm{int}, \, l \ge 2},
\end{align}
where the monopole term depends only on the constant action $L$ (equivalently, on the fixed mean semi-major axis $a$), and may therefore be absorbed into the integrable part of the Hamiltonian; the nontrivial dynamics are contained in the residual interior series.

Since the Keplerian term $-\mu_\E^2/(2L^2)$ depends only on the constant action $L$, it does not affect the reduced dynamics and will be omitted from the decomposition that follows. It is now convenient to separate the perturbing part of the orbit-averaged Hamiltonian into its purely secular and long-periodic components:
\begin{align}
	\label{eq:H_avg_split}
	\overline{\mathcal{H}}
		=
		\mathcal{H}_{\mathrm{sec}}
		+
		\mathcal{H}_{\mathrm{lp}}
		+
		\dot{\Omega}_\M \Gamma_\Omega
		+
		\dot{\omega}_\M \Gamma_\omega.
\end{align}
Here $\mathcal{H}_{\mathrm{sec}}$ collects all harmonics for which the averaged phase
$\overline{\phi}^{\mathsf b}_{lmp\,p^\prime}$ is identically zero, or equivalently independent of the angular variables, while $\mathcal{H}_{\mathrm{lp}}$ contains the remaining orbit-averaged harmonics that still depend on the slow angles $(g,h,\tau_\Omega,\tau_\omega)$.

Equivalently, for each perturber one may write
\begin{align}
	\overline{\mathcal{H}}^{\mathsf b}
		=
		\mathcal{H}^{\mathsf b}_{\mathrm{sec}}
		+
		\mathcal{H}^{\mathsf b}_{\mathrm{lp}},
\end{align}
with
\begin{align}
	\mathcal{H}^{\mathsf b}_{\mathrm{sec}}
		&=
		\sum_{\overline{\phi}^{\mathsf b}_{lmp\,p^\prime} \equiv 0}
		(\cdots),
	\\
	\mathcal{H}^{\mathsf b}_{\mathrm{lp}}
		&=
		\sum_{\overline{\phi}^{\mathsf b}_{lmp\,p^\prime} \not\equiv 0}
		(\cdots).
\end{align}

The purely secular Hamiltonian considered in the following subsection is obtained by retaining only the first of these two contributions. Its quadrupolar truncation then furnishes closed-form apsidal and nodal precession frequencies in the exterior problem, together with the corresponding axisymmetrized interior analogue for the translunar case.

\subsubsection{The Secular and Quadrupolar Approximation}
\label{sec:sec_quad}

The purely secular approximation is obtained by retaining only $\mathcal{H}_{\mathrm{sec}}$ from Eq.~\eqref{eq:H_avg_split} and neglecting the long-periodic contribution $\mathcal{H}_{\mathrm{lp}}$. Truncation at quadrupole order ($l=2$) then yields the closed-form apsidal and nodal precession rates. Since the purely secular Hamiltonian considered below is independent of the lunar angles themselves, the appended linear terms in the conjugate actions $\Gamma_\Omega$ and $\Gamma_\omega$ play no role in the satellite dynamics and will be omitted for brevity.

With the disturbing functions defined as negative potentials, the perturbing secular Hamiltonian may be written as
\begin{align}
	\mathcal{H}_{\mathrm{sec}}
		=
		\mathcal{H}_{\mathrm{sec}}^\M
		+
		\mathcal{H}_{\mathrm{sec}}^\S,
\end{align}
where $\mathcal{H}_{\mathrm{sec}}^{\mathsf b}$ denotes either the exterior-perturber or interior-perturber secular terms (i.e. the negative of the corresponding secular disturbing function), depending on the branch of the problem. 

\paragraph{Exterior perturber ($a_{\mathsf b} > a$).}
At quadrupole order, the perturbing secular contribution for an exterior perturber takes the form
\begin{align}
	\label{eq:ext_sec_hamil}
	\mathcal{H}_{\mathrm{sec,ext}}^{\mathsf b}(G,H;L)
		=
		\frac{\omega^{\mathsf b}_{\mathrm{ext}}(a)\,L}{12}
		\left(
			5 - 3 \frac{G^2}{L^2}
		\right)
		\left(
			1 - 3 \frac{H^2}{G^2}
		\right),
\end{align}
with secular prefactor
\begin{align}
	\label{eq:ext_sec_freq}
	\upwb_{\mathrm{ext}}(a)
		=
		\frac{3}{4}
		\frac{\mu_{\mathsf b}}{\sqrt{\mu_\E}}\,
		K_{\mathsf b}\,
		\frac{a^{3/2}}
		     {a_{\mathsf b}^3 (1 - e_{\mathsf b}^2)^{3/2}},
\end{align}
where
\begin{align}
	K_{\mathsf b}
		=
		1 - \frac{3}{2}\sin^2 I_{\mathsf b}.
\end{align}
Equivalently, this branch follows from the secular Hansen factors
\begin{align}
	\overline{H}_{21}(e)
		&=
		X^{2,0}_{0}(e)
		=
		1 + \frac{3}{2}e^2,
	\\
	\overline{G}_{21}(e_{\mathsf b})
		&=
		X^{-3,0}_{0}(e_{\mathsf b})
		=
		(1 - e_{\mathsf b}^2)^{-3/2},
\end{align}
together with the Kaula inclination function
\begin{align}
	F_{201}(I)
		=
		\frac{3}{4}\sin^2 I - \frac{1}{2}
		=
		\frac{1 - 3 \cos^2 I}{4}.
\end{align}
The corresponding secular precession rates follow from Eq.~\eqref{eq:ext_sec_hamil} through the canonical relations
\begin{align}
	\dot g &= \frac{\partial \mathcal{H}_{\mathrm{sec,ext}}^{\mathsf b}}{\partial G},
	&
	\dot h &= \frac{\partial \mathcal{H}_{\mathrm{sec,ext}}^{\mathsf b}}{\partial H},
\end{align}
with $g=\omega$ and $h=\Omega$, giving
\begin{align}
	\dot{\omega}^{\mathsf b}_{\mathrm{ext}}
		&=
		\frac{\upwb_{\mathrm{ext}}(a)}{2}
		\frac{5 \cos^2 I - 1 + e^2}{\sqrt{1 - e^2}},
	\\
	\dot{\Omega}^{\mathsf b}_{\mathrm{ext}}
		&=
		-
		\upwb_{\mathrm{ext}}(a)
		\frac{1 + \tfrac{3}{2} e^2}{\sqrt{1 - e^2}}
		\cos I.
\end{align}

\paragraph{Interior perturber ($a_{\mathsf b} < a$).}
For the lunar-internal problem, the monopole term $\mu_\M/a$ depends only on the constant action $L$ and may be absorbed into the integrable part of the Hamiltonian; the nontrivial perturbing secular contribution begins with the $l=2$ interior series. Selecting the axisymmetric term $m=0$, $p=1$, and $p^\prime=1$, one obtains
\begin{align}
	\mathcal{H}_{\mathrm{sec,int}}^{\M}(G,H;L)
		=
		\frac{\upwM_{\mathrm{int}}(a)\,L}{6}
		\left(
			\frac{L}{G}
		\right)^3
		\left(
			1 - 3 \frac{H^2}{G^2}
		\right),
\end{align}
where the interior secular prefactor is
\begin{align}
	\label{eq:int_sec_freq}
	\upwM_{\mathrm{int}}(a)
		=
		\frac{3}{4}
		\frac{\mu_\M}{\sqrt{\mu_\E}}\,
		K_\M\,
		\left(
			1 + \frac{3}{2}e_\M^2
		\right)
		\frac{a_\M^2}{a^{7/2}},
\end{align}
with
\begin{align}
	K_\M
		=
		1 - \frac{3}{2}\sin^2 I_\M.
\end{align}
This branch is generated by the interior Hansen factors
\begin{align}
	\overline{\widetilde{H}}_{21}(e)
		&=
		X^{-3,0}_{0}(e)
		=
		(1 - e^2)^{-3/2},
	\\
	\overline{\widetilde{G}}_{21}(e_\M)
		&=
		X^{2,0}_{0}(e_\M)
		=
		1 + \frac{3}{2}e_\M^2,
\end{align}
together with the Kaula inclination factors
\begin{align}
	F_{201}(I)
		&=
		\frac{1 - 3 \cos^2 I}{4},
	\\
	F_{201}(I_\M)
		&=
		\frac{1 - 3 \cos^2 I_\M}{4}
		=
		-\frac{K_\M}{2}.
\end{align}
The corresponding secular precession rates are
\begin{align}
	\dot{\omega}^{\M}_{\mathrm{int}}
		&=
		\frac{\upwM_{\mathrm{int}}(a)}{2}
		\frac{5 \cos^2 I - 1}{(1 - e^2)^2},
	\\
	\dot{\Omega}^{\M}_{\mathrm{int}}
		&=
		-
		\upwM_{\mathrm{int}}(a)
		\frac{\cos I}{(1 - e^2)^2}.
\end{align}

\subsection{The von Zeipel-Lidov-Kozai Mechanism}
\label{sec:lidov}

Until the advent of artificial satellites, most astronomical applications of the restricted three-body problem were conditioned by the principal ``traffic rules'' obeyed by the major bodies of the Solar System: their heliocentric orbits are nearly coplanar and only moderately eccentric. Accordingly, perturbation theory was usually developed as an expansion in small eccentricities and inclinations. The outstanding exception was provided by asteroids and comets, many of which occupy highly eccentric and highly inclined orbits and were found, after decades of study, to be shaped by a rich hierarchy of resonances \citep{cMsD99, aM02, sT23}. With the beginning of the Space Age, comparable dynamical situations were recognized much closer to Earth, where highly eccentric and highly inclined circumterrestrial orbits could undergo similarly dramatic secular and resonant evolution \citep{pM61, mL62, dK62, pM63, mL63, zK67}.

In its classical form, the von Zeipel--Lidov--Kozai (vZLK) mechanism arises for an \emph{interior} test particle perturbed by a distant \emph{exterior} body, after double averaging and truncation at quadrupole order \citep{iS17, tI19}. In the present setting, that classical configuration is realized most cleanly by the Sun, which remains an exterior perturber throughout circumterrestrial space. 

If all angular elements are referred to the ecliptic plane, the solar quadrupole term alone generates the familiar autonomous one-degree-of-freedom Hamiltonian, with stationary solutions at $\omega = \pi/2$ and $3\pi/2$ for inclinations lying between the critical values
\begin{align}
	I_{\mathrm{crit}} \approx 39.2^\circ,
	\qquad
	180^\circ - I_{\mathrm{crit}} \approx 140.8^\circ,
\end{align}
and coupled oscillations in eccentricity and inclination through conservation of the projected angular momentum. In this sense, the solar vZLK mechanism is always present as a possible organizing influence in the distant geocentric problem.

If one further neglects the $5.15^\circ$ inclination of the lunar orbit to the ecliptic, so that the Sun and Moon may both be treated as exterior perturbers in a common reference plane with $K_\S = K_\M = 1$, then the quadrupole truncation of the orbit-averaged exterior Hamiltonian, Eq.~\eqref{eq:H_avg_ext}, yields the familiar idealized lunisolar vZLK Hamiltonian
\begin{align}
	\label{eq:HvZLK}
	\mathcal{H}_{\mathrm{vZLK}}
		=
		-\frac{a^2}{8}
		\left[
			\frac{\mu_\S}{a_\S^3 (1 - e_\S^2)^{3/2}}
			+
			\frac{\mu_\M}{a_\M^3 (1 - e_\M^2)^{3/2}}
		\right]
		\left[
			2 + 3e^2
			-
			3\left(
				1 - e^2 + 5e^2 \sin^2 \omega
			\right)\sin^2 I
		\right],
\end{align}
where the angular elements are again measured with respect to the ecliptic. This autonomous approximation is integrable and possesses the standard first integrals
\begin{align}
	c_0 &= a = \text{constant},
	\\
	c_1 &= (1-e^2)\cos^2 I = \text{constant},
	\\
	c_2 &= e^2\left( \frac{2}{5} - \sin^2 I \sin^2 \omega \right) = \text{constant}.
\end{align}
Its level curves in the $(\omega,\sqrt{1-e^2})$ plane define the classical von Zeipel--Lidov--Kozai diagrams, which furnish a useful idealized backbone for the secularly dominated portion of cislunar phase space.

For the combined lunisolar Hamiltonian, it is convenient to introduce the characteristic quadrupolar frequency
\begin{align}
	\nu_{\mathrm{vZLK}}(a)
		=
		\frac{3}{4}
		\frac{a^{3/2}}{\sqrt{\mu_\E}}
		\left[
			\frac{\mu_\S}{a_\S^3 (1 - e_\S^2)^{3/2}}
			+
			\frac{\mu_\M}{a_\M^3 (1 - e_\M^2)^{3/2}}
		\right].
\end{align}
Away from the separatrix, the corresponding quadrupolar vZLK timescale is therefore of order
\begin{align}
	t_{\mathrm{vZLK}} \sim \nu_{\mathrm{vZLK}}^{-1},
\end{align}
which, in the same normalization as the usual single-perturber estimate \citep{jA15, yWtF23}, may be written as
\begin{align}
	t_{\mathrm{vZLK}}
		\sim
		\frac{16}{15}
		\frac{\sqrt{\mu_\E}}{a^{3/2}}
		\left[
			\frac{\mu_\S}{a_\S^3 (1 - e_\S^2)^{3/2}}
			+
			\frac{\mu_\M}{a_\M^3 (1 - e_\M^2)^{3/2}}
		\right]^{-1}.
\end{align}

That said, the lunar contribution must be interpreted more carefully than the solar one. Inside the Moon's orbit, the lunar disturbing function is formally of exterior-perturber type, but the standard quadrupolar vZLK picture is not generically realized across cislunar space. Classical Lidov--Kozai oscillations are well known to be vulnerable to any additional source of apsidal or nodal precession: in a variety of hierarchical systems, such precession detunes the third-body quadrupolar dynamics and thereby quenches the large-amplitude eccentricity growth of the ideal point-mass problem \citep{bL15, fM20, yWtF23}. In near-Earth space, the relevant additional precession is supplied primarily by Earth's oblateness, which dominates much of the terrestrial domain and suppresses the classical lunar quadrupolar vZLK mechanism long before the Moon would otherwise become the leading secular perturber (cf.~Fig.~\ref{fig:accels}). The point, then, is not that secular structure disappears, but that the classical lunar Lidov--Kozai phase portrait is replaced or overprinted by other precessional dynamics \citep{jDaR16, aR19}. Farther out, as the semi-major-axis ratio ceases to be small and the orbit approaches the lunar distance, the assumptions underlying the classical lunar vZLK approximation degrade rapidly. In particular, for large-$a$ cislunar and translunar trajectories such as \emph{Luna~3} \citep{lS60, vG61, vG62}, lunar close encounters and intermediate-period terms break the doubly averaged dynamics, so that the classical quadrupolar Lidov--Kozai solution is no longer adequate \citep{dA20}. As already emphasized by \citet{lB59}, \citet{pM61}, \citet{mL63}, and \citet{iS63}, Earth oblateness and lunisolar perturbations become comparable near the Laplace radius \citep{rAgC64}; beyond it, higher-order lunar terms grow increasingly important \citep{aR20}.

Thus, for the Moon, the classical vZLK Hamiltonian should be regarded chiefly as an idealized exterior-perturber limit \citep{sTtY14}. It nevertheless remains useful both for building intuition and for the combined lunisolar quadrupole model: if the lunar plane is taken to coincide with the ecliptic, the combined solar and lunar quadrupole torques generate Kozai diagrams whose phase portrait provides a first approximation to the ``secularly dominated'' zone identified later in this Primer. In that sense, the classical vZLK picture still furnishes a meaningful backbone for many highly eccentric cislunar trajectories, even though the real cislunar problem is subsequently modified by lunar inclination, nodal and apsidal precession, higher-order secular terms, and mean-motion commensurabilities. At the same time, it should not be mistaken for a faithful global description of cislunar secular dynamics. In particular, the genuinely \emph{eccentric} lunar Kozai problem belongs to the octupolar and higher-order theory \citep{sN16, cW17, vS18}, where flips, extreme eccentricity excursions, and chaotic secular modulation can arise on timescales much longer than the quadrupole cycle. These effects remain largely unexplored in the circumterrestrial setting, yet they may well be relevant for a nontrivial fraction of the high-eccentricity objects now cataloged in cislunar space (qv.~Fig.~\ref{fig:hills}).

Beyond the Moon, the roles of the two third bodies separate cleanly. The Sun continues to act as an exterior perturber and therefore continues to support the classical quadrupolar vZLK mechanism. The Moon, however, becomes an \emph{interior} perturber, and the secular dynamics enter the complementary ``inverse'' problem of an exterior test particle acted upon by an interior body \citep{fFjL10, tG12, sN17, bVeC18, gdE19, hLyxG24}. At the level of the purely secular axisymmetric quadrupole term derived above, the lunar contribution supplies an inverse-Kozai-type precessional backbone in translunar space. The full inverse Lidov--Kozai dynamics, however, belong to the higher-order secular problem \citep{tI16}: for a circular or nearly circular interior perturber it appears at hexadecapole order, where the outer particle's argument of perigee may librate about $\pm \pi/2$ near inclinations of roughly $63^\circ$ and $117^\circ$, while for eccentric interior perturbers additional octupolar resonances arise. Translunar space therefore admits a genuinely mixed secular architecture, in which the classical solar vZLK mechanism can compete with the interior-lunar inverse-Kozai family.

This distinction is important for the spatiographic program developed here. In cislunar space, the relevant quadrupolar secular scaffold is still predominantly the classical exterior-perturber problem, with the Sun providing the clearest realization and the Moon contributing only in an idealized limit. In translunar space, by contrast, the secular architecture becomes intrinsically hybrid: the Sun organizes the motion through the classical exterior vZLK mechanism, while the Moon contributes through the secular dynamics of an interior perturber. The competition between these two structures is one of the clearest dynamical signatures that translunar motion is not merely an outer continuation of the cislunar problem, but a distinct secular regime.

\subsection{Lunisolar Secular Resonances in Cislunar and Translunar Space}

The preceding development shows that, once the mean anomalies have been averaged out, the long-term dynamics are governed by the orbit-averaged Hamiltonian given by Eq.~\eqref{eq:H_avg_split}. The phase-independent part $\mathcal{H}_{\mathrm{sec}}$ yields the background apsidal and nodal precession, while the resonance web itself is carried by the long-periodic harmonics in $\mathcal{H}_{\mathrm{lp}}$, whose averaged phases evolve only slowly.

For either the exterior or interior branch, the orbit-averaged phase angle, Eq.~\eqref{eq:avg_phase}, has the form
\begin{align}
	\overline{\phi}^{\mathsf b}_{lmp\,p^\prime}
		&=
		(l - 2p)\,\omega 
		+ 
		m\,(\Omega - \Omega_{\mathsf b})
		-
		(l - 2p^\prime)\,\omega_{\mathsf b},
\end{align}
where the integers $(l,m,p,p^\prime)$ are precisely those appearing in the averaged Kaula development. A lunisolar secular resonance occurs when the corresponding slow phase is nearly stationary,
\begin{align}
	\dot{\overline{\phi}}^{\mathsf b}_{lmp\,p^\prime}
		&=
		(l - 2p)\,\dot{\omega} 
		+ 
		m\,(\dot{\Omega} - \dot{\Omega}_{\mathsf b})
		-
		(l - 2p^\prime)\,\dot{\omega}_{\mathsf b}
		\approx 0.
\end{align}
The purely secular subset considered in \textsection\ref{sec:sec_quad} corresponds to the special case
$\overline{\phi}^{\mathsf b}_{lmp\,p^\prime} \equiv 0$; the nontrivial secular resonances arise instead from those long-periodic harmonics for which $\overline{\phi}^{\mathsf b}_{lmp\,p^\prime} \not\equiv 0$ but $\dot{\overline{\phi}}^{\mathsf b}_{lmp\,p^\prime}$ is small.

In cislunar space, both the Moon and Sun contribute through the \emph{exterior}-perturber branch, so that the relevant resonance families are generated by the coefficients 
$\overline{\mathcal{K}}^{\M}_{lmp\,p^\prime}$ and $\overline{\mathcal{K}}^{\S}_{lmp\,p^\prime}$.
At quadrupole order ($l = 2$), the dominant resonances reduce to low-order nodal and apsidal commensurabilities involving the satellite's own precession together with the regression and apsidal rotation of the lunar orbit. If the angular elements are referred to the ecliptic and the solar plane is treated as fixed, then the principal explicit time dependence enters through the lunar angles, yielding resonance families of the form
\begin{align}
	(l - 2p)\,\dot{\omega}
		+
		m\,(\dot{\Omega} - \dot{\Omega}_\M)
		-
		(l - 2p^\prime)\,\dot{\omega}_\M
		\approx 0.
\end{align}
Within this same exterior-perturber setting, the lunar and solar apsidal rates share the same angular dependence and may therefore be combined. For fixed eccentricity, one has
\begin{align}
	\dot{\omega}
		=
		\frac{\upwM_{\mathrm{ext}}(a)+\upwS_{\mathrm{ext}}(a)}{2}
		\frac{5\cos^2 I - 1 + e^2}{\sqrt{1-e^2}},
\end{align}
so that the apsidal-stationary locus $\dot{\omega}=0$ is given by
\begin{align}
	5\cos^2 I = 1-e^2.
\end{align}
For a fixed eccentricity slice, this appears in the $(a,I)$ plane as a horizontal line. These apsidal-stationary loci should not be conflated with the classical quadrupolar vZLK critical inclinations discussed in \textsection\ref{sec:lidov}: they arise here simply as zeros of the combined lowest-order apsidal precession rate.

In translunar space, however, the situation changes qualitatively. The Sun continues to contribute through the exterior branch, but the Moon contributes through the \emph{interior} branch, with coefficients
$\overline{\widetilde{\mathcal{K}}}^{\M}_{lmp\,p^\prime}$ and secular rates furnished by the lunar-internal Hamiltonian derived in \textsection\ref{sec:sec_quad}. The phase structure is formally the same, but the underlying forcing differs: the solar part remains that of a distant exterior perturber, whereas the lunar part is now generated by the interior expansion. The resulting secular web is therefore genuinely hybrid, consisting of a superposition of lunar-interior and solar-exterior secular precession. Accordingly, the solar contribution remains of exterior type while the lunar contribution switches to the interior branch, so the angular factors are no longer identical:
\begin{align}
	\dot{\omega}
		=
		\frac{\upwM_{\mathrm{int}}(a)}{2}
		\frac{5\cos^2 I - 1}{(1-e^2)^2} 
		+
		\frac{\upwS_{\mathrm{ext}}(a)}{2}
		\frac{5\cos^2 I - 1 + e^2}{\sqrt{1-e^2}}.
\end{align}
Setting this equal to zero yields an $a$-dependent apsidal-stationary inclination,
\begin{align}
	\cos^2 I_{\dot{\omega}=0}(a;e)
		=
		\frac{
			\upwM_{\mathrm{int}}(a)
			+
			\upwS_{\mathrm{ext}}(a)(1-e^2)^{5/2}
		}{
			5\!\left[
				\upwM_{\mathrm{int}}(a)
				+
				\upwS_{\mathrm{ext}}(a)(1-e^2)^{3/2}
			\right]
		},
\end{align}
so the translunar branch need not remain horizontal, but may bend with semi-major axis as the relative importance of the solar-exterior and lunar-interior terms changes. In the limit that the interior lunar term dominates just beyond $a_\M$, one recovers
\begin{align}
	\cos^2 I \to \frac{1}{5},
\end{align}
corresponding to
\begin{align}
	I \to 63.4^\circ,
	\qquad
	116.6^\circ.
\end{align}
These are the same characteristic inclinations associated with the inverse-Kozai family for an exterior test particle perturbed by an interior body \citep{fFjL10, tG12, sN17, bVeC18, gdE19, hLyxG24}; in the present quadrupolar, axisymmetric limit they arise simply as zeros of the lunar-interior apsidal precession rate, while the full inverse Lidov--Kozai dynamics belong to the higher-order secular theory discussed in \textsection\ref{sec:lidov}. Once solar forcing overtakes the interior lunar contribution, however, the locus asymptotes back toward the exterior value
\begin{align}
	\cos^2 I \to \frac{1-e^2}{5}.
\end{align}
The continuation of these secular loci across lunar distance should therefore be viewed not as an exact dynamical continuation, but as a lowest-order indication of how the secular scaffold reorganizes as the Moon changes from exterior to interior perturber.

Beyond quadrupole order, the same phase structure persists with $l = 3,4,\ldots$, but the relative importance of the higher harmonics differs between the two perturbers. For the Sun, the octupole correction is weak in the Earth-satellite problem; for the Moon, the octupole and higher terms become increasingly relevant as the orbital radius approaches and exceeds lunar distance. The secular architecture of cislunar and translunar space is therefore not exhausted by the quadrupole approximation, even though the latter provides the natural first scaffold. Accordingly, the secular loci obtained from the lowest-order formulas are best interpreted as schematic spatiographic guides to the evolving secular architecture, rather than as precise local resonance loci in the immediate vicinity of the lunar-distance transition.

\subsection{Lunar and Solar Mean-Motion Resonances}
\label{sec:resonances}

Whereas the secular resonances arise from the slow orbit-averaged phases of the Kaula expansion, mean-motion resonances are generated by the full unaveraged harmonic angles of the third-body disturbing function. For either perturber branch, the relevant phase is given by Eq.~\eqref{eq:phase},
with the same integers $(l,m,p,p^\prime,q,q^\prime)$ that label the Kaula harmonics of \textsection\ref{sec:kaula}. A mean-motion resonance occurs when this angle evolves only slowly,
\begin{align}
	\dot{\phi}^{\mathsf b}_{lmpq\,p^\prime q^\prime}
		&=
		(l - 2p + q)\,n
		-
		(l - 2p^\prime + q^\prime)\,n_{\mathsf b}
		\nonumber\\[3pt]
		&\qquad +
		(l - 2p)\,\dot{\omega}
		-
		(l - 2p^\prime)\,\dot{\omega}_{\mathsf b}
		+
		m\,(\dot{\Omega} - \dot{\Omega}_{\mathsf b})
		\approx 0.
\end{align}

Since the secular frequencies are typically much smaller than the orbital frequencies, the nominal center of the resonance is obtained at leading order from
\begin{align}
	(l - 2p + q)\,n
		-
		(l - 2p^\prime + q^\prime)\,n_{\mathsf b}
		\approx 0.
\end{align}
We denote such a commensurability by $k \! : \! k_{\mathsf b}$, with the first integer referring to the satellite and the second to the perturber; thus the satellite completes $k$ sidereal revolutions in approximately the same inertial time that the perturber completes $k_{\mathsf b}$. Writing $T = 2\pi/n$ and $T_{\mathsf b}=2\pi/n_{\mathsf b}$ for the corresponding sidereal orbital periods, one has
\begin{align}
	k_{\mathsf b} &= l - 2p + q,
	&
	k &= l - 2p^\prime + q^\prime,
\end{align}
so that
\begin{align}
	k_{\mathsf b}\,n - k\,n_{\mathsf b} \approx 0,
	\qquad
	\frac{T}{T_{\mathsf b}} \approx \frac{k_{\mathsf b}}{k}.
\end{align}
The corresponding nominal resonance center is therefore
\begin{align}
    a_{k \, : \, k_{\mathsf b}}
    \approx
    \left(
        \frac{\mu_\E}{\mu_\E+\mu_{\mathsf b}}
    \right)^{1/3}
    \left(
        \frac{k_{\mathsf b}}{k}
    \right)^{2/3}
    a_{\mathsf b}.
\end{align}
For the Moon, this mass factor is simply $(1 - \bar\mu_{\E\M})^{1/3}$, with $\bar\mu_{\E\M}$ the Earth--Moon CR3BP mass ratio, and differs only slightly from unity. Thus the nominal lunar commensurabilities are, to excellent approximation, located at
\begin{align}
    a_{k \, : \, k_\M}
    \approx
    \left(
        \frac{k_\M}{k}
    \right)^{2/3}
    a_\M .
\end{align}
This same nominal condition applies on both sides of lunar distance, but the forcing changes branch across $a = a_\M$. When $a < a_\M$, the Moon acts as an exterior perturber and the resonant harmonics are supplied by the exterior lunar expansion. When $a > a_\M$, the Moon acts as an interior perturber and the same commensurabilities are generated by the lunar-internal expansion. Thus, within the approximation above, the semi-major axis locations of the lunar commensurabilities are unaffected by the switch in disturbing-function expansion, even though the underlying resonant forcing is not. In particular, the translunar lunar problem is not purely tidal: after separation of the monopole term, the leading Earth-centered forcing beyond the Moon already involves the direct-dipole and indirect terms, with the homogeneous $l \ge 2$ interior-Kaula series providing higher-order corrections.

An analogous construction applies to the Sun. Since the Sun is always an exterior perturber for Earth satellites, solar mean-motion resonances arise entirely from the exterior third-body expansion and are described by the same harmonic angle, Eq.~\eqref{eq:phase}, with $\mathsf b = \S$. With the convention adopted above, a $k \! : \! k_\S$ solar commensurability satisfies
\begin{align}
	k_\S\,n - k\,n_\S \approx 0,
\end{align}
and therefore has nominal center
\begin{align}
    a_{k \, : \, k_\S}
    \approx
    \bar\mu_{\S\E}^{1/3}
    \left(
        \frac{k_\S}{k}
    \right)^{2/3}
    a_\S ,
\end{align}
where $\bar\mu_{\S\E}=\mu_\E/(\mu_\S+\mu_\E)$ is the Sun--Earth CR3BP mass ratio. 
Equivalently, in lunar-distance units,
\begin{align}
    \frac{a_{k \, : \, k_\S}}{a_\M}
    \approx
    \bar\mu_{\S\E}^{1/3}
    \left(
        \frac{k_\S}{k}
    \right)^{2/3}
    \frac{a_\S}{a_\M}.
\end{align}
Because the cube-root Sun--Earth mass ratio compensates the large Sun--Earth distance, the first low-order solar commensurabilities fall in the outer translunar problem rather than at heliocentric-scale geocentric distances.

Finally, because both lunar and solar forcing are present in translunar space, one should also expect genuinely mixed three-body commensurabilities. At the level of mean motions, these Laplace-type relations \citep{dNaM98} take schematically the form
\begin{align}
	\kappa\,n - \kappa_\M\,n_\M - \kappa_\S\,n_\S \approx 0,
\end{align}
with $\kappa, \kappa_\M, \kappa_\S \in \mathbb{Z}$, to which one may add slow apsidal and nodal frequencies inherited from the Kaula phase angles. Such resonances are the natural higher-order extension of the separate lunar and solar sequences, and they are especially likely to become relevant in the outer translunar region where low-order exterior lunar and solar commensurabilities coexist.

\subsection{Secular and Resonant Structures in the Curated Catalog}

\begin{figure}[t!]	
	\begin{center}
	\includegraphics[width=0.925\textwidth]{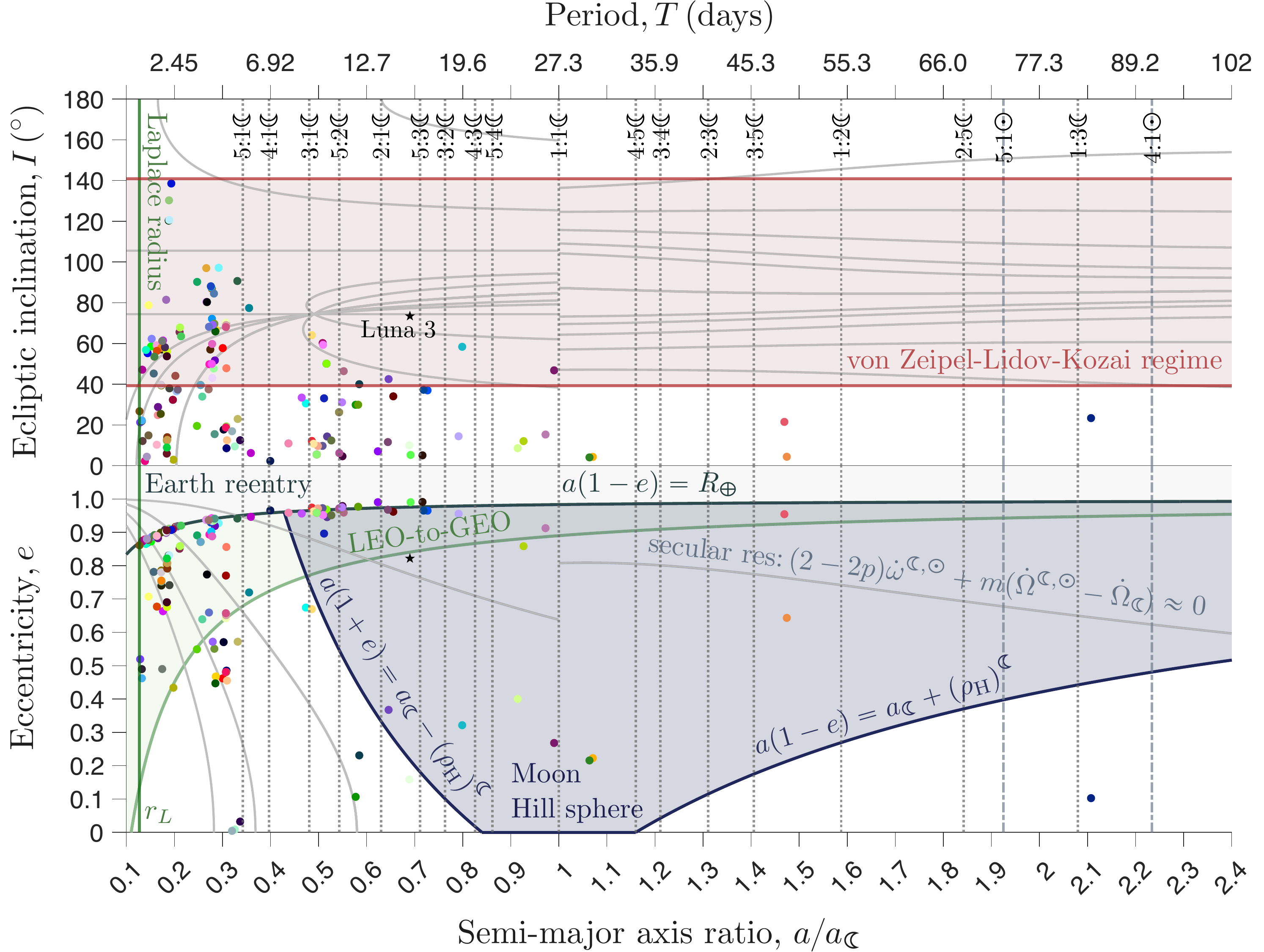}	
	\vskip -0.075in
	\caption{\small 
Snapshot of the historic and current cataloged space objects in the planes of semi-major axis--ecliptic inclination $(a,I)$ ({\it top panel}) and semi-major axis--eccentricity $(a,e)$ ({\it bottom panel}). The {\it colored circles} denote the osculating elements obtained from the latest TLE of each object. The geocentric semi-major axis that marks the dynamical onset of cislunar---the Laplace radius ($r_L$)---is indicated by the {\it vertical line} at $a \sim 0.13\,a_\M$. The approximate loci of the principal lunar mean-motion resonances ({\it vertical dashed lines}) and selected secular structures (other {\it curves}) are overlaid, so that the figure makes visible both the continuous secular web and the discrete resonant ladder that together organize the outer circumterrestrial catalog. The secular {\it curves} are computed from the lowest-order quadrupolar precession formulas at representative fixed slices in eccentricity and inclination, and are intended as schematic spatiographic guides rather than precise local resonance loci, especially in the neighborhood of lunar distance where the exterior and interior lunar branches are stitched together. Objects lying between the delimiting periapsis and apoapsis {\it curves} may enter the Moon's Hill sphere $\left( r_\textrm{H} \right)^\M$, inducing discontinuous jumps in orbital elements \`a la Tisserand's criterion. Luna~3, marked by the $\star$, is discussed further in \cite{dA20}. Objects reaching the Earth-grazing curve $a(1-e)=R_\E$ may reenter the atmosphere. (www.space-track.org. Accessed 1 May 2025.)
	}
	\label{fig:catalog}
	\end{center}
	\vskip -0.075in
	\hspace{2cm}\rule{12.5cm}{0.5pt}
\end{figure}

Figure~\ref{fig:catalog} returns to the geocentric $a$--$e$ catalog projection introduced in Fig.~\ref{fig:hills} and augments it with an accompanying $a$--$I$ view, now with the approximate loci of the predominant secular and mean-motion structures overlaid. The purpose is not to assign a sharp dynamical taxonomy---that synthesis is deferred to Section~\ref{sec:boundaries}---but rather to show how the perturbative framework developed in the preceding subsections already leaves recognizable signatures in the observed distribution of objects.

The top panel displays the catalog in the $( a , I )$ plane, where the overlaid curves represent selected low-order secular loci obtained from the quadrupolar precession formulas of \textsection\ref{sec:sec_quad}, evaluated at a representative eccentricity slice. The bottom panel shows the same objects in the $( a , e )$ plane, where the principal lunar mean-motion commensurabilities appear not merely as reference lines in semi-major axis, but as the first discrete resonant scaffold of the outer circumterrestrial problem. Beyond the secularly dominated inner zone, the cislunar catalog becomes progressively organized by the ladder of low-order lunar period commensurabilities, whose members partition phase space into dynamically distinct bands and culminate in the Earth--Moon gateway near the \res{1}{1}{\M} resonance. Farther out, these lunar resonances are joined by the first solar commensurabilities, so that translunar space begins to exhibit a genuinely mixed lunar--solar resonant architecture. In this form, the figure should be read primarily as a \emph{spatiographic} guide: it identifies where the dominant secular and resonant families are expected to organize the catalog in element space, without implying that the plotted curves furnish exact local resonance centers or separatrices for every object.

This interpretive caution is especially important near lunar distance. The secular curves shown here are based on the lowest-order quadrupolar truncation of the doubly averaged disturbing functions, using the exterior branch for the Sun throughout, the exterior lunar branch in cislunar space, and the axisymmetrized lunar-interior branch in translunar space. That construction captures the first-order reorganization of the secular scaffold as the Moon changes from exterior to interior perturber, but it necessarily stitches together two asymptotic descriptions precisely where each becomes least reliable. Near $a \sim a_\M$, the exterior and interior Legendre expansions both converge slowly, long-period terms omitted from the purely secular model may become important, and the nearby \res{1}{1}{\M} gateway weakens any clean separation between secular and resonant dynamics. The plotted secular curves in that neighborhood should therefore be understood as schematic guides to the changing architecture, not as precise local resonance loci.

With that understanding, several broad features of the catalog become immediately apparent. First, the overwhelming majority of cataloged objects have semi-major axes interior to the Moon's orbit and occupy the broad cislunar band in which slow apsidal and nodal precession still supply the dominant long-term organization. Second, many of these same objects populate a broad high-eccentricity circumterrestrial corridor linking the traditional GEO region to much larger apogee distances, while only a comparatively small subset penetrates deeply into translunar space. Third, the onset of the lunar mean-motion resonance sequence beyond the secularly dominated inner zone shows that the outer circumterrestrial problem is not organized by secular structure alone: discrete period commensurabilities with the Moon emerge as a coequal dynamical framework, furnishing the first resonant partition of cislunar space and supplying the natural bridge from the secularly dominated cislunar interior to the gateway and translunar regimes beyond. In that sense, Fig.~\ref{fig:catalog} offers a first cartographic synthesis of the perturbative structures derived above: not a final dynamical classification, but a map of where the principal secular and resonant organizing mechanisms are expected to reside.

\section{Spatiography: Dynamical Definition of Cislunar, Circumlunar, and Translunar Space}
\label{sec:boundaries}

The spatiographic partition adopted in this Primer---outlined in Table~\ref{tab:spatio}---stands in direct continuity with the early Space-Age work of Hubertus Strughold and Krafft Ehricke. Strughold supplied the general idea of \emph{spatiography} as a geography of space: a topographical description of the region between celestial bodies, using environmental features and orbital characteristics as practical demarcations. Ehricke, in turn, furnished much of the Earth--Moon lexicon and its first dynamical content, treating \emph{terrestrial}, \emph{cislunar}, and \emph{translunar} motion as distinct subdivisions of the broader \emph{circumterrestrial} domain. \emph{Circumlunar} space, by contrast, does not denote a further Earth-centered shell, but the local lunar vicinity itself: a distinct enclave centered on the Moon that cuts across the terrestrial--cislunar--translunar sequence and furnishes the outer boundary of cislunar motion on the near side of the Moon and the inner boundary of translunar motion beyond it.

\begin{table}[p]
    \caption{
        Spatiographic partition of the cislunar and translunar domains by dynamical boundaries, resonances, and tides.
        Here $T=2\pi/n$ denotes the Earth-centered Keplerian sidereal orbital period associated with the listed semi-major axis, reported in days; resonance labels $k \! : \! k_{\mathsf b}$ follow the convention of \textsection\ref{sec:resonances}, so that $k_{\mathsf b} n-k n_{\mathsf b}\approx0$ and $T/T_{\mathsf b}\approx k_{\mathsf b}/k$.
        \label{tab:spatio}
    }
    \begin{tabular*}{0.965\textwidth}{
        p{5.2cm}
        S[table-format=1.2]
        S[table-format=3.2]
        p{6.4cm}
        }      
        
        \toprule
        \addlinespace[1ex]
        
        Dynamical Zone 
        & \multicolumn{1}{c}{\( a / a_\M \)} 
        & \multicolumn{1}{c}{\( T \) [days]} 
        & Description \\
        
        \addlinespace[1.25ex]
	\toprule
        \addlinespace[1ex]
        
        \textbf{Cislunar Lower Bound} & & & \\
            \quad geolunar Laplace radius \( r_L \)
                & 0.13 & 1.24 
                & where lunisolar torques $\approx$ Earth's oblateness 
                \citep{nC01, sT09, aRdS14_ASR} \\
                & & & 
                \( r_L^5 
    		= 2 \mu_\E J_2 R_\E^2 / 
       		 \left( \frac{\mu_\M}{a_\M^3} \frac{1 - \sin^2 i_\M/2}{( 1 - e_\M^2 )^{3/2}} 
        		+ \frac{\mu_\S}{a_\S^3} \frac{1}{( 1 - e_\S^2 )^{3/2}} \right) \)  \\

	\zonebreak
	
	\textbf{Secularly Dominated Zone} & & & \\
            \quad von Zeipel-Lidov-Kozai cycles \& lunar nodal/apsidal secular resonances 
                & {0.13--0.34} & {1.24--5.47}
                & \textit{orbits of 
				OGO \citep{gL63},
				HEOS \citep{aB69},
				Vela \citep{rK73}, 
                			IMP \citep{pB80}, 
				Prognoz \citep{rZ84},
				GEOTAIL \citep{aN94},
				INTERBALL \citep{aG96},
				CXO \citep{mW00},
				XMM-Newton \citep{fJ01}, 
				Cluster \citep{cP01},
				INTEGRAL \citep{nE03},
				\&c.} \\

	\zonebreak

	\textbf{Cislunar Resonant Zone} & & & \\
	\quad resonances interior to Moon's orbit & & & \\[0.5em]
	   \Res{5}{1}{\M} & 0.34 & 5.47  & innermost low-order lunar MMR \\
	   \Res{4}{1}{\M} & 0.40 & 6.84  & \\
	   \Res{3}{1}{\M} & 0.48 & 9.11  & \textit{orbit of IBEX \citep{dM11} \& Tiandu-1} \\
	   \Res{5}{2}{\M} & 0.54 & 10.94 & \\
	   \Res{2}{1}{\M} & 0.63 & 13.67 & \textit{orbit of TESS \citep{gR15}} \\
	   \Res{5}{3}{\M} & 0.71 & 16.41 & \\
	   \Res{3}{2}{\M} & 0.76 & 18.23 & \textit{orbit of DRO-B} \\
	   \Res{4}{3}{\M} & 0.83 & 20.51 & \\
	   \Res{5}{4}{\M} & 0.86 & 21.88 & \\

	\multicolumn{4}{l}{\hspace{0.125em}\rule{0.95\textwidth}{0.3pt}} \\
	\addlinespace[1ex]

        \textbf{Circumlunar Space} & & & \\
            \quad Earth--Moon Lagrange \( \left( L_1 \right)^\M \)
                & 0.84 & 20.94
                & inner zero-velocity surface \\
            \quad Moon's orbit (\res{1}{1}{\M})
                & 1.00 & 27.34
                & lunar semi-major axis, $a_\M$ \\
            \quad Earth--Moon Lagrange \( \left( L_2 \right)^\M \)
                & 1.16 & 34.13
                & outer zero-velocity surface \\

	\zonebreak

	\textbf{Translunar Resonant Zone} & & & \\
	\quad resonances exterior to Moon's orbit & & & \\[0.5em]
    	   \Res{4}{5}{\M} & 1.16 & 34.18  & \\
	   \Res{3}{4}{\M} & 1.21 & 36.46  & \\
	   \Res{2}{3}{\M} & 1.31 & 41.02  & \\
	   \Res{3}{5}{\M} & 1.41 & 45.57  & \\
	   \Res{1}{2}{\M} & 1.59 & 54.69  & \\
	   \Res{2}{5}{\M} & 1.84 & 68.36  & \\
	   \Res{5}{1}{\S} & 1.93 & 73.05  & innermost low-order solar MMR \\
	   \Res{1}{3}{\M} & 2.08 & 82.00  & \\
	   \Res{4}{1}{\S} & 2.23 & 91.31  & \\
	   \Res{1}{4}{\M} & 2.52 & 109.38 & \\
	   \Res{3}{1}{\S} & 2.71 & 121.75 & \\
	   \Res{1}{5}{\M} & 2.92 & 136.72 & outermost low-order lunar MMR \\
	   \Res{5}{2}{\S} & 3.06 & 146.10 & \\
	   \Res{2}{1}{\S} & 3.55 & 182.63 & outermost low-order solar MMR \\             
            
	\zonebreak

	\textbf{Outer Tidal \& SOI Boundaries} & & & \\
    	    \quad lunisolar tidal parity
                 & 1.17 & 34.6
        		& equality of secular (quadrupole) lunar \emph{internal} and solar \emph{external} tidal influence \\
        		& & &
        		\( 
		\left( \dfrac{a_{\rm TP}}{a_\M} \right)^5 = \dfrac{\mu_\M}{\mu_\S} \left( \dfrac{a_\S}{a_\M} \right)^3
		\) 
		\\
		& & & 
		\(
		\hspace{5.0em} \left( 1 - \tfrac{3}{2} \sin^2 i_\M \right) \left( 1 + \tfrac{3}{2} e_\M^2 \right) \left( 1 - e_\S^2 \right)^{3/2} 
		\)
        		\\[0.5em]
            \quad Laplace's patched-conic SOI
                & 2.41 & 102.41 
                & \( \left( r_{\rm SOI} \right)^\E = a_\E \left( \mu_\E / \mu_\S \right)^{2/5} \) \\[0.5em]
            \quad Earth's Hill sphere \( \left( r_H \right)^\E \)        
                & 3.90 & 210.88
                & \( \left( r_H \right)^\E = a_\E \left( \mu_\E / 3 \mu_\S \right)^{1/3} \) \\[0.25em]

        \bottomrule
        
    \end{tabular*}
\end{table}

\citet{kE57} did not impose an exact terrestrial--cislunar boundary. His usage is deliberately qualitative and perturbative. In one formulation, the cislunar satellite ``essentially operates'' inside the lunar orbit beyond about ten Earth radii \citep{kE62a}; in another, he states more explicitly that the transition from terrestrial to cislunar space is difficult to define and depends on the particular requirements of the orbit under consideration \citep{kE62b}. This ambiguity is not a defect of the early taxonomy but a faithful reflection of the underlying dynamics: the change from terrestrial to cislunar motion is not a kinematic crossing of a single geometric shell, but a gradual reordering of the perturbation hierarchy (qv.~Fig.~\ref{fig:accels}). 

The terrestrial--cislunar transition is only the first of the dynamical demarcations that must be clarified. Ehricke's own mature picture already implies a second handoff farther out: cislunar motion is organized principally by the Earth--Moon system, whereas translunar motion becomes progressively governed by the Sun--Earth system. The outer circumterrestrial problem is therefore not uniform beyond the Moon. Its inner part still retains the strong imprint of lunar gateways, encounters, and commensurabilities, while at greater distance the Sun assumes the leading role in setting the stability structure. Nor does translunar space extend indefinitely: beyond the outer Earth-bound domain, motion ceases to be meaningfully circumterrestrial and becomes heliocentrically organized \citep{kE62b, sH61, sH69}. In that sense, the translunar realm is bounded inward by the Earth--Moon $L_2$ gateway and outward not by a single universal shell, but by a sequence of increasingly global demarcations---first tidal, then sphere-of-influence, and finally Hill-stability in character.

Accordingly, our purpose is not to replace Ehricke's usage with an anachronistically rigid definition, but to sharpen these particular transitions within his broader framework. In the specific secular sense relevant here, the Laplace radius furnishes a natural dynamical onset of the cislunar regime: it is the distance at which the characteristic precession driven by the Earth's oblateness becomes comparable to the combined lunar and solar torque. Beyond this point, the motion is no longer usefully ``terrestrial'' in the classical oblateness-dominated sense, and the preferred reference geometry begins to shift away from the equator toward the lunar-orbital and ecliptic planes. The Laplace radius is therefore not presented here as a universal historical definition of cislunar space, but as the first precise dynamical realization of the terrestrial--cislunar transition that Ehricke had already identified in qualitative terms. The spatiographic partition summarized in Table~\ref{tab:spatio} is intended to make both transitions explicit: the inner onset of cislunar space is sharpened through the Laplace radius, while the outer reorganization into translunar dynamics is approximated first by lunisolar tidal parity and then by the progressively weaker Earth-bound limits associated with sphere-of-influence and Hill-type boundaries.

\subsection{Dynamical Transition at the Laplace Radius}
\label{sec:laplace}

In terrestrial space, the dominant conservative perturbation is Earth's oblateness (qv.~Fig.~\ref{fig:accels}). To lowest order, the corresponding secular Hamiltonian is
\begin{align}
    \label{eq:H_sec_J2}
    \mc{H}_\text{sec}^\E (G, H; L)
        =
        \frac{J_2 R_\E^2 \mu_\E^4}{4 L^3}
        \frac{G^2 - 3 H^2}{G^5},
\end{align}
where $J_2$ is the second zonal harmonic coefficient of the geopotential, $R_\E$ is the Earth's mean equatorial radius, and $\mu_\E$ is the terrestrial gravitational parameter. In this inner regime, the apsidal and nodal precession induced by $J_2$ overwhelmingly exceeds the corresponding secular effects of the Moon and Sun, so that the long-term evolution is governed primarily by the oblateness-driven regression of the nodes and rotation of the line of apsides (cf.~Fig.~\ref{fig:accels}).

The associated secular rates follow from $\dot g = \partial \mc{H}_\text{sec}^\E/\partial G$ and $\dot h = \partial \mc{H}_\text{sec}^\E/\partial H$. 
In the terrestrial problem these rates are most naturally written in Earth-equatorial elements, since the equator is the symmetry plane of the oblate geopotential.\footnote{The terrestrial equator is the dynamically preferred reference plane in terrestrial space, where oblateness supplies the leading conservative perturbation \citep{kE62a}. Across the cislunar transition, however, the preferred perturbative reference geometry shifts: lunar forcing is more naturally expressed with respect to the Moon's orbital plane, while solar forcing privileges the ecliptic \citep{kE62b}. The catalog projection in Fig.~\ref{fig:catalog} was therefore plotted using ecliptic inclination to provide a common Earth--Moon--Sun reference geometry across the cislunar and translunar regimes.} Denoting by $I_{\mathrm eq}$ the inclination measured from the terrestrial equator, one recovers the classical expressions \citep{lB59}
\begin{align}
\label{eq:oblateness}
    \dot\omega^\E
    &= \frac{\upwE}{2}\,
       \frac{5\cos^2 I_{\mathrm eq} - 1}{(1 - e^2)^2},
    \\
    \dot\Omega^\E
    &= -\upwE\,
       \frac{\cos I_{\mathrm eq}}{(1 - e^2)^2},
\end{align}
where
\begin{align}
    \upwE
        =
        \frac{3}{2}\sqrt{\mu_\E}\,J_2 R_\E^2 a^{-7/2}.
\end{align}

By contrast, the quadrupolar secular frequencies generated by the lunar and solar disturbing functions increase outward with semi-major axis. Using the exterior-perturber formulas of \textsection\ref{sec:sec_quad}, these are
\begin{align}
    \label{eq:luni}
    \upwM
    &=
    \frac{3}{4}\frac{\mu_\M}{\sqrt{\mu_\E}}
    \frac{1 - \sin^2 I_\M/2}{a_\M^3 (1 - e_\M^2)^{3/2}}
    a^{3/2},
    \\
    \label{eq:solar}
    \upwS
    &=
    \frac{3}{4}\frac{\mu_\S}{\sqrt{\mu_\E}}
    \frac{1}{a_\S^3 (1 - e_\S^2)^{3/2}}
    a^{3/2}.
\end{align}
Thus the terrestrial--cislunar transition may be identified, in the secular sense, with the critical distance at which the combined lunisolar torque becomes comparable to the oblateness-driven precession.

This critical distance is the \emph{Laplace radius}, $r_L$ \citep{rAgC64, pG66, aR69, aD93, nC01, sT09, aRdS14_ASR}, defined in the Earth--Moon--Sun problem by
\begin{align}
    \label{eq:laplace}
    r_L^5
    &=
    a^5 \frac{\upwE}{\upwM + \upwS}
    \nonumber\\
    &=
    \frac{2 \mu_\E J_2 R_\E^2}{
        \dfrac{\mu_\M}{a_\M^3}
        \dfrac{1 - \sin^2 I_\M/2}{(1 - e_\M^2)^{3/2}}
        +
        \dfrac{\mu_\S}{a_\S^3}
        \dfrac{1}{(1 - e_\S^2)^{3/2}}
    },
\end{align}
so that
\begin{align}
    r_L \approx 7.7\,R_\E \approx 0.13\,a_\M .
\end{align}
Here, and throughout this Primer, $R_\E$ and $J_{2,\E}$ are taken from the GGM02 terrestrial gravity-field normalization, with $R_\E=6378.1363~\mathrm{km}$ and $J_{2,\E}=1.08263552549\times10^{-3}$ \citep{bT05}. The mean orbital elements entering the lunisolar terms are taken from \citet{jS94}: 
$a_\M=383397.7725~\mathrm{km}$, 
$e_\M=0.055545526$, 
$I_\M=5.15668983^\circ$, and, for the Sun's apparent geocentric
orbit, 
$a_\S=1.0000010178~\mathrm{au}$ and
$e_\S=0.0167086342$.
The gravitational parameters used in the numerical evaluations are taken from JPL/NAIF SPICE kernels, giving
$\mu_\E=3.986004354360959\times10^5~\mathrm{km^3\,s^{-2}}$,
$\mu_\M=4.902800066163796\times10^3~\mathrm{km^3\,s^{-2}}$, and
$\mu_\S=1.327124400419393\times10^{11}~\mathrm{km^3\,s^{-2}}$.

\begin{figure}[t!]	
	\begin{center}
	\includegraphics[width=0.875\textwidth]{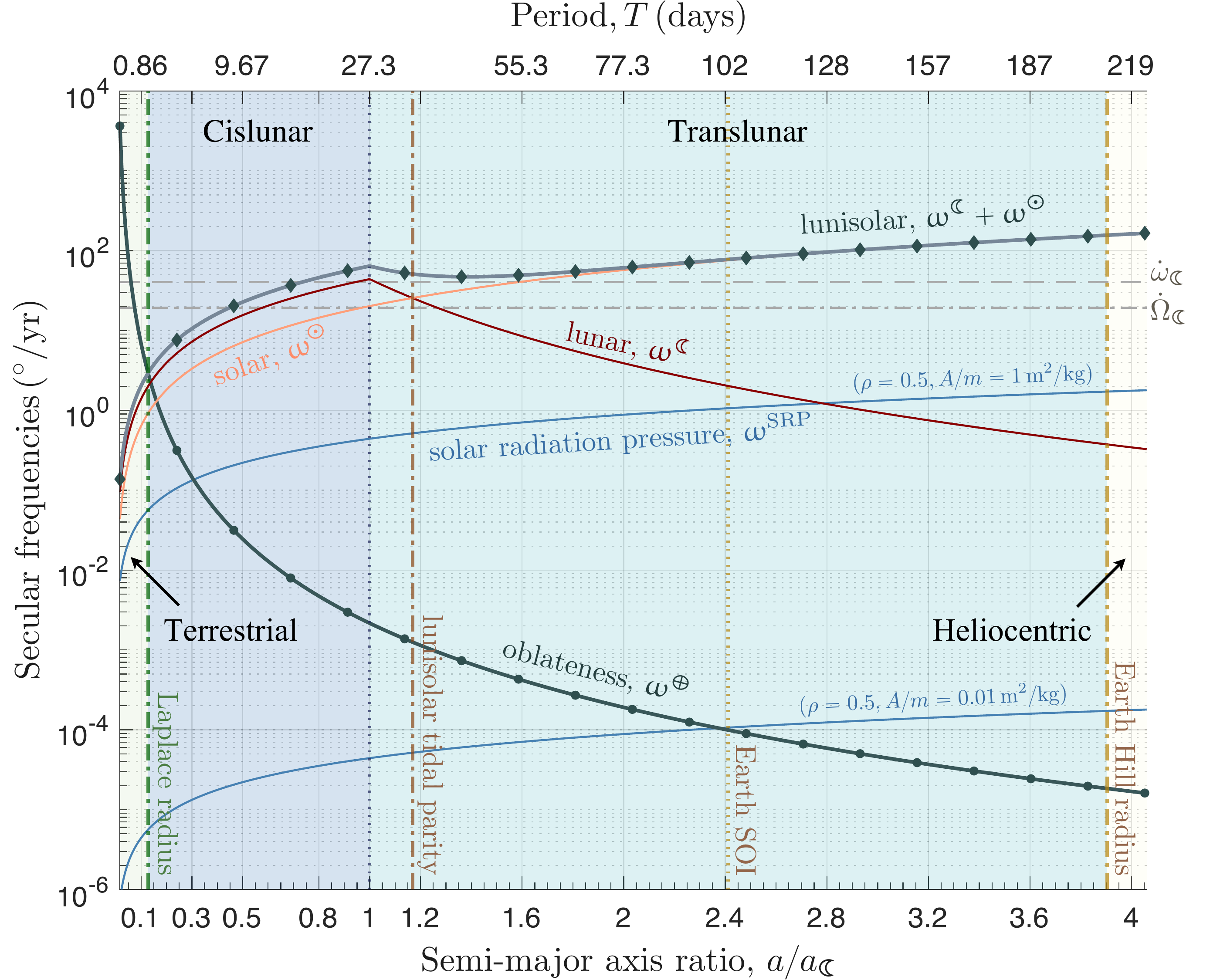}	
	\vskip -0.075in
	\caption{\small 
Secular partition of cislunar space defined by the {\it Laplace radius}, $r_L$, beyond which the characteristic secular apsidal and nodal precession induced by lunisolar perturbations (Eqs.~\eqref{eq:luni} and \eqref{eq:solar}) overtakes that due to Earth oblateness (Eq.~\eqref{eq:oblateness}). Translunar space is then partitioned by the {\it lunisolar tidal parity}, at which the leading-order secular tidal influence on circumterrestrial motion shifts from lunar-dominated to solar-dominated. Also shown are the approximately constant rates of lunar apsidal progression ($\dot\omega_\M \approx 40.68^\circ/\mathrm{yr}$) and nodal regression ($\dot\Omega_\M \approx -19.34^\circ/\mathrm{yr}$), together with the secular solar-radiation-pressure (SRP) precession rates for both low- and high-area-to-mass-ratio objects, parameterized by reflectivity $\rho$ and $A/m$ \citep[qq.v.,][]{aRdS14_ASR,aRdS14_ApJ}.
	}
	\label{fig:laplace}
	\end{center}
	\vskip -0.075in
	\hspace{2cm}\rule{12.5cm}{0.5pt}
\end{figure}

Figure~\ref{fig:laplace} gives a graphical rendering of this balance and, in the present spatiographic framework, furnishes the dynamical lower bound of cislunar space. Beyond $r_L$, atmospheric drag is negligible even for near-Earth grazers \citep{dK62,iS63,bSjC66,gC67,bL72,gJeR76}, while secular solar-radiation-pressure effects remain orders of magnitude weaker than the lunar and solar gravitational torques even for high area-to-mass-ratio objects \citep{aRdS14_CMDA, aRdS14_ASR,aRdS14_ApJ}. The cislunar regime is therefore, to first approximation, the domain in which Earth oblateness has ceased to be the leading secular organizer and the long-term dynamics must instead be understood in explicitly lunisolar terms.

\subsection{Inner Cislunar Zone: The vZLK Effect and Secular Resonances}
\label{sec:secular}

The principal secular architecture of cislunar space has already been identified in \textsection\ref{sec:laplace}: beyond the Laplace radius, apsidal and nodal precession are no longer governed primarily by Earth's oblateness, but by the combined action of the Moon and Sun. In the spatiographic partition of Table~\ref{tab:spatio}, this defines the \emph{secularly dominated zone}, extending from the Laplace transition at $a/a_\M \approx 0.13$ to the onset of the interior lunar mean-motion-resonance sequence at the \res{5}{1}{\M} commensurability near $a/a_\M \approx 0.34$. Within this band, the long-term organization of phase space is set chiefly by commensurabilities among the slow precessional frequencies rather than by period ratios with the Moon or Sun.

At lowest order, the natural backbone of this region is furnished by the quadrupolar lunisolar von Zeipel--Lidov--Kozai Hamiltonian, Eq.~\eqref{eq:HvZLK}, whose level curves in the $( \omega , \sqrt{1-e^2} )$ plane provide an autonomous first portrait of the secular phase space. In this orbit-averaged quadrupolar approximation, the semi-major axis is conserved, the projected angular momentum remains approximately fixed, and the familiar coupled exchange between eccentricity and inclination appears through libration or circulation of the argument of perigee \citep{mL62, yK62}. The corresponding phase portraits are therefore not meant as exact maps of the real cislunar problem, but as the leading-order skeleton on which the more complicated lunisolar secular dynamics are built.

\begin{figure}[t!]	
	\begin{center}
	\includegraphics[width=0.825\textwidth]{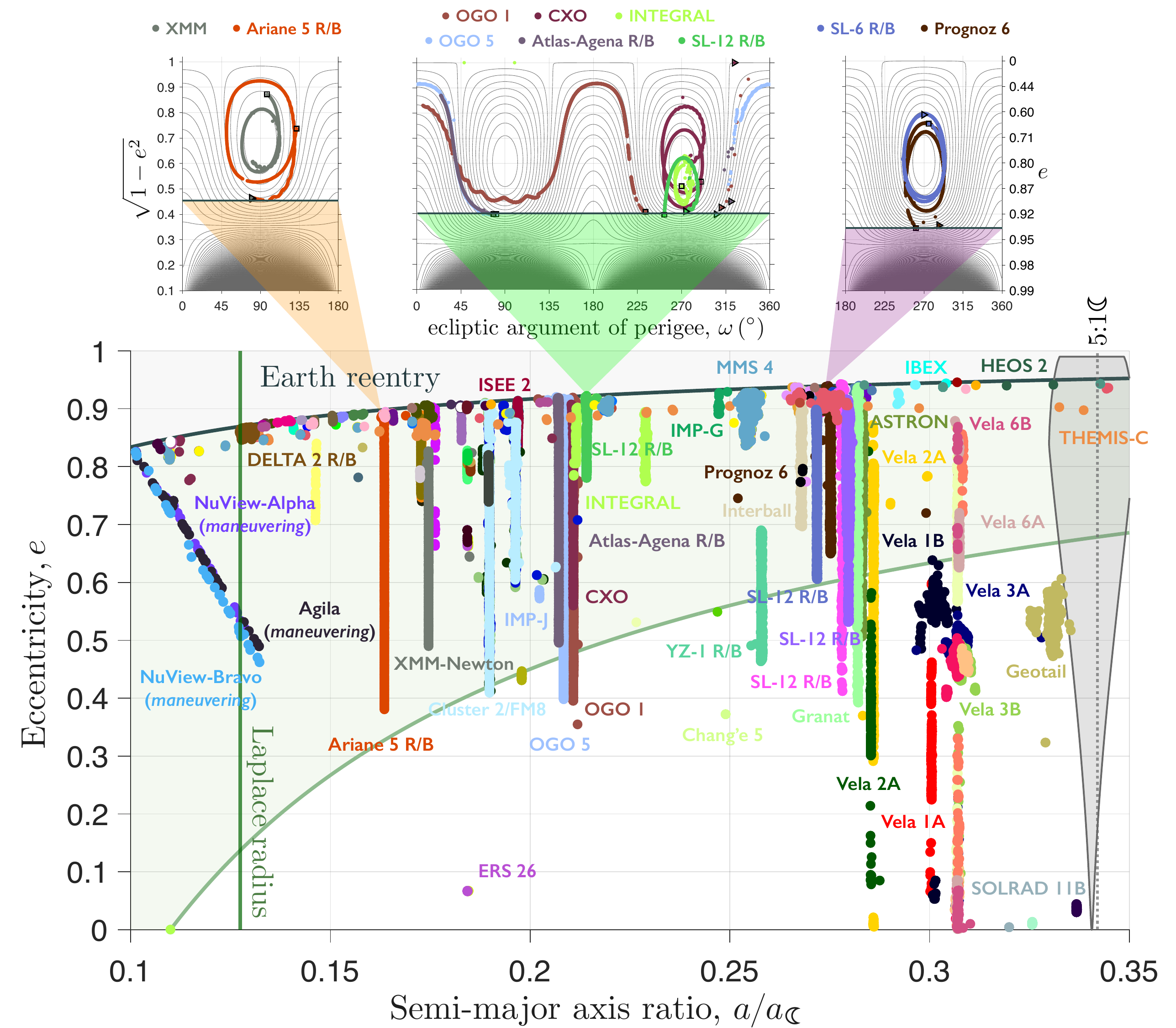}	
	\vskip -0.075in
	\caption{\small 
Representative secular phase-space portraits for objects in the secularly dominated cislunar zone. The \emph{top banner} shows level curves of the idealized orbit-averaged quadrupolar lunisolar von Zeipel--Lidov--Kozai Hamiltonian, Eq.~\eqref{eq:HvZLK}, in the $( \omega , \sqrt{1-e^2} )$ plane at selected semi-major axes, with the corresponding orbit-averaged histories of representative objects overlaid ($\rhd$ denotes the initial state and $\square$ the final state). The \emph{lower panel} displays the osculating-element histories of the 110 objects with at least one valid epoch in the secularly dominated cislunar interval, projected in geocentric $( a , e )$ space relative to the Earth-reentry boundary, the Laplace radius $r_L$, and the \res{5}{1}{\M} commensurability that marks the onset of the interior lunar mean-motion-resonance sequence. The figure shows that the interval $0.13 \lesssim a/a_\M \lesssim 0.34$ is organized primarily by slow lunisolar secular dynamics: the classical vZLK phase portrait provides the leading-order backbone, while the departures of the overlaid trajectories from the idealized level sets reflect the broader web of lunar nodal/apsidal secular resonances and higher-order lunisolar perturbations. Known or suspected maneuvering objects are retained to show catalog occupancy, but their tracks should not be interpreted as ballistic secular evolutions.
	}
	\label{fig:secular}
	\end{center}
	\vskip -0.075in
	\hspace{2cm}\rule{12.5cm}{0.5pt}
\end{figure}

Figure~\ref{fig:secular} shows this structure for selected representative objects drawn from the curated catalog. The banner panels display the level curves of the averaged quadrupolar Hamiltonian at representative semi-major axes within the secularly dominated zone, with the orbit-averaged histories of several xGEO objects overlaid. The lower panel places the same objects in the $( a , e )$ plane relative to the Earth-reentry boundary, the Laplace radius, and the \res{5}{1}{\M} commensurability that marks the onset of the predominant cislunar mean-motion architecture. Read together, the panels make clear that a substantial fraction of the historic and current population occupies precisely this intermediate domain: too distant for the classical oblateness-dominated terrestrial theory, yet still interior to the discrete lunar mean-motion architecture that takes over farther out.

The figure also clarifies the sense in which the vZLK picture should be used here. For some objects, the orbit-averaged trajectories lie close to the idealized quadrupolar level sets, indicating that the classical lunisolar secular Hamiltonian captures an important part of the underlying organization. For others, the traces are visibly distorted, shifted, or broadened relative to that backbone. This is expected. In the real cislunar problem, the secular dynamics are not governed by Eq.~\eqref{eq:HvZLK} alone, but are modulated by the Moon's finite inclination to the ecliptic, the regression of the lunar nodes, the apsidal rotation of the lunar orbit, and higher-order terms in the disturbing function. The result is that the classical vZLK portrait survives not as an exact global theory, but as an idealized first scaffold embedded within a richer web of lunisolar secular resonances.

This distinction matters for the spatiographic partition. The interval $0.13 \lesssim a/a_\M \lesssim 0.34$ is not merely a quiet gap between GEO and the first low-order lunar MMRs. It is a dynamically structured secular province, populated by trajectories whose long-term evolution is controlled primarily by slow lunisolar precession and by resonances among the corresponding apsidal and nodal frequencies. In this sense, the secularly dominated zone forms the natural inner cislunar counterpart to the resonant ladder discussed in the following subsection: here the architecture is continuous and precessional, whereas farther out it becomes increasingly discrete and commensurability-driven.

Historically, this is also the region long exploited by high-altitude scientific missions. Many of the classic astrophysical, magnetospheric, and space-physics observatories listed in Table~\ref{tab:spatio}---including \emph{OGO}, \emph{HEOS}, \emph{GEOTAIL}, \emph{INTERBALL}, \emph{CXO}, \emph{XMM-Newton}, and \emph{INTEGRAL}---occupy or traverse precisely this secular domain. Their presence is a practical reminder that the first dynamical onset of cislunar space is not resonant but secular: before the satellite encounters the strong low-order lunar commensurabilities of the outer cislunar region, it already enters a regime in which the dominant long-term organizing influences are the Moon and Sun.


\subsection{Outer Cislunar Zone: Lunar Mean-Motion Resonances}
\label{sec:cislunar}

Using the notation of \textsection\ref{sec:resonances}, the next province in the spatiographic partition of Table~\ref{tab:spatio} is the \emph{cislunar resonant zone}. It begins where the first low-order \emph{interior} lunar commensurabilities enter the circumterrestrial problem, namely near the \res{5}{1}{\M} resonance at $a/a_\M \simeq 0.34$. Here and in Table~\ref{tab:spatio}, ``low-order'' is used in the spatiographic sense of retaining the principal lunar commensurabilities with $\lvert \, k-k_\M\rvert \leq 4$; the lunar disturbing function of course contains higher-order harmonics, which may structure phase space at smaller semi-major axes or larger eccentricities, but these are not used to define the primary regional partition.

Beyond that point, the outer cislunar phase space is no longer organized chiefly by the continuous secular web discussed in \textsection\ref{sec:secular}, but by a discrete ladder of increasingly prominent low-order period commensurabilities with the Moon. The progression
\begin{center}
	\res{5}{1}{\M}, 
	\res{4}{1}{\M},
	\res{3}{1}{\M},
	\res{5}{2}{\M},
	\res{2}{1}{\M},
	\res{5}{3}{\M},
	\res{3}{2}{\M},
	\res{4}{3}{\M},
	\res{5}{4}{\M}
\end{center}
therefore furnishes the principal resonant scaffold of the outer cislunar domain before the description must pass over to the circumlunar gateway geometry near the co-orbital \res{1}{1}{\M} region.

A useful local estimate of the size and location of these resonances is furnished by the semi-analytical method of \citet{tG06, tG19, tG20}. In spirit, this construction refines the nominal commensurability picture of \textsection\ref{sec:resonances}: one begins from the same resonant angle and period ratio, but instead of retaining only the Keplerian center, one numerically averages the lunar disturbing function at fixed $( e , I , \omega , \Omega )$ over one resonant cycle and thereby reduces the problem to an effective one-degree-of-freedom Hamiltonian in $( a , \sigma )$. The resonant angle may be written in the form
\begin{align}
    \sigma = k_\M \lambda - k \lambda_\M + \gamma,
\end{align}
where $\gamma$ is a slowly varying linear combination of the apsidal and nodal longitudes of the satellite and Moon. The corresponding resonant disturbing function is then
\begin{align}
    \mathcal{R}_{k \, : \, k_\M}(\sigma)
    =
    \frac{1}{2\pi k_\M}
    \int_0^{2\pi k_\M}
        R^\M\!\bigl(\lambda_\M,\lambda(\lambda_\M,\sigma)\bigr)\,
    \mathrm{d}\lambda_\M,
\end{align}
which yields the semi-secular Hamiltonian
\begin{align}
    \mathcal{K}(a,\sigma)
    =
    -\frac{\mu_\E}{2a}
    -
    n_\M \frac{k}{k_\M}\sqrt{\mu_\E a}
    -
    \mathcal{R}_{k \, : \, k_\M}(\sigma).
\end{align}
If $\sigma_{\mathrm s}$ and $\sigma_{\mathrm u}$ denote the stable and unstable equilibria of the averaged resonance, and
\begin{align}
    \Delta \mathcal{R}
    =
    \mathcal{R}_{k \, : \, k_\M}(\sigma_{\mathrm u})
    -
    \mathcal{R}_{k \, : \, k_\M}(\sigma_{\mathrm s}),
\end{align}
then the usual pendulum-like estimate of the half-width is
\begin{align}
    \Delta a_{k \, : \, k_\M}
    \approx
    \sqrt{\frac{8}{3}}\,
    \frac{\sqrt{\Delta \mathcal{R}}}{n},
    \qquad
    n = \sqrt{\frac{\mu_\E}{a_{k \, : \, k_\M}^{\,3}}}.
\end{align}
In this form, Gallardo's construction provides an efficient atlas of approximate libration widths in the $( a , e )$ plane and is therefore well suited to a first cartographic reading of the resonant cislunar belt.

\begin{figure}[t!]	
	\begin{center}
	\includegraphics[width=0.975\textwidth]{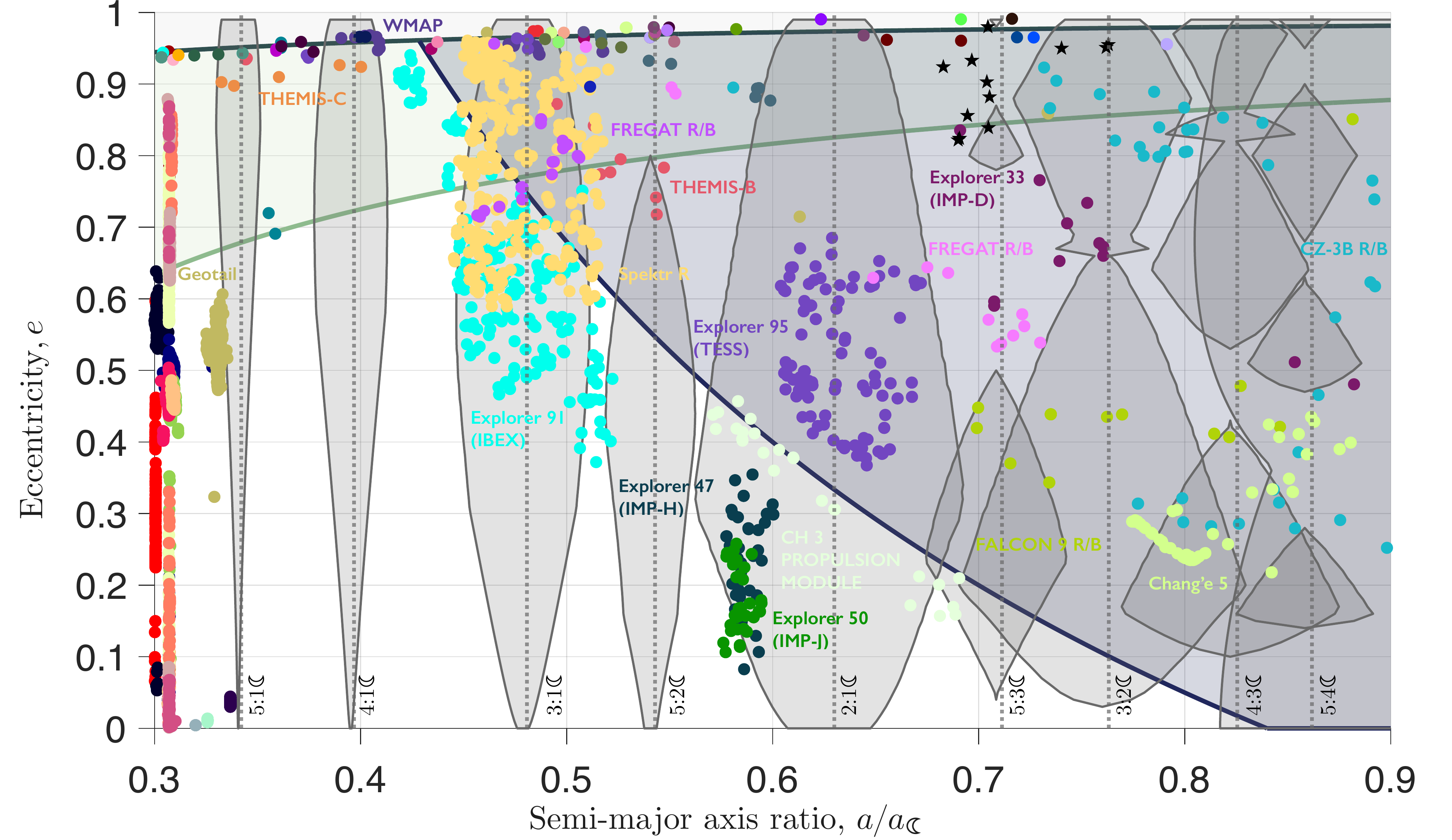}	
	\vskip -0.075in
	\caption{\small 
Representative coplanar Gallardo atlas of the principal \emph{interior} lunar mean-motion resonances, overlaid on the osculating-element histories of the 54 cataloged objects with at least one valid epoch in the cislunar resonant interval in the $( a , e )$ plane. The {\it vertical dotted lines} mark the nominal Keplerian commensurability centers, and the {\it shaded envelopes} indicate the corresponding local semi-analytical libration widths. The {\it shaded} lunar-encounter corridor marks osculating geocentric ellipses whose apoapsis reaches the inner Moon-Hill boundary, $a ( 1 + e ) = a_\M - \left( \rho_H \right)^\M$. The terrestrial-transit band identifies trajectories whose periapsis returns to the inner circumterrestrial domain, bounded by the GEO-crossing curve $a ( 1 - e ) = r_{\mathrm{GEO}}$ and the upper Earth-grazing line $a ( 1 - e ) = R_\E$. Read in conjunction with Table~\ref{tab:spatio}, the figure gives a first cartography of the \emph{cislunar resonant zone}: it shows the onset of the resonant ladder near \res{5}{1}{\M}, its strengthening through the dominant \res{3}{1}{\M} and \res{2}{1}{\M} families, and its accumulation toward the Earth--Moon co-orbital/gateway region near \res{1}{1}{\M}. The {\it overlaid tracks} highlight, in particular, the placement of IBEX near the \res{3}{1}{\M} family and TESS within the \res{2}{1}{\M} domain.
	}
	\label{fig:cislunar_atlas}
	\end{center}
	\vskip -0.075in
	\hspace{2cm}\rule{12.5cm}{0.5pt}
\end{figure}

Figure~\ref{fig:cislunar_atlas} shows a representative coplanar slice of that atlas for the Earth--Moon system, computed using the same mean lunar orbit as above, \citep{jS94}
\[
    a_\M = 383397.7725~\mathrm{km},
    \qquad
    e_\M = 0.055545526,
\]
with the satellite's inclination with respect to the Moon set to zero and a $2 \left( r_H \right)^\M$ cutoff imposed to suppress spurious broadening near close lunar encounters \citep{tG21}. Read in the language of Table~\ref{tab:spatio}, the figure gives a first cartography of the \emph{cislunar resonant zone}: the secularly dominated interior yields near $a/a_\M \simeq 0.34$ to a discrete sequence of interior lunar commensurabilities, whose broadest and dynamically most conspicuous low-order members are the \res{3}{1}{\M} and \res{2}{1}{\M} families \citep{aRbK26}. Higher-order resonances fill much of the intervening large-eccentricity space, while the approach to the \res{1}{1}{\M} co-orbital region signals the breakdown of this local perturbed-Kepler description and the onset of the Earth--Moon gateway regime.

The overlaid object histories are consistent with this reading. TESS \citep{dD14} lies squarely in the domain of the \res{2}{1}{\M} resonance, whereas IBEX \citep{dM11} and Spektr-R \citep{vP14, gZ14, gZ17} probe neighboring higher-order structures. At the same time, the figure also makes clear the limits of the construction: the \res{3}{1}{\M} width does not encompass the full spread of all nearby tracks, and the co-orbital \res{1}{1}{\M} band is well known to be overpredicted when treated in this local width-based manner. Gallardo's atlas should therefore be read not as a global dynamical boundary map, but as a perturbed-Kepler portrait of the dominant \emph{stable} libration islands.

This distinction is important. The nominal Keplerian estimate
\begin{align*}
    a_{k \, : \, k_\M}
    \approx
    \left( \frac{k_\M}{k} \right)^{2/3} a_\M,
\end{align*}
derived in \textsection\ref{sec:resonances}, gives only the first approximation to the resonance location. In the semi-analytical construction, the effective libration centers and widths are obtained instead from the equilibria of the numerically averaged resonant Hamiltonian at fixed $( e , I , \omega , \Omega )$. Consequently, the resonant islands need not be centered exactly on the nominal commensurability line: the full stationary-phase condition contains slow apsidal and nodal contributions omitted in the leading-order period-ratio estimate, and the resulting widths are generally asymmetric in $( a , e )$ space. For this reason, the atlas is best interpreted as a local diagnostic of where resonant structure first becomes important within the cislunar domain, rather than as an exact delimitation of the full chaotic region of influence of each resonance \citep{aRbK26}.

For the purposes of the present spatiographic partition, however, this is precisely the appropriate intermediate description. It is more dynamical than a mere list of nominal commensurabilities, yet less model-dependent than a full global CR3BP resonance atlas. In that sense, Fig.~\ref{fig:cislunar_atlas} marks the point at which the continuous secular province of \textsection\ref{sec:secular} gives way to the discrete lunar commensurability architecture that organizes the outer cislunar domain.

\subsection{Spheres of Gravitational Influence}
\label{sec:soi}

In the patched spatiography of Fig.~\ref{fig:patched}, the next demarcations are no longer set by secular or resonant structure alone, but by the choice of reduced model appropriate to the region of motion. As already suggested by the perturbation hierarchy of Fig.~\ref{fig:accels}, the Earth--Moon environment admits more than one physically meaningful crossover. One may ask where lunar gravity becomes dynamically capable of confining or releasing motion through the CR3BP gateways, or instead where it becomes numerically preferable to replace a geocentric two-body description by a selenocentric one. These are not the same question. In the language of \textsection\ref{sec:perturbed}, the lunar and solar perturbations are represented by disturbing functions whose gradients furnish the corresponding perturbing accelerations; here we ask when those perturbations cease to be secondary in one sense or another. This leads to three related but distinct boundary notions: the Hill radius, the critical Jacobi boundary associated with the opening of the $\left( L_1 \right)^\M$--$\left( L_2 \right)^\M$ necks, and the classical sphere of activity or sphere of influence \citep{kE62b, sH61, sH69, gC64, dSetal20, iC23}. For the present Primer, the Hill/Jacobi viewpoint prepares the definition of \emph{circumlunar} space in \textsection\ref{sec:circumlunar}, whereas the SOI viewpoint remains useful as a patched-conic proxy in the discussion of \emph{translunar} passage in \textsection\ref{sec:translunar}.

\subsubsection{Hill Scales and Jacobi Gateways}

The natural inner gateway scale near the Moon is set by the collinear equilibria of the Earth--Moon circular restricted three-body problem. Let
\begin{align}
    \label{eq:mu_EM}
    \bar\mu_{\E\M} = \frac{\mu_\M}{\mu_\E + \mu_\M},
\end{align}
denote the usual dimensionless Earth--Moon CR3BP mass parameter, i.e., the lunar mass fraction of the two-primary system, and normalize distances by the Earth--Moon separation $a_\M$. In the usual rotating coordinates, the effective potential is
\begin{align}
    \mathcal{U}_{\E\M}(x,y,z)
    =
    \frac{1-\bar\mu_{\E\M}}{r_1}
    +
    \frac{\bar\mu_{\E\M}}{r_2}
    +
    \frac{1}{2}\left(x^2+y^2\right),
\end{align}
where $r_1$ and $r_2$ are the nondimensional distances from the particle to the Earth and Moon respectively (see \textsection\ref{sec:phenomenology}). The collinear libration points are determined by $\partial_x\mathcal{U}_{\E\M}=0$ along the $x$-axis. Writing the location of either lunar collinear point as
\begin{align}
    x = 1 - \bar\mu_{\E\M} \pm \delta,
    \qquad
    \delta \ll 1,
\end{align}
and expanding the equilibrium condition for small $\bar\mu_{\E\M}$, one obtains at leading order
\begin{align}
    3\delta_H^3 - \bar\mu_{\E\M} = 0,
\end{align}
so that
\begin{align}
    \delta_H
    =
    \left(\frac{\bar\mu_{\E\M}}{3}\right)^{1/3}.
\end{align}
Restoring dimensions then gives the familiar lunar Hill scale
\begin{align}
    \label{eq:hill_lunar}
    \left( \rho_H \right)^\M
    =
    a_\M
    \left(
        \frac{\mu_\M}{3(\mu_\E+\mu_\M)}
    \right)^{1/3}
    \simeq
    a_\M
    \left(
        \frac{\mu_\M}{3\mu_\E}
    \right)^{1/3}.
\end{align}
To first order, $\left( L_1 \right)^\M$ and $\left( L_2 \right)^\M$ therefore lie at distances $\pm \left( \rho_H \right)^\M$ from the Moon along the Earth--Moon line. In this sphericalized sense, $\left( \rho_H \right)^\M$ provides the characteristic metric scale of the lunar gateway region \citep{rB56a, bT59, kE59, sH61, pL61}.

Dynamically, however, the true boundary is not a sphere. What matters is the last \emph{closed} lunar lobe of the Earth--Moon zero-velocity surface and its subsequent opening through $\left( L_1 \right)^\M$ and $\left( L_2 \right)^\M$. In the language of modern transport theory, these are the entry and exit bottlenecks through which circumlunar motion communicates with the surrounding cislunar and translunar domains. Long before invariant-manifold language became standard, this topology was already recognized in early Earth--Moon studies. \citet{rB56a} mapped the regions of possible motion and their associated zero-velocity contours in Earth--Moon space. Huang emphasized the role of the \emph{inner contact surface} and its secondary lobe as the limiting closed region around the smaller primary \citep{sH61, sH69}. \citet{bT59}, \citet{vE59}, and \citet{kE59} then exhibited explicit capture-type trajectories threading the narrow necks at $\left( L_1 \right)^\M$ and $\left( L_2 \right)^\M$, with repeated revolutions about the Earth or the Earth--Moon system before entrance into the lunar neighborhood \citep{kE62b}.

This viewpoint is sharpened by Lanzano's \emph{critical Jacobi region}, defined as the largest closed zero-velocity surface surrounding the Moon. In that formulation, a trajectory arriving from Earth may enter the lunar neighborhood provided its Jacobi constant lies below the first critical value associated with the lunar neck, while a sufficiently large impulsive decrease in speed within that region can raise the Jacobi constant so that the new zero-velocity surface closes around the Moon. Capture is then guaranteed in the topological sense: the probe can no longer return to Earth or escape from the lunar neighborhood. \citet{pL61} further obtained an analytical series for the boundary of this critical region, together with the maximum velocity compatible with non-escape from the lunar field. In modern low-energy mission design, this same distinction reappears in the language of temporary gravitational capture and weak lunar residence \citep{hY92, eB93, kY04, rC12, pSS12, fT13, yQ14, pdSS16, kO17, rN19, kOetal19, sA24, sFetal25, sS25}.

\begin{figure}[b!]	
	\begin{center}
	\includegraphics[width=0.925\textwidth]{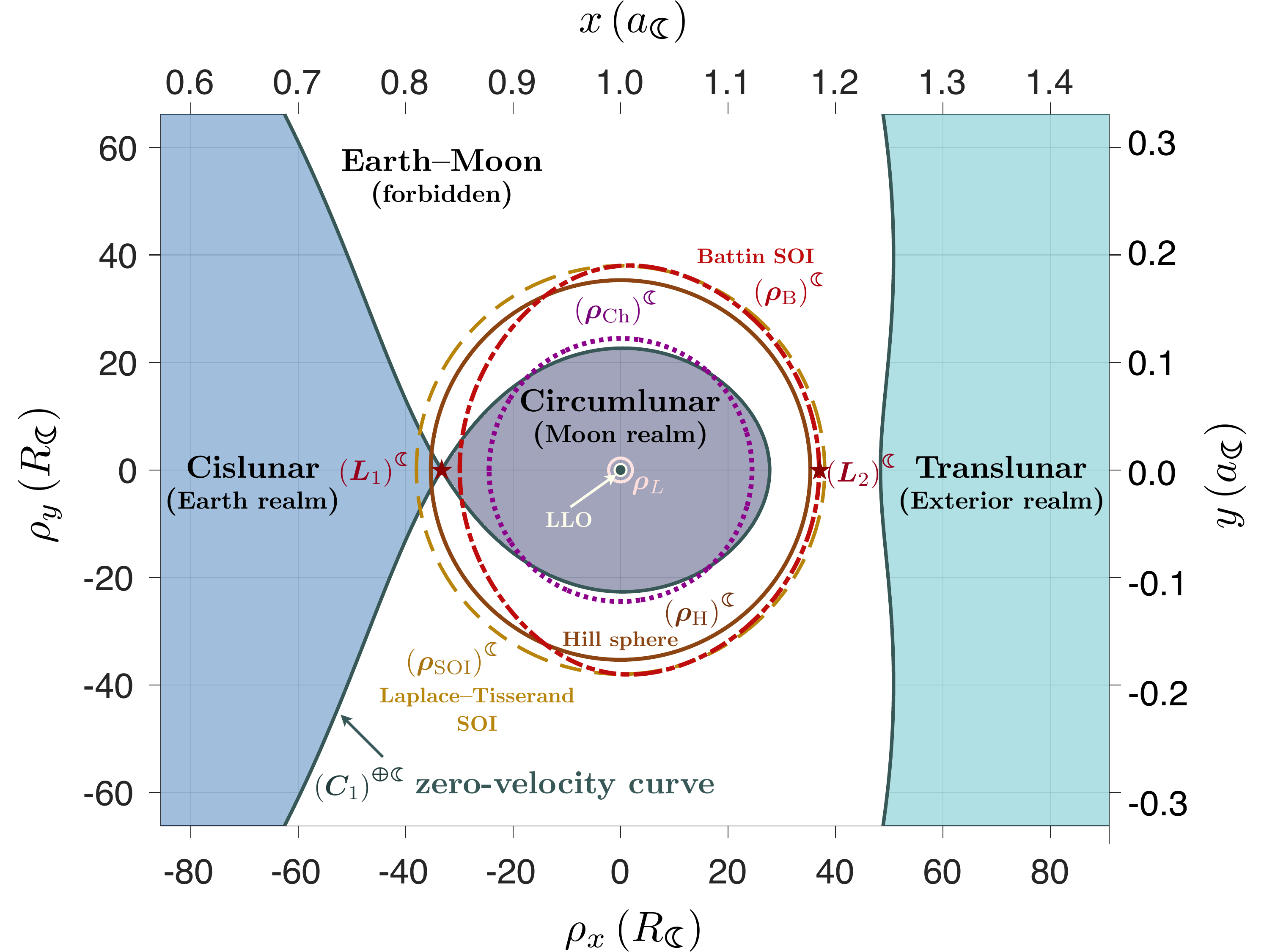}	
	\vskip -0.075in
	\caption{\small 
Local selenocentric spatiography of circumlunar space. The {\it shaded domains} identify the adjacent cislunar, circumlunar, and translunar provinces, with the circumlunar lobe corresponding to the Huang--Lanzano critical Jacobi region about the Moon \citep{sH61, pL61}. The {\it dark curves} show the local Earth--Moon $\left( C_1 \right)^{\E\M}$ zero-velocity boundary associated with the $\left( L_1 \right)^\M$ cislunar--circumlunar gateway; at this Jacobi level, the $\left( L_2 \right)^\M$ circumlunar--translunar gateway remains closed, and the intervening {\it white regions} denote forbidden motion. Superposed on this non-spherical CR3BP boundary are the principal radial proxy scales of the lunar neighborhood: low-lunar orbit, the selenoterrestrial Laplace radius $\rho_L$, the Earth--Moon collinear $\left( L_1 \right)^\M$ and $\left( L_2 \right)^\M$ Lagrange points, the spherical Hill proxy $\left( \rho_H \right)^\M$, and the classical patched-conic SOI radii of Chebotarev, Battin, and Laplace--Tisserand. The exterior terrestrial commensurabilities listed in Table~\ref{tab:circumlunar} are not drawn individually here, but belong to the same local selenocentric partition. The figure is drawn in lunar radii with auxiliary top/right axes in units of the Earth--Moon distance, so as to connect the local circumlunar partition to the broader patched spatiography of Fig.~\ref{fig:patched}.
    }
	\label{fig:circumlunar}
	\end{center}
	\vskip -0.075in
	\hspace{2cm}\rule{12.5cm}{0.5pt}
\end{figure}

The important point for the present spatiography is that circumlunar access is controlled primarily by the opening and closing of the Earth--Moon zero-velocity lobes, not by the crossing of a Euclidean sphere. The spherical Hill and SOI scales remain useful radial proxies, but they are not themselves the topological boundary of circumlunar motion. Figure~\ref{fig:circumlunar} therefore places these radial scales against the local $\left( C_1 \right)^{\E\M}$ zero-velocity boundary, while \textsection\ref{sec:circumlunar} interprets \emph{circumlunar space} as the Moon-centered province embedded between the cislunar--circumlunar and circumlunar--translunar gateway geometry (cf.~Fig.~\ref{fig:patched} and \textsection\ref{sec:spatio}).

For the larger-scale outer Earth-bound problem, the corresponding Sun--Earth Hill scale is obtained analogously by replacing the Moon with the combined Earth--Moon mass and the Earth--Moon distance with the mean Earth--Sun distance:
\begin{align}
    \label{eq:hill_earth}
    \left( r_H \right)^\E
    =
    a_\E
    \left(
        \frac{\mu_\E+\mu_\M}{3\mu_\S}
    \right)^{1/3}
    \simeq
    a_\E
    \left(
        \frac{\mu_\E}{3\mu_\S}
    \right)^{1/3}.
\end{align}
In the patched CR3BP of Ehricke and Huang, this larger lobe furnishes the outermost, stability-based demarcation within which translunar motion remains circumterrestrial before becoming predominantly heliocentric in organization \citep{kE62b, sH69}.

\subsubsection{SOI as Defined by Laplace, Tisserand, Chebotarev, and Battin}

The classical sphere of influence arises from a different construction. In \textsection\ref{sec:perturbed}, the third-body disturbing acceleration is represented at the potential level by the gradient of the disturbing function. Within the patched-conic approximation, one seeks the distance from the Moon at which the perturbation, measured relative to the central attraction, becomes equally important in the competing geocentric and selenocentric two-body descriptions. This defines the Laplace--Tisserand switching criterion, later clarified by Chebotarev and refined for the lunar case by Miller and Battin \citep{jMrB62, jC63, gC64, rB64, wHgM68, dSetal20}.

Let $a_\M$ be the Earth--Moon distance, let $\rho$ be the spacecraft's selenocentric distance from the Moon, and let $\psi_\E$ denote the angle between the Moon--satellite line and the Moon--Earth line. Near the Moon, the terrestrial perturbation on a selenocentric orbit is obtained by expanding the differential terrestrial acceleration about the Moon. To leading order one finds
\begin{align}
    \left\lvert
        \bm a_{\E,\mathrm{pert}}
    \right\rvert
    =
    \frac{\mu_\E}{a_\M^3}\,\rho\,\sqrt{1+3\cos^2\psi_\E}
    +
    O\!\left(\frac{\rho^2}{a_\M^4}\right),
\end{align}
whereas the lunar central attraction is simply $\mu_\M/\rho^2$. The corresponding selenocentric disturbing-to-central ratio is therefore
\begin{align}
    \Pi_{\mathrm{sel}}
    \sim
    \frac{\left\lvert \bm a_{\E,\mathrm{pert}} \right\rvert}{\mu_\M/\rho^2}
    =
    \frac{\mu_\E}{\mu_\M}
    \left(
        \frac{\rho}{a_\M}
    \right)^3
    \sqrt{1+3\cos^2\psi_\E}.
\end{align}

On the other hand, in the competing geocentric description the motion is still primarily Earth-centered, while the Moon acts as a perturbing body. In the transition region one has $r \sim a_\M$, so that the relevant geocentric disturbing-to-central ratio is, to leading order,
\begin{align}
    \Pi_{\mathrm{geo}}
    \sim
    \frac{\mu_\M/\rho^2}{\mu_\E/r^2}
    \simeq
    \frac{\mu_\M/\rho^2}{\mu_\E/a_\M^2}
    =
    \frac{\mu_\M}{\mu_\E}
    \left(
        \frac{a_\M}{\rho}
    \right)^2.
\end{align}
Equating $\Pi_{\mathrm{sel}}$ and $\Pi_{\mathrm{geo}}$ then gives the local Laplace--Tisserand surface of activity
\begin{align}
    \left( \rho_{\mathrm{act}} \right)^\M (\psi_\E)
    =
    a_\M
    \left(
        \frac{\mu_\M}{\mu_\E}
    \right)^{2/5}
    \left(
        1+3\cos^2\psi_\E
    \right)^{-1/10}.
\end{align}
This already shows that the patched-conic boundary is not truly spherical. Its maximum radius occurs at quadrature, $\psi_\E = \pi/2$, yielding the familiar Laplace sphere
\begin{align}
    \left( \rho_{\mathrm{SOI}} \right)^\M
    =
    a_\M
    \left(
        \frac{\mu_\M}{\mu_\E}
    \right)^{2/5},
\end{align}
which is best understood as the maximal spherical proxy of an angle-dependent surface \citep{gC64, dSetal20}.

Chebotarev's clarification is useful because it separates several notions that are often conflated. The \emph{sphere of activity} is the surface just obtained, based on equality of the two perturbation ratios. By contrast, the \emph{gravitational sphere} is defined by direct-force parity,
\begin{align}
    \frac{\mu_\M}{\rho^2}
    \sim
    \frac{\mu_\E}{a_\M^2}
    \qquad\Longrightarrow\qquad
    \left( \rho_{\mathrm{Ch}} \right)^\M
    =
    a_\M
    \left(
        \frac{\mu_\M}{\mu_\E}
    \right)^{1/2}.
\end{align}
The Hill sphere, by contrast, is the stability-based gateway scale already derived in Eq.~\eqref{eq:hill_lunar}, tied to the collinear Earth--Moon bottlenecks at $\left( L_1 \right)^\M$ and $\left( L_2 \right)^\M$. These radii are therefore not interchangeable: each arises from a different approximation and delimits a different aspect of the dynamics \citep{gC64, dSetal20, iC23}.

For the Earth--Moon problem, Battin's importance lies precisely in refusing to suppress the fore--aft asymmetry of the SOI. Retaining the first non-spherical correction in the Laplace criterion, and neglecting only higher powers of $\rho/a_\M$, yields the improved lunar boundary
\begin{align}
    \left( \rho_{\mathrm{B}} \right)^\M (\psi_\E)
    &=
    a_\M
    \left[
        \left(
            \frac{\mu_\M}{\mu_\E}
        \right)^{-2/5}
        \left(
            1+3\cos^2\psi_\E
        \right)^{1/10}
        -
        \frac{2}{5}
        \cos\psi_\E
        \left(
            \frac{1+6\cos^2\psi_\E}{1+3\cos^2\psi_\E}
        \right)
    \right]^{-1},
\end{align}
to first order in $\rho/a_\M$ \citep{rB64, wHgM68}. This form makes explicit the non-symmetric character of the lunar SOI: the boundary is compressed on the Earthward side and expanded in the anti-Earthward direction, rather than being a true sphere. In the present computation, the Battin boundary gives a radius of about $52009$ km ($29.93\,R_\M$) toward the Earth and about $64201$ km ($36.95\,R_\M$) in the anti-Earthward direction, while the familiar spherical Laplace proxy corresponds to about $66010$ km ($37.99\,R_\M$) \citep{rB64, wHgM68, jC63}.\footnote{Here and throughout this Primer, lunar distances expressed in $R_\M$ use the GRAIL lunar gravity-field reference radius ($R_\M=1738.0~\mathrm{km}$) \citep{mZ13}.} 

These scales are shown together with the local Earth--Moon zero-velocity boundary in Fig.~\ref{fig:circumlunar}. Since Battin's correction is directional, the local lunar hierarchy is not strictly a sequence of concentric spheres. At the level of radial proxy scales, however, it may be summarized as the directed ordering
\begin{align}
	\label{eq:em_proxy_hierarchy}
	\left( \rho_{\mathrm{Ch}} \right)^\M
	<
	\left( \rho_{\mathrm{B}} \right)^\M_{\E}
	\lesssim
	\left( \rho_H \right)^\M
	\lesssim
	\left( \rho_{\mathrm{B}} \right)^\M_{-\E}
	\lesssim
	\left( \rho_{\mathrm{SOI}} \right)^\M ,
\end{align}
where the subscripts $\E$ and $-\E$ denote the Earthward and anti-Earthward values of the Battin surface, respectively. This ordering emphasizes that the direct-force parity scale, Hill proxy, and SOI proxies occupy distinct dynamical roles even though several of them cluster within the outer circumlunar neighborhood.

This also explains why early lunar-passage studies adopted different patched-conic or asymptotic constructions according to purpose. Hiller's two-dimensional single-transit treatment used $65983$ km ($37.98\,R_\M$) as a convenient spherical switching boundary \citep{hH62}. Shute's patched-conic study of lunar ejecta, by contrast, employed a reduced asymptotic construction based on a fictitious sphere surrounding the Moon, taken sufficiently large that the motion was near the hyperbolic asymptote, rather than fixing a single universal switching radius \citep{bS66}. Such choices are useful engineering surrogates, but they are not unique dynamical boundaries.

A parallel line of work made clear that patched conics are, strictly speaking, only a heuristic shorthand for the lunar-passage problem. Using singular-perturbation and matched-asymptotic methods, \citet{pLjK63a} first showed in the two-fixed-force-center problem that a uniformly valid Earth-to-Moon trajectory cannot in general be obtained by directly patching an Earth-centered and a Moon-centered Keplerian conic. The corresponding restricted-three-body construction of \citet{pLjK63b} sharpened the same conclusion: the inner lunar hyperbola and the post-encounter Earth-centered branch depend on a first-order correction to the outer approach solution in the small CR3BP mass parameter $\bar\mu_{\E\M}$, defined in Eq.~\eqref{eq:mu_EM}. \citet{pLjK66} extended this matched-asymptotic construction to nonplanar Earth-to-Moon trajectories, and \citet{ySmE67} further generalized it to a more realistic four-body setting including solar forcing and lunar-eccentricity effects. These works do not replace the classical SOI as an engineering switching proxy, but they do show that the actual passage through the lunar neighborhood is more subtle than the direct patching of two conics.

The relation between the SOI and capture should therefore be stated carefully. In the local Earth--Moon problem, a spacecraft may cross a classical lunar SOI surface without ever becoming weakly captured, and conversely a low-energy three-body trajectory may exhibit extended lunar residence governed by the Earth--Moon zero-velocity topology rather than by any spherical switching surface. As shown in Fig.~\ref{fig:circumlunar}, the SOI is best understood here as a patched-conic proxy for passage among adjacent Earth--Moon provinces: from the cislunar Earth realm, through the circumlunar Moon realm, and into the exterior translunar realm of the Earth--Moon scaffold. Modern work has underscored the same point from a different direction: the most suitable patched-conic switching radius can depend on the encounter state and on the metric one seeks to optimize, while temporary-capture problems should be treated directly in the restricted three-body problem rather than forced into a fixed SOI shell \citep{rA08, rN19, iC23}. In that sense, the SOI belongs most naturally to the approximate patched-conic description of \emph{passage} across the local Earth--Moon partition, whereas actual \emph{circumlunar residence} belongs to the gateway language of $\left( L_1 \right)^\M$, $\left( L_2 \right)^\M$, and the critical Jacobi lobes \citep{sH61, pL61}.

\begin{figure}[t!]	
	\begin{center}
	\includegraphics[width=0.925\textwidth]{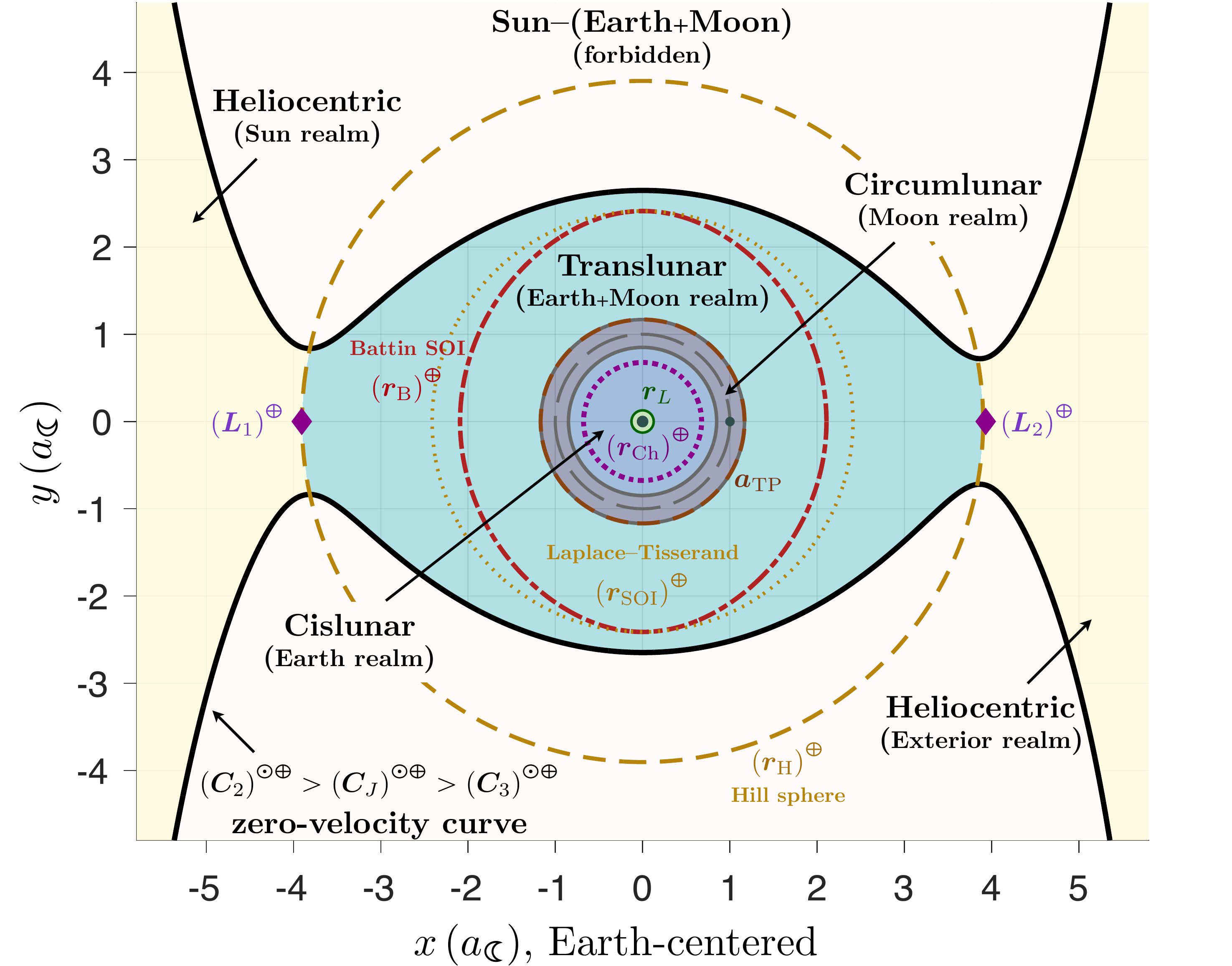}	
	\vskip -0.075in
	\caption{\small 
Earth-centered spatiography of translunar space in the outer Sun--(Earth\texttt{+}Moon) patched restricted problem. The {\it shaded domains} identify the adjacent cislunar, circumlunar, translunar, and heliocentric provinces, with the translunar lobe corresponding to the Earth-associated outer realm bounded by the Sun--(Earth\texttt{+}Moon) zero-velocity geometry and the Earth Hill scale. The {\it dark curves} show the Sun--(Earth\texttt{+}Moon) zero-velocity boundary at a Jacobi level between the second and third critical values, with both heliocentric bottlenecks open and the intervening {\it pale regions} denoting forbidden motion. Superposed on this outer CR3BP boundary are the principal radial proxy scales relevant to translunar passage: the Chebotarev direct-force parity scale $\left( r_{\mathrm{Ch}} \right)^\E$, the lunisolar tidal parity semi-major axis $a_{\mathrm{TP}}$, the first-order Battin SOI surface $\left( r_{\mathrm{B}} \right)^\E$, the spherical Laplace--Tisserand SOI $\left( r_{\mathrm{SOI}} \right)^\E$, and the Earth Hill scale $\left( r_H \right)^\E$. The terrestrial and cislunar subdomains of the Earth realm are shown schematically inside the translunar lobe, with $r_L$ marking the onset of cislunar space; the Moon's orbit and local circumlunar realm are likewise indicated to show where the Earth--Moon gateway structure is embedded within the outer Sun--(Earth\texttt{+}Moon) geometry. The figure is drawn in Earth-centered lunar-distance units, with the Sun in the $-x$ direction and the anti-sunward direction along $+x$, so as to connect the outer translunar partition to the local circumlunar spatiography of Fig.~\ref{fig:circumlunar} and the patched spatiography of Fig.~\ref{fig:patched}.
    }
	\label{fig:translunar}
	\end{center}
	\vskip -0.075in
	\hspace{2cm}\rule{12.5cm}{0.5pt}
\end{figure}

The same hierarchy of proxy scales reappears one level higher in the outer Sun--Earth problem, where it furnishes the corresponding patched-conic context for passage between the outer circumterrestrial realm and heliocentric space. In this description, which corresponds to the outer Sun--(Earth\texttt{+}Moon) geometry of Fig.~\ref{fig:translunar}, the Earth--Moon pair acts, to the accuracy relevant for the large-scale spatiographic partition, as the effective secondary of the Sun--(Earth\texttt{+}Moon) problem: the translunar lobe remains circumterrestrial, but its outer boundary is organized by the Sun--Earth zero-velocity geometry, the Earth SOI proxies, and the Earth Hill scale. Writing $a_\S \equiv a_\E$ for the apparent geocentric solar distance, the spherical Laplace--Tisserand SOI associated with the Earth, or equivalently the Earth--Moon pair to the accuracy relevant here, is
\begin{align}
    \label{eq:soi_earth}
    \left( r_{\mathrm{SOI}} \right)^\E
    &=
    a_\S
    \left(
        \frac{\mu_\E+\mu_\M}{\mu_\S}
    \right)^{2/5}
    \simeq
    a_\S
    \left(
        \frac{\mu_\E}{\mu_\S}
    \right)^{2/5}.
\end{align}

The corresponding Chebotarev direct-force parity scale is
\begin{align}
    \label{eq:chebotarev_earth}
    \left( r_{\mathrm{Ch}} \right)^\E
    &=
    a_\S
    \left(
        \frac{\mu_\E+\mu_\M}{\mu_\S}
    \right)^{1/2}
    \simeq
    a_\S
    \left(
        \frac{\mu_\E}{\mu_\S}
    \right)^{1/2},
\end{align}
In lunar-distance units, this gives $\left( r_{\mathrm{Ch}} \right)^\E \simeq 0.68\,a_\M$, so that the Moon's orbit lies well outside Earth's direct-force gravitational sphere. Chebotarev already emphasized this as an unusual property of the Earth--Moon system: among the satellites of the major planets, the Moon is exceptional in remaining outside the planet's gravitational sphere throughout its orbit, whereas the distant retrograde Jovian satellites he discussed cross the analogous limit only for particular heliocentric or satellite phases \citep{gC64}. The direct-force scale is therefore not the outer edge of Earth-bound motion, but it is a useful reminder that the translunar problem is already strongly conditioned by the solar field before one reaches the classical SOI or Hill scales.

The first-order Battin correction gives the mildly asymmetric Sun--Earth SOI surface
\begin{align}
    \label{eq:battin_earth}
    \left( r_{\mathrm{B}} \right)^\E ( \psi_\S )
    &=
    a_\S
    \left[
        \left(
            \frac{\mu_\E+\mu_\M}{\mu_\S}
        \right)^{-2/5}
        \left(
            1+3\cos^2\psi_\S
        \right)^{1/10}
        -
        \frac{2}{5}
        \cos\psi_\S
        \left(
            \frac{1+6\cos^2\psi_\S}{1+3\cos^2\psi_\S}
        \right)
    \right]^{-1}, 
\end{align}
with $\psi_\S$ measured from the anti-sunward $+x$ direction in the Sun--Earth Hill frame, so that the Sun lies at $\psi_\odot=\pi$. In contrast to the lunar case, the Battin deformation is small for the Sun--Earth mass ratio \citep{rB64, wHgM68}: the sunward and anti-sunward radii differ only weakly, while the quadrature value recovers the spherical Laplace--Tisserand proxy. Thus, at the level of radial proxy scales, the hierarchy is
\begin{align}
    \left( r_{\mathrm{Ch}} \right)^\E
    <
    \left( r_{\mathrm{B}} \right)^\E
    \sim
    \left( r_{\mathrm{SOI}} \right)^\E
    <
    \left( r_H \right)^\E .
\end{align}
This hierarchy is the outer analogue of the circumlunar proxy sequence shown in Fig.~\ref{fig:circumlunar}, but the two sequences differ substantially in scale. Near the Moon, the Chebotarev, Battin, Hill, and Laplace--Tisserand radii all cluster in the outer circumlunar zone, between roughly $0.11\,a_\M$ and $0.17\,a_\M$ from the Moon. In the Sun--Earth analogue, by contrast, the direct-force scale lies inside lunar distance, the secular tidal parity sits just exterior to the Earth--Moon $\left( L_2 \right)^\M$ gateway at $a_{\mathrm{TP}}\simeq1.17,a_\M$ (derived and interpreted in \textsection\ref{sec:tidalparity}), the spherical Laplace--Tisserand SOI lies much farther out at $\left( r_{\mathrm{SOI}} \right)^\E\simeq2.41,a_\M$, and the Earth Hill scale gives the outer stability proxy at $\left( r_H \right)^\E\simeq3.90\,a_\M$. Figure~\ref{fig:translunar} places this outer proxy sequence against the Sun--(Earth\texttt{+}Moon) zero-velocity geometry, thereby completing the large-scale radial and topological boundary picture used in the spatiographic partition of circumterrestrial space summarized in Table~\ref{tab:spatio}. 

These scales do not by themselves define the internal dynamics of the circumterrestrial provinces; rather, they provide the force-parity, patched-conic, and Hill-region context for the remaining stages of the spatiographic construction. The cislunar part of that construction has already been supplied by the preceding two subsections: \textsection\ref{sec:secular} identified the secularly dominated inner cislunar zone, organized by the vZLK effect and associated lunisolar secular resonances, while \textsection\ref{sec:cislunar} identified the outer cislunar resonant zone, structured by the ladder of interior lunar mean-motion commensurabilities. The next subsection turns to the Moon's immediate dynamical neighborhood, where the local Earth--Moon gateway geometry governs the transition between Earth-centered passage and lunar-vicinity residence; \textsection\ref{sec:tidalparity} then identifies the secular lunisolar parity scale just outside the Earth--Moon gateway; and \textsection\ref{sec:translunar} finally resolves the outer province into its exterior lunar and interior solar mean-motion resonances and mixed lunisolar secular architecture.

\subsection{Circumlunar Space: A Local Enclave of the Earth--Moon System}
\label{sec:circumlunar}

The lunar neighborhood admits its own local spatiography when recast in selenocentric distance. In this local view, the Earth enters simply as an external orbit at $\rho = a_\M \simeq 383398$ km $\simeq 220.67\,R_\M$, rather than as the central scale of the partition. Figure~\ref{fig:circumlunar} shows the principal radial boundary scales of the local circumlunar problem: low-lunar orbit, the selenoterrestrial Laplace radius $\rho_L$, the Earth--Moon gateway distances $\left( L_1 \right)^\M$ and $\left( L_2 \right)^\M$, the spherical Hill proxy $\left( \rho_H \right)^\M$, and the patched-conic SOI scales discussed in \textsection\ref{sec:soi}. Table~\ref{tab:circumlunar} complements this radial picture by collecting the principal selenocentric boundaries, resonances, and gateway scales that stratify the same circumlunar domain, in the spirit of the older distinction---already clear in Ehricke's classification of cislunar, translunar, and lunar-satellite motion---between the broader Earth-bound problem and the Moon's immediate dynamical neighborhood \citep{kE62a, kE62b}.

Historically, the circumlunar problem was already recognized as dynamically distinct from both the near-Earth and the merely patched-conic lunar-passage problems \citep{rB56b}. Ehricke's two-volume \emph{Space Flight} already makes that separation explicit at the level of both taxonomy and dynamics: cislunar and translunar satellites remain subclasses of the broader circumterrestrial problem, whereas the lunar satellite is treated as a distinct local class associated with motion in the Moon's immediate vicinity under strong terrestrial perturbation \citep{kE62a, kE62b}. In a perturbed-Hamiltonian treatment of stable and unstable lunar orbits, \citet{yK63} found that the Earth's perturbation becomes comparable to the Moon's departure from sphericity already near $\rho \sim 2\,R_\M$. \citet{aR68a, aR68b} then developed a full disturbing-function theory for a lunar orbiter under the combined action of the Moon's figure, the Earth, and the Sun, explicitly retaining the indirect solar contribution transmitted through the Earth--Moon motion. In that formulation, the Earth's perturbation was shown to overtake the lunar figure already at heights of order $1700$ km, while the Sun remained much smaller over the same range. \citet{gG70} continued this line with a semi-analytic lunar-satellite theory in selenocentric canonical variables, incorporating terrestrial, solar, and nonspherical lunar effects. From a different direction, \citet{wB71} treated near-parabolic trajectories between the Moon and the cislunar libration-point neighborhood, emphasizing the transition between local circumlunar motion and the gateway geometry of the Earth--Moon problem. Taken together, these works show that the lunar neighborhood had already acquired its own local dynamical identity well before the modern language of weak stability boundaries, temporary capture, and invariant manifolds.

The innermost transition is the selenoterrestrial Laplace radius, where the secular torque due to the Moon's oblateness becomes comparable to the combined quadrupolar tides of the Earth and Sun. To estimate this dominant local crossover scale, we adopt an axisymmetric small-obliquity approximation in which the external terrestrial and solar torques are averaged about the lunar equator while the lunar and terrestrial orbital eccentricities are retained. This gives
\begin{align}
    \rho_L^5
    \approx
    \frac{
        2 \mu_\M J_{2,\M} R_\M^2
    }{
        \dfrac{\mu_\E}{a_\M^3 ( 1 - e_\M^2 )^{3/2}}
        +
        \dfrac{\mu_\S}{a_\S^3 ( 1 - e_\S^2 )^{3/2}}
    }.
\end{align}
Using the lunar value of $J_{2,\M} \simeq 2.0322 \times 10^{-4}$ adopted by \citet{eLcE24}, one finds
\begin{align}
    \rho_L \simeq 3846~\mathrm{km} \simeq 2.21\,R_\M,
\end{align}
corresponding to an altitude of about $2109$ km above the lunar surface and a local circular Keplerian period of about $0.25$ days, consistent with the earlier estimates of \citet{yK63} and \citet{aR68a}.

Inside this boundary the lunar figure still provides the dominant non-central structure, and the traditional low-orbit questions of critical inclination, frozen behavior, and impact avoidance are naturally posed selenocentrically \citep{yK63, zK98, tE05, tNpG18}. That dominance, however, weakens rapidly with altitude. Roy showed that the Moon's bulge remains the principal perturbation only out to about $600$ km above the surface, and exceeds the Earth's perturbation only to roughly $1600$ km \citep{aR68a}. Modern comparisons using contemporary gravity models \citep{mZ13} place the same transition in a more systematic hierarchy: at very low altitude many harmonics compete, above about $100$ km the problem becomes usefully secular, above about $500$ km the Earth begins to compete seriously with the dominant low-degree lunar harmonics, and by about $2065$ km the Earth's tidal acceleration overtakes the second-degree lunar potential terms, very close to the present $\rho_L$ estimate \citep{bDS06, tNpG18, eLcE24}. That local dynamical replacement is reflected in the geometry of the orbit families themselves. In axisymmetric models one may still speak of frozen-orbit structures in the usual zonal sense \citep{zK98}. But once the non-axisymmetric harmonics are retained, the longitude of node enters essentially and genuine frozen orbits no longer exist in general; the natural organizing objects are instead periodic orbits in the full two-degree-of-freedom secular problem \citep{jV70, dSt98, sTetal14, aNetal15}. This is one reason the circumlunar region cannot be described adequately by a single spherical boundary or a single asymptotic switching surface: already in Ehricke's treatment, lunar-satellite motion appears as a local problem of motion near the smaller primary, while Kozai, Roy, and later secular theories show how rapidly the internal hierarchy of perturbations changes with selenocentric distance \citep{kE62b, yK63, aR68a, gG70, dSt98, sTetal14, aNetal15}. The gateway viewpoint of \textsection\ref{sec:soi} remains indispensable at the outer edge, but the interior of the region is itself resonantly and secularly structured.

\begin{table}[t!]
    \caption{
        Local spatiographic partition of the circumlunar neighborhood in selenocentric distance. The quoted periods are the circular Keplerian periods associated with the listed radii.
        \label{tab:circumlunar}
    }
    \begin{tabular*}{0.965\textwidth}{
        p{5.2cm}
        S[table-format=2.2]
        S[table-format=2.2]
        p{6.4cm}
        }

        \toprule
        \addlinespace[1ex]

        Local Dynamical Zone
        & \multicolumn{1}{c}{$\rho / R_\M$}
        & \multicolumn{1}{c}{$T$ [days]}
        & Description \\

        \addlinespace[1.25ex]
        \toprule
        \addlinespace[1ex]

        \textbf{Inner Circumlunar Zone} & & & \\
            \quad low-lunar orbit ($h = 100$ km)
                & 1.06 & 0.08
                & representative close lunar orbiter; lunar figure dominates strongly \\
            \quad selenoterrestrial Laplace radius $\rho_L$
                & 2.21 & 0.25
                & where the combined Earth--Sun quadrupolar tide $\approx$ the lunar $J_2$ torque \\
                & & &
                $\rho_L^5
                =
                2 \mu_\M J_{2,\M} R_\M^2
                \Big/
                \left(
                    \dfrac{\mu_\E}{a_\M^3 (1-e_\M^2)^{3/2}}
                    +
                    \dfrac{\mu_\S}{a_\S^3 (1-e_\S^2)^{3/2}}
                \right)$ \\

        \zonebreak

        \textbf{Circumlunar Resonant Zone} & & & \\
            \quad exterior terrestrial resonances & & & \\[0.5em]
            \Res{8}{1}{\E}  & 12.73 & 3.42  & innermost 7th-order terrestrial commensurability \\
            \Res{7}{1}{\E}  & 13.92 & 3.91  & \\
            \Res{6}{1}{\E}  & 15.43 & 4.56  & \\
            \Res{5}{1}{\E}  & 17.42 & 5.47  & \\
            \Res{9}{2}{\E}  & 18.69 & 6.08  & \\
            \Res{4}{1}{\E}  & 20.22 & 6.84  & \\
            \Res{7}{2}{\E}  & 22.10 & 7.81  & \\
            \Res{10}{3}{\E} & 22.83 & 8.20  & \\
            \Res{3}{1}{\E}  & 24.49 & 9.11  & \\
            \Res{8}{3}{\E}  & 26.49 & 10.25 & \\
            \Res{5}{2}{\E}  & 27.65 & 10.94 & \\
            \Res{7}{3}{\E}  & 28.96 & 11.72 & \\
            \Res{9}{4}{\E}  & 29.67 & 12.15 & \\
            \Res{2}{1}{\E}  & 32.09 & 13.67 & \\
            \Res{9}{5}{\E}  & 34.42 & 15.19 & \\
	    \Res{7}{4}{\E}  & 35.08 & 15.63 & outermost terrestrial resonance interior to $\left( \rho_H \right)^\M$ \\

        \zonebreak

        \textbf{Gateway \& SOI Boundaries} & & & \\
            \quad Chebotarev grav. sphere $\left( \rho_{\mathrm{Ch}} \right)^\M$
                & 24.47 & 9.11
                & direct-force parity with the Earth \\
            \quad Battin SOI $\left( \rho_{\mathrm{B}} \right)^\M$ (Earthward)
                & 29.93 & 12.32
                & first-order asymmetric patched-conic boundary toward the Earth \\
            \quad Earth--Moon Lagrange $\left( L_1 \right)^\M$
                & 33.31 & 14.46
                & inner zero-velocity bottleneck \\
            \quad Moon's Hill sphere $\left( \rho_H \right)^\M$
                & 35.32 & 15.79
                & spherical proxy for the circumlunar gateway scale \\
            \quad Battin SOI $\left( \rho_{\mathrm{B}} \right)^\M$ (anti-Earthward)
                & 36.95 & 16.90
                & first-order asymmetric patched-conic boundary away from the Earth \\
            \quad Earth--Moon Lagrange $\left( L_2 \right)^\M$
                & 37.04 & 16.95
                & outer zero-velocity bottleneck \\
            \quad Laplace patched-conic SOI $\left( \rho_{\mathrm{SOI}} \right)^\M$
                & 37.99 & 17.61
                & spherical Laplace--Tisserand proxy \\

        \bottomrule

    \end{tabular*}
\end{table}

Table~\ref{tab:circumlunar} summarizes this local selenocentric partition. For a Moon-centered satellite with selenocentric semi-major axis $\left( a \right)^\M$ in exterior commensurability with the apparent terrestrial motion, the nominal centers follow from
\begin{align}
    \label{eq:terrestrial_circumlunar_mmr}
    \left( a_{k \, : \, k_\E} \right)^\M
    &\simeq 
    \bar\mu_{\E\M}^{1/3}
    \left(
        \frac{k_\E}{k}
    \right)^{2/3}
    a_\M,
\end{align}
with $\bar\mu_{\E\M}$ defined in Eq.~\eqref{eq:mu_EM} and $a_\M$ denoting the Earth--Moon separation, here viewed as the radius of the Earth's apparent selenocentric orbit. Here the resonance label $k \! : \! k_\E$ follows the convention of \textsection\ref{sec:resonances}: the satellite completes $k$ selenocentric revolutions in approximately the same inertial time that the Earth completes $k_\E$ apparent revolutions about the Moon. Beyond the selenoterrestrial Laplace radius there appears a dense sequence of exterior terrestrial commensurabilities, beginning near $\left( a \right)^\M \simeq 12.73\,R_\M$ for the \res{8}{1}{\E} resonance and extending outward to $\left( a \right)^\M \simeq 35.08\,R_\M$ for the \res{7}{4}{\E} resonance, just inside the Hill proxy. Although Earth perturbations are of course a classical ingredient of circumlunar theory, the literature surveyed here appears to treat them chiefly through secular effects, frozen-orbit structure, critical-inclination theory, periodic-orbit families, or gateway/capture geometry. To our knowledge, these terrestrial resonances have not previously been tabulated explicitly as a local spatiographic partition of circumlunar space. They do not replace the gateway interpretation of circumlunar motion, but they do refine it by showing that the lunar neighborhood contains its own internal resonant stratification before the Earth--Moon necks are reached.

At the outer edge, the local selenocentric and patched-conic boundaries sit in the same broad zone but answer different questions. In the present computation, $\left( L_1 \right)^\M \simeq 57868$ km $\simeq 33.31\,R_\M$ and $\left( L_2 \right)^\M \simeq 64347$ km $\simeq 37.04\,R_\M$, while the spherical Hill proxy is $\left( \rho_H \right)^\M \simeq 61364$ km $\simeq 35.32\,R_\M$. The corresponding SOI radii lie nearby but are not identical: $\left( \rho_{\mathrm{Ch}} \right)^\M \simeq 42499$ km $\simeq 24.47\,R_\M$, the first-order Battin boundary $\left( \rho_{\mathrm{B}} \right)^\M$ spans roughly $52009$ km ($29.93\,R_\M$) Earthward to $64201$ km ($36.95\,R_\M$) anti-Earthward, and the spherical Laplace proxy gives about $66010$ km ($37.99\,R_\M$). Thus the circumlunar gateway region is not only non-spherical in the CR3BP sense, as emphasized in \textsection\ref{sec:soi}, but also non-spherical even at the patched-conic level.

This distinction matters physically. The low-lunar-orbit problem remains dominated by the Moon's gravity field and its strong tesseral structure, so that impact, frozen behavior, and long-term eccentricity growth are governed primarily by local lunar harmonics rather than by any Euclidean switching shell \citep{zK98, bDS06, aNetal15}. By contrast, once one approaches the outer circumlunar zone, the Earth becomes the dominant non-Keplerian perturber, and the local picture must be completed by the gateway geometry at $\left( L_1 \right)^\M$ and $\left( L_2 \right)^\M$, together with the near-parabolic and weak-capture passages already visible in the early literature \citep{wB71}. Modern work has only sharpened this picture: high-order gravity models replace exact frozen orbits by small-amplitude periodic families \citep{aNetal15}, while secular lifetime studies show that re-entry and long residence in the middle-altitude circumlunar zone are governed chiefly by the interplay between mascon-driven structure and a small number of dominant secular resonances, most notably the $2g$ resonance \citep{eLcE24}. In that sense, circumlunar space is best understood not as a mere local annex of circumterrestrial space, but as a dynamically self-organized lunar province embedded within the wider Earth--Moon system.

\subsection{Lunisolar Tidal Parity as a Secular Translunar Marker}
\label{sec:tidalparity}

In the patched spatiography of Figs.~\ref{fig:patched} and~\ref{fig:circumlunar}, the topological entry into the translunar province is the circumlunar--translunar gateway associated with the exterior Earth--Moon neck near $\left(L_2\right)^\M$. The lunisolar tidal parity considered here is a different kind of demarcation. It is not a gateway, a zero-velocity boundary, or a patched-conic switching surface, but a secular tidal crossover lying just exterior to the Earth--Moon gateway region. In the Earth-centered, orbit-averaged description, it marks the distance at which the leading quadrupolar secular influence on circumterrestrial motion shifts from lunar-internal to solar-external dominance (qv.~Fig.~\ref{fig:laplace} and Table~\ref{tab:spatio}).

The relevant secular framework has already been developed in \textsection\ref{sec:perturbed}. Beyond lunar distance, the Moon enters the disturbing function as an \emph{interior} perturber, while the Sun remains an \emph{exterior} perturber. At quadrupole order, the lunar and solar secular contributions therefore scale differently with semi-major axis. In consequence, the lunar interior quadrupolar contribution weakens outward beyond lunar distance, while the solar exterior tide becomes progressively more important and eventually dominates. Using the leading-order coefficients given in \textsection\ref{sec:secular}, the parity condition is obtained by equating the amplitudes of these two quadrupolar contributions. This gives
\begin{align}
    \left(
        \frac{a_{\mathrm{TP}}}{a_\M}
    \right)^5
    =
    \frac{\mu_\M}{\mu_\S}
    \left(
        \frac{a_\S}{a_\M}
    \right)^3
    \left(
        1
        -
        \frac{3}{2}\sin^2 i_\M
    \right)
    \left(
        1
        +
        \frac{3}{2}e_\M^2
    \right)
    \left(
        1
        -
        e_\S^2
    \right)^{3/2}.
\end{align}
With the constants adopted here, this yields
\begin{align}
    a_{\mathrm{TP}} \simeq 1.17\,a_\M,
\end{align}
corresponding to a Keplerian period of about $34.6$ days.

The construction is analogous in spirit to the geolunar Laplace radius of Eq.~\eqref{eq:laplace}: both identify a secular crossover by equating the leading quadrupolar contributions that organize long-term motion. The analogy, however, is only formal. The Laplace radius marks a comparatively clean transition between terrestrial oblateness and lunisolar torques, whereas the translunar parity considered here compares two third-body secular terms in a region already complicated by the Earth--Moon gateway geometry and by the slow convergence of the lunar interior expansion. 

The meaning of this scale is therefor limited but useful. It does \emph{not} imply that the motion beyond $a_{\mathrm{TP}}$ suddenly ceases to be organized by the Earth--Moon problem, nor that the Moon becomes dynamically irrelevant immediately outside parity. Nor should its numerical proximity to the Earth--Moon $\left( L_2 \right)^\M$ distance be assigned topological significance. The two constructions answer different questions: the $\left( L_2 \right)^\M$ neck is a zero-velocity boundary of the Earth--Moon CR3BP, whereas $a_{\mathrm{TP}}$ is an orbit-averaged quadrupolar balance between lunar-internal and solar-external secular terms in an Earth-centered disturbing function. In that restricted sense, it identifies an early outer circumterrestrial scale at which the secular organization of translunar motion begins to shift from lunar-dominated to solar-dominated. It is also not the radius at which the instantaneous solar third-body acceleration first exceeds the lunar third-body acceleration. At the acceleration level of Fig.~\ref{fig:accels}, the lunar branches remain comparable to, and in some geometries larger than, the solar branches over a substantially broader interval beyond lunar distance; the solar perturbation overtakes the lunar contribution only farther out, with the crossover depending on syzygy/quadrature geometry and on whether the Moon-facing or anti-Moon branch is being compared.

Several caveats are therefore important. First, the parity condition is only a lowest-order secular quadrupolar balance. The translunar secular problem is intrinsically hybrid (qv.~\textsection\ref{sec:secular}): the Moon contributes through the \emph{interior} branch of the disturbing function, while the Sun contributes through the \emph{exterior} branch, so the angular structures of the two averaged terms are not identical. Second, near lunar distance the Moon's interior expansion is not in the deep hierarchical limit; higher-order Legendre terms, long-period terms, and the nearby co-orbital gateway geometry can all matter. Thus the parity condition should be interpreted as a schematic secular crossover scale, not as a precise local resonance surface or dynamical boundary. Third, the result is orbit-averaged and says nothing directly about short-period encounters, lunar flybys, zero-velocity topology, or the transport geometry associated with the Earth--Moon necks.

Accordingly, the lunisolar tidal parity should be understood not as the first boundary of translunar space, but as an early \emph{secular marker} within a broader sequence of translunar demarcations. It signals the point at which the outer circumterrestrial problem begins, in the orbit-averaged quadrupolar sense, to be governed more by the Sun--Earth hierarchy than by the Earth--Moon hierarchy, with the Moon acting increasingly as an interior perturbation. In that restricted sense, it usefully complements the gateway definition of translunar access supplied by the patched CR3BP, while remaining distinct from the larger sphere-of-influence and Hill-type outer scales considered below.

\subsection{The Translunar Realm and Exterior Resonances}
\label{sec:translunar}

Beyond the Moon lies a distinct outer circumterrestrial province in which motion remains Earth-bound, but is no longer organized principally by the Earth--Moon problem alone. In the spatiographic framework adopted here, this is the \emph{translunar} realm: the domain exterior to the lunar orbit in which the Moon acts as an \emph{interior} perturber while the Sun remains an \emph{exterior} perturber. The boundary scales developed in \textsection\ref{sec:soi}---the circumlunar--translunar gateway, the lunisolar tidal parity, the sphere-of-influence proxies, and the Earth Hill scale---provide the radial and topological setting for this province (Table~\ref{tab:spatio}; Fig.~\ref{fig:translunar}). The task now is to resolve its internal dynamical architecture. The result is not merely a geometric extension of cislunar space, but a mixed secular--resonant region whose organizing structure differs qualitatively from that of the inner Earth--Moon environment \citep{kE62a, kE62b, mA76}.

Historically, this outer regime was already implicit in Ehricke's mature cislunar--translunar picture of circumterrestrial flight \citep{kE62a, kE62b}, which distinguished the Earth--Moon-controlled region from the more remote outer Earth-bound domain in which solar control grows progressively more important. In modern terms, translunar space begins not simply ``beyond the Moon'' in a kinematic sense, but where the outer circumterrestrial problem must be described by the combined action of an interior lunar perturber and an exterior solar perturber. Its inner access is tied first to the lunar gateway: as discussed in \textsection\ref{sec:soi}, the opening of the Earth--Moon neck at $\left( L_2 \right)^\M$ provides the natural dynamical exit from circumlunar residence. But the translunar region is not thereby immediately solar-dominated. Rather, there follows an overlapping outer circumterrestrial zone in which lunar encounters, exterior lunar resonances, solar secular forcing, and eventually solar commensurabilities all coexist.

The first secular handoff within this outer province is the \emph{lunisolar tidal parity} presented in \textsection\ref{sec:tidalparity}, where the lowest-order lunar \emph{internal} quadrupolar contribution and the solar \emph{external} quadrupolar contribution become comparable. As already emphasized there, this parity is only a schematic secular marker: near lunar distance, both the exterior and interior Legendre expansions converge slowly, long-period terms omitted from the purely quadrupolar truncation may become important, and the nearby \res{1}{1}{\M} gateway weakens any clean separation between secular and resonant organization \citep{mA76}. It nevertheless identifies the scale at which the orbit-averaged translunar problem begins to look less like an Earth--Moon hierarchy and more like a Sun--Earth hierarchy with the Moon retained as an interior perturbation.


The resonant structure beyond the Moon is equally distinctive. Using the nominal mean-motion commensurability condition of \textsection\ref{sec:resonances}, the first low-order \emph{exterior} lunar resonances appear immediately outside the lunar gateway, beginning with \res{4}{5}{\M}, \res{3}{4}{\M}, \res{2}{3}{\M}, \res{3}{5}{\M}, and \res{1}{2}{\M}, and continuing outward through \res{1}{5}{\M}. Farther out, these are joined by the first low-order \emph{interior} solar commensurabilities. Here, as in the cislunar resonant zone, ``low-order'' is used in the spatiographic sense of retaining a compact set of principal commensurabilities, rather than attempting an exhaustive resonance census: Table~\ref{tab:spatio} therefore lists the solar sequence \res{5}{1}{\S}, \res{4}{1}{\S}, \res{3}{1}{\S}, \res{5}{2}{\S}, and \res{2}{1}{\S}, together with their Keplerian centers. This truncation is deliberately conservative. The next solar tier begins already with the fifth-order \res{6}{1}{\S} commensurability near $a/a_\M\simeq1.70$, followed by a denser sequence that interleaves with the exterior lunar resonances throughout the translunar domain. Such higher-order solar commensurabilities may be dynamically important---for example, in producing small stable pockets or fine resonant texture---but they are not used here to define the primary spatiographic partition. Their early appearance nevertheless underscores the central distinction: unlike the cislunar resonant zone, where the principal architecture is supplied by the Moon alone, translunar space is organized by a genuinely coupled lunisolar resonant web.

This is one reason the translunar realm is better compared, in broad dynamical spirit, to the trans-Neptunian region than to the asteroid belt. In the trans-Neptunian problem, an exterior population is structured by mean-motion commensurabilities with an interior giant planet; here, the outer circumterrestrial domain is likewise partitioned first by exterior lunar resonances with the interior Moon, before being further threaded by solar commensurabilities as the Sun--Earth hierarchy intrudes. The analogy is not exact, but it is suggestive: the translunar realm is the outer resonant fringe of circumterrestrial motion, where an interior secondary reshapes the architecture of weakly bound bodies far beyond its own orbit \citep{cB94, oWcM97, mB00, dNrR00, dNrR01, gV05, rMzC23}.

\begin{figure}[t!]	
	\begin{center}
	\includegraphics[width=0.975\textwidth]{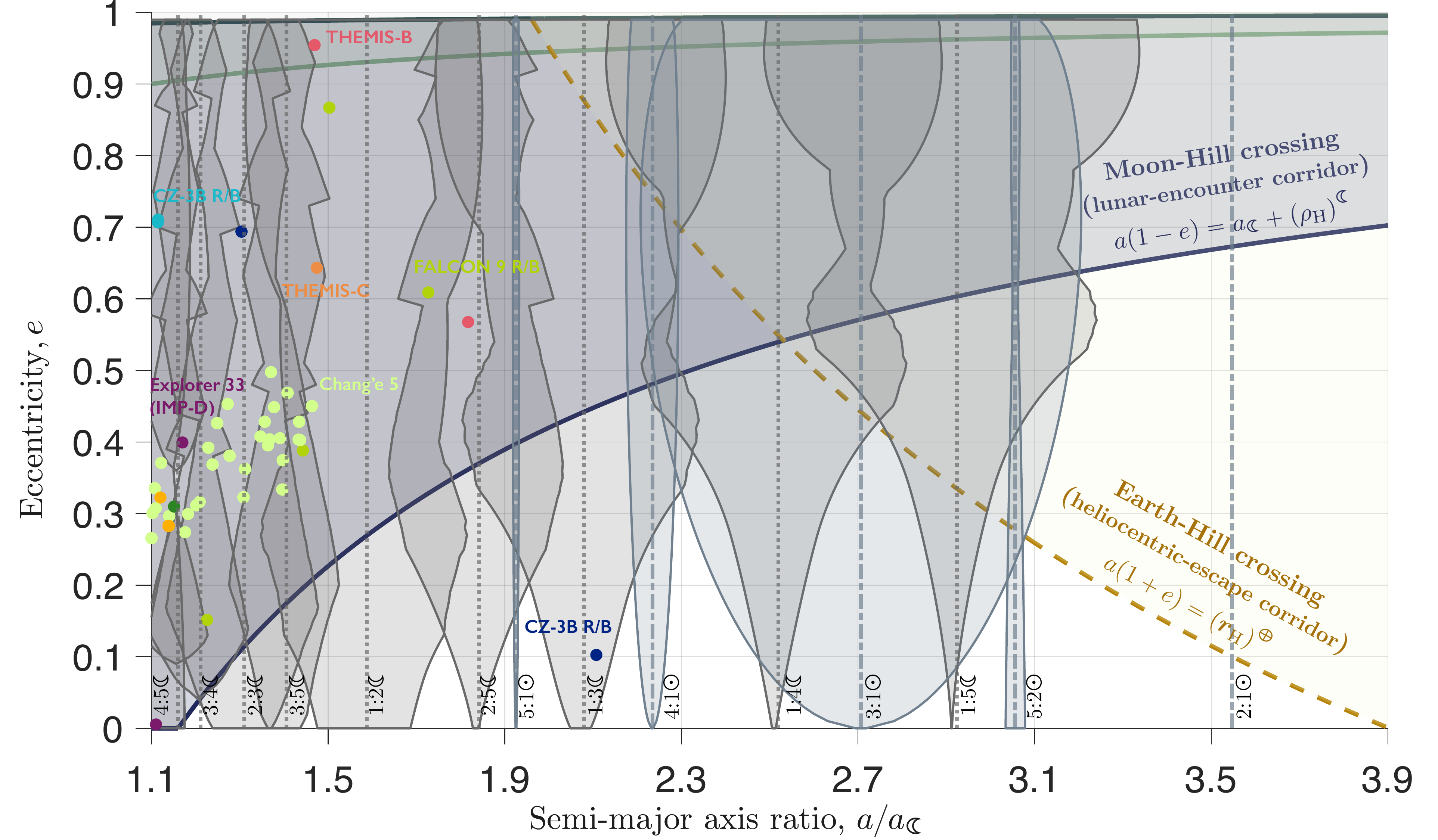}	
	\vskip -0.075in
	\caption{\small 
Resonant atlas of the translunar realm in the $( a , e )$ plane. {\it Grey contours} show the semi-analytical widths predicted by Gallardo's algorithm for the exterior lunar and interior solar mean-motion resonances. {\it Colored points} show the six curated catalog and transient objects occupying or crossing this domain. {\it Vertical} guides mark the nominal Keplerian centers of the low-order commensurabilities listed in Table~\ref{tab:spatio}. The {\it shaded} crossing corridors mark osculating geocentric ellipses whose periapsis reaches the Moon-Hill encounter region, $a ( 1 - e ) = a_\M + \left( \rho_H \right )^\M$, or whose apoapsis reaches the Earth-Hill boundary, $a ( 1 + e ) = \left( r_H \right )^\E$; these curves are element-space crossing diagnostics rather than physical surfaces. The additional GEO-crossing curve, $a ( 1-e ) = r_{\mathrm{GEO}}$, marks trajectories whose periapsis returns to the inner terrestrial operational domain. The formal semi-analytical width of the \res{2}{1}{\S} resonance is omitted because its nominal center lies close to the Earth Hill boundary, where the averaged geocentric width estimate is no longer a reliable indicator of a stable libration domain. 
	}
	\label{fig:translunar_atlas}
	\end{center}
	\vskip -0.075in
	\hspace{2cm}\rule{12.5cm}{0.5pt}
\end{figure}



Figure~\ref{fig:translunar_atlas} shows a representative coplanar slice of the corresponding atlas for the translunar realm. The contours show the local semi-analytical libration widths obtained from Gallardo's averaged-Hamiltonian construction, while the vertical guides mark the nominal Keplerian centers of the exterior lunar and interior solar commensurabilities listed in Table~\ref{tab:spatio}. Read in the language of that table, the figure gives a first cartography of the \emph{translunar resonant zone}: the circumlunar gateway gives way immediately to exterior lunar resonances beyond the Moon's orbit, while progressively farther out these are interleaved with the first interior solar commensurabilities. The result is not a single ladder of lunar resonances, as in the outer cislunar zone, but a mixed lunisolar resonant web superposed on the secular handoff and outer SOI/Hill-scale demarcations discussed in \textsection\ref{sec:soi}.

The translunar atlas should, however, be read with greater caution than its cislunar counterpart. Immediately outside the lunar gateway, exterior lunar resonances coexist with possible close approaches to the Moon, so the osculating geocentric elements can vary rapidly over the same timescale on which the averaged resonant Hamiltonian assumes them to be fixed. The close-approach filtering used in the width calculation should therefore be understood simply as a device for suppressing spurious broadening where the averaged disturbing-function description is no longer locally reliable. Farther outward, the same fixed-element assumption becomes delicate because the geocentric description approaches the outer circumterrestrial fringe, where the Earth SOI and Hill scales mark the gradual loss of a clean Earth-bound hierarchy. For this reason, Fig.~\ref{fig:translunar_atlas} does not show the formal semi-analytical width of the \res{2}{1}{\S} solar commensurability. Its nominal center lies close enough to the Earth Hill boundary that the averaged geocentric width returned by the atlas is not physically meaningful as a stable resonant domain. More generally, the plotted solar structures are best interpreted as formal averaged-Hamiltonian diagnostics of where solar resonant organization enters the outer circumterrestrial problem, rather than as guaranteed stable libration islands all the way to the heliocentric transition.

Even with these caveats, Fig.~\ref{fig:translunar_atlas} makes clear why translunar space cannot be treated as a simple geometric extension of cislunar space. The atlas strongly suggests that much of the translunar domain is pervaded by broad resonance overlap. The exterior lunar resonances broaden outward, the interior solar resonances intrude before the Earth Hill scale is reached, and the intervening region is crossed by the secular tidal parity, the Earth SOI, and the outer Hill boundary. Thus the region beyond the Moon appears less like a set of isolated islands than like a mixed and, in places, globally chaotic outer circumterrestrial sea: a sparsely populated, weakly bound province in which lunar and solar commensurabilities, secular forcing, close lunar encounters, and eventual heliocentric escape all compete. At present, the curated catalog contains only a small resident population in this domain, together with a larger transient set that crosses in and out of it; the sparseness of known occupants underscores how little explored the region remains dynamically. This reading is qualitatively consistent with the broader exterior-resonance literature, where low-order exterior commensurabilities are known to admit asymmetric libration centers, bifurcated resonant structure, and extended chaotic layers \citep{cB94, oWcM97, gV05, rMzC23, aRetal25}.

The translunar architecture also interfaces naturally with the weak-capture and transport literature. Resonance transitions in the Earth--Moon restricted problem, together with weak capture near the Moon and ballistic escape from the Earth--Moon system, show that the translunar region is not merely a passive outer shell but an active transport corridor linking cislunar, circumlunar, and heliocentric behaviors \citep{jKjL68, rFaK73, fT08, eB08}. Sun-perturbed extensions of the three-body problem sharpen the same point: once solar forcing is admitted, the outer circumterrestrial region is threaded by families of transfers, escapes, and temporary captures that are best understood in the language of bicircular or more general restricted four-body dynamics rather than in that of purely patched conics \citep{hY92, eB93, kY04, rC12, pSS12, fT13, yQ14, pdSS16, kO17, sA24, sFetal25, sS25}. In that sense, translunar space is the dynamical borderland of the Earth--Moon system \citep{kTtK25}: still circumterrestrial in the broad sense, yet already structured by the same exterior-resonance and chaotic-transport phenomena that govern weakly bound small bodies elsewhere in celestial mechanics.
    
\section{Restricted Multi-Body Dynamical Systems Theory}
\label{sec:restricted}

\subsection{Phenomenological Models for the Restricted Problem}
\label{sec:phenomenology}

While the perturbed Hamiltonian formulation of \textsection\ref{sec:perturbed} permits the calculation of resonant half-width excursions in $(a, e, i)$ and furnishes a local characterization of the dominant commensurabilities (see Figs.~\ref{fig:cislunar_atlas} and \ref{fig:translunar_atlas}), the full MMR problem is more naturally interpreted on restricted-problem backbones. In that sense, the averaged Hamiltonian and the restricted problem play complementary roles rather than competing ones: the former supplies resonant centers, equilibria, and first-order width estimates within a locally valid reduced-dimensional model, whereas the latter realizes the corresponding structures in the non-reduced phase space as global objects---periodic and quasi-periodic invariant objects, separatrices, chaotic layers, and transport channels. In the patched-CR3BP view introduced in \textsection\ref{sec:spatio}, the relevant scaffold is not fixed once and for all: the Earth--Moon restricted problem organizes cislunar and circumlunar transport, while the Sun--Earth restricted problem supplies the outer translunar, secondary-region description \citep{sH60, sH61, kE62b, sH69, kTtK25}.

For a patched restricted-problem representation of circumterrestrial space, it is useful not to privilege the Earth--Moon barycenter at the outset. Let bodies $1$ and $2$ denote the two massive primaries that define the local restricted-problem scaffold, and let body $3$ denote the remaining massive body retained as a perturbation. In a frame rotating with the instantaneous angular velocity $\bm{\Omega}_{12}$ of the $1$--$2$ line and with origin at the $1$--$2$ barycenter, the motion of a massless particle may be written as \citep{lMjK67,dS98}
\begin{align}
	\label{eq:patched_R4BP}
	\ddot{\rb}
	+ \dot{\bm{\Omega}}_{12}\times\rb
	+ 2\bm{\Omega}_{12}\times\dot{\rb}
	+ \bm{\Omega}_{12}\times
    \left(
        \bm{\Omega}_{12}\times\rb
    \right)
	= \frac{\partial U_{12,3}}{\partial \rb},
\end{align}
where
\begin{align}
	U_{12,3}(\rb,t) & =
	{{\cal G}(M_1+M_2)}
	\left[
	\frac{1-\bar\mu_{12}}{\| \rb-\rb_1 \|}
	+
	\frac{\bar\mu_{12}}{\| \rb-\rb_2 \|}
	\right] \nonumber \\
	& \quad +
	{\cal G}M_3
	\left[
	\frac{1}{\| \rb-\rb_3 \|}
	-
	\frac{\rb\cdot\rb_3}{\| \rb_3 \|^3}
	\right],
\end{align}
with
\begin{align}
	\bar\mu_{12}=\frac{M_2}{M_1+M_2}.
\end{align}
Here $\rb_1$, $\rb_2$, and $\rb_3$ are measured from the $1$--$2$ barycenter. The first two terms define the instantaneous restricted-problem backbone, while the final bracket is the direct-minus-indirect contribution of the third massive body; the barycentric analogue of the body-centered disturbing-function form introduced in \textsection\ref{sec:disturbing}. Thus the indirect term accounts for the acceleration of the chosen barycentric origin by body $3$.

The two limiting branches used in the patched-CR3BP spatiography are obtained by the assignments
\begin{align}
	(1,2;3) &= (\E,\M;\S),
	&&\text{Earth--Moon scaffold with the Sun as an exterior perturber}, \\
	(1,2;3) &= (\S,\E;\M),
	&&\text{Sun--Earth scaffold with the Moon as an interior perturber}.
\end{align}
The first branch is the natural cislunar--circumlunar model: the rotating frame is tied to the Earth--Moon pair and the Sun supplies the dominant exterior perturbation. The second branch is the natural outer-translunar model: the rotating frame is tied to the Sun--Earth pair, while the Moon is no longer part of the scaffold pair but is retained as an interior perturbing body. Read in this way, the two branches are not separate theories for separate subjects, but asymptotic descriptions of different parts of the same circumterrestrial phase space.

This change of scaffold should not be interpreted as a sharp switch at lunar distance. The inner translunar problem---especially near the Earth--Moon exterior gateway and throughout the strongly Moon-perturbed portion of the translunar lobe---remains most naturally described as an Earth--Moon restricted problem with the Sun as an exterior perturbation. Farther out, as the Sun--Earth hierarchy becomes organizing and the motion approaches the Earth-centered secondary region of the Sun--Earth problem, the Sun--Earth scaffold with the Moon as an interior perturbation becomes the more faithful asymptotic description. The handoff is therefore dynamical rather than kinematic: it is set not by a single geometric shell, but by the gradual reordering of the dominant transport and secular structures. Locating and quantifying that handoff between the inner- and outer-translunar descriptions is reserved for a more detailed treatment elsewhere.

A concrete cartographic signature of this handoff is furnished by the Sun--Earth Jacobi constant. In the Sun--Earth CR3BP---the $(1,2;3)=(\S,\E;\M)$ branch of the patched formulation above with the lunar perturbation suppressed---the autonomous equations admit the usual Jacobi integral. The critical value $C_J^{\S\E}=C_1^{\S\E}$, evaluated at the Sun--Earth $\left( L_1 \right)^\E$ Lagrange point, defines the zero-velocity level at which the Earth-centered lobe first opens toward the Sun. Orbits whose Sun--Earth Jacobi value lies below this critical level are energetically able, in the CR3BP sense, to access the Sun--Earth $\left( L_1 \right)^\E$ gateway and hence the broader heliocentric environment (see Fig.~\ref{fig:translunar}). The corresponding $C_J^{\S\E}=C_2^{\S\E}$ level opens the anti-sunward $\left( L_2 \right)^\E$ bottleneck, so that the higher-energy Sun--(Earth\texttt{+}Moon) level represented schematically in Fig.~\ref{fig:translunar} displays both heliocentric necks open. For the element-space fate maps, however, the relevant signature is the existence of a Sun--Earth zero-velocity boundary in the patched description, not exact coincidence with either critical level. This level-set structure therefore acts, in the patched description, as a natural dynamical handoff surface between the two CR3BP branches. Its trace in geocentric orbital-element space appears as a sharply defined, geometrically smooth boundary in Sun-perturbed fate-classified astro-cartographies (see \textsection\ref{sec:cartographies})---a feature absent in the Earth--Moon-only model, where the conserved Jacobi integral belongs instead to the $(1,2;3)=(\E,\M;\S)$ branch and carries no Sun--Earth gateway boundary.

For comparison with the Earth--Moon spatiography, the Sun--Earth branch is most usefully viewed in a secondary-region, or Hill, frame about the Earth and scaled by the Earth--Moon distance rather than by the full Sun--Earth separation. In the notation above, this local coordinate may be written
\begin{align}
	\bm{\xi}
	=
	\frac{\rb-\rb_2}{a_{\E\M}},
	\qquad (1,2;3)=(\S,\E;\M),
\end{align}
where $\rb_2=\rb_\E$ in the Sun--Earth branch.

For the cislunar specialization, $(1,2;3)=(\E,\M;\S)$, Eq.~\eqref{eq:patched_R4BP} reduces to the Earth--Moon rotating-frame model with solar perturbations.  In this case $\bar\mu_{12}=\bar\mu_{\E\M}$, $\bm{\Omega}_{12}$ is the instantaneous angular velocity of the Earth and Moon about each other, and $\rb$ is measured from the Earth--Moon barycenter.  Under the additional assumptions of circular Earth--Moon motion and prescribed solar motion, this reduces to the simplified bicircular restricted four-body problem \citep{aJetal20}. More generally, however, the formulation does not require the two scaffold primaries to move on circular, or even Keplerian, orbits. The attraction of either primary may also be replaced by a higher-fidelity gravity-field expansion when the application warrants it.

Although this more general form is useful for exposing the connection between the Earth--Moon and Sun--Earth descriptions, many applications begin instead with the circular restricted three-body problem (CR3BP). It is therefore not the full cislunar model, but the autonomous reference problem against which more elaborate four-body and ephemeris formulations should be read. In the cislunar branch, the CR3BP neglects the Sun, treats the Earth and Moon as point masses, and constrains the Earth--Moon primary pair to uniform circular motion.  Thus $\dot{\bm{\Omega}}_{\E\M}=0$ and, after the usual nondimensionalization by the Earth--Moon distance and the inverse Earth--Moon mean motion, one may take $\bm{\Omega}_{\E\M}=\hat{\bm{z}}$. Equation~\eqref{eq:patched_R4BP} then reduces to
\begin{align}
	\label{eq:CR3BP}
	\rb''
	+ 2 \hat{\bm{z}}\times\rb'
	+ \hat{\bm{z}}\times
	\left(
		\hat{\bm{z}}\times\rb
	\right)
	=
	\frac{\partial U_{\E\M}}{\partial \rb},
\end{align}
where $()'$ denotes differentiation with respect to the nondimensional time $\tau=n_{\E\M}t$, equivalently the mean anomaly of the circular Earth--Moon motion, and
\begin{align}
	U_{\E\M}(\rb)
	=
	\frac{1-\bar\mu_{\E\M}}{\| \rb+\bar\mu_{\E\M}\hat{\bm{x}}\|}
	+
	\frac{\bar\mu_{\E\M}}{\| \rb-(1-\bar\mu_{\E\M})\hat{\bm{x}}\|}.
\end{align}
Here the Earth and Moon occupy the fixed nondimensional locations $-\bar\mu_{\E\M}\hat{\bm{x}}$ and $(1-\bar\mu_{\E\M})\hat{\bm{x}}$, respectively, in the rotating Earth--Moon barycentric frame.

Equivalently, moving the centrifugal term to the right-hand side gives the standard effective-potential form
\begin{align}
	\rb'' 
    + 
    2 \hat{\bm{z}}\times\rb'
	=
	\frac{\partial \mathcal{U}_{\E\M}}{\partial \rb},
	\qquad
	\mathcal{U}_{\E\M}(\rb)
	=
	U_{\E\M}(\rb)
	+
	\frac{1}{2}\|\hat{\bm{z}}\times\rb\|^2 .
\end{align}

Let $\vb=\rb'$ denote the nondimensional velocity in the rotating frame, with $v=\|\vb\|$.  This autonomous system possesses the Jacobi integral
\begin{align}
	J(\rb,\vb)
	=
	\frac{1}{2}v^2-\mathcal{U}_{\E\M}(\rb),
\end{align}
or, equivalently, the classical Jacobi constant
\begin{align}
	C_J=-2J=2\mathcal{U}_{\E\M}(\rb)-v^2 .
\end{align}
The stationary points of $\mathcal{U}_{\E\M}$ are the five Lagrange equilibria, whose associated Hill-region topology, gateway openings, and invariant-manifold structures underlie much of modern cislunar and translunar mission design.

An approximation to the Jacobi constant in orbital-element space with respect to the Moon is the Tisserand parameter ($T_\M \approx C_J$), defined as
\begin{align}
    \label{eq:tisserand_moon}
    T_\M
    =
    \frac{a_\M}{a}
    +
    2 \cos I^\M
    \sqrt{
    \frac{a}{a_\M}
    \left( 1-e^2 \right)
    },
\end{align}
where $a$ and $e$ are the osculating geocentric semi-major axis and eccentricity, and $I^\M$ is the inclination of the satellite orbit measured with respect to the lunar orbital plane. In this form, $T_\M$ is the orbital-element analogue of the Jacobi constant: it maps the Hill-region accessibility constraints of the CR3BP encoded by $C_J$; in the absence of maneuvers, dissipative forces, or strong departures from the restricted-problem approximation, motion is constrained to remain near a constant-$T_\M$ surface in $( a , e , I^\M )$ space \citep{jL86, wMlB22}. The coplanar contours used in the catalog projections in Fig.~\ref{fig:hills} correspond to the specialization $I^\M = 0$.

The CR3BP exposes, without explicit time dependence, the resonant periodic orbits, quasi-periodic invariant tori, separatrices, Poincar\'e sections, Hill-region openings, and invariant manifolds that organize the onset of transport and chaos \citep[qq.v.][]{cC68, McGehee1969, jL85, eB93, eBjM95, cSeO97, wK00, wK01, gG04, eD07, sR07, eB08, mB10, dGY13, aU15, kO19a, kO19b, vdO20, wK22, sS25, sFsG25, nB26}. In this sense it supplies the global phase-space geometry that underlies the local resonant atlas of \textsection\ref{sec:perturbed}: the averaged Hamiltonian identifies where resonant structure first appears, while the CR3BP reveals how that structure is embedded in a wider network of stable islands, unstable periodic or quasi-periodic invariant objects, tube dynamics, whiskered tori, and chaotic transport \citep{fT08, mVkH14a, mVkH14b, dVP23, bMkW23, dHdS23, bKaR26, sP26}. The more elaborate elliptic restricted three-body problem (ER3BP), R4BP, and ephemeris models are then best understood as successive tests of the persistence, deformation, or operational relevance of those organizing structures.

It is worth emphasizing that these four-body refinements are not a recent invention. Already in the early Space-Age literature, the Earth--Moon--Sun problem was treated through a hierarchy of restricted models lying between the CR3BP and the full ephemeris problem. Huang's ``very restricted'' four-body model idealized the Sun--Earth--Moon geometry by prescribing two nested circular motions and examined the resulting zero-velocity surfaces \citep{sH60}. Shortly thereafter, the bicircular restricted four-body problem was developed explicitly as a periodic solar perturbation of the Earth--Moon restricted problem, with particular attention to periodic motions near the libration points and to the persistence, displacement, or destruction of the familiar three-body structures \citep{jdV64, jCetal64, jD65, lW66, rKlC67, jCetal68, bTbS68}. In parallel, several authors studied the effect of the Sun on the triangular points, including near-resonant, nonlinear, and stability questions that anticipate much of the modern invariant-object language \citep{hSwH64, jdV64, jD65, lW66, rKlC67, hS68, rKlC68, aKjB70}. Matched-asymptotic and limiting formulations were also pursued for Earth-to-Moon motion and for Hill-type four-body reductions, including models in which the lunar motion was represented by de~Pont\'ecoulant theory or by Hill's variational orbit \citep{ySmE67, lMjK67, pB77}. Thus the BCR4BP, Hill-type restricted four-body reductions, and patched restricted-problem viewpoints all have substantial antecedents in the early Space Age, although this lineage is often obscured in contemporary discussions.

The later literature may be read as a revival and refinement of this same phenomenological hierarchy. The ballistic-capture and weak-stability-boundary perspective reasserted the practical importance of the Sun--Earth--Moon four-body geometry for low-energy lunar transfer design \citep{hY92, eB93, kY04, pSS12, fT13}. The patched or coupled restricted-problem approach then supplied a dynamical-systems interpretation: exterior low-energy transfers are naturally organized by the Sun--Earth restricted problem, while lunar capture and cislunar transport are organized by the Earth--Moon restricted problem \citep{rC12, pdSS16, kOetal17, bPjM25}. In this language, the bicircular problem (BCR4BP) remains the simplest time-periodic Sun-perturbed model, the quasi-bicircular problem (QBCP) restores coherence in the prescribed motion of the primaries, and the Hill restricted four-body problem (HR4BP) supplies a coherent Hill-limit model retaining both the direct and indirect solar effects \citep{dS98, mA02, aLcB08, jBG15}. Recent work has returned to the local invariant structures of these models---near $L_1$, $L_2$, $L_3$, the triangular points, and NRHO-like families---using non-autonomous normal forms, center-manifold reductions, invariant tori, Lagrangian coherent structures, and manifold continuation \citep{aJbN20, aJetal20, kB20, jR21, kO21, dVP23, lPetal23, lPetal24, sA24, bPkH24, lP25, sFetal25, sS25, rS26, rLetal26}. The models adopted in the literature should therefore be understood not as competing approximations, but as members of a single explanatory hierarchy: each sacrifices some fidelity in order to expose a particular organizing mechanism of the Earth--Moon--Sun system.

In this sense, the CR3BP supplies the autonomous reference problem; the BCR4BP adds a time-periodic solar perturbation to the Earth--Moon restricted problem; the QBCP modifies the prescribed primary motion to recover greater dynamical coherence; the HR4BP introduces the solar tide through a Hill-limit description of the Earth--Moon pair; and patched or coupled CR3BP models use different restricted problems in the regions where each is dynamically prevalent. The full ephemeris model is then the target model for validation and mission realization, not the first instrument of explanation. Section~\ref{sec:cartographies} takes the next step in this hierarchy: the numerically constructed MEGNO and fate maps show which of these backbone structures survive, broaden, or break into sticky transport layers in the Earth--Moon and Earth--Moon--Sun circumterrestrial problem surveyed here.

\subsection{Resonant Dynamics and the Onset of Chaos in Cislunar and Translunar Space}
\label{sec:resonant_dynamics}

The resonance skeleton has already been exposed by the perturbative treatment of \textsection\ref{sec:perturbed}: the Gallardo-style atlas identifies the principal lunar and solar commensurabilities, and Table~\ref{tab:spatio} organizes their spatiographic placement across the cislunar and translunar provinces. The role of the restricted-problem dynamics is therefore not to redetermine the nominal resonance centers, but to embed that skeleton in the global phase-space geometry \citep{wK22}. Whereas the perturbed Hamiltonian formulation supplies local resonance centers, approximate half-widths, and a first-order hierarchy of resonant importance, the restricted-problem description furnishes the associated periodic and quasi-periodic invariant objects, Poincar\'e sections, separatrices, chaotic layers, and invariant manifolds. The two descriptions should not be identified term by term: the equilibria and widths of the averaged resonant Hamiltonian are local shadows of the full phase-space structure, not one-to-one substitutes for the periodic-orbit families and their global manifolds. The CR3BP and its elliptical and Sun-perturbed extensions therefore complement the local Hamiltonian atlas with a genuinely global portrait of how resonances organize confinement, transport, and the onset of chaos.

\begin{figure*}[t!]
	\begin{center}
	\includegraphics[width=0.85\textwidth]{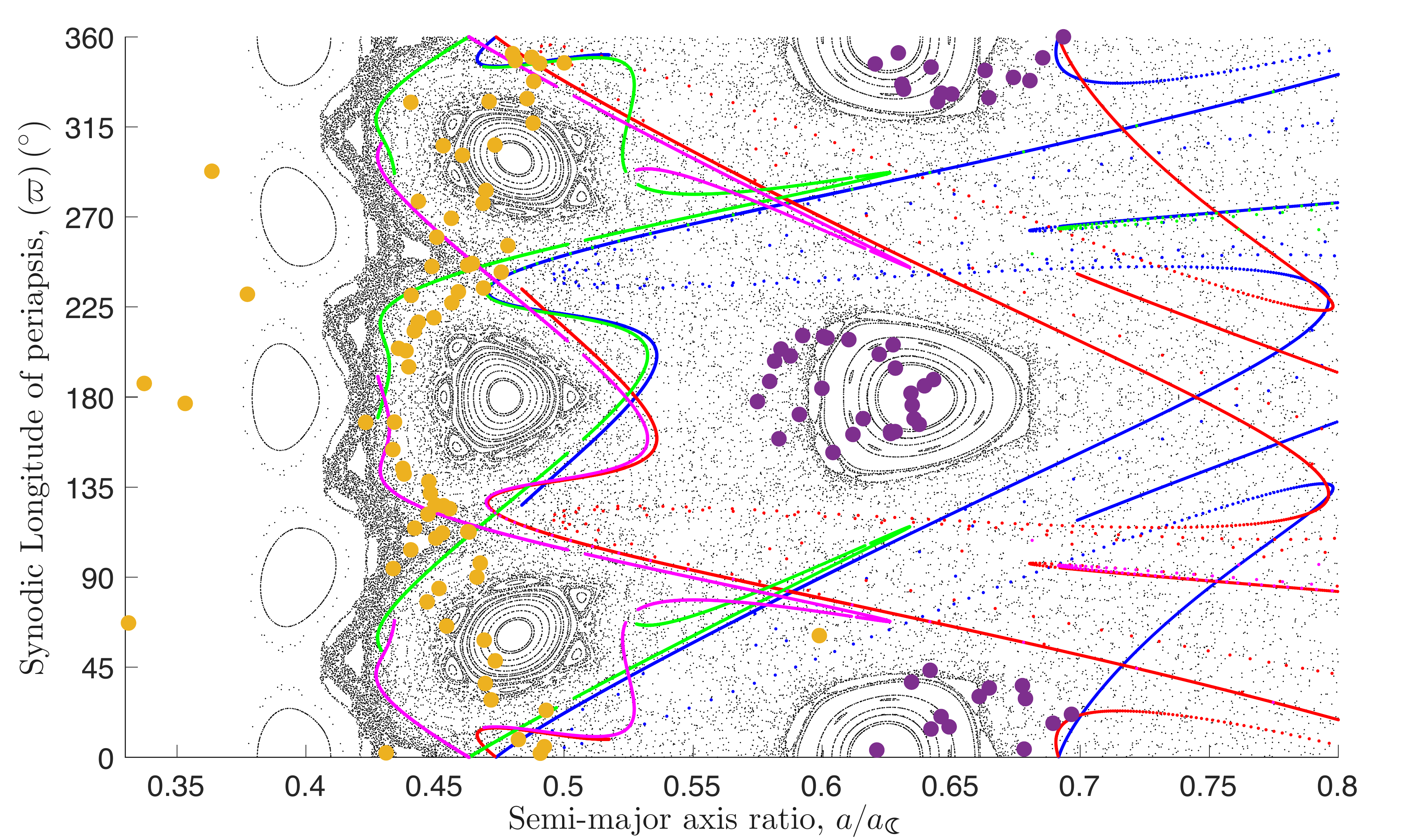}
	\vskip -0.075in
	\caption{\small 
Perigee Poincar\'e section of the Earth--Moon CR3BP showing the global resonant geometry in $( \varpi , a )$ at a fixed Jacobi constant. {\it Gray points}: phase-space sample revealing resonant islands and chaotic layers. {\it Colored curves}: separatrices traced from unstable resonant periodic orbits (weak/strong branches); intersections highlight heteroclinic links. Filled markers indicate elliptic island centers; open markers denote hyperbolic saddles. The separatrix extent defines each resonance's zone of influence and reveals corridor connections across the section. Perigee crossings from JPL Horizons ephemerides for Spektr--R and TESS are overlaid.
	}
	\label{fig:periapsis_poincare}
	\end{center}
	\vskip -0.075in
	\hspace{2cm}\rule{12.5cm}{0.5pt}
\end{figure*}

Beyond geosynchronous orbit and interior to the lunar distance, lunar mean-motion resonances generate alternating stable and unstable resonant structures \citep{fT08, eB08}. In the planar circular restricted problem, this organization is seen most directly on surfaces of fixed Jacobi constant through Poincar\'e maps at osculating perigee: stable islands are centered on elliptic resonant fixed points, unstable resonant periodic orbits appear as hyperbolic fixed points from which separatrices emanate, and larger chaotic zones are threaded and bounded by segments of the associated stable and unstable manifolds (Fig.~\ref{fig:periapsis_poincare}). The resulting global resonance zones and chaotic layers are generally broader than those predicted by local perturbed-Keplerian estimates \citep{aRbK26}, indicating that the CR3BP captures transport channels and heteroclinic connections among neighboring resonance regions that are not resolved by a purely local Hamiltonian treatment. This is precisely why the Gallardo atlas is best read as a local cartographic guide to where resonant structure first becomes important, rather than as a global delimitation of each resonance's full region of influence.

The same construction extends naturally to the translunar realm. There, Poincar\'e maps at osculating apogee expose exterior $1\!:\!n$ lunar resonances, whose geometry differs from the interior case: stable asymmetric libration zones coexist with weak and strong symmetric unstable resonance zones. The invariant manifolds of the unstable resonant orbits provide separatrices that delimit their regions of influence and, through the $\left( L_1 \right)^\M$--$\left( L_2 \right)^\M$ tube geometry, mediate transitions between exterior and interior resonant domains \citep{aRetal25}. Together with the heteroclinic connections among the interior $4\!:\!1$, $3\!:\!1$, and $2\!:\!1$ resonances and the lunar $\left( L_1 \right)^\M$ region, these structures define a network of ballistic pathways by which the semimajor axis of a spacecraft may change substantially under lunar perturbations alone \citep{eB08, yL17, hLbX18, cP24, bKaR26, sP26}.\footnote{A concrete mission-scale illustration is provided by ESA's SMART-1 transfer \citep{gRetal02}, whose low-thrust trajectory closely shadowed many of these natural resonant and gateway corridors, even though the mission design itself did not explicitly rely on a manifold-based construction \citep{bKaRa26_ISSFD}.} In the translunar problem, however, this global resonant geometry must also be read together with residence and escape behavior: the same phase-space region may support prompt escape, long sticky residence, or repeated gateway passage, which is why the astro-cartographies of \textsection\ref{sec:cartographies} combine variational indicators with explicit fate diagnostics rather than relying on MEGNO alone.

A complementary criterion for the onset of \emph{global} chaos in the outer circumterrestrial problem is provided by the Chirikov resonance-overlap condition \citep{bC79, aLmL92, aM02}. In its classical planetary form, this mechanism gives rise to a chaotic zone around the orbit of a secondary: the accumulation and overlap of neighboring first-order mean-motion resonances destroy intervening invariant barriers, producing a broad annulus in which initially nearby particles are scattered, collide, escape, or survive only in protected co-orbital islands \citep{jW80, mH93, jBbG98, sM15, tG19, iS20, tDiS21}. In the present Earth--Moon setting, this language should be read dynamically rather than literally. The issue is not the clearing of a material disk, but the loss of local resonant confinement and the opening of large-scale transport across the outer circumterrestrial phase space.

In Fig.~\ref{fig:cislunar_atlas}, the principal interior lunar resonances remain, over much of the cislunar resonant zone, recognizable as distinct bands; appreciable crowding appears mainly near the outer end of the interior ladder, as the resonances approach the cislunar--circumlunar gateway at $\left( L_1 \right)^\M$, with the \res{4}{3}{\M} and \res{5}{4}{\M} commensurabilities appearing as the last clearly marked interior lunar resonances before the circumlunar transition. In the translunar atlas of Fig.~\ref{fig:translunar_atlas}, by contrast, the exterior lunar resonances and the first interior solar resonances form a much more entangled overlap web across the translunar domain. The onset of large-scale stochasticity is therefore not confined to a narrow anti-Earthward strip immediately outside the Moon, but is better understood as a broad outer-circumterrestrial chaotic zone produced by the overlap of exterior lunar resonances and interior solar commensurabilities, and framed radially by the lunisolar tidal-parity scale, the Earth SOI, and the Hill boundary. The corresponding Poincar\'e maps at osculating apogee make this transition explicit: below the overlap threshold, coherent resonant islands persist; above it, the stochastic layer connects continuously to the outer escape boundary and individual trajectories diffuse across the full chaotic region \citep{aRetal25}. Quantitative validation of this overlap onset against the inner edge of the outer chaotic sea identified in the astro-cartographies is deferred to \textsection\ref{sec:cartographies}.

\subsection{Stability Structure and Bifurcations of Periodic-Orbit Families}
\label{sec:stability_bifurcations}

In the planar circular restricted setting, the periodic-orbit families generated by the cislunar and translunar restricted problems provide the natural organizing objects that sit between the local resonant atlas of \textsection\ref{sec:boundaries} and the global transport maps of \textsection\ref{sec:cartographies}. As parameters such as the Jacobi constant, resonance ratio, or continuation amplitude are varied, these families undergo the familiar sequence of stability changes, symmetry breaking, and branch bifurcations that populate the reduced phase space with stable periodic orbits surrounded by islands of stability, as well as unstable periodic orbits and their associated stable and unstable manifolds. In this sense, the onset of chaos is not a featureless loss of regularity, but the cumulative consequence of repeated bifurcation and manifold interactions leading to resonance overlap. The practical consequence is especially important for cislunar and translunar astrodynamics: broad stable regions, gateway layers, and sticky chaotic corridors are all rooted in the stability structure of the underlying periodic families, and should therefore be read as the global restricted-problem architecture associated with the same resonant skeleton already visible in the local Gallardo-style Hamiltonian portraits. This viewpoint is also the natural place to connect the perturbed Hamiltonian and restricted-problem languages. In the averaged resonant problem, one speaks of stable and unstable equilibria of the reduced Hamiltonian; in the restricted problem, those same local structures continue into families of periodic solutions whose stability determines the surrounding phase-space organization. The correspondence is not exact, especially near the Earth--Moon gateway and in the strongly perturbed translunar region, but it is close enough to justify speaking of the local resonant equilibria as the first shadows of the global periodic-orbit architecture.

The Earth--Moon CR3BP supports a rich atlas of periodic orbit families, both resonant and non-resonant, whose classification and global structure have been mapped in detail \citep{rB68, kH01, eD07, sBnB23}. The collinear libration points each seed a planar Lyapunov family parameterized by the Jacobi constant: the $\left( L_1 \right)^\M$ and $\left( L_2 \right)^\M$ families provide the innermost organizing structures of the Earth--Moon gateway region, while the $\left( L_3 \right)^\M$ family generates planar periodic orbits about the exterior collinear point on the far side of Earth from the Moon. As the Jacobi constant decreases, equivalently as the three-body energy level rises, the planar Lyapunov families at $\left( L_1 \right)^\M$ and $\left( L_2 \right)^\M$ each undergo a vertical bifurcation at the \emph{vertical critical orbit}, where the linearized out-of-plane frequency resonates with the in-plane frequency and the monodromy eigenvalues for the out-of-plane motion pass through unity. This bifurcation spawns the three-dimensional halo families (north and south), whose high-inclination limiting members near $\left( L_2 \right)^\M$---the near-rectilinear halo orbits (NRHOs)---have been identified as the reference architecture for the lunar Gateway and related mission concepts \citep{rFaK73, kH84, eZ20}. More generally, relaxing the coplanar restriction lifts the planar periodic-orbit skeleton into a spatial architecture of vertical families, halo branches, quasi-periodic Lissajous-type tori, and their associated invariant manifolds \citep{gG04}. Distinct from the libration-point families, the \emph{distant retrograde orbit} (DRO) family consists of large retrograde periodic orbits about the Moon in the rotating Earth--Moon frame and maintains high linear stability across a wide range of Jacobi constants, extending continuously from the outer cislunar region into the inner translunar regime \citep{cB17, kO21}. The resonant symmetric and asymmetric periodic-orbit families then populate the commensurability skeleton of the atlas in a layered manner. In Poincar\'e-section views of the Earth--Moon planar problem, these families appear together with the libration-point and retrograde families as a common architecture of elliptic islands, surrounding quasi-periodic tori, separatrix transitions, and stability changes as the Jacobi constant is varied \citep{rB68, fT08, pDrK10, aRbK26, bKaR26}. Retaining lunar eccentricity in the elliptic restricted problem does not erase this skeleton, but changes its parametrization: the Jacobi integral is lost, the equations become explicitly time dependent, and only commensurable members of the circular families continue as periodic solutions at fixed eccentricity, with their geometry and stability followed through eccentricity continuation \citep{rB69, fFmL18}.

The complementary outer-boundary problem is organized by the collinear libration-point families of the Sun--Earth restricted problem, interpreted at the coarse spatiographic scale of Fig.~\ref{fig:translunar} as the Sun--(Earth\texttt{+}Moon) system. In this outer CR3BP, the Earth--Moon pair behaves as the effective secondary, and the Sun--Earth $\left( L_1 \right)^\E$ and $\left( L_2 \right)^\E$ neighborhoods provide the two heliocentric bottlenecks through which translunar motion communicates with the surrounding solar realm. As in the Earth--Moon problem, the collinear points seed planar Lyapunov, vertical Lyapunov, halo, and Lissajous-type families and tori, together with stable and unstable invariant manifolds that separate transit from non-transit motion and furnish low-energy pathways between dynamically distinct realms \citep{kHhP93, rFetal04, kHmK06, dGY13, bPjM25}. In this setting, the Earth--Moon $\left( L_1 \right)^\M$ and $\left( L_2 \right)^\M$ families organize the local cislunar--circumlunar and circumlunar--translunar gateways, while the Sun--Earth $\left( L_1 \right)^\E$ and $\left( L_2 \right)^\E$ families organize the translunar--heliocentric handoff. In higher-fidelity Sun--Earth--Moon models, this outer restricted-problem architecture is further modulated by explicit time dependence: the autonomous Sun--Earth periodic families continue as quasi-periodic or time-dependent analogues, and their connections acquire a dependence on the lunar phase and on close lunar encounters. Nevertheless, the same manifold geometry remains the organizing skeleton for low-cost passage between the Sun--Earth and Earth--Moon dynamical systems \citep{rLetal26}.

Viewed through the spatiographic partition of Table~\ref{tab:spatio}, these families do not belong to a single region by pointwise location alone. The Earth--Moon $\left( L_1 \right)^\M$ Lyapunov, halo, and related quasi-periodic families are best interpreted as cislunar--circumlunar gateway structures: they are rooted at the Earthward lunar gateway and organize passage between the outer cislunar resonant zone and the Moon's local dynamical neighborhood. The corresponding Earth--Moon $\left( L_2 \right)^\M$ families are translunar--circumlunar gateway structures, attached to the exterior lunar gateway and more directly exposed to the Sun-perturbed outer circumterrestrial environment (Fig.~\ref{fig:circumlunar}). The Sun--Earth $\left( L_1 \right)^\E$ and $\left( L_2 \right)^\E$ families, by contrast, belong to the outer translunar boundary of the Earth-bound problem: their manifolds govern passage through the heliocentric bottlenecks of the Sun--(Earth\texttt{+}Moon) zero-velocity geometry shown schematically in Fig.~\ref{fig:translunar}. The $\left( L_3 \right)^\M$ family occupies a different role, furnishing an exterior Earth--Moon collinear branch on the anti-lunar side of Earth rather than a Moon-encircling or gateway family. The DRO family, finally, is most naturally read as a circumlunar stability structure in the rotating frame: its members encircle the Moon retrograde to the Earth--Moon rotation, while the larger members extend across the neighboring cislunar and translunar partitions. Thus the same periodic-orbit atlas supplies not only stability classes, but also the transport-theoretic refinement of the cislunar, circumlunar, translunar, and heliocentric provinces introduced in \textsection\ref{sec:boundaries}.

The stability of each periodic orbit is characterized by the eigenvalue structure of its \emph{monodromy matrix}---the linearized Poincar\'e return map after one full period \citep{rB68}. For planar orbits in the CR3BP, the non-trivial eigenvalue pair $\{\lambda, \lambda^{-1}\}$ defines the stability index 
\begin{align}   
    \nu = \frac{1}{2}(\lambda + \lambda^{-1}).
\end{align}
When $|\nu| < 1$ the orbit is \emph{elliptic} and surrounded by resonant tori on which nearby trajectories librate; when $|\nu| > 1$ the orbit is \emph{hyperbolic}, possessing stable and unstable manifolds that generate chaotic transport and, near the Earth--Moon $\left( L_1 \right)^\M$ and $\left( L_2 \right)^\M$, the tube structures that mediate gateway passage. The stability boundary $|\nu| = 1$ is precisely the bifurcation locus at which new family branches may appear. For three-dimensional families, an independent eigenvalue pair must be tracked for the out-of-plane direction simultaneously: an orbit that is elliptic in-plane may be simultaneously hyperbolic out-of-plane, and the fully \emph{complex-unstable} regime---characteristic of large-amplitude halo orbits and certain resonant families---generates diffusion through the Arnold web of the three-degree-of-freedom system \citep{aLmL92, aM02}. In the Earth--Moon halo family, the stability character transitions from the elliptic NRHO arc (near the bifurcation point) to hyperbolic and eventually complex-unstable as the orbit amplitude grows toward the Moon, partitioning the family into a mission-relevant stable arc and a transport-active unstable arc.

Each stability change along a periodic-orbit family entails a specific geometric reorganization of the surrounding phase space. A \emph{tangent} (fold) bifurcation creates or annihilates a stable--unstable periodic orbit pair: the elliptic member seeds a new island chain in the Poincar\'e section, while the hyperbolic member contributes separatrices and stable and unstable manifolds. A \emph{period-doubling} bifurcation, occurring when $\nu = -1$, spawns a period-doubled daughter family whose own subsequent bifurcations can generate still higher-period resonant islands, contributing to the island-in-island hierarchy characteristic of near-integrable Hamiltonian systems. Such bifurcations are therefore important features of the periodic-orbit atlas, but they are not the sole origin of the resonant hierarchy. Many of the dominant cislunar and translunar resonant periodic orbits are more naturally understood as finite-$\mu$ continuations of specially phased subharmonic Keplerian orbits, with folds, stability changes, and symmetry breaking delimiting their existence and reshaping the nearby phase space. The perigee Poincar\'e section in Fig.~\ref{fig:periapsis_poincare} illustrates the resulting restricted-problem geometry: elliptic resonant centers are embedded within islands of quasi-periodic motion, while hyperbolic resonant orbits generate separatrices whose stable and unstable manifolds bound chaotic layers and form heteroclinic links across neighboring resonance regions \citep{eB08, fT08, hLbX18, aRbK26, bKaR26, sP26}. The broad chaotic layers that emerge in the outer cislunar and translunar regions consequently reflect the combined effects of resonance overlap, separatrix splitting, and manifold intersections associated with hyperbolic periodic orbits and invariant tori, rather than a bifurcation sequence alone.

The stability type of each family member should therefore leave a recognizable imprint on the MEGNO astro-cartographies of \textsection\ref{sec:cartographies}, where the MEGNO diagnostic is defined in Eq.~\eqref{eq:MEGNO}. Near a stable periodic orbit, surrounding invariant tori support quasi-periodic trajectories and may appear in element space as coherent low-MEGNO pockets, bands, or filaments. Near an unstable family member, exponential divergence of nearby trajectories produces elevated MEGNO, while the associated separatrix webs and manifold intersections broaden the corresponding chaotic layers. The transition from stable to unstable along a family is therefore expected to appear in the maps not as an abrupt geometric boundary, but as a degradation from coherent low-MEGNO structure to diffuse high-MEGNO texture, with stability changes and bifurcation acting as one organizing source of that transition. In practice, the DRO family---highly stable across a wide range of Jacobi constants---is expected to generate a broad and easily identified low-chaos band; the stable arcs of the lunar-gateway halo families should produce compact pockets of regular motion near the circumlunar gateways; and the resonant periodic-orbit branches of the outer cislunar and translunar zones should appear as discrete filaments that thin, fragment, or dissolve as resonance overlap becomes dominant. Section~\ref{sec:cartographies} examines which of these restricted-problem stability signatures survive, broaden, or disappear when the idealized Earth--Moon architecture is embedded in the full Sun--Earth--Moon environment.

\subsection{Earth--Moon Cyclers and Collision Orbits}
\label{sec:cyclers_collisions}

The preceding discussion emphasizes resonant islands, chaotic layers, and manifold-mediated transport. A closely related, but less frequently foregrounded, part of the same CR3BP landscape is furnished by repeated-encounter periodic orbits: trajectories that return successively to the neighborhoods of both primaries. In the Earth--Moon problem these are naturally interpreted as cyclers when they provide recurring access to the terrestrial and lunar vicinities without remaining permanently attached to either body. They are also closely connected to the older theory of collision and close-encounter periodic orbits, in which passage through, or arbitrarily near, one of the primaries is treated as a singular limiting geometry of the restricted problem. In either interpretation, such orbits are not merely curiosities in the old atlases of periodic solutions. They are periodic manifestations of the same phase-space organization already visible in the resonant and gateway problems, and therefore candidate geometries for communications, navigation, search-and-rescue, space-domain awareness, logistics, and staging architectures distributed across geolunar space \citep{aBsR26}.
 
This viewpoint also has a much older pedigree than is usually acknowledged in modern cislunar mission design. \citet{rN59} studied periodic orbits of a planetoid passing close to two gravitating masses, explicitly treating the repeated close passage by both primaries as a distinguished class of restricted-problem motion. Shortly thereafter, \citet{sH62} introduced ideal moon-probe orbits that enclose the Earth and Moon and pass near both bodies repeatedly, so that observations made near the Moon could be returned to the Earth's vicinity; \citet{sHcW63} then computed two planar families of such Earth--Moon periodic orbits. \citet{rA63} proved the existence of periodic solutions passing near both masses, while \citet{jKjL68} obtained matched-asymptotic descriptions of Earth--Moon periodic orbits that come close to both primaries and exhibit prescribed synodic commensurability. \citet{rHbW68} compared such Earth--Moon restricted orbits with rotating Keplerian orbits, making explicit how far the three-body families depart from their two-body progenitors while remaining interpretable in an orbital-element language. This early literature already contains the essential idea of an Earth--Moon cycler: a periodic, repeated-access trajectory generated by the lunar perturbation rather than by station-keeping about a libration point.

The operational one-pass counterpart of these repeated-access families is the free-return lunar flyby. In its ideal form, a translunar injection places the spacecraft on an Earth--Moon transit for which the lunar encounter itself supplies the passive gravity assist onto a transearth return, so that, absent lunar orbit insertion or another major maneuver, Earth return remains feasible after the initial boost. Schwaniger's Apollo-era free-return study made this geometry explicit in the Earth--Moon restricted problem, classifying symmetric free returns into circumlunar cases, with far-side periselenum, and cislunar cases, with near-side periselenum, and noting that one such free-return case closes as a periodic trajectory \citep{aS63}. The Apollo figure-eight free-return architecture should be read in this same lineage: not as a cycler in the repeated-access sense, but as the one-pass mission analogue of the close-encounter geometry that Arenstorf and related authors regularized into periodic Earth--Moon families \citep{rA63}. Modern cycler studies make this bridge explicit by embedding Apollo-like figure-eight free-return segments within monthly Earth--Moon cyclers, including Arenstorf-type four-leaf-clover or big-loop geometries \citep{aGbA15}.


Modern treatments have recast this older material in dynamical-systems language. A useful modern distinction is between high-energy, near-Keplerian cyclers and lower-energy gateway-mediated cyclers. The first class is organized by $p\!:\!q$ resonant circumterrestrial orbits corrected for lunar flybys, together with their three-dimensional CR3BP generalizations; it remains close to the perturbed-Keplerian language of resonant orbital elements \citep{aBdA24}. The second class, already visible in earlier orbit atlases and in Davidson's periodic transition-orbit construction, is organized by the homoclinic and manifold structure associated with the Earth--Moon collinear gateways \citep{mD64}. \citet{jC10} synthesized this distinction in the planar CR3BP, emphasizing high-energy resonant cyclers on the one hand and lower-energy $\left( L_1 \right)^\M$-associated homoclinic-type cyclers on the other. In that formulation, however, the latter class is naturally treated as an outgrowth of unstable homoclinic structure rather than as a stable family in its own right. Recent work by \citet{sRmRT25} therefore adds an important refinement: stable, low-energy prograde Earth--Moon cyclers can also exist in this gateway-mediated class, providing natural repeated access to both primaries without requiring station-keeping about a libration point. This distinction is useful not only for mission design but also for the internal logic of the present Primer: high-energy cyclers lie nearer the local resonant-orbital-element language of \textsection\ref{sec:perturbed}, whereas low-energy cyclers belong more transparently to the gateway, manifold, and realm-transition geometry developed in the present section. Subsequent infrastructure-oriented studies have continued to exploit both sides of this bridge, using resonant and gateway-mediated cycler families to provide repeatable access patterns, regular phasing opportunities, and distributed coverage of geolunar space \citep{yL20}.

A second, still less familiar, lineage runs through periodic collision and consecutive close-encounter orbits. These are not operational impact trajectories in the usual sense; rather, they are limiting objects obtained by allowing a repeated-encounter family to pass through collision in a regularized model. H\'enon's study of orbits that encounter the Earth twice already made the connection between consecutive collisions and the classical periodic solutions of the restricted three-body problem, and noted that the construction carries over to the Earth--Moon setting \citep{mH68}. \citet{dH77} later used consecutive-collision orbits as generating objects for stable close-encounter periodic solutions, thereby bridging the periodic-collision literature and the practical cycler problem.  In the Earth--Moon context, \citet{aPrB94, aPrB96} further developed H\'enon-type transfer and regularized formulations, showing how collision regularization supplies a computationally useful way to continue and classify transfer-like repeated-encounter solutions.

The importance of these collision families is conceptual as much as computational.  In a regularized phase space, collision is not simply a singular endpoint at which the dynamics cease to be meaningful; it becomes a limiting geometric object whose stable and unstable sets organize nearby close-encounter motion. Periodic collision orbits may therefore be read as the singular skeleton of repeated flyby families, just as unstable resonant periodic orbits serve as the skeleton of chaotic resonance zones. From this perspective, cyclers, near-collision families, and resonant transport are different faces of the same global Earth--Moon phase-space architecture: some provide regular repeated access to the terrestrial and lunar neighborhoods, some form collision- and escape-limiting structures, and some lie close enough to chaotic layers to be highly maneuverable with small control inputs.


This perspective also clarifies how repeated-access structures should be read in the astro-cartographies of \textsection\ref{sec:cartographies}. A stable cycler or near-cycler family, if it persists beyond the idealized CR3BP setting, would appear only indirectly: its surrounding invariant tori would register as coherent low-MEGNO pockets, bands, or filaments embedded within the surrounding fate structure. Nearby unstable or near-collision branches, by contrast, would be associated with elevated variational growth, separatrix texture, and fate-map signatures of escape, impact, temporary capture, or repeated gateway passage. The cartographies therefore provide a coarse diagnostic for where repeated-access families, close-encounter structures, or their remnants may remain dynamically relevant, but such signatures are not by themselves cycler-family identifications; rather, they mark regions where periodic-orbit continuation or targeted recurrence analysis would be dynamically well motivated.

This section has developed the theoretical scaffolding for the circumterrestrial problem in the restricted multi-body setting: a phenomenological account of the two patched-CR3BP branches and their handoff, including the Sun--Earth Jacobi constant as its dynamical signature; a global resonance portrait in which the Chirikov overlap criterion links the semi-analytical atlas to the onset of large-scale chaos; and an account of periodic-orbit stability, bifurcations, collision limits, and cycler families that encode regular access patterns within the outer chaotic sea. The function of the next section is to render this scaffolding visible in numerically constructed element-space cartographies. The MEGNO and fate-class maps of \textsection\ref{sec:cartographies} translate these theoretical structures---resonant islands and separatrices, chaotic seas, gateway and escape channels, Sun--Earth zero-velocity boundaries, and possible repeated-access remnants---into spatially resolved signatures across the $(a,e)$ plane. They are not periodic-orbit catalogs in their own right; rather, they test how the restricted-problem architecture persists, broadens, or reorganizes in the elliptic Earth--Moon and Sun-perturbed Earth--Moon--Sun models, with the results calibrated against the spatiographic boundaries of Table~\ref{tab:spatio} and interpreted through the fate diagnostics described in \textsection\ref{sec:fate_classes}.

\section{Astro-Cartographies from the Laplace Radius to the Hill Sphere}
\label{sec:cartographies}

\subsection{MEGNO Cartography with REBOUND}
\label{sec:megno_rebound}

To reveal the fine structure of circumterrestrial phase space, we complement the semi-analytical portraits of the preceding sections with direct numerical astro-cartographies based on the \emph{mean exponential growth factor of nearby orbits} (MEGNO). It is therefore the diagnostic used here to translate the spatiographic and resonant architecture synthesized in \textsection\ref{sec:boundaries}, together with the stability and chaos signatures anticipated in \textsection\ref{sec:restricted}, into element-space dynamical maps. Introduced by \citet{pCcS00} and further developed by \citet{pC03}, MEGNO is a variational chaos indicator closely related to the largest Lyapunov characteristic number, but with substantially faster practical convergence in distinguishing ordered from stochastic motion. If $\delta(t)$ denotes the norm of the tangent-vector separation between two nearby trajectories, then
\begin{align}
    \label{eq:MEGNO}
	Y(t) = \frac{2}{t}\int_{0}^{t} \frac{\dot{\delta}(s)}{\delta(s)}\, s\, ds,
    \qquad
    \overline{Y}(t) = \frac{1}{t}\int_{0}^{t} Y(s)\, ds .
\end{align}
For quasi-periodic trajectories, $Y(t)$ oscillates about $2$ and $\overline{Y}(t)\rightarrow 2$; for chaotic motion, both $Y$ and $\overline{Y}$ grow asymptotically linearly, with slope proportional to the maximal Lyapunov exponent \citep{pCcS00, pC03, tH10, mM11}. In practice, this makes MEGNO especially well suited to dynamical maps, because regular islands, resonance borders, thin chaotic layers, and sticky transport channels can often be separated over integration intervals far shorter than those needed for a clean direct estimate of the Lyapunov time \citep{tH10, mM11, gC22}.

The numerical integrations underlying the maps are carried out with the open-source \texttt{REBOUND} $N$-body package \citep{hR12}, using its adaptive, high-order \texttt{IAS15} integrator \citep{hR15}. This choice is motivated by the dynamical character of the problem. The circumterrestrial phase space surveyed here contains strongly eccentric trajectories, repeated close approaches to the Moon, temporary capture into gateway neighborhoods, and resonant evolutions whose sensitivity to initial conditions can be extreme. \texttt{IAS15} was developed precisely for such settings: it is a 15th-order Gau{\ss}--Radau scheme with adaptive step-size control, designed to handle close encounters and high-eccentricity motion while maintaining errors near machine precision over very long integrations \citep{hR15}. Modern resonance-map studies using \texttt{REBOUND} likewise exploit this combination of flexibility and accuracy when constructing MEGNO cartographies of sensitive three-body phase space \citep{gC22}. In the present context, it furnishes a natural numerical backbone for maps extending from the geolunar Laplace radius to the Earth Hill sphere.

The integration spans adopted for the maps are not uniform across circumterrestrial space, but are matched to the dynamical provinces identified by the spatiographic partition of Table~\ref{tab:spatio} and refined, for numerical cartography, in Table~\ref{tab:megno_timescales} of Appendix~\ref{app:megno}. The secularly dominated cislunar and cislunar resonant domains admit shorter production maps because even a $19$-yr window covers hundreds to thousands of orbital revolutions. The translunar domains, by contrast, require substantially longer horizons to separate prompt escape from delayed escape, sticky residence, long-lived resonance sticking, and regular finite-time structures such as quasiperiodic pockets or orderly escape channels. The regime limits quoted there should be understood as \emph{working astro-cartographic boundaries}: they are chosen from the separatrix-edge envelope of the bounding resonance, and thus are intended to delimit the principal dynamical province of each map rather than to mark an exact global discontinuity in phase space. In consequence, adjacent regimes are allowed to overlap deliberately near their ends, reflecting the fact that the circumterrestrial problem passes between secular, resonant, gateway, and escape-dominated behavior through broad transitional belts rather than through infinitesimally sharp frontiers. The cartographic timescale is therefore treated here not as a purely numerical convenience, but as a dynamical quantity that must be matched to regime.

\subsection{Fate Classes and Survival Diagnostics}
\label{sec:fate_classes}

MEGNO alone, however, does not exhaust the dynamical information carried by the outer circumterrestrial problem. Once the maps extend into the circumlunar gateway and translunar provinces, trajectories with similar MEGNO values may correspond to qualitatively different outcomes. A low-MEGNO trajectory may remain bounded and quasiperiodic over the full integration window, escape in an orderly, nearly regular manner, or terminate early by collision before substantial variational growth has accumulated. Conversely, trajectories with elevated MEGNO may escape promptly through the outer circumterrestrial boundary, remain Earth-centered for decades while wandering through a sticky resonant corridor, or terminate after chaotic close-encounter evolution. In such regions, a pure regular-versus-chaotic coloring can collapse physically distinct classes of motion into the same numerical bucket. That limitation is not a defect of MEGNO itself, but a reflection of the fact that the circumlunar and translunar regimes are mixed transport borderlands rather than homogeneous dynamical seas.

For this reason, each initial condition in the presented maps is supplemented by a compact set of \emph{fate diagnostics}. In addition to the MEGNO value, we record whether the trajectory remains Earth-bound over the adopted integration window or crosses beyond the Earth Hill sphere, the first escape time when such escape occurs, Earth-reentry and lunar-impact flags, the minimum geocentric and selenocentric distances attained, and, where relevant, entries into the lunar Hill region. These diagnostics are not used to produce a separate statistical census here; rather, they provide the categorical layer needed to interpret the maps dynamically. Terminal events such as Earth reentry and lunar impact are assigned as geometric outcomes regardless of the accumulated MEGNO value, while non-terminal and escaping trajectories are further distinguished by their variational character. Thus the fate layer separates bounded quasiperiodic residence from bounded chaotic or sticky residence, orderly escape from chaotic escape, and terminal collision or reentry from both. Two additional categories---bounded-unclassified and escape-unclassified---capture trajectories whose geometric outcome is known but whose regular or chaotic character is not cleanly resolved within the adopted integration window. In the inner cislunar regime such a classification is mostly secondary, because regular islands and chaotic layers are already well separated by MEGNO itself. By contrast, in the circumlunar gateway and translunar regimes, it becomes essential: low MEGNO may correspond to bounded regular motion, orderly escape, or early termination by collision, while high MEGNO may correspond to sticky residence, chaotic escape, or terminal close-encounter evolution.

The resulting astro-cartographies should therefore be read as \emph{two-layer} maps. The first layer is variational and measures local dynamical regularity through MEGNO. The second layer is geometric and classificatory: it records whether the corresponding trajectory remains Earth-bound, escapes through the outer circumterrestrial boundary, reenters Earth, impacts the Moon, or undergoes repeated lunar-neighborhood passage, and then uses the MEGNO value to distinguish regular or orderly motion from chaotic or sticky transport when the geometric outcome alone is not decisive. Read together, these two layers recover the structured spatiography argued for throughout this Primer. The cislunar domain is not merely a place where resonant islands sit within a chaotic background, and the translunar domain is not merely a uniformly unstable exterior shell. Rather, both are populated by families of motion whose dynamical character must be described simultaneously in terms of resonance, regularity, transport, and fate.

\subsection{Spatiographic Dynamical Maps}
\label{sec:spatio_maps}

\begin{figure}[htp!]
	\begin{center}
	\includegraphics[width=0.495\textwidth]{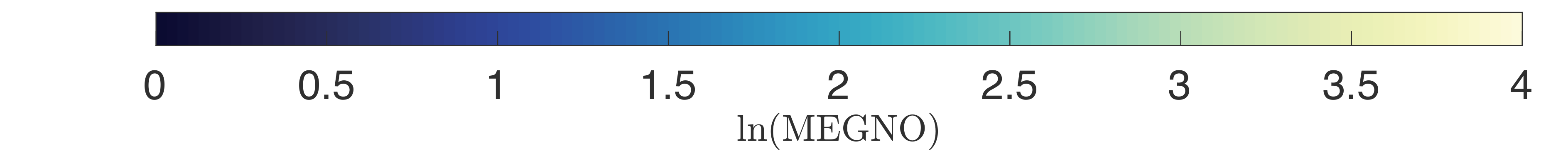}
	\includegraphics[width=0.495\textwidth]{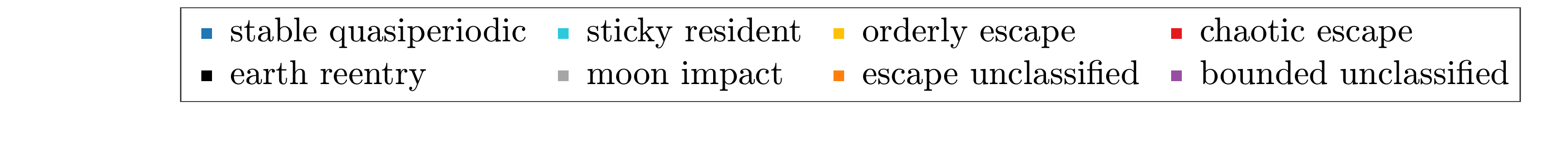}
	\includegraphics[width=0.495\textwidth]{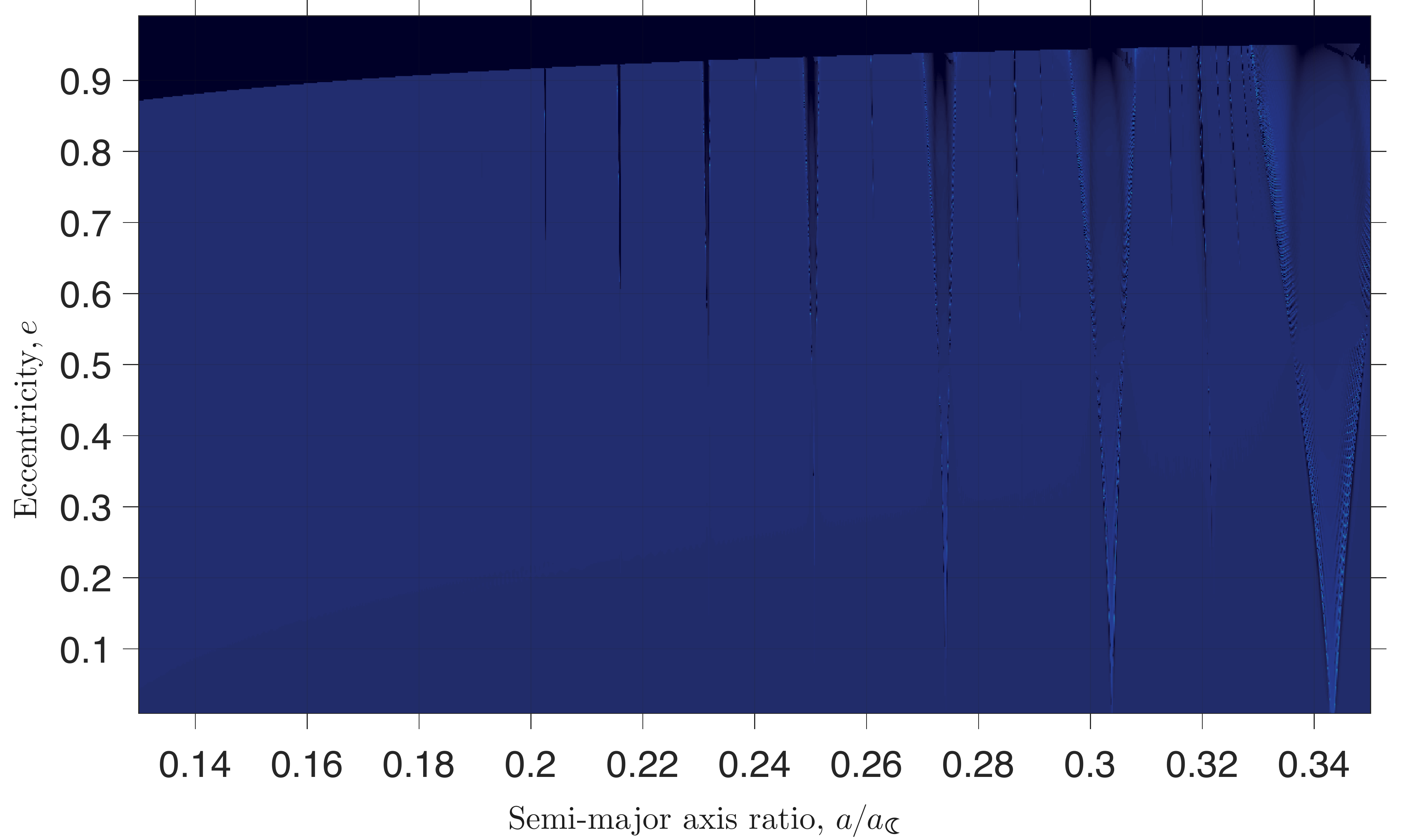}
	\includegraphics[width=0.495\textwidth]{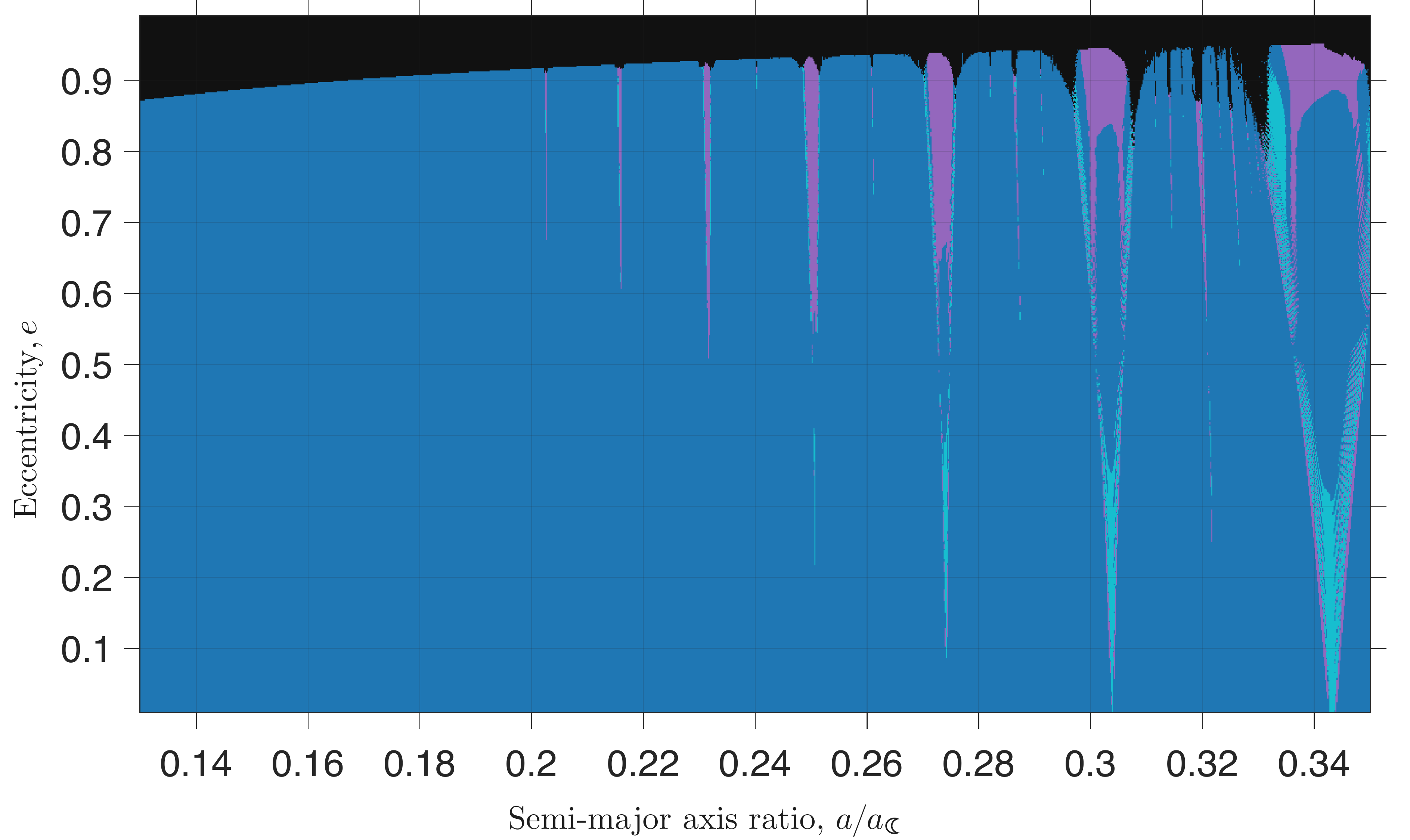}
	\includegraphics[width=0.495\textwidth]{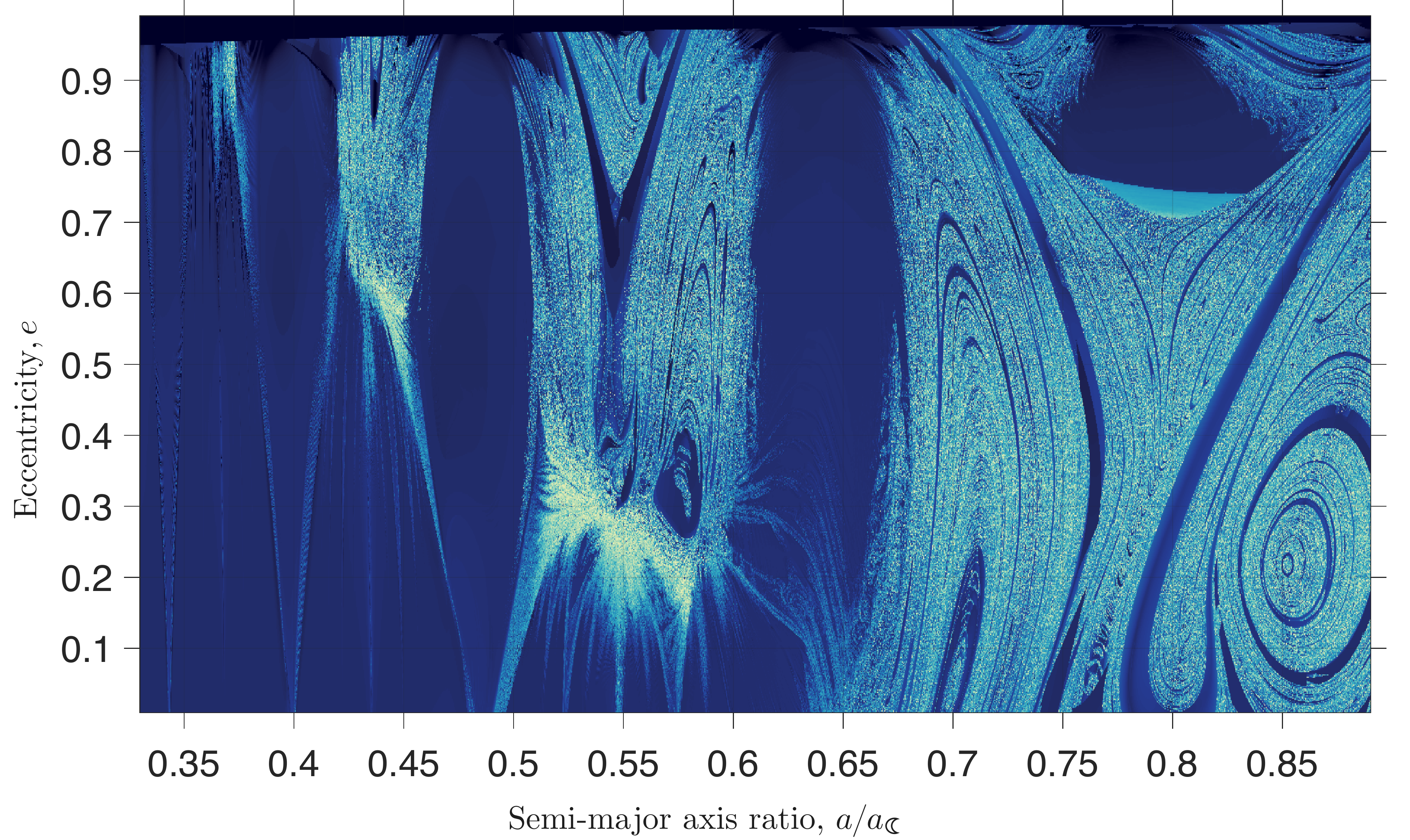}
	\includegraphics[width=0.495\textwidth]{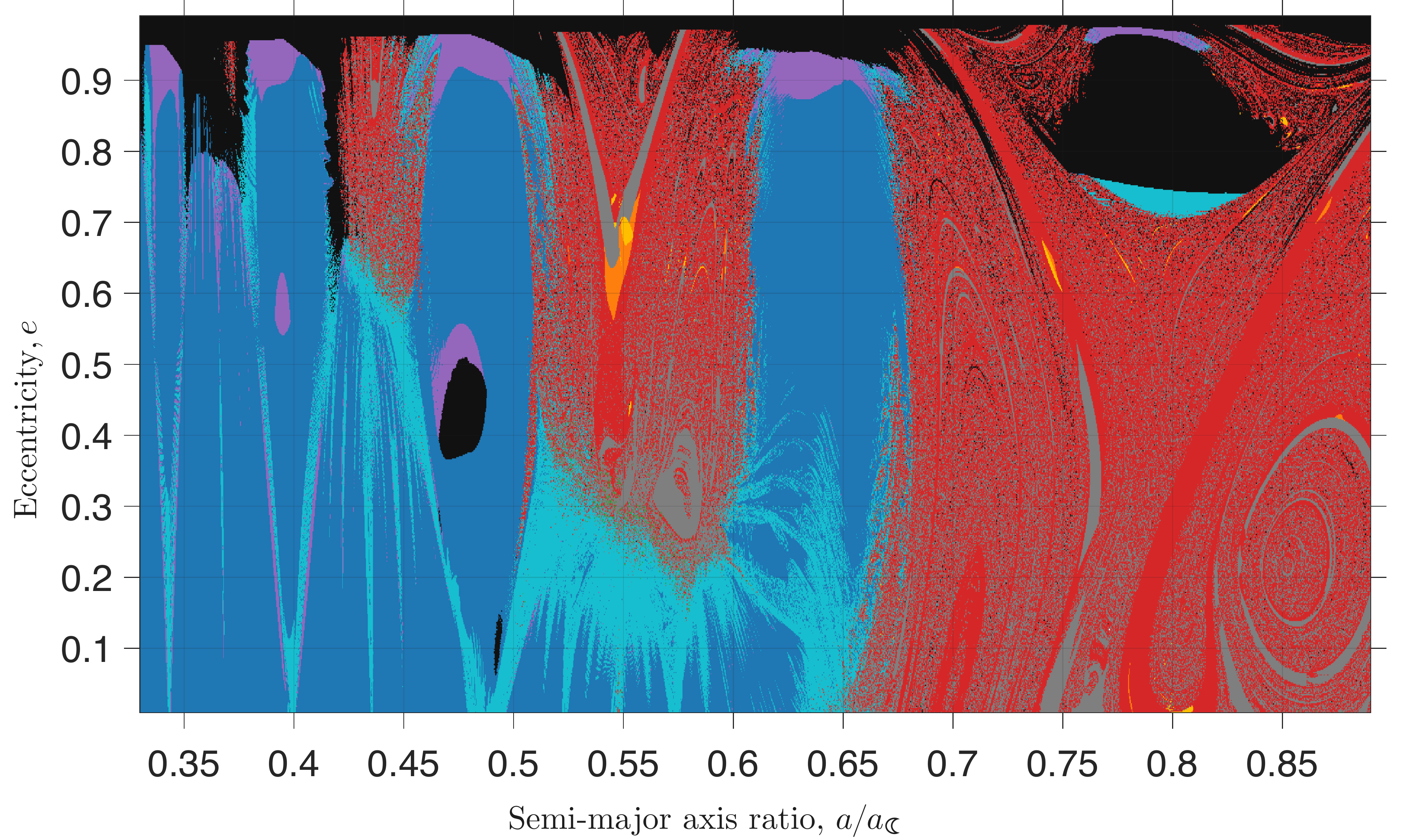}
	\includegraphics[width=0.495\textwidth]{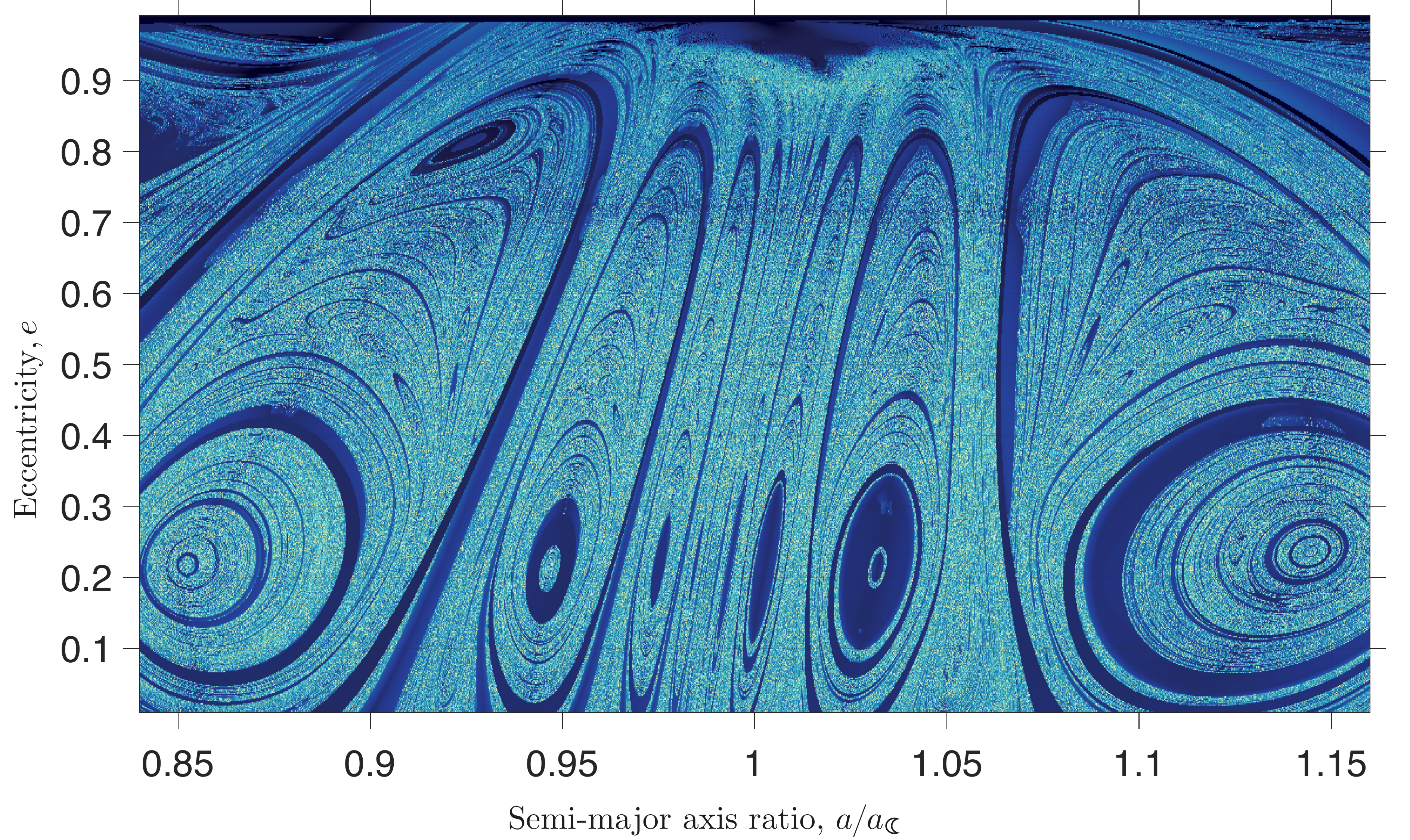}
	\includegraphics[width=0.495\textwidth]{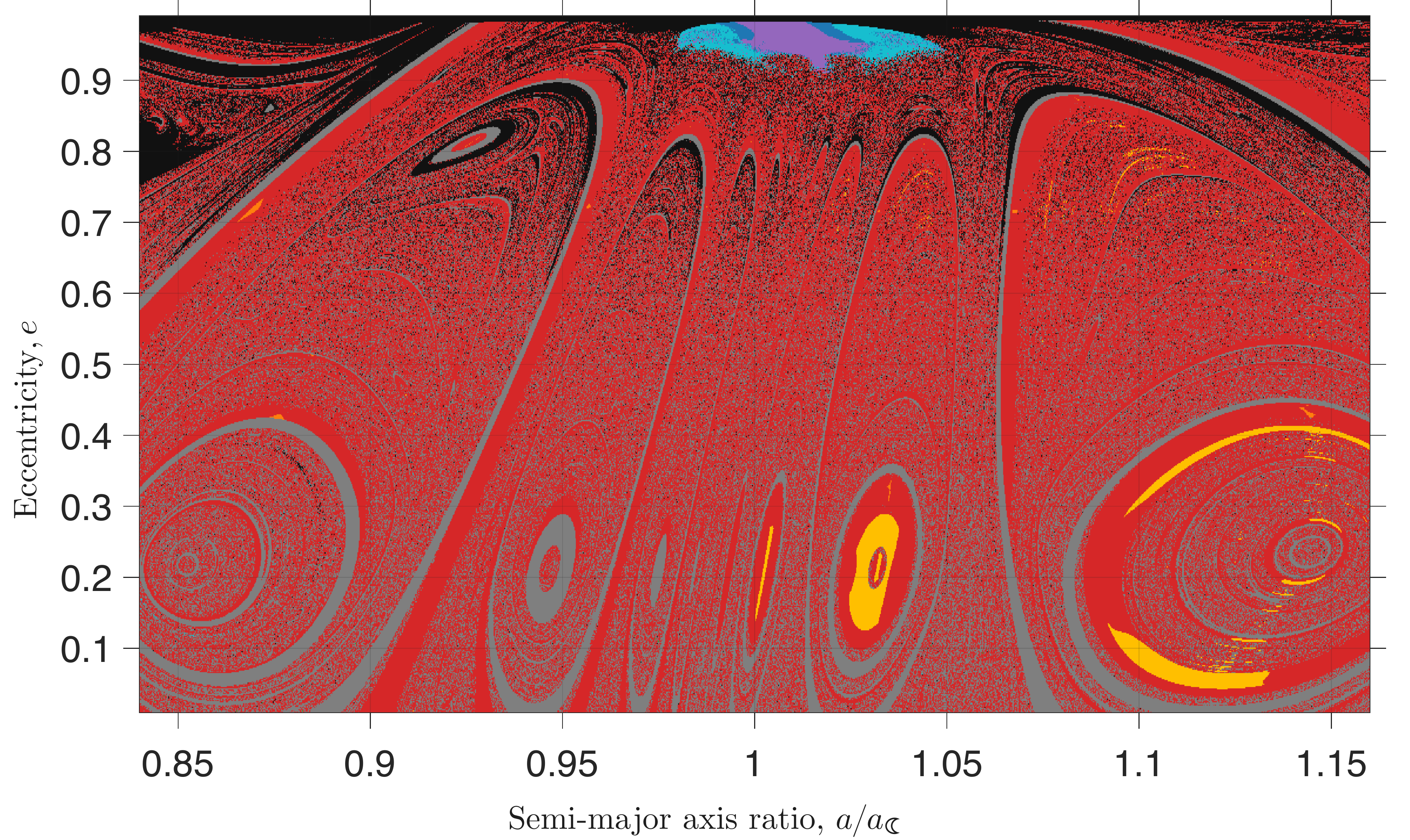}
	\vskip -0.075in
	\caption{\small
Earth--Moon (EM) two-layer astro-cartographies for the cislunar domain---secular-cislunar (SC), cislunar-resonant (CR), and circumlunar-gateway (CG) regimes---computed from the 2027 August~2 total solar eclipse epoch with $(\Omega,\omega,M)=(311.07^\circ,355.84^\circ,0^\circ)$ and inclination set to the lunar orbital plane at epoch. {\it Left column}: MEGNO maps rendered with the Van~Gogh colormap (dark navy: $\bar{Y}\approx2$, ordered; white--yellow: strongly chaotic). {\it Right column}: fate maps; color coding follows the legend and the class definitions of \textsection\ref{sec:fate_classes}. Read together, the two layers expose the transition from a nearly uniform secular background (SC) through the resonant archipelago (CR) to the globally chaotic but island-threaded circumlunar gateway (CG).
        }
	\label{fig:megno_fates_EM_cis}
	\end{center}
	\vskip -0.075in
	\hspace{2cm}\rule{12.5cm}{0.5pt}
\end{figure}

\begin{figure}[htp!]
	\begin{center}
	\includegraphics[width=0.495\textwidth]{megno_banner}
	\includegraphics[width=0.495\textwidth]{fates_banner}
	\includegraphics[width=0.495\textwidth]{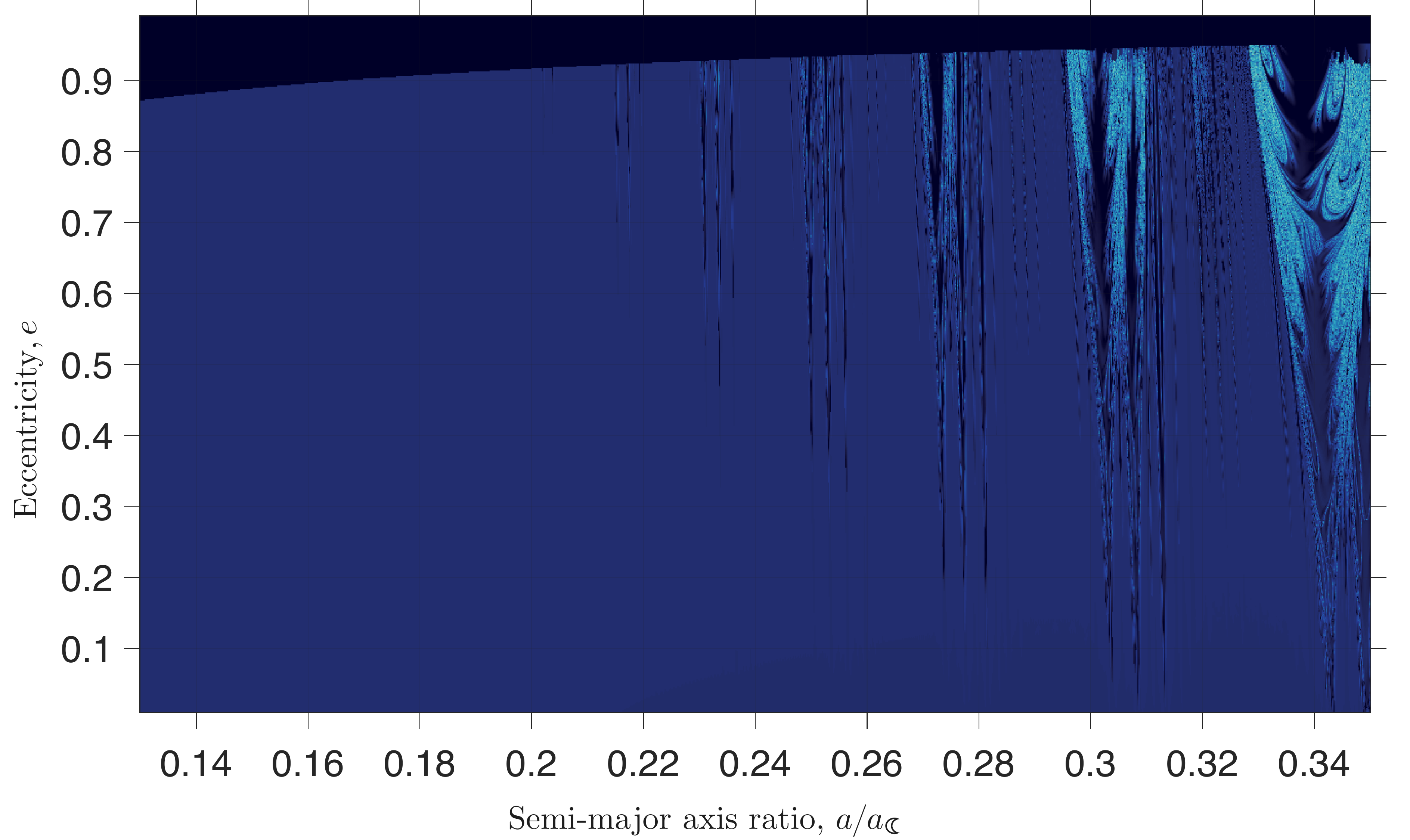}
	\includegraphics[width=0.495\textwidth]{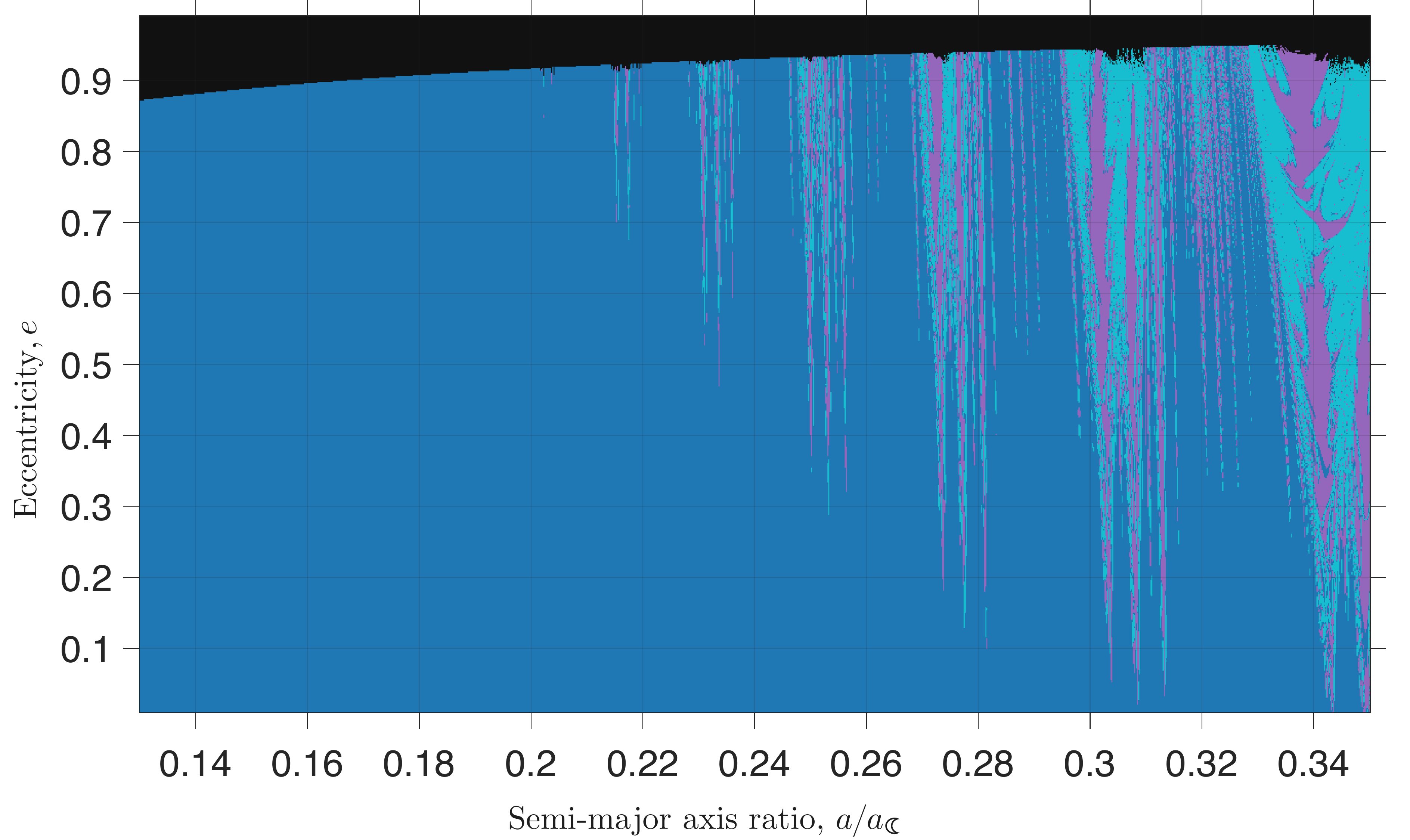}
	\includegraphics[width=0.495\textwidth]{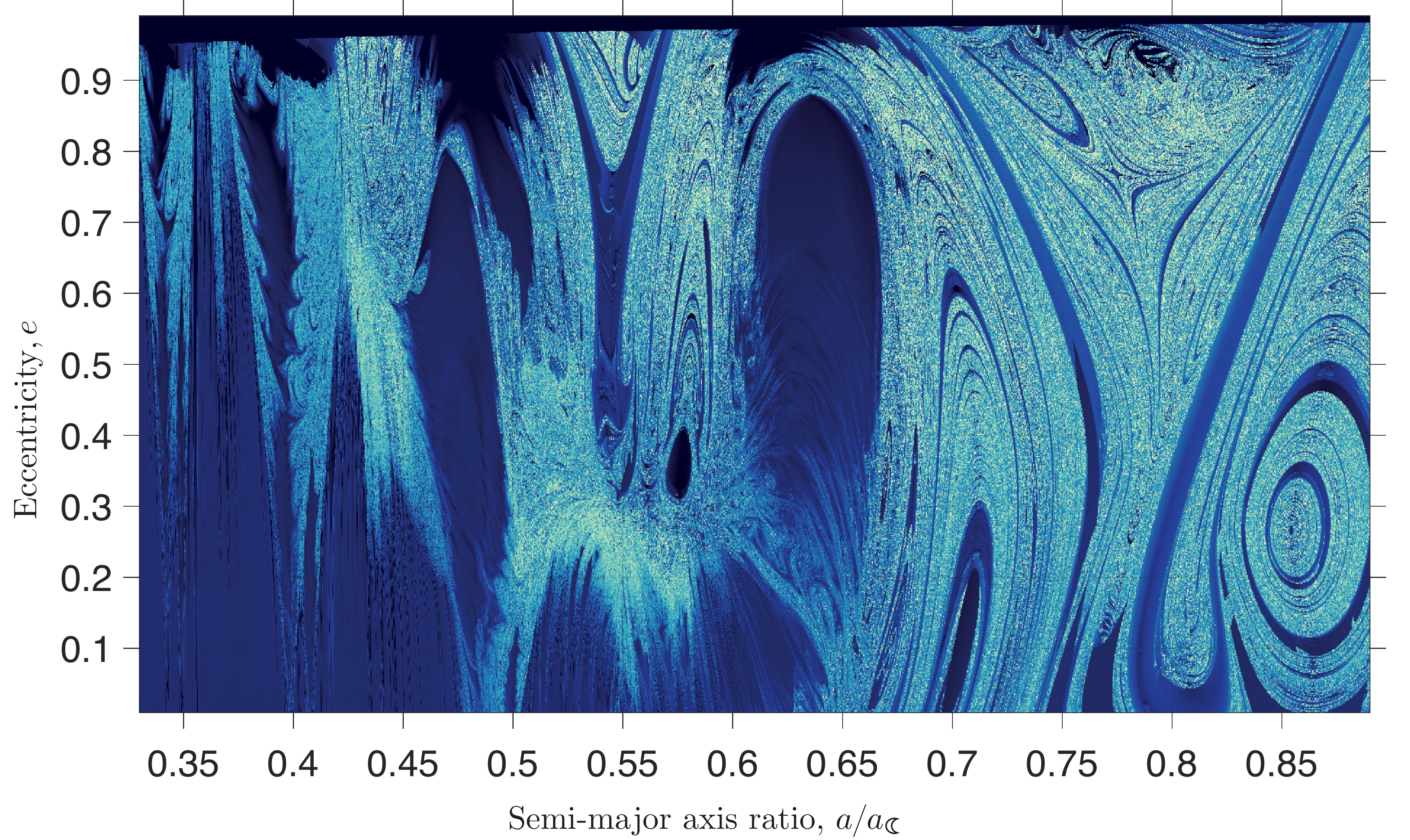}
	\includegraphics[width=0.495\textwidth]{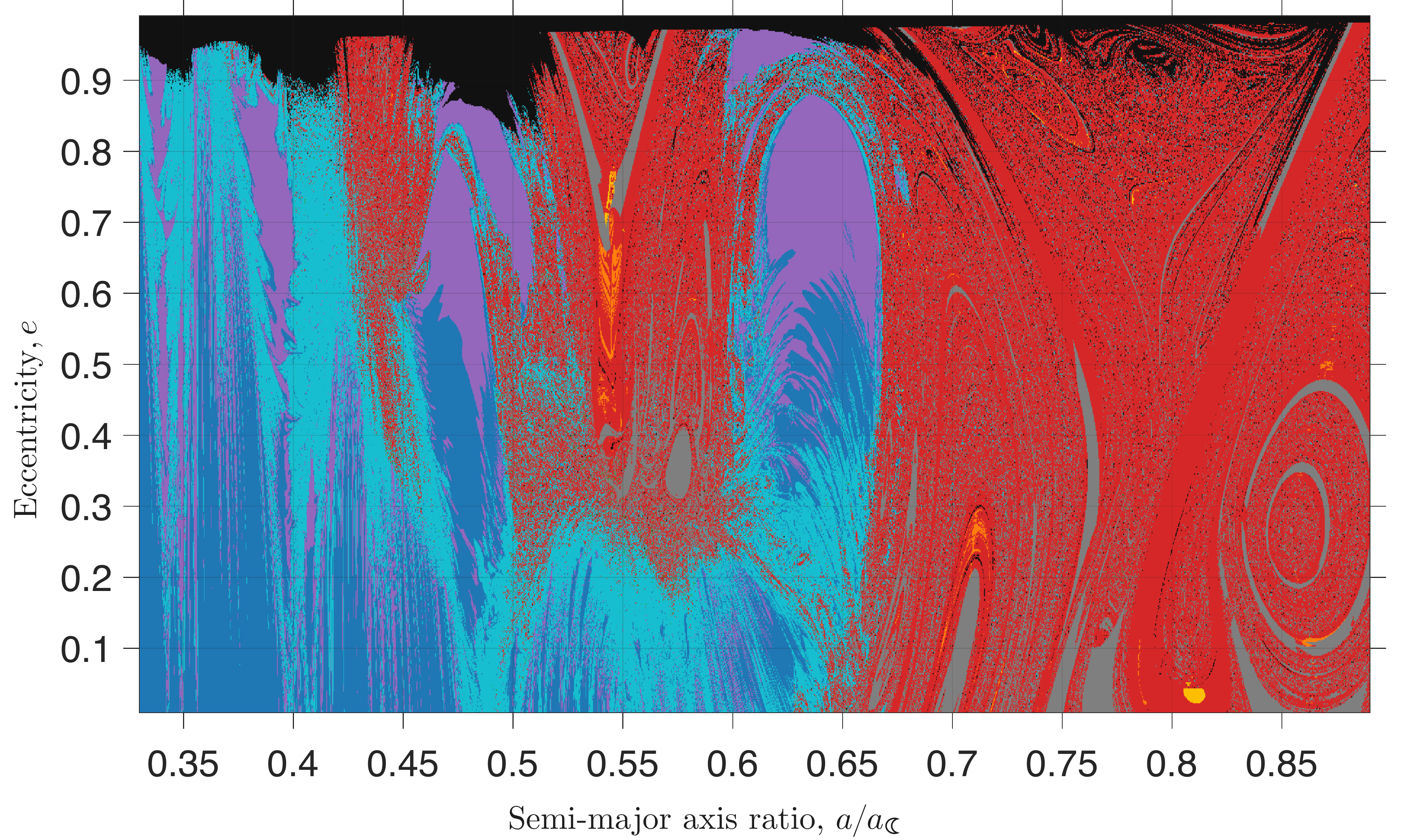}
	\includegraphics[width=0.495\textwidth]{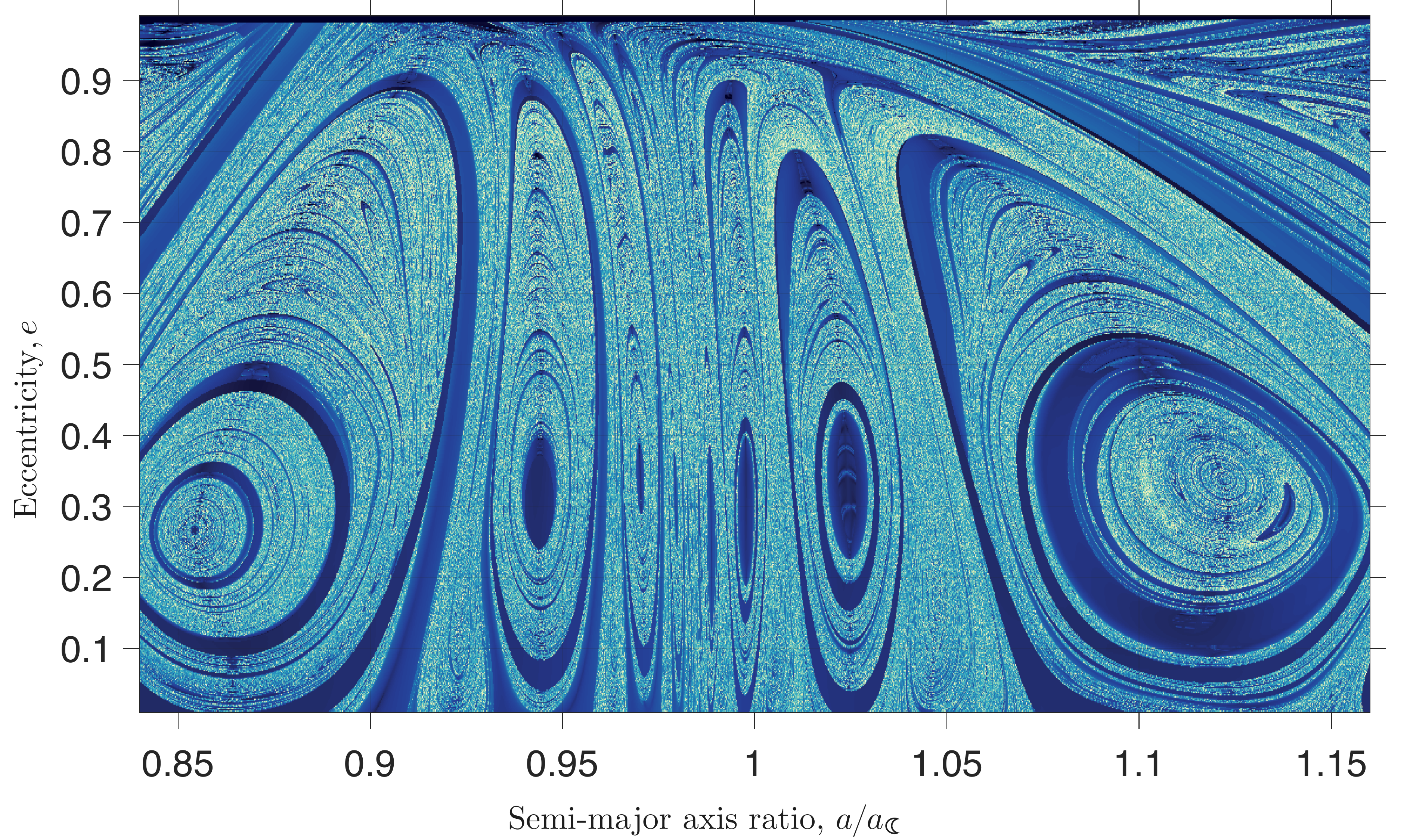}
	\includegraphics[width=0.495\textwidth]{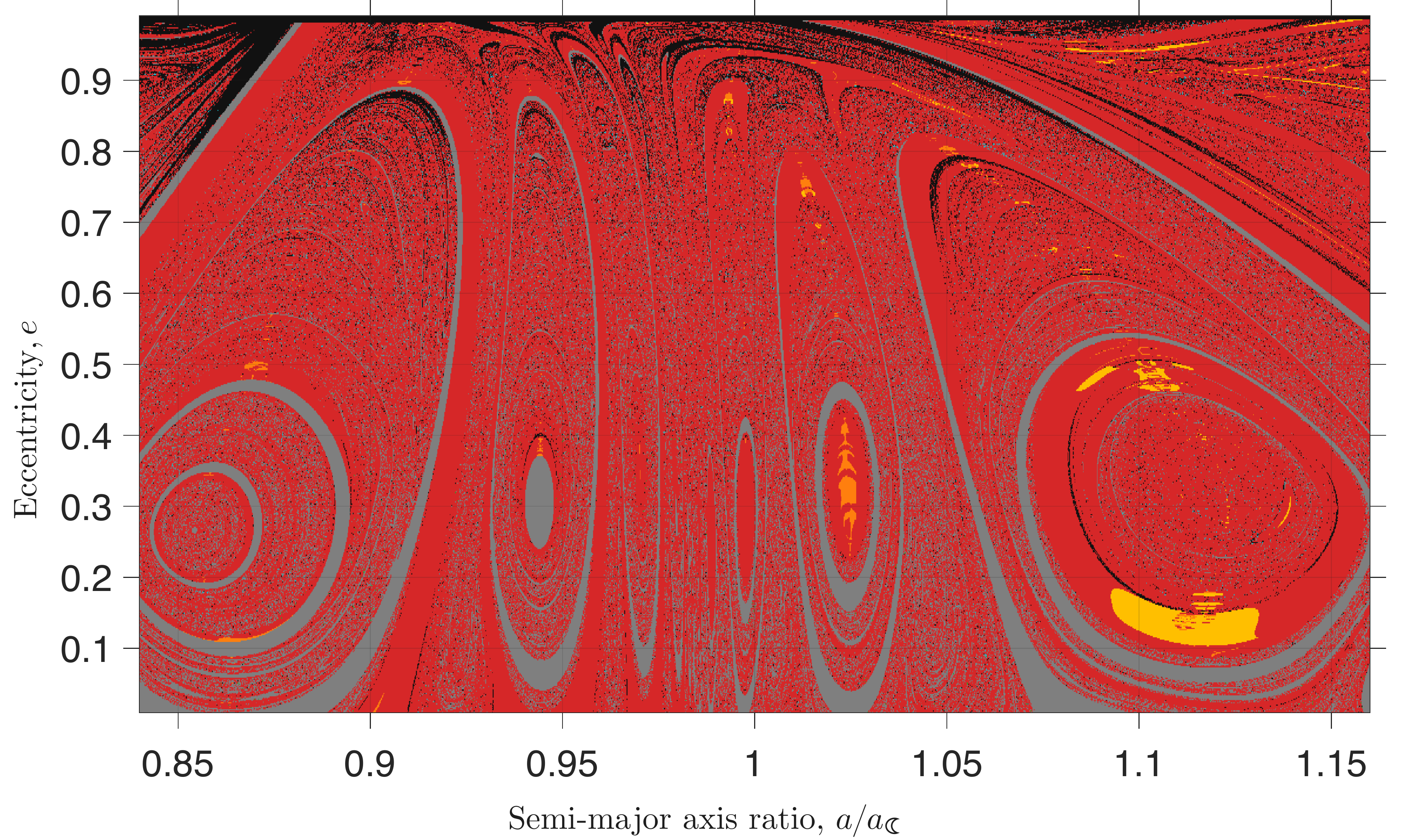}
	\vskip -0.075in
	\caption{\small
Earth--Moon--Sun (EMS) two-layer astro-cartographies for the SC, CR, and CG regimes, for the same epoch and initial angles as Fig.~\ref{fig:megno_fates_EM_cis}. Fate class color coding follows the legend shown at top; class definitions are given in \textsection\ref{sec:fate_classes}. Direct comparison with the corresponding EM panels reveals the qualitative impact of solar forcing on the secular, resonant, and gateway structure of the cislunar domain.
        }
	\label{fig:megno_fates_EMS_cis}
	\end{center}
	\vskip -0.075in
	\hspace{2cm}\rule{12.5cm}{0.5pt}
\end{figure}

\begin{figure}[htp!]
	\begin{center}
	\includegraphics[width=0.495\textwidth]{megno_banner}
	\includegraphics[width=0.495\textwidth]{fates_banner}
	\includegraphics[width=0.495\textwidth]{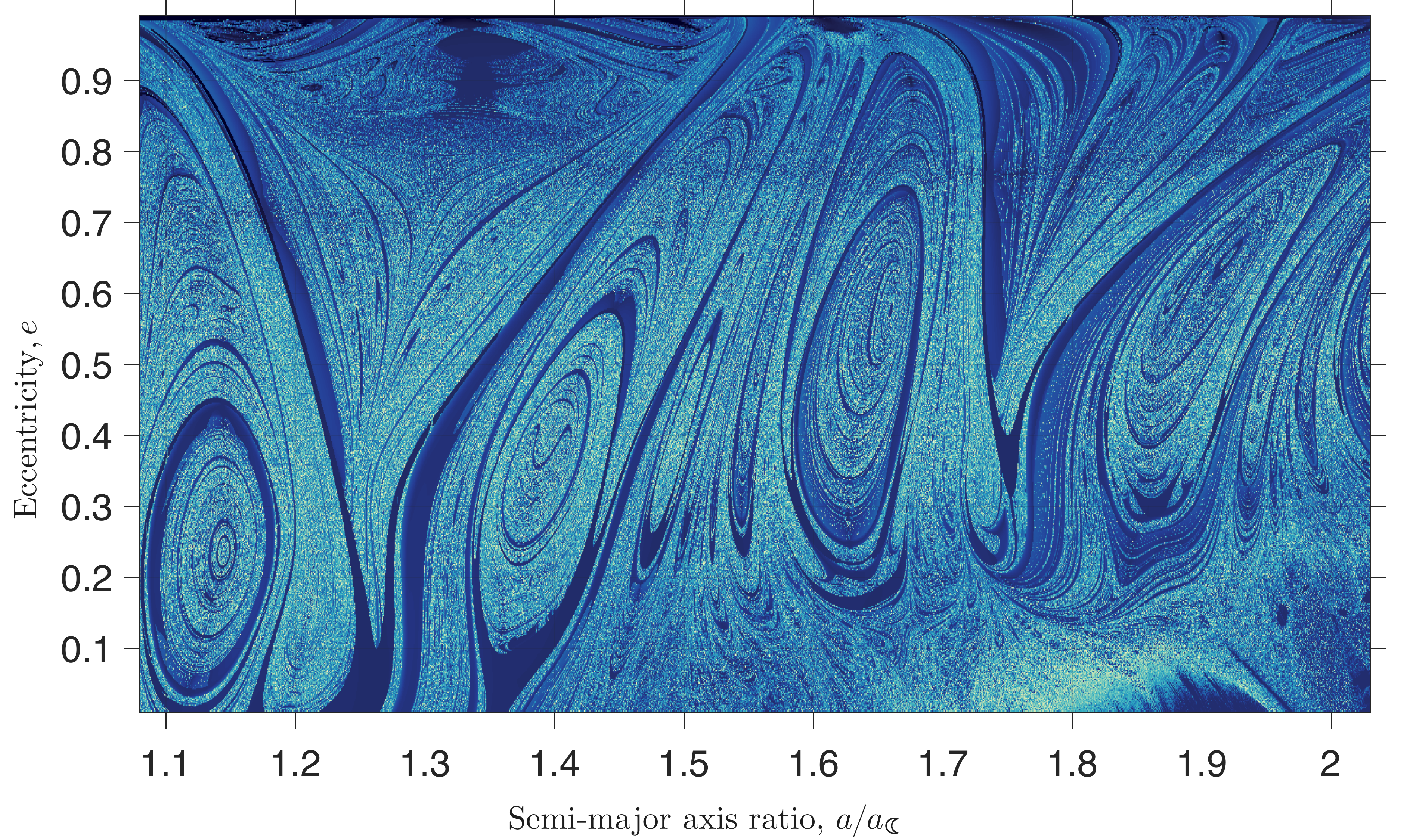}
	\includegraphics[width=0.495\textwidth]{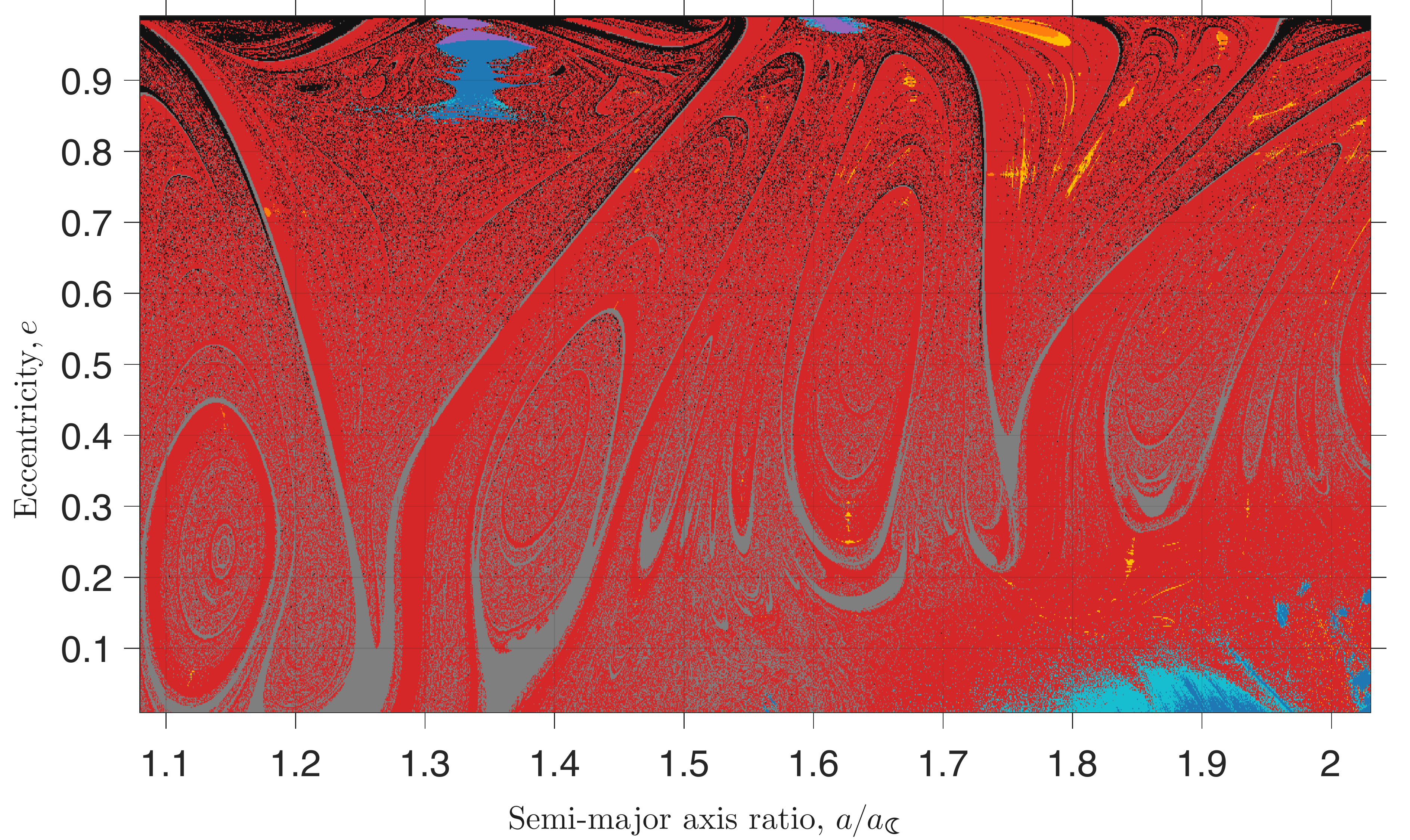}
	\includegraphics[width=0.495\textwidth]{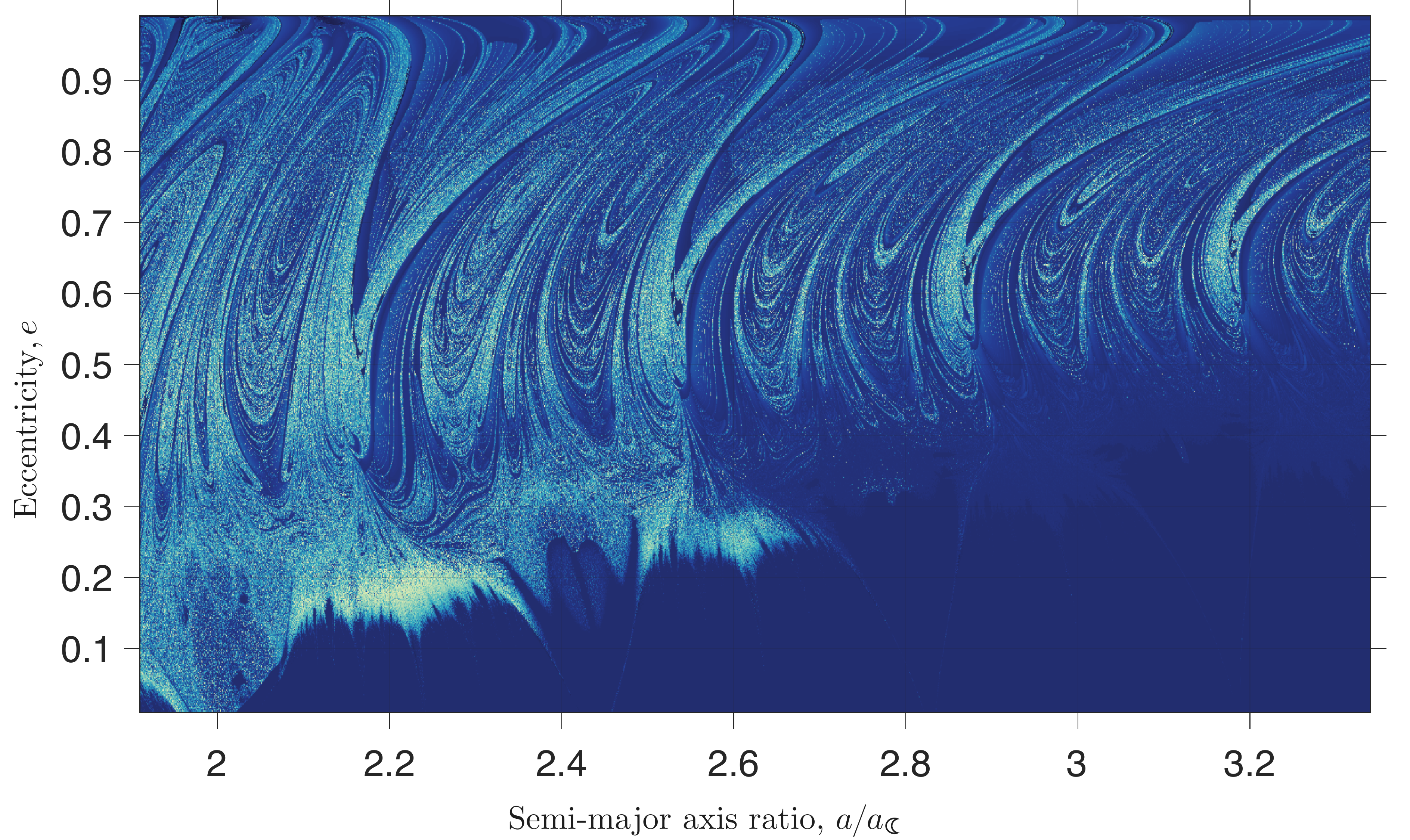}
	\includegraphics[width=0.495\textwidth]{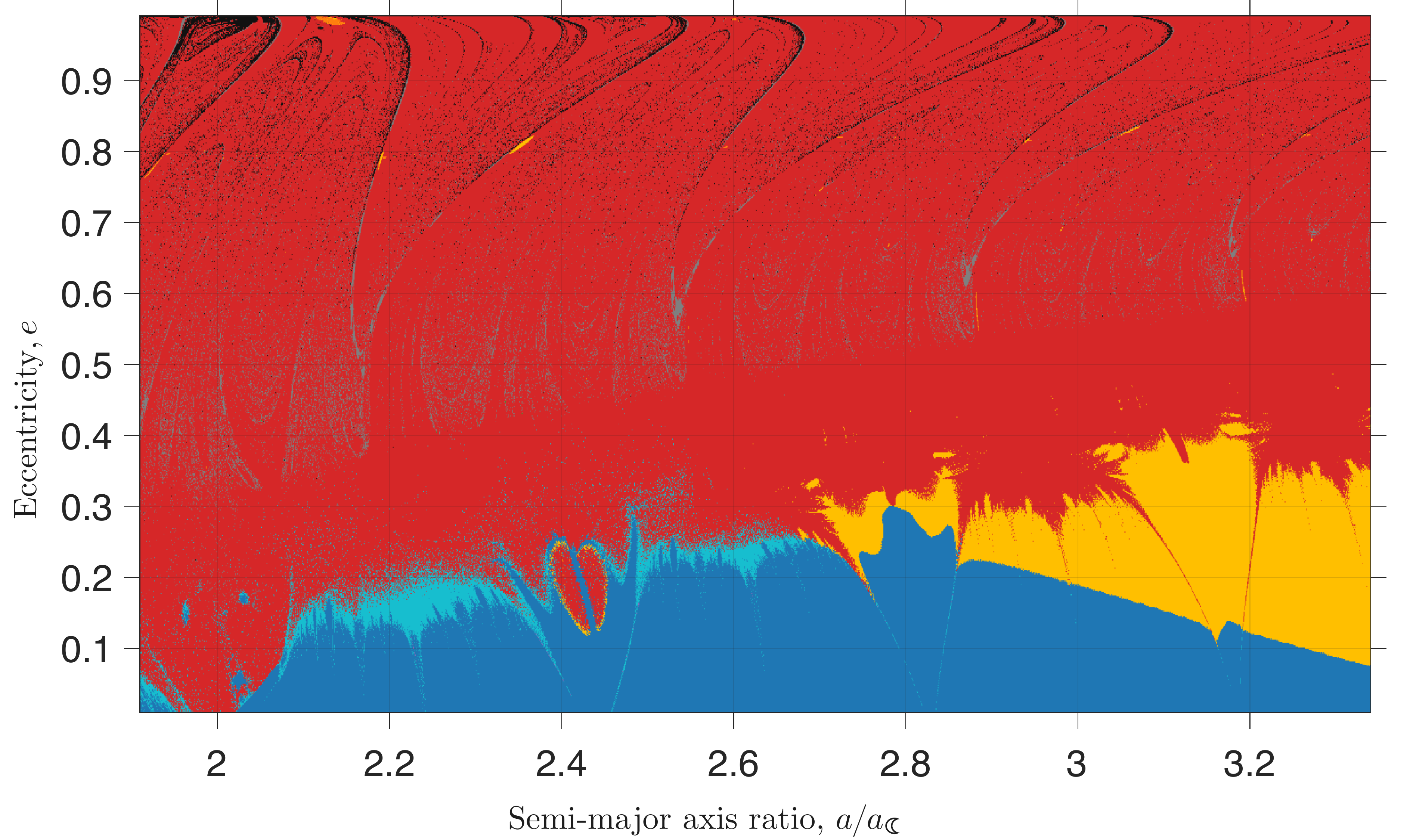}
	\includegraphics[width=0.495\textwidth]{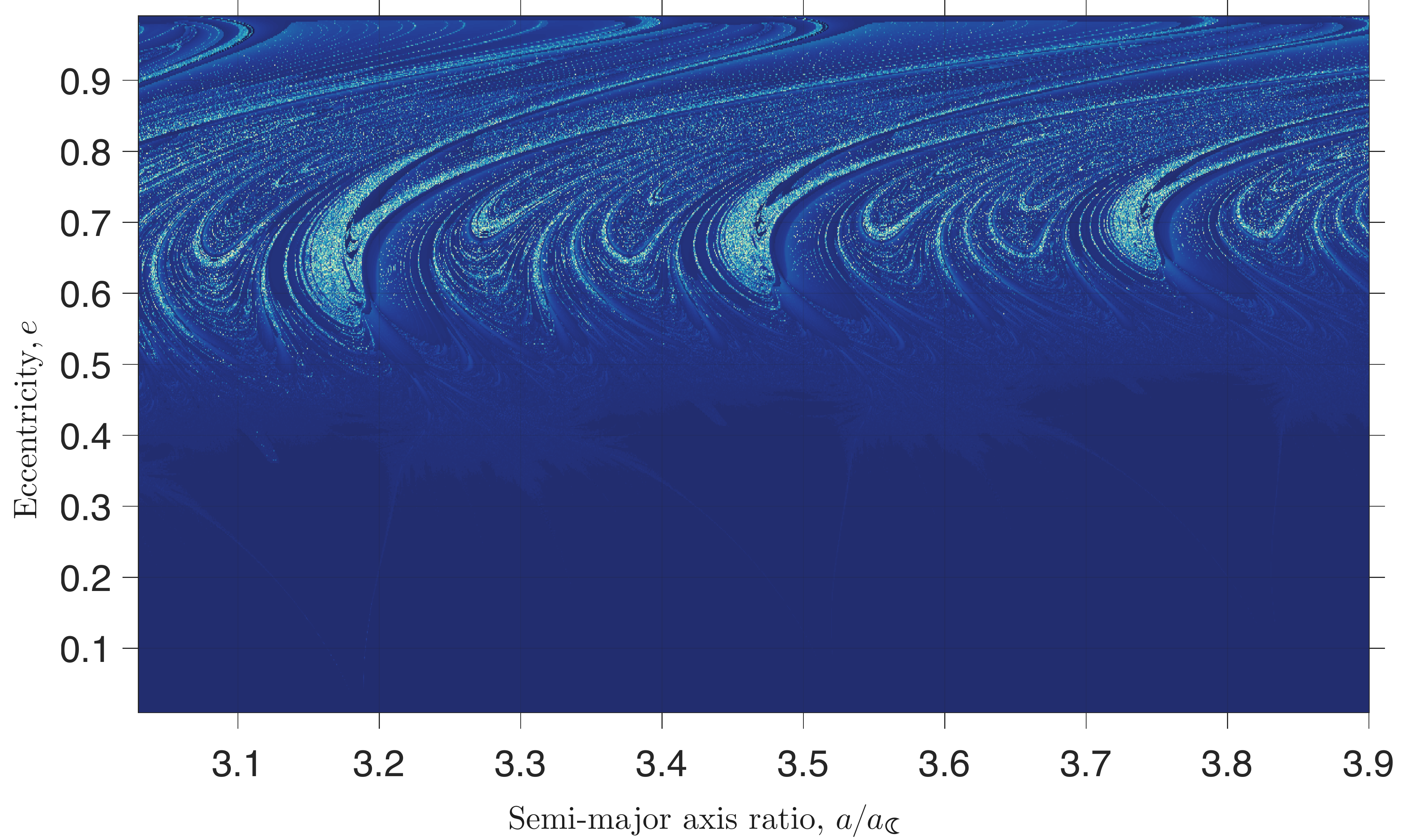}
	\includegraphics[width=0.495\textwidth]{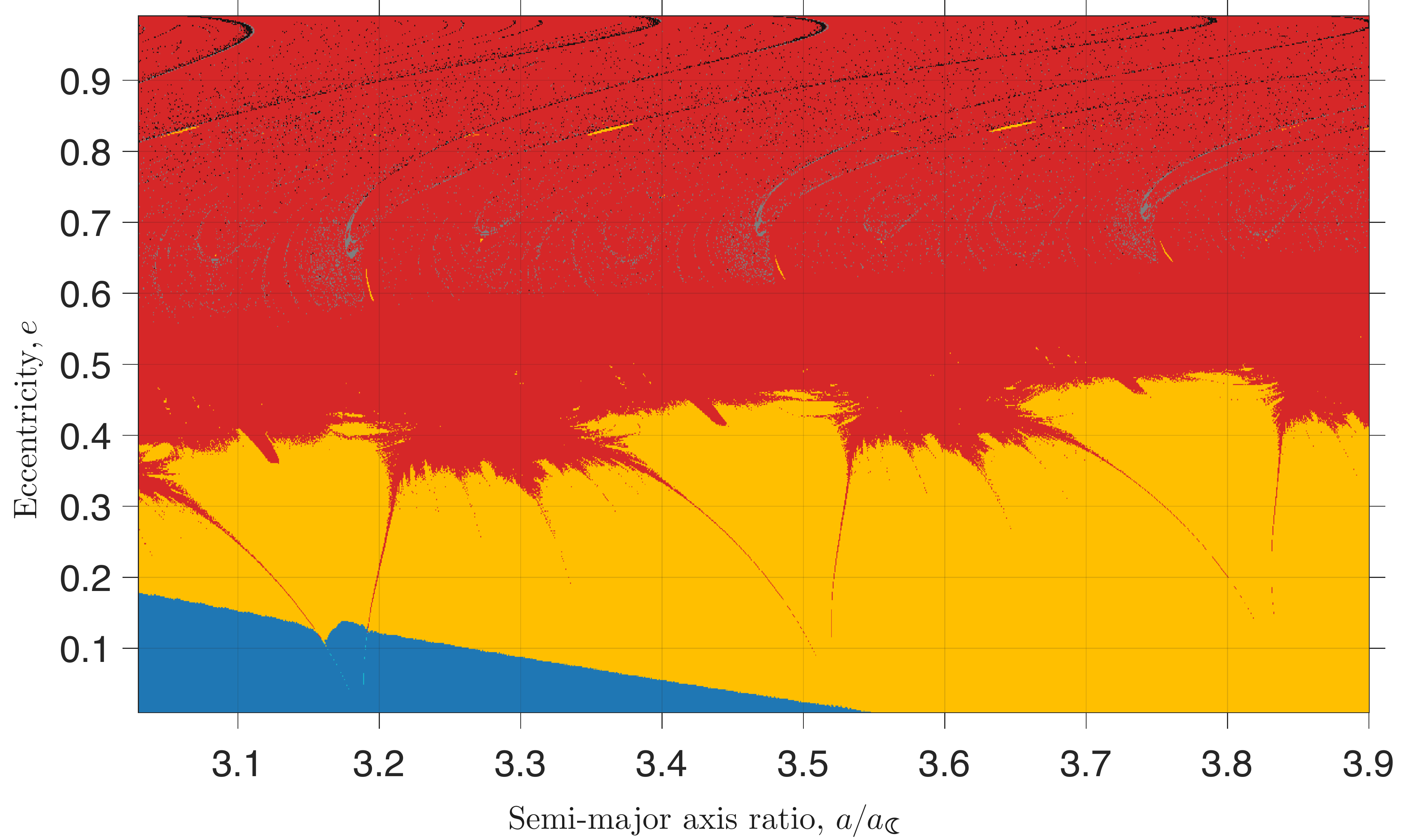}
	\vskip -0.075in
	\caption{\small
Earth--Moon (EM) two-layer astro-cartographies for the translunar domain---inner translunar (IT), outer translunar (OT), and translunar fringe (TF) regimes---for the same epoch and initial angles as Fig.~\ref{fig:megno_fates_EM_cis}. The IT map shows a globally chaotic background punctuated by exterior lunar resonance island chains; the OT and TF maps exhibit the defining shelf-plus-sea structure of the outer circumterrestrial problem, in which a low-eccentricity regular domain is separated by a complex boundary from a chaotic escape sea threaded by resonant filaments tracing exterior lunar commensurabilities.
        }
	\label{fig:megno_fates_EM_trans}
	\end{center}
	\vskip -0.075in
	\hspace{2cm}\rule{12.5cm}{0.5pt}
\end{figure}

\begin{figure}[htp!]
	\begin{center}
	\includegraphics[width=0.495\textwidth]{megno_banner}
	\includegraphics[width=0.495\textwidth]{fates_banner}
	\includegraphics[width=0.495\textwidth]{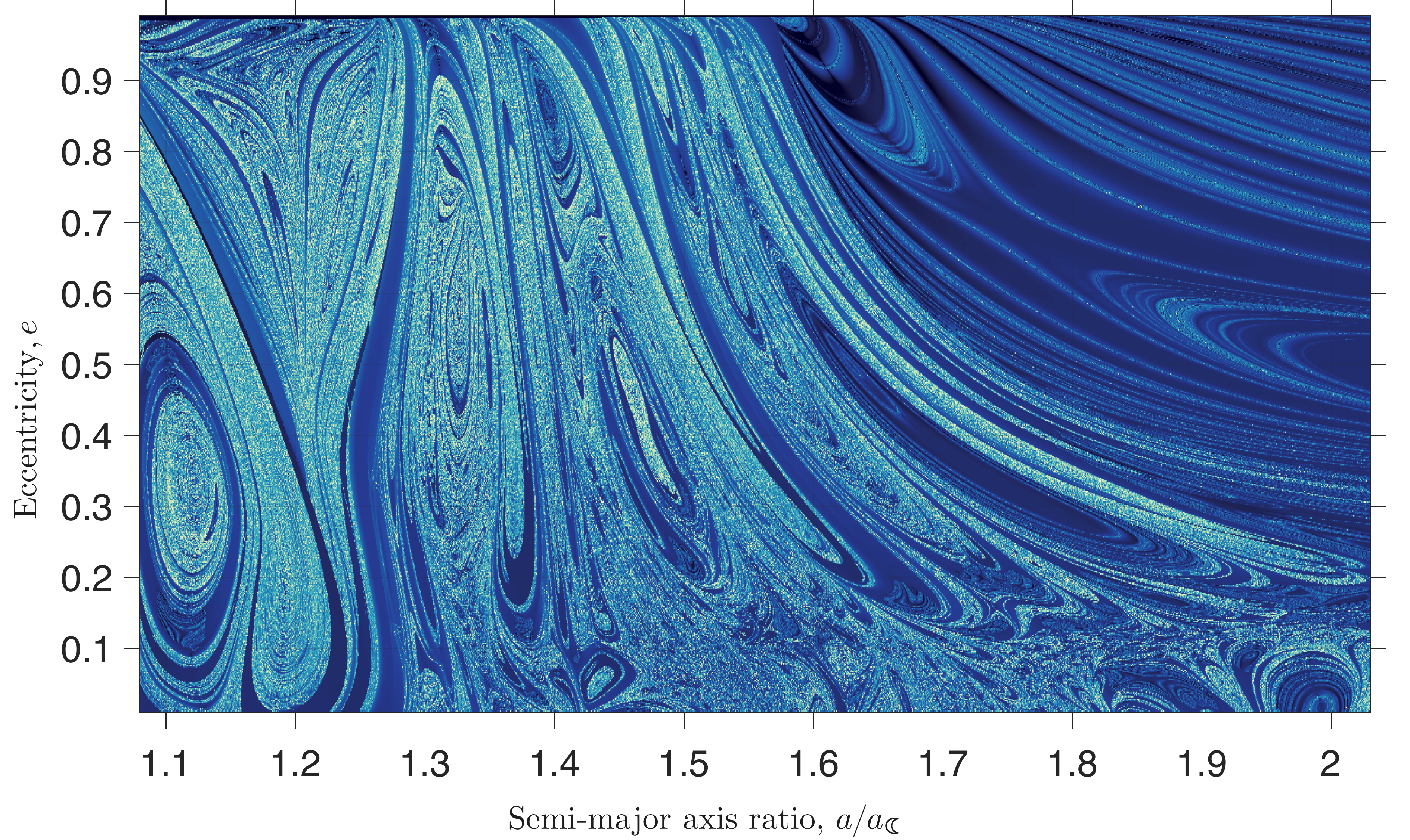}
	\includegraphics[width=0.495\textwidth]{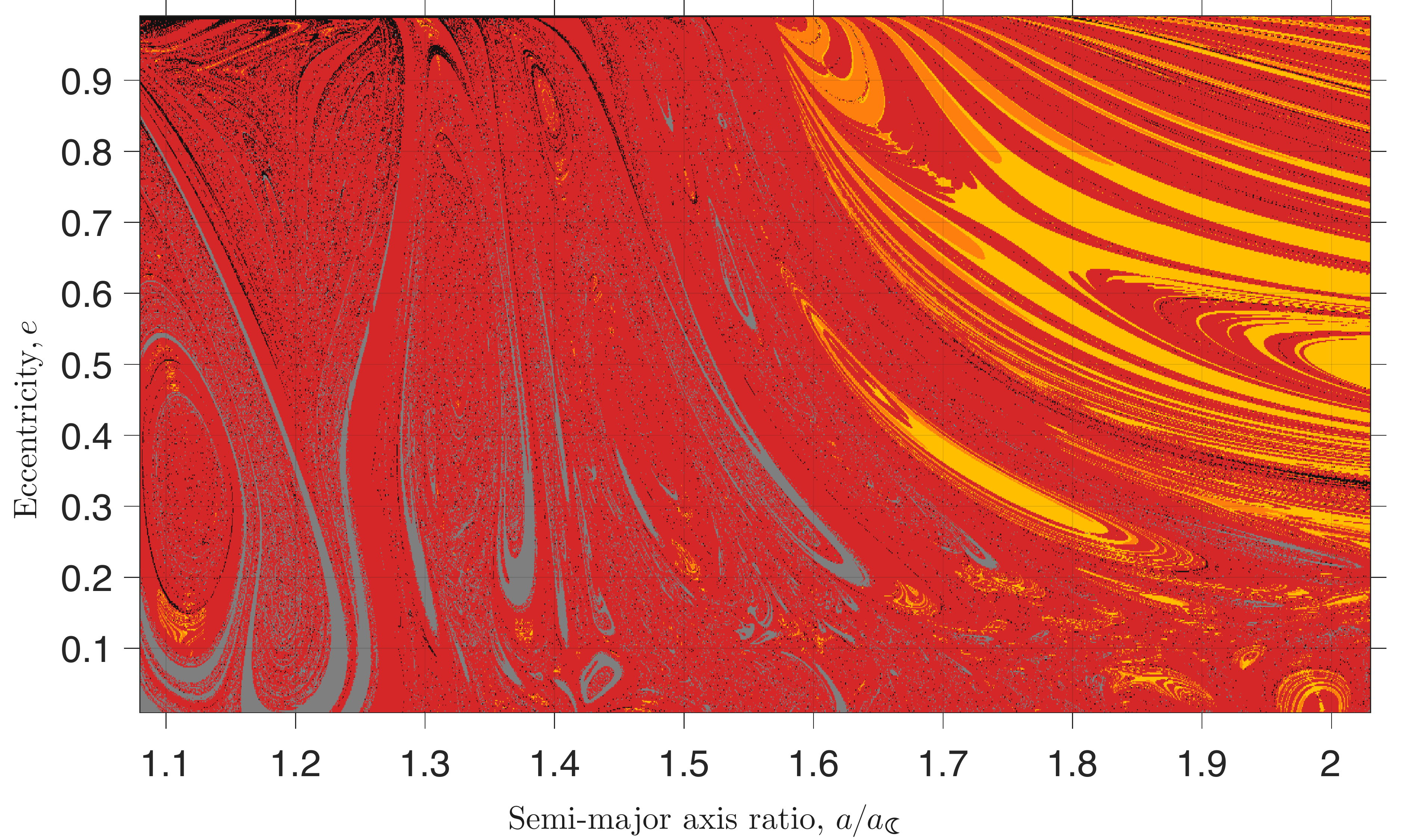}
	\includegraphics[width=0.495\textwidth]{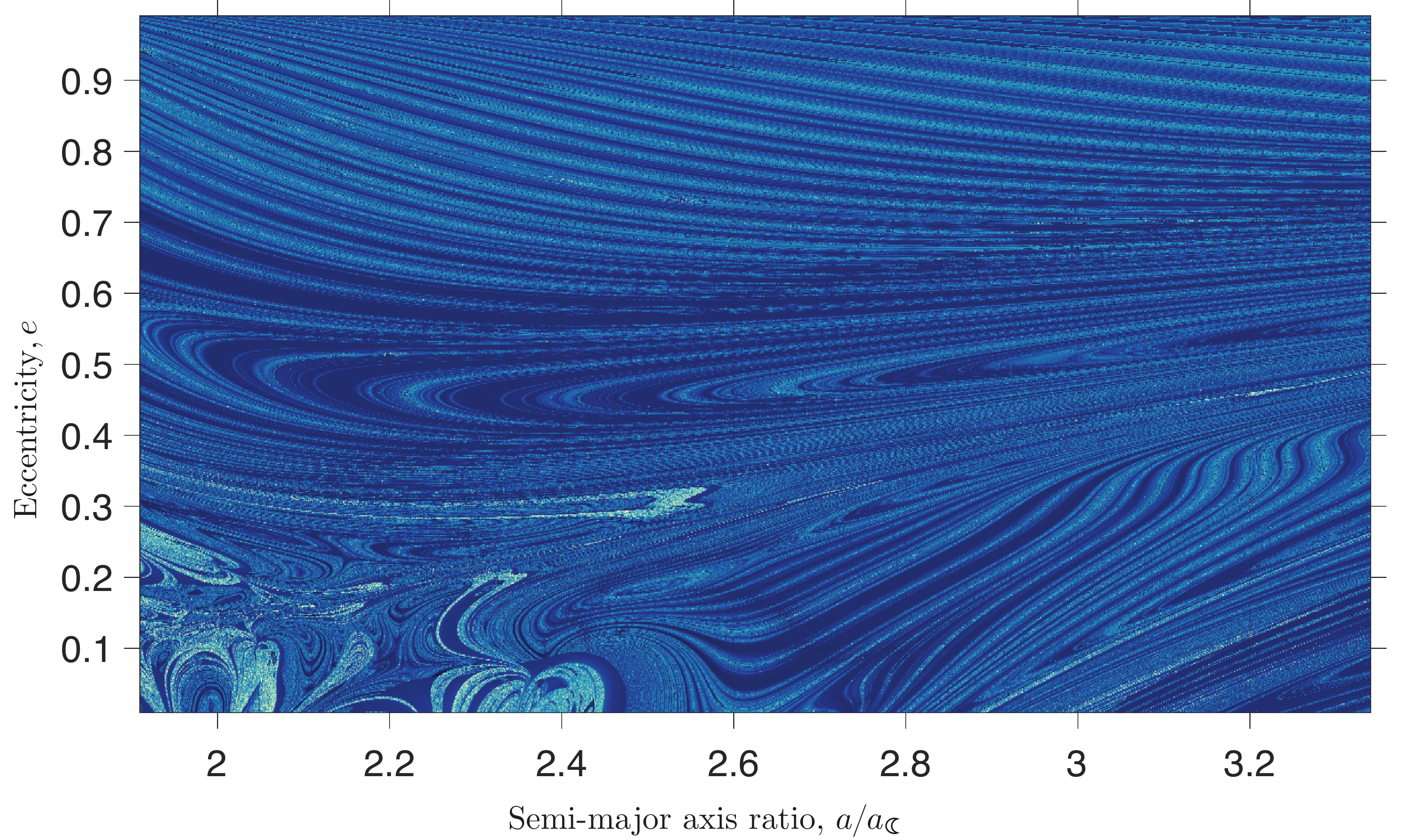}
	\includegraphics[width=0.495\textwidth]{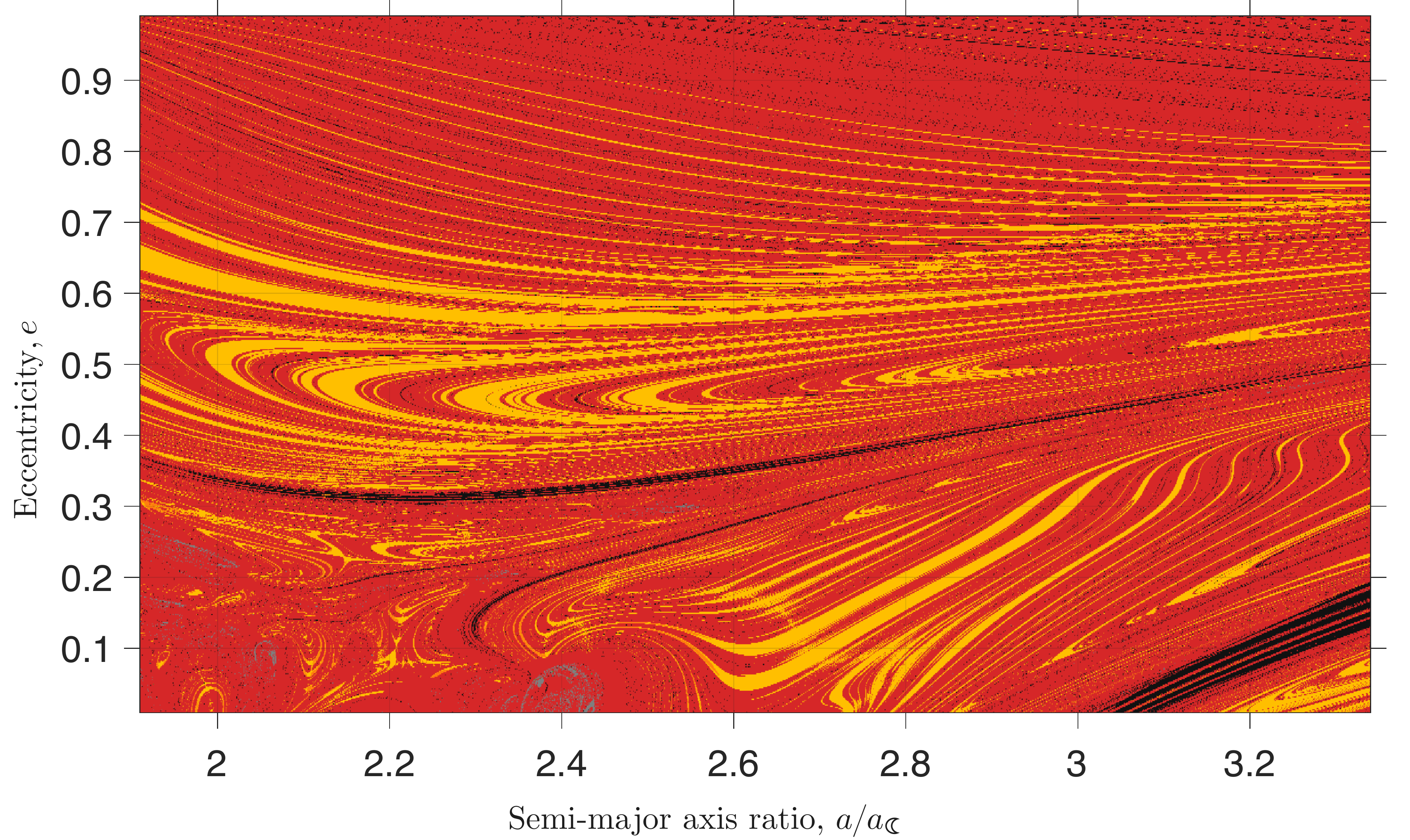}
	\includegraphics[width=0.495\textwidth]{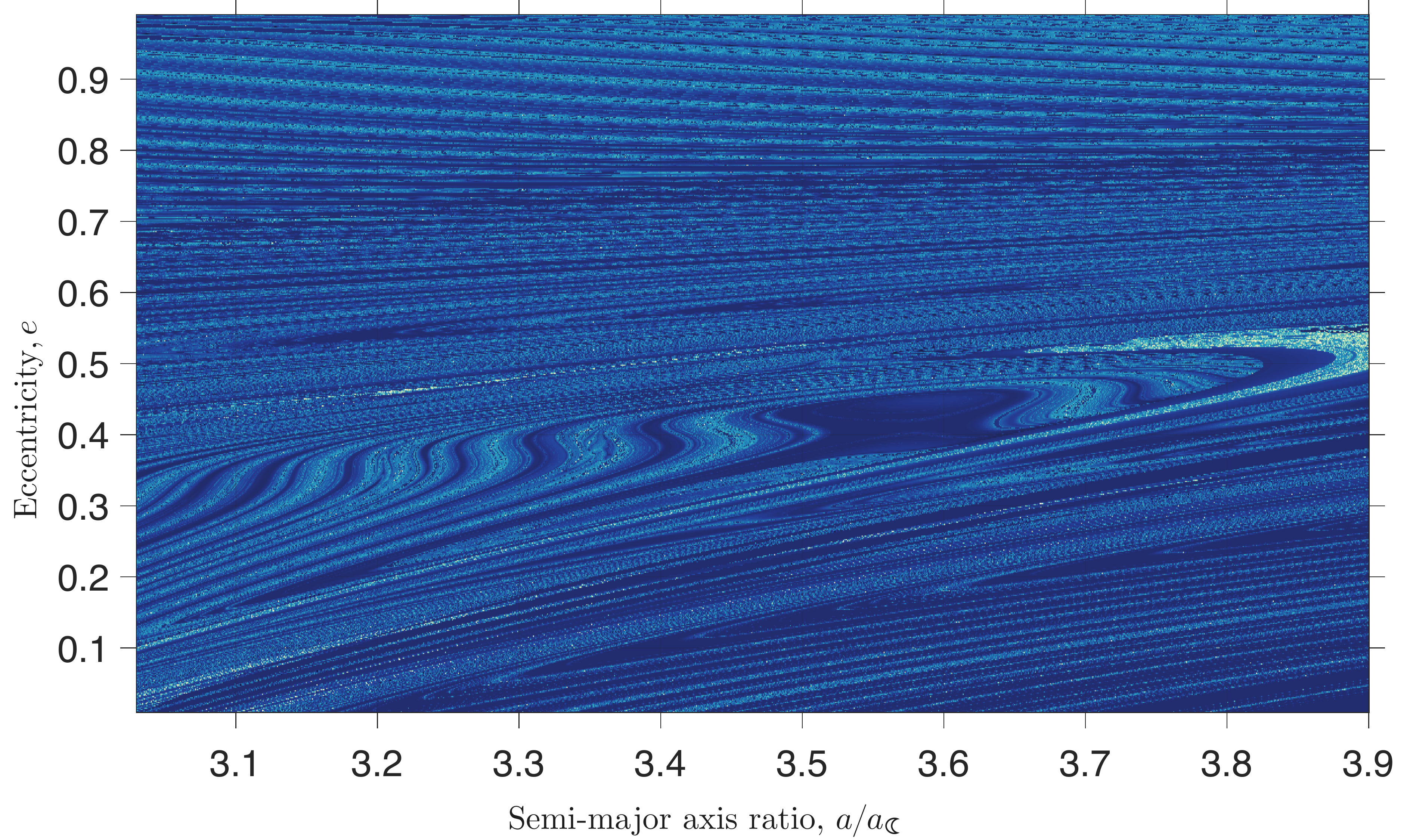}
	\includegraphics[width=0.495\textwidth]{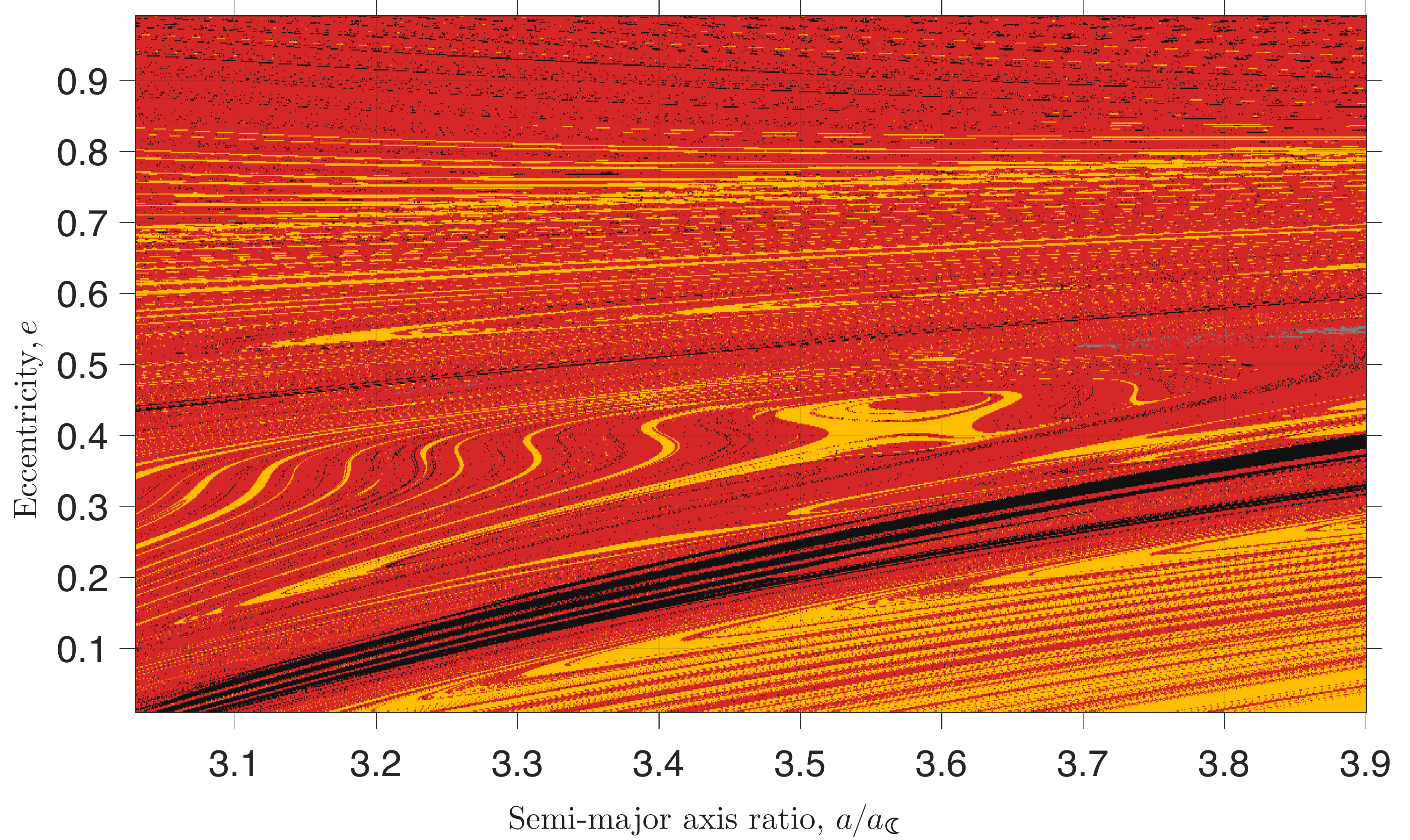}	
	\vskip -0.075in
	\caption{\small
Earth--Moon--Sun (EMS) two-layer astro-cartographies for the IT, OT, and TF regimes, for the same epoch and initial angles as Fig.~\ref{fig:megno_fates_EM_cis}. Fate class color coding follows the legend shown at top; class definitions are given in \textsection\ref{sec:fate_classes}. Direct comparison with the corresponding EM panels reveals how solar forcing reorganizes the outer circumterrestrial architecture.
        }
	\label{fig:megno_fates_EMS_trans}
	\end{center}
	\vskip -0.075in
	\hspace{2cm}\rule{12.5cm}{0.5pt}
\end{figure}

The two-layer astro-cartographies of Figs.~\ref{fig:megno_fates_EM_cis}--\ref{fig:megno_fates_EMS_trans} present $( a , e )$ dynamical maps of circumterrestrial space from the geolunar Laplace transition to the Earth Hill sphere. The maps follow the spatiographic partition of Table~\ref{tab:spatio}, but, for numerical cartography, the broad cislunar, circumlunar, and translunar provinces are refined into the six map domains summarized in Table~\ref{tab:megno_timescales}. All integrations are initialized at the total solar eclipse epoch of 2027 August~2 (UTC 10:06:37). Throughout, the initial angular elements are fixed at $\Omega = 311.07^\circ$, $\omega = 355.84^\circ$, and $M = 0^\circ$, with inclination set to the Moon's orbital plane at epoch. This is the anti-aligned apsidal convention: $\Omega=\Omega_\M$ and $\omega=\omega_\M+180^\circ$ place the test-particle line of apsides anti-parallel to the Moon's apsidal line, while $M=0^\circ$ initializes the orbit at periapsis. In the cislunar resonant regime in particular, this choice aligns the numerical maps with the same periapsis-based phase slice used in the Gallardo atlas and in the periapsis Poincar\'e maps of \citet{aRbK26}, permitting direct comparison among the three representations while still providing a consistent global cut through the full circumterrestrial survey.

Two dynamical configurations are shown. The Earth--Moon (EM) model is an elliptic, point-mass Earth--Moon problem initialized from JPL Horizons, with the spacecraft treated as a massless test particle. It therefore isolates the lunar resonant and gateway architecture in a time-dependent geocentric setting, rather than in the autonomous circular restricted three-body problem. The Earth--Moon--Sun (EMS) model then restores the solar point mass, again initialized from the same epoch, and may be regarded for the present purpose as a Sun-perturbed elliptic Earth--Moon restricted problem, or equivalently as an ephemeris-style restricted four-body model at the level of the dominant point-mass perturbations. The resulting comparison is therefore a test of architectural persistence: from the low-inclination secular cislunar interior, through the resonant and gateway zones, and out into the outer translunar and fringe regimes, the maps show where the lunar elliptic baseline is sufficient, where it is merely suggestive, and where solar forcing reorganizes the fate structure of circumterrestrial motion.

In each panel pair, the left column shows the MEGNO cartography rendered with the ``Van~Gogh'' colormap---dark navy for low-MEGNO, near-quasiperiodic motion, cyan for weakly chaotic or slowly diverging motion, and white-yellow for strongly chaotic variational growth---while the right column shows the companion fate classification for the same grid of initial conditions. The primary fate classes used in the interpretation below are defined in \textsection\ref{sec:fate_classes}: {\it stable quasiperiodic} (navy), {\it sticky resident} (cyan), {\it orderly escape} (gold), {\it chaotic escape} (red), {\it earth reentry} (black), {\it moon impact} (gray), {\it escape unclassified} (orange), and {\it bounded unclassified} (purple). The full numerical classifier also retains auxiliary lunar-neighborhood labels, but these play no central role in the present $( a , e )$ cartographic interpretation. Read together, the MEGNO and fate layers expose both the local variational character of an initial condition and its larger transport role within circumterrestrial phase space.

\paragraph{Secular-cislunar (SC) zone \textnormal{(\(0.13 \lesssim a/a_\M \lesssim 0.35\))}.}
In the low-inclination slice adopted here, the EM SC MEGNO map is dominated by a nearly uniform deep-navy field. This shows that the chosen section through the inner circumterrestrial phase space---beyond the Laplace transition of \textsection\ref{sec:laplace} but still interior to the first low-order lunar commensurability used to define the cislunar resonant zone---is overwhelmingly regular over the integration window. Since the maps are sampled at low ecliptic inclination ($I\approx I_\M\approx5^\circ$), classical Kozai--Lidov oscillations and much of the quadrupole-level apsidal structure are strongly suppressed. The observed regularity therefore reflects the near-integrability of this low-inclination, phase-fixed slice over much of the sampled eccentricity range, rather than a full cartographic imprint of the secular scaffold described in \textsection\ref{sec:secular}. The visible departures from the quasiperiodic background are thin resonant filaments at isolated commensurability locations. These appear already by $a/a_\M \sim 0.20$, well before the nominal \res{5}{1}{\M} boundary, and are naturally interpreted as the higher-order resonant texture anticipated in \textsection\ref{sec:cislunar}: dynamically real, but not part of the low-order scaffold used to define the principal spatiographic partition. They remain sparse through the inner part of the SC zone, but thicken progressively toward $a/a_\M\approx0.30$--$0.35$, where the influence of the \res{5}{1}{\M} family begins to disturb the background and marks the transition into the cislunar resonant zone.

The EM companion fate map confirms this reading. Away from the Earth-grazing boundary, stable quasiperiodic motion (navy) overwhelmingly dominates, indicating that most non-terminal initial conditions in this slice remain Earth-bound and regular over the full integration window. A broad black reentry band marks the Earth-grazing condition $a ( 1 - e ) \le R_\E$, sweeping downward in $a$ as $e \rightarrow 1$. Along the thin resonant filaments seen in the MEGNO layer, the fate map reveals narrow threads of sticky residents (cyan) and bounded-unclassified trajectories (purple): the first indication that bounded irregular or slowly convergent behavior enters the circumterrestrial problem already near the outer edge of the secularly dominated zone, even before the cislunar resonant structure becomes globally conspicuous.


The EMS comparison (Fig.~\ref{fig:megno_fates_EMS_cis}) shows that, in this particular low-inclination slice, solar forcing does not destroy the basic inner-cislunar architecture, but it does refine it in a dynamically meaningful way. The broad regular background persists, and the overall fate structure remains dominated by stable quasiperiodic motion together with Earth reentry along the grazing boundary. The principal EMS signature is instead concentrated on the thin resonant filaments: the comparatively clean EM strands are no longer single isolated lines, but are resolved into fine multiplet-like substructure, suggestive of solar sidebands or mixed lunisolar commensurabilities attached to the underlying lunar resonances. These splittings are visible already well inside the nominal \res{5}{1}{\M} boundary, including the higher-order filamentary texture near $a/a_\M \sim 0.2$, and they thicken toward the outer edge of the SC zone. This behavior is consistent with the interpretation emphasized in \textsection\ref{sec:secular}: the present slice does not directly expose the full lunisolar secular web, but samples a low-inclination section for which the dominant secular architecture remains comparatively quiet. The SC comparison therefore serves a useful baseline role. It shows that the Sun is not yet architecturally decisive in the sense of reorganizing the full map into an escape-dominated or resonance-overlap sea; rather, in this zone it first appears as a fine-structure perturbation of an otherwise regular low-inclination secular background. A quantitative theory of the sideband spacing, based on the mixed lunar--solar resonance conditions, is left for future work.

\paragraph{Cislunar-resonant (CR) zone \textnormal{($0.33 \lesssim a/a_\M \lesssim 0.89$)}.}
The transition into the CR zone is immediate and dramatic. The nearly featureless secular background gives way to a dense resonant archipelago in which large dark-navy island chains are embedded within an increasingly bright chaotic sea. The principal low-order lunar families predicted by the Gallardo-style atlas of \textsection\ref{sec:resonances}, and whose global phase-space structure is furnished by the periapsis Poincar\'e section of Fig.~\ref{fig:periapsis_poincare}, appear as coherent island-and-separatrix structures at the expected locations: the \res{5}{1}{\M} family ($a/a_\M\approx0.34$) as a narrow V-shaped resonant envelope at the left edge of the map; the \res{4}{1}{\M} family ($\approx0.40$) as a pair of elongated arches flanked by thin chaotic layers; the \res{3}{1}{\M} family ($\approx0.48$) as a prominent island complex with a well-developed dark interior and a broad low-eccentricity separatrix fan; and the \res{2}{1}{\M} family ($\approx0.63$) as the most expansive stable island in the cislunar domain, spanning a wide eccentricity range consistent with its dominant width in the Gallardo atlas and with the phase portraits of \citet{aRbK26}. The \res{3}{2}{\M} neighborhood ($a/a_\M \approx 0.76$) is weaker but still traceable: in the Gallardo atlas it appears as an hourglass-like, double-silhouette envelope, of which mainly the upper high-eccentricity portion appears in the MEGNO map as a dark, low-variational pocket bordered by brighter filamentary layers. Between the major islands, the inter-resonance background brightens progressively with increasing $a$, tracing the onset of the Chirikov-type resonance-overlap condition discussed in the preceding section. Adjacent lunar families begin to impinge upon one another, their chaotic layers merge, and the dark elliptic cores that dominate the inner resonant ladder give way to filamentary MEGNO texture, sticky corridors, and increasingly connected chaotic transport. The result is not an abrupt disappearance of ordered motion, but a progressive fragmentation of the resonant archipelago: beyond the \res{3}{2}{\M} neighborhood, no comparably broad, long-lived island family survives in the present slice, and the remaining quasiperiodic structure appears only as narrower and more fragile remnants embedded in the growing chaotic sea approaching the circumlunar gateway.

The companion fate map makes this structure dynamically legible. Stable quasiperiodic trajectories (navy) occupy the dark MEGNO island interiors and mark the long-lived survival cores of the cislunar resonant architecture. Around them, broad fans of sticky residents (cyan) spread across regions where trajectories remain Earth-bound over the integration window but exhibit elevated variational growth. These fans are especially conspicuous around the \res{3}{1}{\M} island, whose low-eccentricity flank opens into a wide separatrix-like boundary region. Their morphology is consistent with the broad separatrix regions and inter-island transport channels identified in \textsection\ref{sec:resonant_dynamics}. Terminal outcomes are not confined to marginal high-eccentricity edges. The black Earth-reentry class includes the expected high-$e$ grazing cap, but also localized pockets embedded within the resonant architecture, most visibly near the \res{3}{1}{\M} complex. The gray lunar-impact class likewise appears as coherent impact domains interwoven with the resonant and chaotic texture, rather than appearing only as a boundary effect near the lunar-crossing limit. Thus, the fate layer shows that the CR zone is not merely a competition among quasiperiodic islands, sticky residence, and chaotic escape: resonant transport also feeds substantial collision basins. This distinction is important for interpreting the MEGNO layer, since terminal trajectories can display low or moderate finite-time variational growth when impact, reentry, or escape occurs before long-term divergence has time to accumulate. Farther outward, chaotic escape (red) becomes increasingly prevalent; in and beyond the \res{3}{2}{\M} neighborhood, the broad stable cores have largely disappeared, and the remaining low-MEGNO structure is increasingly fragmented by impact and escape channels within the cislunar--circumlunar transport region.

The EMS cislunar-resonant maps (Fig.~\ref{fig:megno_fates_EMS_cis}, middle pair) show that solar forcing is not a minor correction but a qualitatively important perturbation throughout this zone. In the MEGNO layer, the principal lunar commensurability locations remain legible, but the island interiors are less cleanly separated from their surroundings than in the Earth--Moon baseline. The inter-resonance background is substantially brighter and more filamentary, indicating stronger local instability and more pervasive resonance overlap. The \res{3}{1}{\M} and \res{2}{1}{\M} neighborhoods still retain recognizable low-MEGNO cores, but these cores are surrounded by broad mottled cyan and light-blue separatrix texture rather than by sharply defined dark elliptic interiors. Farther outward, the \res{3}{2}{\M} and adjacent outer-CR structures are embedded in a nearly continuous bright transport field, with only narrow dark filaments and small low-variational pockets surviving inside the broader chaotic web.

The fate layer clarifies the dynamical consequence of this variational erosion. Stable-quasiperiodic regions (navy) persist only as reduced cores or narrow remnants, while much of the resonant area that remains Earth-bound is classified instead as sticky residence (cyan) or bounded-unclassified motion (purple). This replacement is especially prominent across the \res{3}{1}{\M} and \res{2}{1}{\M} families, where the nominal resonant skeleton survives but its bounded component is no longer dominated by clean quasiperiodic residence. The outer half of the CR zone is also more strongly escape dominated, with chaotic escape (red) occupying much of the space between and beyond the surviving resonant remnants. Moon-impact structure (gray), by contrast, is less dominant than in the EM baseline, indicating that solar forcing does not simply add instability locally, but redirects part of the collision-fed resonant transport into escape. In other words, the dominant lunar resonances persist under solar forcing, but their surrounding architecture is reorganized: the same nominal resonant locations survive, yet the balance between regular confinement, bounded chaotic residence, and escape is appreciably altered. The patched spatiography of \textsection\ref{sec:spatio} is therefore not merely a conceptual refinement; it is required to understand the actual fate structure of the outer cislunar domain.

\paragraph{Circumlunar-gateway (CG) zone \textnormal{($0.84 \lesssim a/a_\M \lesssim 1.16$)}.}
By the MEGNO diagnostic, the circumlunar-gateway map is the most globally chaotic of the six regime-split cartographies. Its background is predominantly cyan and light blue, indicating that most of this band is strongly perturbed even where the finite-time variational growth is not maximal. The mapped initial conditions are geocentric orbits whose osculating semi-major axes straddle the lunar orbital distance; at these scales, repeated close lunar encounters become common, and the confining invariant structure that protects much of the inner cislunar population has largely broken down. This is the cartographic signature of the gateway geometry described in \textsection\ref{sec:circumlunar}: once the $\left( L_1 \right)^\M$--$\left( L_2 \right)^\M$ passages are effectively accessible, transport between Earth-bound, lunar-vicinity, and escaping trajectories becomes dynamically intrinsic rather than exceptional. What survives are isolated low-MEGNO islands and filaments embedded in the scattering web, rather than broad resonant continents.


The broad resonant islands of the CR zone have therefore given way to an island-threaded scattering web. Dark low-MEGNO structures still appear, but only as isolated ovals, loops, caps, and narrow filaments embedded in the gateway sea. Near the inner edge of the gateway, a wrapped island complex at $a/a_\M \simeq 0.86$--$0.88$ and $e \simeq 0.2$--$0.3$ marks the place where the outermost interior lunar resonances, especially the \res{4}{3}{\M} and \res{5}{4}{\M} families, feed into the circumlunar-gateway architecture. Near the co-orbital center, several slender loop-like structures between $a/a_\M \simeq 0.95$ and $1.05$ bracket the \res{1}{1}{\M} commensurability from either side, while a distinct high-eccentricity low-MEGNO cap near $a/a_\M \simeq 1$ and $e \gtrsim 0.9$ marks an upper co-orbital feature close to the lunar-crossing boundary. The clustering of the principal low-MEGNO islands near $e \simeq 0.2$--$0.3$ is consistent with the combined Tisserand and resonant-family geometry of the gateway. In the coplanar limit, Eq.~\eqref{eq:tisserand_moon} gives $T_\M = a_\M / a + 2 \sqrt{ ( a / a_\M ) ( 1 - e^2 ) }$; at $a = a_\M$, even the circular value is only $T_\M = 3$, below the first Earth--Moon neck-opening threshold. Thus the CG zone lies naturally in an open-gateway topology (see Fig.~\ref{fig:hills}). Within that open region, however, the locations of the surviving low-MEGNO centers occur where the resonant families themselves cross this open region without being immediately erased by close-encounter scattering. The outermost interior and first exterior lunar resonance envelopes shown in Figs.~\ref{fig:cislunar_atlas} and \ref{fig:translunar_atlas} cross the CG $a$-window at approximately these eccentricities. Below this band, the near-co-orbital dynamics are already strongly mixed by the gateway geometry; above it, lunar-encounter, impact, and escape channels increasingly dominate. Farther outward, a larger wrapped island-and-separatrix structure near $a/a_\M \simeq 1.12$ forms the most prominent exterior low-MEGNO feature before the map passes fully into the translunar resonant zone. These structures are therefore not arranged as a new confining architecture, but appear as finite-time low-variational pockets embedded within the dominant close-encounter transport layer generated by the circumlunar gateway.

The companion fate map shows why the low-MEGNO structures must be interpreted with care. The CG zone is dominated by chaotic escape (red), indicating that most initial conditions leave the Earth--Moon Hill region within the integration window. This escape sea is not structureless, however: coherent lunar-impact domains (gray) thread through the same island-and-separatrix geometry that appears as organized loops and filaments in the MEGNO layer. Earth reentry (black) occupies the expected high-eccentricity grazing band, while compact patches of orderly escape (gold) appear in and around several of the low-variational structures, especially on the exterior side of the gateway. Stable-quasiperiodic residence (navy) is nearly absent as a macroscopic fate class; the only clearly bounded low-MEGNO population is confined to a narrow near-parabolic cap close to the co-orbital region. Thus the EM gateway is not simply a chaotic sea with a few stable islands. It is a mixed transport layer in which low-variational motion, lunar-impact basins, Earth-reentry channels, and escape pathways are geometrically interleaved.

The EMS gateway maps preserve the same underlying gateway skeleton, but solar forcing changes the population of that skeleton. In the MEGNO layer, the principal low-variational loops and island-like structures remain recognizable, although their placement, thickness, and internal texture are modified. The fate layer shows the more important dynamical consequence: chaotic escape (red) expands, the lunar-impact contribution (gray) is reduced, and the already small bounded signatures of the EM gateway are essentially erased. Solar forcing therefore does not create an entirely new gateway architecture. Rather, it changes the phasing and timing of repeated lunar encounters, diverting part of the EM impact and sticky-residence population into escape channels. This explains why the gross morphology of the gateway survives the addition of the Sun, while the fate balance shifts toward more pervasive escape. The circumlunar gateway is therefore the first region in the cislunar survey where the patched Earth--Moon and Sun-perturbed descriptions must be read together: the lunar gateway geometry supplies the scattering skeleton, but the Sun alters how that skeleton is populated by impact, bounded residence, and escape.

\paragraph{Inner-translunar (IT) zone \textnormal{($1.08 \lesssim a/a_\M \lesssim 2.03$)}.}
Beyond the $\left( L_2 \right)^\M$ gateway, the EM inner-translunar map enters a phase space qualitatively distinct from both the cislunar resonant archipelago and the circumlunar-gateway scattering layer. The MEGNO layer is globally bright and filamentary, but it is not featureless: it is threaded by a sequence of dark low-variational island chains associated with the exterior lunar resonance ladder. These structures have a different morphology from the arched interior-resonance islands of the CR zone. They appear as broader, more distorted, and often more strongly wound island-and-separatrix complexes, consistent with the exterior lunar resonant families discussed in \textsection\ref{sec:resonant_dynamics}, where asymmetric librations and unstable symmetric branches organize the local phase portrait. The innermost structures enter immediately beyond the gateway near the \res{4}{5}{\M} and \res{3}{4}{\M} neighborhoods, including a large low-MEGNO complex just outside the gateway boundary near $a/a_\M \simeq 1.15$. Additional chains extend across the \res{2}{3}{\M}, \res{3}{5}{\M}, \res{1}{2}{\M}, and \res{2}{5}{\M} neighborhoods, with the \res{1}{2}{\M} region near $a/a_\M \simeq 1.59$ forming one of the most conspicuous outer island systems. The map therefore suggests a hierarchy of embedded low-variational pockets and surrounding unstable layers, consistent with the bifurcation sequences described in \textsection\ref{sec:stability_bifurcations}. The result is a resonance-threaded transport web rather than a broad stable domain: a chaotic background punctuated by coherent exterior lunar island chains, separatrix layers, and increasingly sheared low-MEGNO filaments toward the outer edge of the IT zone, where the first low-order solar commensurability begins to intrude.

The fate layer shows that these low-MEGNO structures are primarily transport organizers rather than long-term confinement domains. Chaotic escape (red) dominates most of the IT zone, confirming that the opening through $\left( L_2 \right)^\M$ leads rapidly into global loss for a large fraction of the sampled initial conditions. Coherent lunar-impact domains (gray) follow many of the same wrapped structures visible in the MEGNO layer, showing that the exterior resonant skeleton also feeds collision channels. Orderly escape (gold) appears in compact arcs and patches along selected low-variational structures, particularly in the outer half of the panel, where trajectories can leave the Earth--Moon Hill region without first developing large finite-time variational growth. Stable-quasiperiodic residence is not a prominent fate class in this regime, and the small sticky or bounded regions that do appear are confined to limited pockets, most visibly near the low-eccentricity outer edge. Thus the EM IT zone is not simply the continuation of the cislunar resonant archipelago beyond the Moon. Its principal island chains act less as permanent reservoirs than as organized transport conduits, embedded in a resonance-threaded escape layer where low-MEGNO islands, lunar-impact basins, and orderly or chaotic escape channels are geometrically intertwined.

The EMS comparison shows that solar forcing preserves the exterior-resonance skeleton but reorganizes its expression. In the MEGNO layer, the inner exterior lunar resonances remain recognizable, but the outer half of the IT zone is transformed into broader, more oblique, and more coherently sheared low-variational sweeps. This is consistent with the interpretation developed above: the lunar exterior resonances continue to supply the primary scaffold, while the Sun introduces an additional slow forcing that changes the phasing of repeated lunar encounters and begins to couple the IT dynamics to the mixed lunisolar resonance structure of the outer translunar realm. The fate layer records the corresponding redistribution of outcomes. Chaotic escape remains the dominant background, but orderly and unclassified escape pathways become more organized and more extensive, while the lunar-impact structures are displaced and partially fragmented. Solar forcing therefore does not erase the inner-translunar resonant architecture; rather, it deforms the exterior lunar skeleton and changes how that skeleton is populated by impact, temporary residence, and escape. In this sense, the IT zone is the first translunar regime in which the Earth--Moon resonant problem remains visibly legible but no longer sufficient by itself: the Sun begins to control the preferred geometry of escape through the same resonant channels.

\paragraph{Outer-translunar (OT) zone \textnormal{($1.91 \lesssim a/a_\M \lesssim 3.34$)}.}
The EM outer-translunar map presents one of the clearest topological contrasts in the full circumterrestrial survey. The MEGNO layer separates into a deep-navy low-eccentricity shelf and, above it, a bright filamentary chaotic sea divided from the shelf by a complex tongue-like boundary. Unlike the circumlunar gateway, this regime is no longer dominated by immediate close lunar scattering. The low-eccentricity shelf corresponds to geocentric ellipses whose periapses remain exterior to the lunar orbital region and whose apoapses, over much of the shelf, remain inside the Earth-Hill boundary. In the EM problem, this leaves the exterior lunar resonance ladder as the principal organizing structure rather than as a rapid collision trigger. Above the shelf, the map is threaded by curved dark filaments and nested tongue-like structures associated with the exterior lunar commensurabilities cataloged in the Gallardo atlas and discussed in \textsection\ref{sec:resonant_dynamics}. These filaments do not form broad stable continents, but they remain dynamically legible as organized resonant traces embedded within the surrounding stochastic layer.

The fate layer refines this interpretation in an important way. The low-eccentricity shelf is not merely a region of finite-time low MEGNO: it contains a genuine stable-quasiperiodic population (navy), concentrated at low eccentricity and moderate-to-large translunar semi-major axes. Just above this shelf, a narrow sticky-resident band (cyan) marks the loss of clean variational regularity along the upper stability boundary. Chaotic escape (red) dominates the high-eccentricity sea, while orderly escape (gold) appears primarily along selected resonant filaments and near the outer, low-eccentricity edge of the zone, where the apoapsis approaches or exceeds the Earth-Hill boundary. Lunar-impact domains (gray) are comparatively minor in this regime, and Earth-reentry appears mainly at high eccentricity. Thus the OT zone is not simply an escape sea with a misleadingly regular MEGNO shelf. In the EM model it contains a low-eccentricity quasiperiodic shelf, bounded above by a sticky transition layer and surrounded by a resonance-threaded escape sea. The two-layer representation is nevertheless essential, because the same low-MEGNO diagnostic marks two different behaviors: stable residence within the shelf, and organized escape along selected resonant transport channels.

The EMS comparison shows that this EM shelf is not robust to solar forcing. Once the Sun is restored, the low-eccentricity stable-quasiperiodic domain is eliminated, the sticky boundary is greatly weakened, and the simple shelf--sea partition is replaced by broad diagonal and curvilinear low-MEGNO bands that cut across the OT panel. These structures are consistent with the mixed lunar--solar character of the outer translunar regime: the exterior lunar resonances remain part of the scaffold, but interior solar commensurabilities and the solar quadrupolar tide now participate directly in organizing the transport. The fate layer records the corresponding redistribution of outcomes. Chaotic escape remains the dominant background, orderly escape becomes much more extensive, lunar impact is strongly reduced, and Earth reentry increases through moderate-eccentricity pathways rather than only through the high-eccentricity grazing boundary. Solar forcing therefore does not merely displace the EM shelf boundary. It changes the character of the OT zone from an Earth--Moon exterior-resonance shelf with surrounding escape channels into a lunisolar transport web in which the same region that was quasiperiodically confined in the EM model is drained into organized escape, chaotic escape, or solar-assisted Earth-crossing trajectories.

\paragraph{Translunar-fringe (TF) zone \textnormal{($3.03 \lesssim a/a_\M \lesssim 3.90$)}.}
In the EM model, the translunar fringe carries the outer-translunar shelf--sea structure into the immediate neighborhood of the Earth Hill boundary, but the shelf is now compressed by the finite Hill aperture. The MEGNO layer contains a broad deep-navy low-eccentricity domain whose upper edge is scalloped by tongue-like intrusions, while the higher-eccentricity sea is threaded by repeating curved filaments associated with the outermost exterior lunar commensurabilities. The relevant geometric constraint is the Earth-Hill apocenter condition $a ( 1 + e ) \approx \left( r_H \right)^\E$: as $a$ approaches $\left( r_H \right) ^\E \simeq 3.9\, a_\M$, only progressively lower eccentricities can remain wholly inside the nominal circumterrestrial Hill volume. The dark shelf is therefore the terminal low-variational remnant of the EM circumterrestrial problem, bounded above not by a sharp analytic curve, but by a resonance- and escape-mediated transition layer. Above this shelf, the repeating dark arcs in the MEGNO layer remain dynamically legible as exterior lunar-resonance traces, but they are increasingly embedded in a background that is already close to global escape.

The fate layer shows that the EM fringe is not a single regular reservoir. Its lowest-eccentricity base contains a narrow stable-quasiperiodic population (navy), concentrated near the inner part of the fringe and disappearing before the Hill boundary is reached. Above it, much of the low-to-moderate eccentricity shelf is classified as orderly escape (gold): these trajectories satisfy the escape criterion by the final epoch while retaining near-regular finite-time variational behavior over the full integration span. The high-eccentricity sea is dominated by chaotic escape (red), with thin resonant threads and small orderly-escape patches tracing the same filamentary skeleton visible in the MEGNO layer; lunar-impact and Earth-reentry classes remain minor and are concentrated mostly at higher eccentricity. Thus the EM translunar fringe is not simply an escape sea, nor is it a stable extension of the outer-translunar shelf. It is a terminal shelf in which a small quasiperiodic base, a broad orderly-escape sheet, and a chaotic exterior-resonance sea coexist immediately inside the Earth-Hill boundary. 

The EMS fringe map makes the transition to a genuinely lunisolar outer boundary explicit. Once solar forcing is included, the MEGNO layer is qualitatively reorganized: the simple EM shelf--sea partition is replaced by a dense system of oblique and nearly horizontal bands spanning the full $( a , e )$ domain, and a prominent low-variational lens near $( a/a_\M , e ) \approx ( 3.5 , 0.45 )$ stands out as the most organized feature in the map. Its location is consistent with the neighborhood of the \res{2}{1}{\S} interior solar commensurability cataloged in Table~\ref{tab:spatio} and Fig.~\ref{fig:translunar_atlas}, while the surrounding field remains structured but broadly escaping. The fate layer shows the corresponding redistribution of outcomes: the stable-quasiperiodic EM base is eliminated, the orderly-escape sheet contracts into structured bands and lenses, and chaotic escape becomes the dominant background. At the same time, Earth reentry (black) appears as a broad diagonal corridor through the lower and middle eccentricity range, indicating a solar-assisted return channel rather than the high-eccentricity grazing boundary seen in the EM map. The TF zone therefore marks the outermost stage of the circumterrestrial transition: the last Earth--Moon shelf is not merely eroded by the Sun, but reorganized into a weakly bound lunisolar transport layer whose principal outcomes are orderly escape, chaotic escape, and solar-assisted Earth return, with the \res{2}{1}{\S} commensurability imprinting a distinct island-like order within the otherwise escaping fringe.

Taken together, the six regime maps provide the numerical cartography that the preceding analytical, semi-analytical, and restricted-problem sections could only foreshadow. The transition from the secularly dominated interior to the resonant cislunar domain near $a/a_\M \approx 0.35$ is cartographically sharp. The resonant archipelago of the cislunar zone agrees in broad outline with the Gallardo atlas and with the island topology of the periapsis Poincar\'e sections, while the fate maps add the indispensable second layer of transport directionality: they distinguish true stable confinement from sticky residence, orderly escape, chaotic escape, and terminal reentry or impact. The circumlunar gateway then marks the destruction of broad confining structure and the onset of globally accessible chaotic transport, interrupted only by isolated regular survivors. Farther out, the translunar regimes reveal that exterior lunar resonances first organize transport locally and then give way to a broader mixed lunar--solar architecture in which the Sun becomes the dominant large-scale sculptor of escape. The EM-to-EMS comparison is therefore not merely quantitative. It demonstrates that solar forcing is itself an architectural ingredient of circumterrestrial dynamics, one that alters the balance among confinement, sticky residence, and escape in ways the Earth--Moon restricted problem alone cannot capture. This is precisely why the patched spatiography of \textsection\ref{sec:spatio}, together with full EMS integrations, is required for a faithful dynamical geography of cislunar and translunar space.

\section{Conclusions}

The central result of this paper is that the Earth--Moon environment is not a single vast beyond-GEO regime, but a stratified circumterrestrial geography whose provinces are distinguished by different organizing mechanisms. A geocentric high-eccentricity orbit interior to the Moon, a lunar-vicinity trajectory threaded through the $\left( L_1 \right)^\M / \left( L_2 \right)^\M$ gateway, and a weakly bound Earth-centered orbit well beyond lunar distance do not belong to one and the same dynamical class simply because all lie outside the traditional terrestrial satellite bands. Any useful description of the region must therefore distinguish where secular dynamics dominate, where discrete lunar commensurabilities organize the phase space, where gateway transport destroys broad confinement, and where solar forcing becomes an architectural element rather than a perturbative correction.

The spatiographic partition developed here makes those distinctions explicit. The onset of cislunar dynamics is sharpened through the Laplace transition, which marks the loss of the classical oblateness-dominated terrestrial picture. Beyond it lies a secularly dominated inner cislunar zone, followed by an outer cislunar region structured by interior lunar mean-motion resonances. Near the Moon, circumlunar space appears not as a mere continuation of Earth-centered motion, but as a distinct enclave organized by the gateway geometry of the Earth--Moon problem and by the Moon's sphere of influence. Beyond lunar distance, translunar space is shown to be neither a simple extension of cislunar dynamics nor a mere prelude to heliocentric escape, but an outer circumterrestrial province in which the Moon and Sun jointly sculpt a mixed secular and resonant architecture.

Several methodological conclusions follow. First, perturbative theory, resonance atlases, restricted-problem dynamics, and direct numerical cartography are not competing descriptions, but mutually necessary ones. Orbit-averaged theory supplies the secular scaffold; semi-analytical atlases identify the local resonant structures; periodic orbits, bifurcations, and separatrices reveal the global phase-space skeleton; and the MEGNO and fate-class maps show which of those structures survive in higher-fidelity Earth--Moon and Earth--Moon--Sun models, and how they govern confinement, sticky residence, impact, capture, and escape. Second, the comparison between the Earth--Moon and Earth--Moon--Sun cartographies confirms that solar forcing is not merely a quantitative perturbation on an otherwise lunar architecture. In the outer cislunar, gateway, and translunar regimes, it is a qualitative architectural ingredient that reshapes the balance among regular motion, bounded chaotic residence, and organized or chaotic escape.

A broader historical point also emerges. The modern umbrella use of \emph{cislunar} is sociolinguistically understandable, but dynamically imprecise. The early Space Age literature had already glimpsed much of the taxonomy and much of the restricted-problem structure needed to describe the Earth--Moon environment more faithfully. What was missing was not raw insight, but synthesis: a framework that could place historical terminology, perturbation theory, resonance structure, gateway geometry, and modern ephemeris cartography into one coherent picture. That spatiographic synthesis has been the aim of this Primer.

The same geography also frames the scientific and stewardship value of the region. Cislunar space was recognized early not only as a route to the Moon, but as a laboratory for measuring Earth’s extended atmosphere, magnetic environment, electron-density structure, charged-particle and dust populations, and lunar-vicinity fields and exospheres \citep{sS59, vE60, jP62, wA71}. In his expansive geolunar vision, \citet{kE81, kE85} gave this operational geography its most ambitious form, treating the Earth--Moon system as an integrated arena for transportation, resource use, industrial development, and eventual settlement, with the lunar surface, circumlunar orbit, circumterrestrial markets, and Earth itself forming coupled parts of a larger operational economy. Modern mission concepts sharpen that view: Sun--Earth sub-$\left( L_1 \right)^\E$ platforms extend the space-weather-monitoring architecture at the heliocentric boundary of the geospace environment; Earth--Moon resonant orbits can provide long-lived vantage points for magnetospheric science; cislunar resource-staging concepts can support in-situ resource utilization and deep-space exploration architectures; cislunar constellations can provide communication, navigation, domain awareness, and exploration logistics; and fragmentation studies show that debris generated in this region may not remain local or intuitively predictable \citep{dMetal15, nLetal25, jLetal25, yHsCyW26, nB23, pG23, aB25}. Thus, the same dynamical structure that enables access, residence, and transfer also defines the constraints under which exploration, science, and long-term sustainability must be planned.

If future activity beyond GEO is to include long-lived infrastructure, lunar-vicinity operations, translunar navigation, and effective catalog maintenance for artificial satellites and debris, then the Earth--Moon environment cannot be treated as a single operational shell. Stability, observability, predictability, transport, and disposal are all regime dependent. The circumterrestrial geography recovered here is intended as a step toward that more precise view: one that treats the Earth--Moon environment neither as a blank slate nor as a rhetorical frontier, but as an old dynamical country whose provinces, passages, and borders can now be drawn with greater clarity.

\section*{Acknowledgments}

This research made use of NASA's Astrophysics Data System and the open-source \texttt{REBOUND} (N)-body integration package \citep{hR12, hR15}. The authors are deeply grateful to the UC San Diego Interlibrary Loan and Document Delivery staff for invaluable assistance in procuring many manuscripts and out-of-circulation articles. A.J. Rosengren thanks Tabar'e Gallardo (University of the Republic, Uruguay) for guidance on his semi-analytical method and Di Wu (Embry-Riddle Aeronautical University) for his initial implementation of the Gallardo codes. A.J. Rosengren also thanks Binyamin J. Stivi (Advanced Space) for producing an early rendition of Fig.~\ref{fig:synodic} during their shared Summer 2025 period at the 15th Space Surveillance Squadron, and Lamberto Dell'Elce (Universit{\'e} C{\^o}te d'Azur and Inria Sophia Antipolis M{\'e}diterran{\'e}e) for providing the fast-Fourier-transform-based numerical averaging scheme used to produce the top panel of Fig.~\ref{fig:secular}.

A.J. Rosengren and S.D. Ross acknowledge support from the Air Force Office of Scientific Research (AFOSR) under Grant No.~FA9550-24-1-0194.

\begin{appendices}

\section{Etymological Timeline of Earth--Moon--Sun Spatial Terms}
\label{app:glossary}

\begingroup

\small
\setlength{\tabcolsep}{6pt}
\renewcommand{\arraystretch}{1.15}
\arrayrulecolor{tblRule}

\newcommand{\RowSep}{\addlinespace[0.5ex]\midrule\addlinespace[0.5ex]}

\begin{longtable}{%
    p{2.4cm}
    p{2.4cm}
    p{10.0cm}
    }

    \caption{\color{tblText}%
        Etymological glossary of Earth--Moon--Sun spatial terms. Within this lexicographic timeline, shaded rows indicate terms that
        are widely used in modern astrodynamics and spaceflight discourse.
        \label{tab:ems_lexicon}
    } \\[-1.5ex]

    \toprule
    \addlinespace[0.5ex]

    \rowcolor{tblHeader}
    {\bf\color{tblHeadText} Term} &
    {\bf\color{tblHeadText} Earliest Use Date} &
    {\bf\color{tblHeadText} Meaning (incl. historical/literary/philosophical/astronomical use)} \\

    \addlinespace[0.75ex]
    \midrule
    \addlinespace[0.5ex]
    \endfirsthead

    \toprule
    \addlinespace[0.75ex]

    \rowcolor{tblHeader}
    {\bf\color{tblHeadText} Term} &
    {\bf\color{tblHeadText} Earliest Use Date} &
    {\bf\color{tblHeadText} Meaning (incl. historical/literary/philosophical/astronomical use)} \\

    \addlinespace[0.75ex]
    \midrule
    \addlinespace[0.5ex]
    \endhead

    \midrule
    \addlinespace[0.5ex]
    \multicolumn{3}{r}{\small\textit{(continued on next page)}}\\
    \addlinespace[0.5ex]
    \endfoot

    \bottomrule
    \endlastfoot

    \rowcolor{tblBase}
    {\bf\color{tblTerm} sublunar} & 1598 &
        Existing/situated {\bf beneath the Moon}; by extension {\bf of this world}, the earthly realm
        (classically opposed to a higher celestial realm). In historical/philosophical contexts:
        the {\bf mutable}, {\bf perishable} sphere of human affairs and material change (Aristotelian--medieval
        cosmology), often used morally/literarily to evoke {\bf impermanence} and ``worldliness.'' \\

    \RowSep

    \rowcolor{tblBase}
    {\bf\color{tblTerm} sublunary} & 1601 &
        Variant adjective/noun form of sublunar. Used historically in cosmology and literature for what
        pertains to the {\bf terrestrial}, {\bf changeable} world below the lunar sphere; often used contrastively
        with higher/ideal realms. \\

    \RowSep

    \rowcolor{tblBase}
    {\bf\color{tblTerm} superlunary} & 1614 &
        Situated {\bf above or beyond the Moon}; belonging to a {\bf higher world} (celestial/divine).
        In literary/philosophical use: connotes {\bf transcendence}, an exalted or ``otherworldly''
        quality, sometimes explicitly opposed to {\it sublunary}. \\

    \RowSep

    \rowcolor{tblBase}
    {\bf\color{tblTerm} translunary} & 1627 &
        Existing/situated {\bf beyond} or above the (supposed) sphere of the Moon (now chiefly historical/literary).
        In philosophical/literary contexts: ``beyond ordinary experience,'' {\bf heavenly}/{\bf visionary} or otherworldly.
        (Distinct from later astronautical/technical senses.) \\

    \RowSep

    \rowcolor{tblBase}
    {\bf\color{tblTerm} supralunary} & 1635 &
        {\bf Above the Moon}; a celestial or higher-region sense closely related to {\it superlunary}/{\it supralunar}
        in early cosmology. In historical writing: used within the old ``two-region'' universe (sub-/supra-lunar distinction),
        and figuratively for what is elevated beyond earthly concerns. \\

    \RowSep

    \rowcolor{tblAccent}
    {\bf\color{tblTerm} heliocentric} & 1655 &
        {\bf Sun-centered}. Historically: (i) astronomical quantities referred to the Sun as center (positions/longitudes/
        latitudes ``as viewed from the Sun''), and (ii) the Copernican {\bf heliocentric system}. Often used as shorthand for a
        paradigm shift in cosmology/scientific worldview. \\

    \RowSep

    \rowcolor{tblAccent}
    {\bf\color{tblTerm} geocentric} & 1664 &
        {\bf Earth-centered}. Historically: (i) a person/position supporting Earth as the universe's center (obsolete sense),
        and (ii) in astronomy, measured/considered {\bf with reference to Earth} (Earth as observational center).
        Often used in intellectual history to contrast with heliocentric framing. \\

    \RowSep

    \rowcolor{tblAccent}
    {\bf\color{tblTerm} lunisolar} & 1675 &
        Pertaining to the {\bf mutual relations} of the {\bf Sun} and {\bf Moon}, or resulting from their combined action;
        early attestations occur in calendrical/chronological contexts as {\it luni-solar} (e.g., {\it luni-solar cycle}).
        Later technical uses include {\it lunisolar year}, {\it lunisolar precession}, and {\it lunisolar perturbations}. \\

    \RowSep

    \rowcolor{tblBase}
    {\bf\color{tblTerm} supralunar} & 1719 &
        Revolving about/situated {\bf above the Moon} (essentially {\it supralunary}; also with figurative extension).
        In later literary usage, can serve as a metaphorical ``higher realm'' contrasted with {\it sublunar} life. \\

    \RowSep

    \rowcolor{tblAccent}
    {\bf\color{tblTerm} circumterrestrial} & 1728 &
        Situated/moving {\bf round the Earth}. In early scientific writing used descriptively for phenomena
        pertaining to Earth's vicinity (often paired with {\it circumsolar}). Later compatible with technical orbital
        usage: Earth-surrounding space, motion, or phenomena. \\

    \RowSep

    \rowcolor{tblBase}
    {\bf\color{tblTerm} superlunar} & 1742 &
        Variant of {\it superlunary}; beyond/above the Moon, but often encountered historically in {\bf literary}
        or elevated figurative contexts. Typically carries an air of {\bf exaltation}, ``higher sphere'' sensibility,
        or distance from mundane concerns. \\

    \RowSep

    \rowcolor{tblAccent}
    {\bf\color{tblTerm} translunar} & 1791 &
        {\bf Passing beyond/across the Moon} (older, often literary/historical). In modern technical use it becomes spaceflight-adjacent,
        but its earliest attestation is a pre-spaceflight {\bf figurative/higher-realm} usage. \\

    \RowSep

    \rowcolor{tblAccent}
    {\bf\color{tblTerm} selenocentric} & 1852 &
        Having relation to the {\bf center of the Moon}; considered as seen/estimated from the Moon's center.
        A technical astronomical/positional term that later becomes standard in dynamical/coordinate-frame discussions. \\

    \RowSep

    \rowcolor{tblAccent}
    {\bf\color{tblTerm} circumlunar} & 1871 &
        Situated/occurring {\bf round the Moon} (in the Moon's vicinity). Early astronomical usage appears
        in eclipse/corona discussions (e.g., ``circumlunar matter''); later widely used for mission profiles
        and trajectories (lunar flyby or ``around-the-Moon'' voyages). \\

    \RowSep

    \rowcolor{tblAccent}
    {\bf\color{tblTerm} cislunar} & 1872 &
        On this {\bf side of the Moon} (Earthward; near-side in a relational sense). First attested in eclipse--corona
        discourse to describe phenomena inferred to lie {\bf between Earth and Moon}/{\bf within the Moon's orbit},
        contrasted with {\it lunar} or {\it solar} origin (and with extra-terrestrial interpretations more broadly). Later
        extended in astronautics and policy usage to the broader Earth--Moon domain (especially ``cislunar space''). \\

    \RowSep
    
    \rowcolor{tblAccent}
    {\bf\color{tblTerm} spatiography} & 1958 &
        Early astronautical term for a ``geography of space'': a {\bf topographical description of extra-atmospheric space},
        especially the region {\bf between celestial bodies}, intended to support orientation, navigation, and the classification
        of space operations. In Strughold's formulation, spatiography concerns the ``empty'' space between bodies (distinct from
        \emph{planetography}, which describes conditions on planetary bodies). \\

    \RowSep

    \rowcolor{tblAccent}
    {\bf\color{tblTerm} transearth} & 1963 &
        Astronautics: pertaining to the {\bf Earth-return leg} following a lunar encounter or departure, esp.\ in
        \textit{transearth trajectory}, \textit{transearth injection (TEI)}, and \textit{transearth coast}.
        Commonly paired with \textit{translunar injection (TLI)} for the outbound leg; in free-return and related lunar-transfer geometries,
        both phases are naturally ``trans-'' in the literal sense ({\bf across/beyond}) because the corresponding capture/return maneuvers are
        typically executed near perilune on the {\bf lunar far-side} portion of the approach hyperbola, where direct Earth line-of-sight is unavailable. \\

\end{longtable}

\endgroup

\clearpage 

\section{Spatiographic Regimes, Integration Spans, and Numerical Cartography Guidance}
\label{app:megno}

\begin{table}[h!]
    \caption{
        Recommended integration spans for the regime-split astro-cartographies of circumterrestrial space, organized by the spatiographic partition of Table~\ref{tab:spatio}. The regime limits are working astro-cartographic boundaries set by the Gallardo atlas predictions, i.e., by the separatrix-edge envelope of the bounding resonance, so adjacent zones are allowed to overlap deliberately near their ends. The quoted orbital-revolution counts are approximate and correspond to the recommended integration span in each regime, evaluated across the listed semi-major-axis range.
        \label{tab:megno_timescales}
    }

    \begin{tabular*}{1.0\textwidth}{
        @{\extracolsep{\fill}}
        >{\raggedright\arraybackslash}p{4.2cm}
        >{\centering\arraybackslash}p{1.3cm}
        >{\centering\arraybackslash}p{1.2cm}
        >{\centering\arraybackslash}p{2.0cm}
        >{\raggedright\arraybackslash}p{5.2cm}
    }

        \toprule
        \addlinespace[1ex]

        Dynamical Zone
        & {$a/a_\M$}
        & {Span [yr]}
        & {Revolutions}
        & Description \\

        \addlinespace[1.25ex]
        \toprule
        \addlinespace[1ex]

        \textbf{Secularly Dominated Cislunar Zone} & & & & \\
        \quad geolunar Laplace radius to inner edge of \mbox{$5\!:\!1_{\M}$}
            & 0.13--0.35 & 19 & 5419--1227
            & MEGNO cartography of the vZLK backbone together with lunar nodal and apsidal secular resonances \\

        \addlinespace[0.8ex]
        \multicolumn{5}{@{}l}{\hspace{0.8em}\rule{0.965\textwidth}{0.3pt}}\\
        \addlinespace[1ex]

        \textbf{Cislunar Resonant Zone} & & & & \\
        \quad Gallardo-predicted interior lunar resonance belt up to the co-orbital/gateway envelope
            & 0.33--0.89 & 19 & 1340--303
            & MEGNO cartography of the interior lunar resonance ladder, including stable islands, separatrix layers, and surrounding chaotic bands \\

        \addlinespace[0.8ex]
        \multicolumn{5}{@{}l}{\hspace{0.8em}\rule{0.965\textwidth}{0.3pt}}\\
        \addlinespace[1ex]

        \textbf{Circumlunar Gateway Zone} & & & & \\
        \quad co-orbital/gateway envelope spanning \((L_1)^\M\), \mbox{$1\!:\!1_{\M}$}, and \((L_2)^\M\)
            & 0.84--1.16 & 19 & 330--203
            & Two-layer cartography: MEGNO plus gateway-crossing, lunar-impact, Earth-reentry, and temporary-residence diagnostics \\

        \addlinespace[0.8ex]
        \multicolumn{5}{@{}l}{\hspace{0.8em}\rule{0.965\textwidth}{0.3pt}}\\
        \addlinespace[1ex]

        \textbf{Inner Translunar Resonant Zone} & & & & \\
        \quad Gallardo-predicted exterior lunar resonance belt from \mbox{$4\!:\!5_{\M}$} through \mbox{$2\!:\!5_{\M}$}, with first solar intrusion near \mbox{$5\!:\!1_{\S}$}
            & 1.08--2.03 & 38 & 453--176
            & Two-layer cartography of exterior lunar resonances and first low-order solar commensurabilities, with survival/escape diagnostics \\

        \addlinespace[0.8ex]
        \multicolumn{5}{@{}l}{\hspace{0.8em}\rule{0.965\textwidth}{0.3pt}}\\
        \addlinespace[1ex]

        \textbf{Outer Translunar Resonant Zone} & & & & \\
        \quad mixed Gallardo-predicted exterior lunar and interior solar resonance belt, from \mbox{$5\!:\!1_{\S}$} through \mbox{$1\!:\!5_{\M}$}
            & 1.91--3.34 & 57 & 289--125
            & MEGNO plus first-escape time and sticky-residence classification in the mixed lunisolar borderland \\

        \addlinespace[0.8ex]
        \multicolumn{5}{@{}l}{\hspace{0.8em}\rule{0.965\textwidth}{0.3pt}}\\
        \addlinespace[1ex]

        \textbf{Translunar Fringe} & & & & \\
        \quad outer fringe from inner edge of \mbox{$5\!:\!2_{\S}$} to Earth's Hill sphere
            & 3.03--3.90 & 57 & 144--99
            & Survival/escape cartography; MEGNO retained only as a secondary indicator \\

        \bottomrule

    \end{tabular*}
\end{table}

\end{appendices}

\bibliographystyle{spbasic}
\bibliography{primer_refs}

@BOOK{ rB64,
  author  = "R. H. Battin",
  title = "Astronautical guidance",
  year = "1964",
  publisher = "McGraw-Hill, Inc.",
  address = "New York, NY"
  }

@BOOK{ kE62a,
  author  = "K. A. Ehricke",
  title = "Space Flight: Flight I. Environment and Celestial Mechanics",
  year = "1962",
  publisher = "D. Van Nostrand Company, Inc.",
  address = "Princeton",
  series = "Principles of Guided Missile Design"
  }

@BOOK{ kE62b,
  author  = "K. A. Ehricke",
  title = "Space Flight: Flight II. Dynamics",
  year = "1962",
  publisher = "D. Van Nostrand Company, Inc.",
  address = "Princeton",
  series = "Principles of Guided Missile Design"
  }

@ BOOK{ vE69,
  author = "V. A. Egorov",
  year = "1969",  
  title = "Three-Dimensional Lunar Trajectories",
  address = "Jerusalem", 
  publisher = "Israel Program for Scientific Translations"  
  }

@ BOOK{ gH99,
  author = "G. Haller",
  year = "1999",  
  title = "Chaos Near Resonance",
  address = "New York", 
  publisher = "Springer-Verlag"  
  }

@ BOOK{ wK22,
  author = "W. S. Koon and M. W. Lo and J. E. Marsden and S. D. Ross",
  year = "2022",  
  title = "Dynamical Systems, the Three-Body Problem and Space Mission Design",
  publisher = "Marsden Books, ISBN 978-0-615-24095-4"
  }

@ BOOK{ aLmL92,
  author = "A. J. Lichtenberg and M. A. Lieberman",
  year = "1992",  
  title = "Regular and Chaotic Dynamics",
  address = "New York",   
  publisher = "Springer-Verlag",
  edition = "2"  
  }

@ BOOK{ aM02,
  author = "A. Morbidelli",
  year = "2002",  
  title = "Modern Celestial Mechanics: Aspects of Solar System Dynamics",
  address = "London",   
  publisher = "Taylor \& Francis"
  }

@ BOOK{ cMsD99,
  author = "C. D. Murray and S. F. Dermott",
  year = "1999",  
  title = "Solar System Dynamics",
  address = "Cambridge",   
  publisher = "Cambridge University Press"
  }

@ BOOK{ rZ84,
  author = "R. Z. Sagdeev",
  year = "1984",  
  title = "Investigations of Solar Activity and the Prognoz Space System",
  address = "Moscow",   
  publisher = "Izdatel'stvo Nauka"
  }

@ BOOK{ iS17,
  author = "I. I. Shevchenko",
  year = "2017",  
  title = "The Lidov--Kozai Effect - Applications in Exoplanet Research and Dynamical Astronomy",
  address = "Cham",   
  publisher = "Springer"
  }

@ BOOK{ sT23,
  author = "S. Tremaine",
  year = "2023",  
  title = "Dynamics of Planetary Systems",
  address = "Princeton",   
  publisher = "Princeton University Press"
  }

@ INCOLLECTION{ wA71,
  author = "W. M. Alexander and C. W. Arthur and J. L. Bohn",
  year = "1971",
  title = "{Lunar Explorer 35} and {OGO 3}: dust particle measurements in selenocentric and cislunar space from 1967 to 1969",
  booktitle = "Space Research XI",
  editor = "K. Ya. Kondratyev and M. J. Rycroft and C. Sagan",
  address = "Berlin",
  publisher = "Akademie-Verlag",
  pages = "279--285"
  }

@ INCOLLECTION{ mB10,
  author = "M. Bell{\' o} and G. G{\' o}mez and J. J. Masdemont",
  year = "2010",   
  title = "Invariant manifolds, {Lagrangian} trajectories and space mission design",
  booktitle = "Space Manifold Dynamics: Novel Spaceways for Science and Exploration",  
  editor = "E. Perozzi and S. Ferraz-Mello",
  address = "New York",   
  publisher = "Springer",
  pages = "1--96"
  }

@ INCOLLECTION{ jdV64,
  author = "J. P. {de Vries}",
  year = "1964",
  title = "The {Sun's} perturbing effect on motion near a triangular {Lagrange} point",
  booktitle = "XIII$^{th}$ International Astronautical Congress Varna 1962",
  editor = "N. Boneff and I. Hersey",
  address = "Vienna",
  publisher = "Springer-Verlag",
  pages = "432--450"
  }

@INCOLLECTION{ kE56,
  author = "K. A. Ehricke",
  year = "1956",
  title = "Astronautical and space-medical research with automatic satellites",
  booktitle = "Earth Satellites as Research Vehicles",
  address = "Lancaster, Pa.",
  publisher = "Journal of the Franklin Institute",
  pages = "25--66"
  }

@ INCOLLECTION{ kE58a,
  author = "K. A. Ehricke",
  year = "1958",
  title = "Instrumented comets---{Astronautics} of solar and planetary probes",
  booktitle = "VIII$^{th}$ International Astronautical Congress Barcelona 1957",
  editor = "F. Hecht",
  address = "Vienna",
  publisher = "Springer-Verlag",
  pages = "74--126"
  }

@INCOLLECTION{ kE58b,
  author = "K. A. Ehricke",
  year = "1958",
  title = "Error Analysis of Single and Two-Force Field Spacecraft Orbits",
  booktitle = "Ten Steps into Space",
  address = "Lancaster, Pa.",
  publisher = "Journal of the Franklin Institute",
  pages = "114--149"
  }

@ INCOLLECTION{ kE59,
  author = "K. A. Ehricke",
  year = "1959",   
  title = "Cislunar orbits",
  booktitle = "Orbit Theory",  
  editor = "G. Birkhoff and R. E. Langer",
  address = "Providence, RI",   
  publisher = "American Mathematical Society",
  pages = "48--74"
    }

@INCOLLECTION{ kE85,
  author = "K. A. Ehricke",
  year = "1985",
  title = "Lunar industrialization and settlement---Birth of polyglobal civilization",
  booktitle = "Lunar Bases and Space Activities of the 21st Century",
  editor = "W. W. Mendell",
  address = "Houston",
  publisher = "Lunar and Planetary Institute",
  pages = "827--855"
  }

@INCOLLECTION{ gEaD61,
  author = "G. A. Ellis and A. C. Diana",
  year = "1961",
  title = "The use of the {Earth--Moon} libration points as terminals for space stations",
  booktitle = "Propulsion Space Science and Space Exploration",
  editor = "C. T. Morrow and L. D. Ely and M. R. Smith",
  address = "New York",
  publisher = "Academic Press",
  pages = "427--445"
  }

@ INCOLLECTION{ vE58,
  author = "V. A. Egorov",
  year = "1958",
  title = "Certain problems of {Moon} flight dynamics",
  booktitle = "The Russian Literature of Satellites, Part I",
  editor = "A. V. Shpol'skii",
  address = "New York",
  publisher = "International Physical Index, Inc.",
  pages = "107--174"
  }

@ INCOLLECTION{ bE99,
  author = "B. \'Erdi",
  year = "1999",   
  title = "Dynamics of satellites with multi-day periods",
  booktitle = "The Dynamics of Small Bodies in the Solar System",  
  editor = "B.A. Steves and A.E. Roy",
  address = "Dordrecht",   
  publisher = "Kluwer Academic Publishers",
  pages = "303--307"
  }

@ INCOLLECTION{ hH62,
  author = "H. Hiller",
  year = "1962",
  title = "A generalized study of two-dimensional trajectories of a vehicle in {Earth--Moon} space",
  booktitle = "Space Flight and Re-Entry Trajectories",
  editor = "P. A. Libby",
  address = "Vienna",
  publisher = "Springer-Verlag",
  pages = "82--105"
  }

@ INCOLLECTION{ aKjB70,
  author = "A. A. Kamel and J. V. Breakwell",
  year = "1970",
  title = "Stability of motion near {Sun}-perturbed {Earth--Moon} triangular libration points",
  booktitle = "Periodic Orbits, Stability and Resonances",
  editor = "G. E. O. Giacaglia",
  address = "Dordrecht-Holland",
  publisher = "D. Reidel",
  pages = "82--90"
  }

@ INCOLLECTION{ pL61,
  author = "P. Lanzano",
  year = "1961",
  title = "Applications of the {Jacobi} integral of celestial mechanics to the terminal guidance of space probes",
  booktitle = "XI$^{th}$ International Astronautical Congress Stockholm 1960",
  editor = "C. W. P. Reutersw{\"a}rd",
  address = "Vienna",
  publisher = "Springer-Verlag",
  pages = "114--124"
  }

@ INCOLLECTION{ mL63,
  author = "M. L. Lidov",
  year = "1963",   
  title = "On the approximated analysis of the orbit evolution of artificial satellites",
  booktitle = "Dynamics of Satellites",  
  editor = "M. Roy",
  address = "Berlin",   
  publisher = "Springer-Verlag",
  pages = "168--179"
  }

@ INCOLLECTION{ pM63,
  author = "P. Musen",
  year = "1963",   
  title = "On long range effects in the motion of artificial satellites",
  booktitle = "Dynamics of Satellites",  
  editor = "M. Roy",
  address = "Berlin",   
  publisher = "Springer-Verlag",
  pages = "21--33"
  }

@ INCOLLECTION{ wHgM68,
  author = "P. Musen and R. K. Squires",
  year = "1968",   
  title = "Orbital mechanics",
  booktitle = "Introduction to Space Science",  
  editor = "W.N. Hess and G.D. Mead",
  address = "New York",   
  publisher = "Gordon and Breach",
  pages = "529--554", 
  edition = "2"
  }

@ INCOLLECTION{ iS63,
  author = "I. I. Shapiro",
  year = "1963",   
  title = "The prediction of satellite orbits",
  booktitle = "Dynamics of Satellites",  
  editor = "M. Roy",
  address = "New York",   
  publisher ="Academic Press",
  pages = "257--312"
  }

@ INCOLLECTION{ sS59,
  author = "S. F. Singer",
  year = "1959",
  title = "Scientific problems in cislunar space and their exploration with rocket vehicles",
  booktitle = "IX$^{th}$ International Astronautical Congress Amsterdam 1958",
  editor = "F. Hecht",
  address = "Vienna",
  publisher = "Springer-Verlag",
  pages = "904--913"
  }

@ INCOLLECTION{ hS59,
  author = "H. Strughold",
  year = "1959",   
  title = "Definitions and subdivisions of space---{Bioastronautical} aspect",
  booktitle = "First Colloquium on the Law of Outer Space",  
  editor = "A. G. Haley and W. Heinrich",
  address = "Vienna",   
  publisher = "Springer-Verlag",
  pages = "110--113"
  }

@ INCOLLECTION{ mV84,
  author = "M. E. Vaucher",
  year = "1984",
  title = "Geographical parameters for military doctrine in space and the defense of the space-based enterprise",
  booktitle = "International Security Dimensions of Space",
  editor = "U. Ra'anan and R. L. Pfaltzgraff, Jr.",
  address = "Hamden, CT",
  publisher = "Archon Books",
  pages = "32--46"
  }

@ INCOLLECTION{ jV70,
  author = "J. Vagners",
  year = "1970",
  title = "On the long-term evolution of lunar satellite orbits",
  booktitle = "Periodic Orbits, Stability and Resonances",
  editor = "G. E. O. Giacaglia",
  address = "Dordrecht-Holland",
  publisher = "D. Reidel",
  pages = "304--313"
  }

@ INCOLLECTION{ mW00,
  author = "M. C. Weisskopf and H. D. Tananbaum and L. P. {Van Speybroeck} and S. L. O'Dell",
  year = "2000",   
  title = "{Chandra X-Ray Observatory (CXO): Overview}",
  booktitle = "XRay Optics, Instruments, and Missions III",  
  editor = "J. E. Truemper and B. Aschenbach",
  address = "Munich",  
  publisher = "SPIE",   
  volume = "2012",
  pages = "2--16"
  }

@ INCOLLECTION{ lW66,
  author = "L. E. Wolaver",
  year = "1966",   
  title = "Effect of initial configurations on libration point motion",
  booktitle = "Methods in Astrodynamics and Celestial Mechanics",  
  editor = "R. L. Duncombe and V. G. Szebehely",
  address = "New York",   
  publisher ="Academic Press",
  pages = "75--99"
  }

@ ARTICLE{ rAgC64,
  author = "R. R. Allan and G. E. Cook",
  year = "1964",   
  title = "The long-period motion of the plane of a distant circular orbit",
  journal = "Proceedings of the Royal Society A",
  volume = "280",
  pages = "97--109"
  }

@ ARTICLE{ dA20,
  author = "D. Amato and R. Malhotra and V. Sidorenko and A. J. Rosengren",
  year = "2020",   
  title = "Lunar close encounters compete with the circumterrestrial {Lidov--Kozai} effect: {The} dynamical demise of {Luna 3}",
  journal = "Celestial Mechanics and Dynamical Astronomy",
  volume = "132",
  pages = "35 (18 pp.)"
  }

@ ARTICLE{ sA24,
  author = "S. An and M. Liu and H. Li and F. Wu",
  year = "2024",   
  title = "Five-impulse low-energy {Earth}--{Moon} transfer using manifolds",
  journal = "Advances in Space Research",
  volume = "73", 
  pages = "201--224"
  }

@ ARTICLE{ mA02,
  author = "M. A. Andreu",
  year = "2002",   
  title = "Dynamics in the center manifold around {L2} in the quasi-bicircular problem",
  journal = "Celestial Mechanics and Dynamical Astronomy",
  volume = "84",
  pages = "105--133"
  }

@ ARTICLE{ jA15,
  author = "J. M. O. Antognini",
  year = "2015",   
  title = "Timescales of {Kozai--Lidov} oscillations at quadrupole and octupole order in the test particle limit", 
  journal = "Monthly Notices of the Royal Astronomical Society",
  volume  = "452",
  pages = "3610--3619",
  }

@ ARTICLE{ rA08,
  author = "R. A. N. Araujo and O. C. Winter and A. F. B. A. Prado and R. {Vieira Martins}",
  year = "2008",   
  title = "Sphere of influence and gravitational capture radius: a dynamical approach", 
  journal = "Monthly Notices of the Royal Astronomical Society",
  volume  = "391",
  pages = "675--684",
  }

@ ARTICLE{ rA63,
  author = "R. F. Arenstorf",
  year = "1963",   
  title = "Existence of periodic solutions passing near both masses of the restricted three-body problem", 
  journal = "AIAA Journal",
  volume  = "1",
  pages = "238--240",  
  }

@ ARTICLE{ mA76,
  author = "M. E. Ash",
  year = "1976",   
  title = "Doubly averaged effect of the {Moon} and {Sun} on a high altitude {Earth} satellite orbit", 
  journal = "Celestial Mechanics",
  volume  = "14",
  pages = "209--238",
  }

@ ARTICLE{ bB24,
  author = "B. {Baker-McEvilly} and S. Bhadauria and D. Canales and C. Frueh",
  year = "2024",   
  title = "A comprehensive review on cislunar expansion and space domain awareness", 
  journal = "Progress in Aerospace Sciences",
  volume  = "147",
  pages = "101019 (16 pp.)"
  }

@ ARTICLE{ pB77,
  author = "P. G. D. Barkham and V. J. Modi and A. C. Soudack",
  year = "1977",   
  title = "Asymptotic solutions in the many-body problem: {Part II: Periodic} orbits in four-body systems",
  journal = "Celestial Mechanics",
  volume = "15",
  pages = "5--20"
  }

@ ARTICLE{ cB94,
  author = "C. Beaug\'e",
  year = "1994",   
  title = "Asymmetric librations in exterior resonances",
  journal = "Celestial Mechanics and Dynamical Astronomy",
  volume = "60",
  pages = "225--248"
  }

@ ARTICLE{ eB93,
  author = "E. A. Belbruno and J. K. Miller",
  year = "1993",   
  title = "Sun-perturbed {Earth-to-Moon} transfers with ballistic capture",
  journal = "Journal of Guidance, Control, and Dynamics",
  volume = "16",
  pages = "770--775"
  }

@ ARTICLE{ eB08,
  author = "E. A. Belbruno and F. Topputo and M. Gidea",
  year = "2008",   
  title = "Resonance transitions associated to weak capture in the restricted three-body problem",
  journal = "Advances in Space Research",
  volume = "42",
  pages = "1330--1351"
  }

@ ARTICLE{ cB17,
  author = "C. Bezrouk and J. S. Parker",
  year = "2017",   
  title = "Long term evolution of distant retrograde orbits in the {Earth-Moon} system",
  journal = "Astrophysics and Space Science",
  volume = "362",
  pages = "176 (11 pp.)"
  }

@ ARTICLE{ aBdA24,
  author = "D. Binder and D. Arnas",
  year = "2024",   
  title = "Reliable and repeatable transit through cislunar space using 2:1 resonant spatial orbits",
  journal = "Journal of Guidance, Control, and Dynamics",
  volume = "47",
  pages = "1973--1979"
  }

@ ARTICLE{ aB25,
  author = "A. Black and C. Frueh",
  year = "2025",   
  title = "Fragmentation characterization in the circular restricted three body problem for cislunar space domain awareness",
  journal = "Advances in Space Research",
  volume = "75",
  pages = "1177--1204"
  }

@ ARTICLE{ wB71,
  author = "W. B. Blair",
  year = "1971", 
  title = "Analytical investigation of near-parabolic lunar trajectories between {Moon} and cislunar libration point",
  journal = "AIAA Journal",
  volume = "9",
  pages = "2437--2442"
  }

@ ARTICLE{ lB59,
  author = "L. Blitzer",
  year = "1959", 
  title = "Lunar-solar perturbations of an {Earth} satellite",
  journal = "American Journal of Physics",
  volume = "27",
  pages = "634--645"
  }

@ ARTICLE{ eBjM95,
  author = "E. M. Bollt and J. D. Meiss",
  year = "1995",   
  title = "Targeting chaotic orbits to the {Moon} through recurrence",
  journal = "Physics Letters A",
  volume = "204",
  pages = "373--378"
  }

@ ARTICLE{ sBnB23,
  author = "S. Bonasera and N. Bosanac",
  year = "2023",   
  title = "Computing natural transitions between tori near resonances in the {Earth--Moon} system",
  journal = "Journal of Guidance, Control, and Dynamics",
  volume = "46",
  pages = "443--454"
  }

@ ARTICLE{ nB23,
  author = "N. R. Boone and R. A. Bettinger",
  year = "2023",   
  title = "Debris collision risk analysis following simulated cislunar spacecraft explosions",
  journal = "Journal of Spacecraft and Rockets",
  volume = "60",
  pages = "668--684"
  }

@ ARTICLE{ kB20,
  author = "K. K. Boudad and K. C. Howell and D. C. Davis",
  year = "2020",   
  title = "Dynamics of synodic resonant near rectilinear halo orbits in the bi-circular four-body problem",
  journal = "Advances in Space Research",
  volume = "66",
  pages = "2194--2214"
  }

@ ARTICLE{ aB69,
  author = "A. Bonetti and G. Moreno and S. Cantarano and others",
  year = "1969",   
  title = "Solar-wind observations with satellite {ESRO HEOS-1} in {December} 1968",
  journal = "Il Nuovo Cimento",
  volume = "64",
  pages = "307--323"
  }

@ ARTICLE{ nB26,
  author = "N. Bosanac",
  year = "2026",   
  title = "Clustering natural trajectories in the {Earth}--{Moon} circular restricted three-body problem",
  journal = "The Journal of the Astronautical Sciences",
  volume = "73",
  pages = "2 (41 pp.)"
  }

@ ARTICLE{ jBG15,
  author = "J. Burgos-Garc\'ia and M. Gidea",
  year = "2015",   
  title = "Hill's approximation in a restricted four-body problem",
  journal = "Celestial Mechanics and Dynamical Astronomy",
  volume = "122",
  pages = "117--141"
  }

@ ARTICLE{ jBbG98,
  author = "J. A. Burns and B. J. Gladman",
  year = "1998",   
  title = "Dynamically depleted zones for {Cassini}'s safe passage beyond {Saturn}'s rings",
  journal = "Planetary and Space Science",
  volume = "46",
  pages = "1401--1407"
  }

@ ARTICLE{ nC01,
  author = "N. {Callegari, Jr.} and T. Yokoyama",
  year = "2001",   
  title = "Some aspects of the dynamics of fictitious {Earth's} satellites",
  journal = "Planetary and Space Science",
  volume = "49",
  pages = "35--46"
  }

@ ARTICLE{ gC22,
  author = "G. A. Carit\'a and A. C. Signor and M. H. M. Morais",
  year = "2022",
  title = "A numerical study of the 1/2, 2/1, and 1/1 retrograde mean motion resonances in planetary systems",
  journal = "Monthly Notices of the Royal Astronomical Society",
  volume = "515",
  pages = "2280--2292"
  }

@ ARTICLE{ jC10,
  author = "J. Casoliva and J. M. Mondelo and B. F. Villac and others",
  year = "2010",   
  title = "Two classes of cycler trajectories in the {Earth--Moon} system",
  journal = "Journal of Guidance, Control, and Dynamics",
  volume = "33",
  pages = "1623--1640"
  }

@ ARTICLE{ rC12,
  author = "R. Castelli",
  year = "2012",   
  title = "Regions of prevalence in the coupled restricted three-body problem approximation",
  journal = "Communications in Nonlinear Science and Numerical Simulation",
  volume = "2012",
  pages = "1804--816"
  }

@ ARTICLE{ iC23,
  author = "I. Cavallari and C. Grassi and G. F. Gronchi and G. Ba\`{u} and G. B. Valsecchi",
  year = "2023",   
  title = "A dynamical definition of the sphere of influence of the {Earth}",
  journal = "Communications in Nonlinear Science and Numerical Simulation",
  volume = "119",
  pages = "107091 (26 pp.)"
  }

@ ARTICLE{ aCaR17,
  author = "A. Celletti and C. Gale\c{s} and G. Pucacco and A. J. Rosengren",
  year = "2017",   
  title = "Analytical development of the lunisolar disturbing function and the critical inclination secular resonance",
  journal = "Celestial Mechanics and Dynamical Astronomy",
  volume = "127",
  pages = "259--283"
  }

@ ARTICLE{ gC64,
  author = "G. A. Chebotarev",
  year = "1964",   
  title = "Gravitational spheres of the major planets, {Moon}, and {Sun}", 
  journal = "Soviet Astronomy",
  volume  = "7",
  pages = "618--622",
  }

@ ARTICLE{ bC79,
  author = "B. V. Chirikov",
  year = "1979",   
  title = "A universal instability of many-dimensional oscillator systems",
  journal = "Physics Reports",
  volume = "52",
  pages = "263--379"
  }

@ ARTICLE{ pCcS00,
  author = "P. Cincotta and C. Sim\'{o}",
  year = "2000",   
  title = "Simple tools to study global dynamics in non-axisymmetric galactic potentials - {I}",
  journal = "Astronomy and Astrophysics Supplement",
  volume = "147",
  pages = "205--228"
  }

@ ARTICLE{ pC03,
  author = "P. Cincotta and C. Giordano and C. Sim\'{o}",
  year = "2003",   
  title = "Phase space structure of multi-dimensional systems by means of the mean exponential growth factor of nearby orbits",
  journal = "Physica D",
  volume = "182",
  pages = "151--178"
  }

@ ARTICLE{ mC22,
  author = "M. Cinelli",
  year = "2022",   
  title = "Inner third-body perturbations including the inclination and eccentricity of the perturbing body",
  journal = "Monthly Notices of the Royal Astronomical Society",
  volume = "517",
  note = "3904--3915"
  }

@ ARTICLE{ mC24,
  author = "M. Cinelli",
  year = "2024",   
  title = "Frozen orbits with inner planar perturbing body up to triakontadipole level of approximation",
  journal = "Acta Astronautica",
  volume = "214",
  note = "72--82"
  }

@ ARTICLE{ gC60,
  author = "G. Colombo",
  year = "1960",
  title = "Sui satelliti del sistema terra-luna",
  journal = "Rendiconti dell'Accademia Nazionale dei Lincei",
  series = "Serie VIII",
  volume = "28",
  pages = "169--172"
  }

@ ARTICLE{ gC61,
  author = "G. Colombo",
  year = "1961",
  title = "The stabilization of an artificial satellite at the inferior conjunction point of the {Earth--Moon} system",
  journal = "Smithsonian Astrophysical Observatory Special Report",
  volume = "80",
  pages = "1--13"
  }

@ ARTICLE{ cC68,
  author = "C. C. Conley",
  year = "1968",
  title = "Low energy transit orbits in the restricted three-body problem",
  journal = "SIAM Journal on Applied Mathematics",
  volume = "16",
  pages = "732--746"
  }

@ ARTICLE{ gC67,
  author = "G. E. Cook and D. W. Scott",
  year = "1967",   
  title = "Lifetimes of satellites in large-eccentricity orbits",
  journal = "Planetary and Space Science",
  volume = "15",
  pages = "1549--1556"
  }

@ ARTICLE{ jC63,
  author = "J. W. Crenshaw",
  year = "1963",   
  title = "Sphere of influence in patched-conic methods", 
  journal = "AIAA Journal",
  volume  = "1",
  pages = "2168--2170",
  }

@ ARTICLE{ jCetal64,
  author = "J. Cronin and P. B. Richards and L. H. Russell",
  year = "1964",   
  title = "Some periodic solutions of a four-body problem", 
  journal = "Icarus",
  volume  = "3",
  pages = "423--428",
  }

@ ARTICLE{ jCetal68,
  author = "J. Cronin and P. B. Richards and I. S. Bernstein",
  year = "1968",   
  title = "Some periodic solutions of a four-body problem. {II}", 
  journal = "Icarus",
  volume  = "9",
  pages = "281--290",
  }

@ ARTICLE{ jD65,
  author = "J. M. A. Danby",
  year = "1965",  
  title = "Inclusion of extra forces in the problem of three bodies",
  journal = "The Astronomical Journal",
  volume = "70", 
  pages = "181--189"
  }

@ ARTICLE{ jD22,
  author = "J. Daquin and E. Legnaro and I. Gkolias and C. Efthymiopoulos",
  year = "2022",  
  title = "A deep dive into the $2 g + h$ resonance: {Separatrices}, manifolds and phase-space structure of navigation satellites",
  journal = "Celestial Mechanics and Dynamical Astronomy",
  volume = "134", 
  pages = "6 (31 pp.)"
  }

@ ARTICLE{ jDaR16,
  author = "J. Daquin and A. J. Rosengren and E. M. Alessi and F. Deleflie and G. B. Valsecchi and A. Rossi",
  year = "2016",  
  title = "The dynamical structure of the {MEO} region: {Long}-term stability, chaos, and transport",
  journal = "Celestial Mechanics and Dynamical Astronomy",
  volume = "124", 
  pages = "335--366"
  }

@ ARTICLE{ mD64,
  author = "M. C. Davidson",
  year = "1964",
  title = "Numerical examples of transition orbits in the restricted three body problem",
  journal = "Astronautica Acta",
  volume = "10",
  pages = "309--313"
  }

@ARTICLE{ gdE19,
  author = "G. C. {de El{\'i}a} and M. Zanardi and A. Dugaro and S. Naoz",
  year = "2019",
  title = "The inverse {Lidov-Kozai} resonance for an outer test particle due to an eccentric perturber",
  journal = "Astronomy and Astrophysics",
  volume = "627",
  pages = "A17 (15 pp.)"
  }

@ ARTICLE{ vdO20,
  author = "V. M. {de Oliveira} and P. A. {Sousa-Silva} and I. L. Caldas",
  year = "2020",   
  title = "Order-chaos-order and invariant manifolds in the bounded planar {Earth--Moon} system",
  journal = "Celestial Mechanics and Dynamical Astronomy",
  volume = "132",
  pages = "51 (17 pp.)"
  }

@ ARTICLE{ bDS06,
  author = "B. {De Saedeleer}",
  year = "2006",   
  title = "Analytical theory of a lunar artificial satellite with third body perturbations",
  journal = "Celestial Mechanics and Dynamical Astronomy",
  volume = "95",
  pages = "407--423"
  }

@ ARTICLE{ pdSS16,
  author = "P. A. {de Sousa-Silva} and M. O. Terra",
  year = "2016",   
  title = "A survey of different classes of {Earth-to-Moon} trajectories in the patched three-body approach",
  journal = "Acta Astronautica",
  volume = "123",
  pages = "340--349"
  }

@ ARTICLE{ tDiS21,
  author = "T. V. Demidova and I. I. Shevchenko",
  year = "2021",   
  title = "Evolution of planetary chaotic zones in planetesimal disks",
  journal = "Astronomy Letters",
  volume = "47",
  pages = "771--781"
  }

@ ARTICLE{ dD14,
  author = "D. J. Dichmann and R. Lebois and J. P. {Carrico, Jr.}",
  year = "2014",   
  title = "Dynamics of orbits near 3:1 resonance in the {Earth-Moon} system",
  journal = "The Journal of the Astronautical Sciences",
  volume = "60", 
  pages = "51--86"
  }

@ ARTICLE{ aD93,
  author = "A. R. Dobrovolskis",
  year = "1993",   
  title = "The {Laplace} planes of {Uranus} and {Pluto}",
  journal = "Icarus",
  volume = "105",
  pages = "400--407"
  }

@ ARTICLE{ eD07,
  author = "E. J. Doedel and V. A. Romanov and R. C. Paffenroth and others",
  year = "2007",   
  title = "Elemental periodic orbits associated with the libration points in the circular restricted 3-body problem",
  journal = "International Journal of Bifurcation and Chaos",
  volume = "17",
  pages = "2625--2677"
  }

@ ARTICLE{ pDrK10,
  author = "P. Dutt and R. K. Sharma",
  year = "2010",   
  title = "Analysis of periodic and quasi-periodic orbits in the {Earth--Moon} system",
  journal = "Journal of Guidance, Control, and Dynamics",
  volume = "33",
  pages = "1010--1017"
  }

@ ARTICLE{ kE81,
  author = "K. A. Ehricke",
  year = "1981",   
  title = "A soci-economic evaluation of the lunar environment and resources. {I.} {Principles} and overall system strategy",
  journal = "Acta Astronautica",
  volume = "8", 
  pages = "1389--1405"
  }

@ ARTICLE{ nE03,
  author = "N. A. Eismont and A. V. Ditrikh and G. Janin and others",
  year = "2003",   
  title = "Orbit design for launching {INTEGRAL} on the {Proton/Block-DM} launcher",
  journal = "Astronomy and Astrophysics",
  volume = "411", 
  pages = "L37--L41"
  }

@ ARTICLE{ tE05,
  author = "T. A. Ely",
  year = "2005",   
  title = "Stable constellations of frozen elliptical inclined lunar orbits", 
  journal = "The Journal of the Astronautical Sciences",
  volume  = "53",
  pages = "301--316",
  }

@ ARTICLE{ tEkH97,
  author = "T. A. Ely and K. C. Howell",
  year = "1997",   
  title = "Dynamics of artificial satellite orbits with tesseral resonances 
  including the effects of luni-solar perturbations.",
  journal = "International Journal of Dynamics and Stability of Systems",
  volume = "12",
  pages = "243--269"
  }

@ ARTICLE{ vE59,
  author = "V. A. Egorov",
  year = "1959",   
  title = "On the question of capture in the restricted circular three-body problem", 
  journal = "Iskusstvennye Sputniki Zemli",
  volume  = "3",
  pages = "3--12",
  }

@ ARTICLE{ cP01,
  author = "C. P. Escoubet and M. Fehringer and M. Goldstein",
  year = "2001",   
  title = "Introduction: {The Cluster} mission", 
  journal = "Annales Geophysicae",
  volume  = "19",
  pages = "1197--1200",
  }

@ ARTICLE{ vE60,
  author = "V. R. Eshleman and P. B. Gallagher and R. C. Barthle",
  year = "1960",   
  title = "Radar methods of measuring the cislunar electron density",
  journal = "Journal of Geophysical Research",
  volume = "65",
  pages = "3079--3086"
  }

@ ARTICLE{ fFjL10,
  author = "F. Farago and J. Laskar",
  year = "2010",   
  title = "High-inclination orbits in the secular quadrupolar three-body problem",
  journal = "Monthly Notices of the Royal Astronomical Society",
  volume = "401",
  note = "1189--1198"
  }

@ ARTICLE{ rFaK73,
  author = "R. W. Farquhar and A. A. Kamel",
  year = "1973",   
  title = "Quasi-periodic orbits about the translunar libration point",
  journal = "Celestial Mechanics",
  volume = "7",
  pages = "458--473"
  }

@ ARTICLE{ rFetal04,
  author = "R. W. Farquhar and D. W. Dunham and Y. Guo and J. V. McAdams",
  year = "2004",   
  title = "Utilization of libration points for human exploration in the {Sun--Earth--Moon} system and beyond",
  journal = "Acta Astronautica",
  volume = "55",
  pages = "687--700"
  }

@ ARTICLE{ fFmL18,
  author = "F. Ferrari and M. Lavagna",
  year = "2018",   
  title = "Periodic motion around libration points in the elliptic restricted three-body problem",
  journal = "Nonlinear Dynamics",
  volume = "93",
  pages = "453--462"
  }

@ ARTICLE{ sFsG25,
  author = "S. Fu and S. Gong",
  year = "2025",   
  title = "Escape criterion for restricted three-body problem",
  journal = "Advances in Space Research",
  volume = "75",
  pages = "6376--6393"
  }

@ ARTICLE{ sFetal25,
  author = "S. Fu and Y. Liang and D. Wu and S. Gong",
  year = "2025",   
  title = "Low-energy {Earth--Moon} transfers with lunar ballistic capture based on {Lagrangian} coherent structures in a four-body model",
  journal = "Advances in Space Research",
  volume = "75",
  pages = "4993--5013"
  }

@ ARTICLE{ aG96,
  author = "A. A. Galeev and Y. I. Gal'Perin and L. M. Zelenyi",
  year = "1996",   
  title = "The {INTERBALL} project to study solar-terrestrial physics",
  journal = "Kosmicheskie Issledovaniya", 
  volume = "34", 
  pages = "339--362"
  }

@ ARTICLE{ tG06,
  author = "T. Gallardo",
  year = "2006",   
  title = "Atlas of the mean motion resonances in the {Solar System}",
  journal = "Icarus",
  volume = "184",
  pages = "29--38"
  }

@ ARTICLE{ tG19,
  author = "T. Gallardo",
  year = "2019",   
  title = "Strength, stability and three dimensional structure of mean-motion resonances in the {Solar System}",
  journal = "Icarus",
  volume = "317",
  pages = "121--134"
  }

@ ARTICLE{ tG20,
  author = "T. Gallardo",
  year = "2020",   
  title = "Three-dimensional structure of mean-motion resonances beyond {Neptune}",
  journal = "Celestial Mechanics and Dynamical Astronomy",
  volume = "132",
  pages = "9 (26 pp.)"
  }

@ ARTICLE{ tG12,
  author = "T. Gallardo and G. Hugo and P. Pais",
  year = "2012",   
  title = "Survey of {Kozai} dynamics beyond {Neptune}",
  journal = "Icarus",
  volume = "220",
  pages = "392--403"
  }

@ ARTICLE{ tG21,
  author = "T. Gallardo and C. Beaug\'e and C. A. Giuppone",
  year = "2021",   
  title = "Semianalytical model for planetary resonances: {Application} to planets around single and binary stars",
  journal = "Astronomy and Astrophysics",
  volume = "646",
  pages = "A148 (14 pp.)"
  }

@ ARTICLE{ gG70,
  author = "G. E. O. Giacaglia and J. P. Murphy and T. L. Felsentreger",
  year = "1970",   
  title = "A semi-analytic theory for the motion of a lunar satellite",
  journal = "Celestial Mechanics",
  volume = "3",
  pages = "3--66"
  }

@ ARTICLE{ gG04,
  author = "G{\'o}mez and G., Koon, W. S. and  Lo, M. W. and Marsden, J. E. and Masdemont, J. and Ross, S. D.",
  year = "2004",   
  title = "Connecting orbits and invariant manifolds in the spatial restricted three-body problem",
  journal = "Nonlinearity",
  volume = "17",
  pages = "1571--1606"
  }

@ ARTICLE{ pG66,
  author = "P. Goldreich",
  year = "1966",   
  title = "History of the lunar orbit",
  journal = "Reviews of Geophysics",
  volume = "4",
  pages = "411--439"
  }

@ ARTICLE{ vG61,
  author = "V. T. Gontkovskaya and G. A. Chebotarev",
  year = "1961",   
  title = "The motion of the space probe {Lunik III}",
  journal = "Soviet Astronomy",
  volume = "5",
  pages = "91--94"
  }

@ ARTICLE{ vG62,
  author = "V. T. Gontkovskaya and G. A. Chebotarev",
  year = "1962",   
  title = "Lunar and solar perturbations of {Lunik III}",
  journal = "Soviet Astronomy",
  volume = "5",
  pages = "728--732"
  }

@ ARTICLE{ pG23,
  author = "P. Guardabasso and D. K. Skoulidou and L. Bucci and others",
  year = "2023",   
  title = "Analysis of accidental spacecraft break-up events in cislunar space",
  journal = "Advances in Space Research",
  volume = "72",
  pages = "1550--1569"
  }

@ ARTICLE{ mGkH25,
  author = "M. Gupta and K. C. Howell",
  year = "2025",   
  title = "Leveraging {Earth--Moon} resonant orbits for access throughout the cislunar region",
  journal = "The Journal of the Astronautical Sciences",
  volume = "72",
  pages = "42 (39 pp.)"
  }

@ ARTICLE{ yHsCyW26,
  author = "Y. He and S. Cui and Y. Wang",
  year = "2026",   
  title = "A review of cislunar constellation design and optimization",
  journal = "Advances in Astronautics",
  volume = "9",
  pages = "1--31"
  }

@ ARTICLE{ mH68,
  author = "M. H\'enon",
  year = "1968",   
  title = "Sur les orbites interplan\'etaires qui rencontrent deux fois la terre",
  journal = "Bulletin astronomique",
  volume = "3",
  pages = "377--402"
  }

@ ARTICLE{ dHdS23,
  author = "D. B. Henry and D. J. Scheeres",
  year = "2023",   
  title = "Quasi-periodic orbit transfer design via whisker intersection sets",
  journal = "Journal of Guidance, Control, and Dynamics",
  volume = "46",
  pages = "1929--1944"
  }

@ ARTICLE{ tH10,
  author = "T. C. Hinse and A. A. Christou and J. L. A. Alvarellos and K. Go\'zdziewski",
  year = "2010",   
  title = "Application of the {MEGNO} technique to the dynamics of {Jovian} irregular satellites",
  journal = "Monthly Notices of the Royal Astronomical Society",
  volume = "2",
  pages = "837--857"
  }

@ ARTICLE{ dH77,
  author = "D. L. Hitzl",
  year = "1977",   
  title = "Generating orbits for stable close encounter periodic solutions of the restricted problem",
  journal = "AIAA Journal",
  volume = "15",
  pages = "1410--1418"
  }

@ ARTICLE{ mH93,
  author = "M. J. Holman and J. Wisdom",
  year = "1993",   
  title = "Dynamical stability in the outer {Solar System} and the delivery of short period comets",
  journal = "The Astronomical Journal",
  volume = "105",
  pages = "1987--1999"
  }

@ ARTICLE{ kH01,
  author = "K. C. Howell",
  year = "2001",   
  title = "Families of orbits in the vicinity of the collinear libration points",
  journal = "The Journal of the Astronautical Sciences",
  volume = "49",
  pages = "107--125"
  }

@ ARTICLE{ kH84,
  author = "K. C. Howell and J. V. Breakwell",
  year = "1984",   
  title = "Almost rectilinear halo orbits",
  journal = "Celestial Mechanics",
  volume = "32",
  pages = "29--52"
  }

@ ARTICLE{ kHmK06,
  author = "K. C. Howell and M. Kakoi",
  year = "2006",   
  title = "Transfers between the {Earth--Moon} and {Sun--Earth} systems using manifolds and transit orbits",
  journal = "Acta Astronautica",
  volume = "59",
  pages = "367--380"
  }

@ ARTICLE{ kHhP93,
  author = "K. C. Howell and H. J. Pernicka",
  year = "1993",   
  title = "Stationkeeping method for libration point trajectories",
  journal = "Journal of Guidance, Control, and Dynamics",
  volume = "16",
  pages = "151--159"
  }

@ ARTICLE{ sH61,
  author = "S. -S. Huang",
  year = "1961",   
  title = "Some dynamical properties of natural and artificial satellites",
  journal = "The Astronomical Journal",
  volume = "66",
  pages = "157--159"
  }

@ ARTICLE{ sH62,
  author = "S. -S. Huang",
  year = "1962",   
  title = "Preliminary study of orbits of interest for {Moon} probes",
  journal = "The Astronomical Journal",
  volume = "67",
  pages = "304--310"
  }

@ ARTICLE{ sH69,
  author = "S. -S. Huang",
  year = "1969",   
  title = "A hypothetical four-body problem and its applications",
  journal = "Vistas in Astronomy",
  volume = "10",
  pages = "113--124"
  }

@ ARTICLE{ sHcW63,
  author = "S. -S. Huang and C. {Wade, Jr.}",
  year = "1963",   
  title = "Preliminary study of orbits of interest for {Moon} probes. {II}",
  journal = "The Astronomical Journal",
  volume = "68",
  pages = "388--391"
  }

@ ARTICLE{ tI16,
  author = "T. Ito",
  year = "2016",   
  title = "High-order analytic expansion of disturbing function for doubly averaged circular restricted three-body problem",
  journal = "Advances in Astronomy",
  volume = "2016", 
  pages = "8945090 (23 pp.)"
  }

@ ARTICLE{ tI19,
  author = "T. Ito and K. Ohtsuka",
  year = "2019",   
  title = "The {Lidov-Kozai} oscillation and {Hugo von Zeipel}",
  journal = "Monographs on Environment, Earth and Planets",
  volume = "7",
  pages = "1--113"
  }

@ ARTICLE{ gJeR76,
  author = "G. Janin and E. A. Roth",
  year = "1976",   
  title = "Decay of a highly eccentric satellite",
  journal = "Celestial Mechanics",
  volume = "14", 
  pages = "141--149"
  }

@ ARTICLE{ fJ01,
  author = "F. Jansen and D. Lumb and B. Altieri and others",
  year = "2001",   
  title = "{XMM-Newton} observatory: {I}. {The} spacecraft and operations",
  journal = "Astronomy and Astrophysics",
  volume = "365", 
  pages = "L1--L6"
  }

@ ARTICLE{ aJ99,
  author = "\`A. Jorba",
  year = "1999",   
  title = "A methodology for the numerical computation of normal forms, centre manifolds and first integrals of {Hamiltonian} systems",
  journal = "Experimental Mathematics",
  volume = "8",
  pages = "155--195"
  }

@ ARTICLE{ aJbN20,
  author = "\`A. Jorba and B. Nicol\'as",
  year = "2020",   
  title = "Transport and invariant manifolds near $L_3$ in the {Earth--Moon} bicircular model",
  journal = "Communications in Nonlinear Science and Numerical Simulation",
  volume = "89",
  pages = "105327 (19 pp.)"
  }

@ ARTICLE{ aJjV98,
  author = "\`A. Jorba and J. Villanueva",
  year = "1998",   
  title = "Numerical computation of normal forms around some periodic orbits of the restricted three-body problem",
  journal = "Physica D",
  volume = "114",
  pages = "197--229"
  }

@ ARTICLE{ aJetal20,
  author = "{\'A}. Jorba and M. Jorba-Cusc{\'o} and J. Rosales",
  year = "2020",   
  title = "The vicinity of the {Earth--Moon} $L_1$ point in the bicircular problem",
  journal = "Celestial Mechanics and Dynamical Astronomy",
  volume = "132",
  pages = "11 (25 pp.)"
  }

@ ARTICLE{ jKjL68,
  author = "J. Kevorkian and J. E. Lancaster",
  year = "1968",   
  title = "An asymptotic solution for a class of periodic orbits of the restricted three-body problem",
  journal = "The Astronomical Journal",
  volume = "73",
  pages = "791--806"
  }

@ ARTICLE{ rK73,
  author = "R. W. Klebesadel and I. B. Strong and R. A. Olson",
  year = "1973",   
  title = "Observations of gamma-ray bursts of cosmic origin",
  journal = "The Astrophysical Journal", 
  volume = "182",
  pages = "L85--L88"
  }

@ ARTICLE{ wKeB58,
  author = "W. B. Klemperer and E. T. Benedikt",
  year = "1958",   
  title = "Selenoid satellites",
  journal = "Astronautica Acta", 
  volume = "4",
  pages = "25--30"
  }

@ ARTICLE{ dK62,
  author = "D. G. King-Hele",
  year = "1962",   
  title = "The contraction of satellite orbits under the influence of air drag. {III.} {High}-eccentricity orbits $(0.2 \leq e < 1)$",
  journal = "Proceedings of the Royal Society A",
  volume = "267",
  pages = "541--557"
  }

@ ARTICLE{ zK98,
  author = "Z. Kne\v{z}evi\'{c} and A. Milani",
  year = "1998",   
  title = "Orbit maintenance of a lunar polar orbiter",
  journal = "Planetary and Space Science",
  volume = "46",
  pages = "1605--1611"
  }

@ ARTICLE{ rKlC67,
  author = "R. Kolenkiewicz and L. Carpenter",
  year = "1967",   
  title = "Periodic motion around the triangular libration point in the restricted problem of four bodies",
  journal = "AIAA Journal",
  volume = "6",
  pages = "1301--1304"
  }

@ ARTICLE{ rKlC68,
  author = "R. Kolenkiewicz and L. Carpenter",
  year = "1968",   
  title = "Stable periodic orbits about the {Sun}-perturbed {Earth--Moon} triangular points",
  journal = "AIAA Journal",
  volume = "6",
  pages = "1301--1304"
  }

@ ARTICLE{ wK00,
  author = "Koon, W. S. and Lo, M. W. and Marsden, J. E. and Ross, S. D.",
  year = "2000",   
  title = "Heteroclinic connections between periodic orbits and resonance transitions in celestial mechanics",
  journal = "Chaos: An Interdisciplinary Journal of Nonlinear Science",
  volume = "10",
  pages = "427--469"
  }

@ ARTICLE{ wK01,
  author = "Koon, W. S. and Lo, M. W. and Marsden, J. E. and Ross, S. D.",
  year = "2001",   
  title = "Low energy transfer to the {Moon}",
  journal = "Chaos: An Interdisciplinary Journal of Nonlinear Science",
  volume = "81",
  pages = "63--73"
  }

@ ARTICLE{ zK67,
  author = "Z. Kopal",
  year = "1967",   
  title = "Perturbations of the orbits of artificial satellites by an attraction of external bodies",
  journal = "Icarus",
  volume = "6",
  pages = "298--314"
  }

@ ARTICLE{ dK25,
  author = "D. A. Koplow",
  year = "2025",   
  title = "Pave outer space and put up a parking lot: {Lagrange} points should be the common heritage of mankind",
  journal = "Michigan Journal of International Law",
  volume = "46",
  pages = "403--461"
  }

@ ARTICLE{ yK62,
  author = "Y. Kozai",
  year = "1962",   
  title = "Secular perturbations of asteroids with high inclination and eccentricity",
  journal = "The Astronomical Journal",
  volume = "67",
  pages = "591--598"
  }

@ ARTICLE{ yK63,
  author = "Y. Kozai",
  year = "1963",   
  title = "Motion of a lunar orbiter",
  journal = "Publications of the Astronomical Society of Japan",
  volume = "3",
  pages = "301--312"
  }

@ ARTICLE{ wK62,
  author = "W. M. Kaula",
  year = "1962",   
  title = "Development of the lunar and solar disturbing functions for a close satellite",
  journal = "The Astronomical Journal",
  volume = "67",
  pages = "300--303"
  }

@ ARTICLE{ bKaR26,
  author = "B. Kumar and A. Rawat and A. J. Rosengren and S. D. Ross",
  year = "2026",
  title = "Cislunar resonant transport and heteroclinic pathways: {From} 3:1 to 2:2 to {L1}",
  journal = "Advances in Space Research",
  volume = "77", 
  pages = "3815--3843"
  }

@ ARTICLE{ pLjK63a,
  author = "P. A. Lagerstrom and J. Kevorkian",
  year = "1963",   
  title = "Matched-conic approximation to the two fixed force-center problem",
  journal = "The Astronomical Journal",
  volume = "68",
  pages = "84--92"
  }

@ ARTICLE{ pLjK63b,
  author = "P. A. Lagerstrom and J. Kevorkian",
  year = "1963",   
  title = "{Earth-to-Moon} trajectories in the restricted three-body problem",
  journal = "Journal de M\'ecanique",
  volume = "2",
  pages = "189--218"
  }

@ ARTICLE{ pLjK66,
  author = "P. A. Lagerstrom and J. Kevorkian",
  year = "1966",   
  title = "Nonplanar {Earth-to-Moon} trajectories in the restricted three-body problem",
  journal = "AIAA Journal",
  volume = "4",
  pages = "149--152"
  }

@ ARTICLE{ jLgB10,
  author = "J. Laskar and G. Bou{\'e}",
  year = "2010",
  title = "Explicit expansion of the three-body disturbing function for arbitrary eccentricities and inclinations",
  journal = "Astronomy and Astrophysics",
  volume = "522",
  pages = "A60 (11 pp.)"
  }

@ ARTICLE{ jLetal25,
  author = "J. Lee and K. {Sun Park} and K.-J. Hwang and others",
  year = "2025",   
  title = "Long-term {Earth} magnetosphere science orbit with {Earth}--{Moon} resonance orbit",
  journal = "Advances in Space Research",
  volume = "76",
  pages = "3515--3527"
  }

@ ARTICLE{ eLcE23,
  author = "E. Legnaro and C. Efthymiopoulos",
  year = "2023",   
  title = "A detailed dynamical model for inclination-only dependent lunisolar resonances. {Effect} on the ``eccentricity growth'' mechanism",
  journal = "Advances in Space Research",
  volume = "72",
  pages = "2460--2480"
  }

@ ARTICLE{ eLcE24,
  author = "E. Legnaro and C. Efthymiopoulos",
  year = "2024",   
  title = "Secular dynamics and the lifetimes of lunar artificial satellites under natural force-driven orbital evolution",
  journal = "Acta Astronautica",
  volume = "225",
  pages = "768--787"
  }

@ARTICLE{ hLyxG24,
  author = "H. Lei and Y. -X. Gong",
  year = "2024",
  title = "Dynamical structures of misaligned circumbinary planets under hierarchical three-body systems",
  journal = "Monthly Notices of the Royal Astronomical Society",
  volume = "532",
  pages = "1580--1597"
  }

@ ARTICLE{ hLbX18,
  author = "H. Lei and B. Xu",
  year = "2018",   
  title = "Resonance transition periodic orbits in the circular restricted three-body problem",
  journal = "Astrophysics and Space Science",
  volume = "363",
  pages = "70 (11 pp.)"
  }

@ ARTICLE{ aLcB08,
  author = "A. M. Leiva and C. B. Briozzo",
  year = "2008",   
  title = "Extension of fast periodic transfer orbits from the {Earth--Moon} {RTBP} to the {Sun--Earth--Moon} quasi-bicircular problem",
  journal = "Celestial Mechanics and Dynamical Astronomy",
  volume = "101",
  pages = "225--245"
  }

@ ARTICLE{ cLwM11,
  author = "C. Levit and W. Marshall",
  year = "2011",   
  title = "Improved orbit prediction using two-line elements",
  journal = "Advances in Space Research",
  volume = "47",
  pages = "1107--1115"
  }

@ARTICLE{ dL14,
  author = "D. Li and J. -L. Zhou and H. Zhang",
  year = "2014",
  title = "Analytical theories for near coplanar and polar circumbinary orbits",
  journal = "Monthly Notices of the Royal Astronomical Society",
  volume = "437",
  pages = "3832--3841"
  }

@ ARTICLE{ rLetal26,
  author = "R. Li and J. J. Masdemont and Z. Zhu and C. Gao",
  year = "2026",   
  title = "The {Sun--Earth} heteroclinics in restricted four-body non-autonomous models",
  journal = "Communications in Nonlinear Science and Numerical Simulation",
  volume = "152",
  pages = "109356"
  }

@ ARTICLE{ yL17,
  author = "Y. Liang and M. Xu and S. Xu",
  year = "2017",   
  title = "The cislunar polygonal-like periodic orbit: {Construction}, transition and its application", 
  journal = "Acta Astronautica",
  volume  = "133",
  pages = "282--301"
  }

@ ARTICLE{ yL20,
  author = "Y. Liang and M. Xu and K. Peng and S. Xu",
  year = "2020",   
  title = "A cislunar in-orbit infrastructure based on $p:q$ resonant cycler orbits", 
  journal = "Acta Astronautica",
  volume  = "170",
  pages = "539--551"
  }

@ ARTICLE{ mL62,
  author = "M. L. Lidov",
  year = "1962",   
  title = "The Evolution of Orbits of Artificial Satellites of Planets Under the 
  Action of Gravitational Perturbations of External Bodies",
  journal = "Planetary and Space Science",
  volume = "9",
  pages = "719--759"
  }

@ARTICLE{ bL15,
  author = "B. Liu and D. J. Mu{\~n}oz and D. Lai",
  year = "2015",
  title = "Suppression of extreme orbital evolution in triple systems with short-range forces",
  journal = "Monthly Notices of the Royal Astronomical Society",
  volume = "447",
  pages = "747--764"
  }

@ ARTICLE{ jL86,
  author = "J. J. F. Liu and J. Segrest and V. Szebehely",
  year = "1986",   
  title = "Orbit mechanics of deep space probes",
  journal = "The Journal of the Astronautical Sciences",
  volume = "34",
  pages = "171--187"
  }

@ ARTICLE{ jL85,
  author = "J. Llibre and R. Martinez and C. Sim\'o",
  year = "1985",   
  title = "Tranversality of the invariant manifolds associated to the {Lyapunov} family of periodic orbits near {L2} in the restricted three-body problem",
  journal = "Journal of Differential Equations",
  volume = "58",
  pages = "104--156"
  }

@ ARTICLE{ bL72,
  author = "B. E. Lowrey",
  year = "1972",   
  title = "Ephemeris of a highly eccentric orbit: {Explorer} 28",
  journal = "Celestial Mechanics",
  volume = "5",
  pages = "107--125"
  }

@ ARTICLE{ gL63,
  author = "G. H. Ludwig",
  year = "1963",   
  title = "The orbiting geophysical observatories",
  journal = "Space Science Reviews",
  volume = "2",
  pages = "175--218"
  }

@ ARTICLE{ nLetal25,
  author = "N. Lugaz and N. Al-Haddad and B. Zhuang and others",
  year = "2025",   
  title = "The need for a sub-{L1} space weather research mission: {Current} knowledge gaps on coronal mass ejections",
  journal = "Space Weather",
  volume = "23",
  pages = "e2024SW004189 (18 pp.)"
  }

@ ARTICLE{ gMpG20,
  author = "G. Marcus and P. Gurfil",
  year = "2020",   
  title = "Inner third-body perturbations",
  journal = "Celestial Mechanics and Dynamical Astronomy",
  volume = "132",
  pages = "31 (17 pp.)"
  }

@ ARTICLE{ rMzC23,
  author = "R. Malhotra and Z. Chen",
  year = "2023",   
  title = "Non-perturbative investigation of low-eccentricity exterior mean motion resonances",
  journal = "Monthly Notices of the Royal Astronomical Society",
  volume = "521",
  note = "1253--1263"
  }

@ ARTICLE{ rM13,
  author = "R. A. Mardling",
  year = "2013",   
  title = "New developments for modern celestial mechanics -- {I. General} coplanar three-body systems. {Application} to exoplanets",
  journal = "Monthly Notices of the Royal Astronomical Society",
  volume = "435",
  note = "2187--2226"
  }

@ ARTICLE{ wMlB22,
  author = "W. Martens and L. Bucci",
  year = "2022",   
  title = "Double {Tisserand} graphs for low-energy lunar transfer design",
  journal = "Frontiers in Space Technologies", 
  volume = "3", 
  pages = "920456 (12 pp.)"
  }

@ ARTICLE{ fM20,
  author = "F. Marzari",
  year = "2020",   
  title = "Ring dynamics around an oblate body with an inclined satellite: the case of {Haumea}",
  journal = "Astronomy and Astrophysics",
  volume = "643", 
  pages = "A67 (6 pp.)"
  }

@ ARTICLE{ dMetal15,
  author = "D. D. Mazanek and R. G. Merrill and J. R. Brophy and R. P. Mueller",
  year = "2015",   
  title = "{Asteroid Redirect Mission} concept: A bold approach for utilizing space resources",
  journal = "Acta Astronautica",
  volume = "117",
  pages = "163--171"
  }

@ ARTICLE{ bMkW23,
  author = "B. McCarthy and K. Howell",
  year = "2023",   
  title = "Construction of heteroclinic connections between quasi-periodic orbits in the three-body problem",
  journal = "The Journal of the Astronautical Sciences",
  volume = "70",
  note = "24 (16 pp.)"
  }

@ ARTICLE{ mB00,
  author = "M. D. Melita and A. Brunini",
  year = "2020",   
  title = "Comparative study of mean-motion resonances in the trans-{Neptunian} region",
  journal = "Icarus",
  volume = "147", 
  pages = "205--219"
  }

@ ARTICLE{ mM11,
  author = "M. F. Mestre and P. M. Cincotta and C. M. Giordano",
  year = "2011",   
  title = "Analytical relation between two chaos indicators: {FLI and MEGNO}",
  journal = "Monthly Notices of the Royal Astronomical Society",
  volume = "414", 
  pages = "L100--103"
  }

@ ARTICLE{ jM63,
  author = "J. E. Michaels",
  year = "1963",   
  title = "Design analysis of {Earth}--lunar trajectories: {Launch} and transfer characteristics",
  journal = "AIAA Journal",
  volume = "1",
  note = "1342--1350"
  }

@ ARTICLE{ dM11,
  author = "D. J. McComas and J. P. {Carrico, Jr.} and B. Hautamaki and others",
  year = "2011",   
  title = "A new class of long-term stable lunar resonance orbits: 
  {Space} weather applications and the {Interstellar Boundary Explorer}",
  journal = "Space Weather",
  volume = "9",
  note = "S11002 (9 pp.)"
  }

@ ARTICLE{ lMjK67,
  author = "L. Mohn and J. Kevorkian",
  year = "1967",   
  title = "Some limiting cases of the restricted four-body problem",
  journal = "The Astronomical Journal",
  volume = "72",
  pages = "959--963"
  }

@ ARTICLE{ sM15,
  author = "S. Morrison and R. Malhotra",
  year = "2015",
  title = "Planetary chaotic zone clearing: destinations and timescales",
  journal = "The Astrophysical Journal",
  volume = "799",
  pages = "41 (8 pp)"
  }

@ ARTICLE{ pM61,
  author = "P. Musen",
  year = "1961",   
  title = "On the Long-Period Lunar and Solar Effects on the Motion of an Artificial Satellite, 2",
  journal = "Journal of Geophysical Research",
  volume = "66",
  pages = "2797--2805"
  }

@ ARTICLE{ sN16,
  author = "S. Naoz",
  year = "2016",
  title = "The eccentric {Kozai-Lidov} effect and its applications",
  journal = "Annual Review of Astronomy and Astrophysics",
  volume = "54",
  pages = "441--489"
  }

@ ARTICLE{ sN17,
  author = "S. Naoz and G. Li and M. Zanardi and others",
  year = "2017",
  title = "The eccentric {Kozai--Lidov} mechanism for outer test particle",
  journal = "The Astronomical Journal",
  volume = "154",
  pages = "18 (11 pp.)"
  }

@ ARTICLE{ rN19,
  author = "R. B. Negri and A. Sukhanov and A. F. B. {de Almeida Prado}",
  year = "2019",   
  title = "Lunar gravity assists using patched-conics approximation, three and four body problems",
  journal = "Advances in Space Research", 
  volume = "64", 
  pages = "42--63"
  }

@ ARTICLE{ dNaM98,
  author = "D. Nesvorn{\'y} and A. Morbidelli",
  year = "1998",   
  title = "Three-body mean motion resonances and the chaotic structure of the asteroid belt",
  journal = "The Astronomical Journal",
  volume = "116",
  pages = "3029--3037"
  }

@ ARTICLE{ dNrR00,
  author = "D. Nesvorn{\'y} and F. Roig",
  year = "2000",   
  title = "Mean motion resonances in the trans-neptunian region part {I: The} 2:3 Resonance with {Neptune}",
  journal = "Icarus",
  volume = "148",
  pages = "282--300"
  }

@ ARTICLE{ dNrR01,
  author = "D. Nesvorn{\'y} and F. Roig",
  year = "2001",   
  title = "Mean motion resonances in the transneptunian region part {II: The} 1:2, 3:4, and weaker resonances",
  journal = "Icarus",
  volume = "150",
  pages = "104--123"
  }

@ ARTICLE{ rN59,
  author = "R. R. Newton",
  year = "1959",   
  title = "Periodic orbits of a planetoid passing close to two gravitating masses",
  journal = "Smithsonian Contribution to Astrophysics",
  volume = "3",
  pages = "69--78"
  }

@ ARTICLE{ tNpG18,
  author = "T. Nie and P. Gurfil",
  year = "2018",   
  title = "Lunar frozen orbits revisited",
  journal = "Celestial Mechanics and Dynamical Astronomy",
  volume = "130",
  pages = "61 (35 pp.)"
  }

@ ARTICLE{ aN94,
  author = "A. Nishida",
  year = "1994",   
  title = "The {GEOTAIL} mission",
  journal = "Geophysical Research Letters", 
  volume = "21", 
  pages = "2871--2873"
  }

@ ARTICLE{ aNetal15,
  author = "A. Noullez and K. Tsiganis and S. Tzirti",
  year = "2015",   
  title = "Satellite orbits design using frequency analysis",
  journal = "Advances in Space Research", 
  volume = "56", 
  pages = "163--175"
  }

@ ARTICLE{ kO17,
  author = "K. Onozaki and H. Yoshimura and S. D. Ross",
  year = "2017",   
  title = "Tube dynamics and low energy {Earth}--{Moon} transfers in the 4-body system",
  journal = "Advances in Space Research", 
  volume = "60",
  pages = "2117--2132"
  }

@ ARTICLE{ kO19a,
  author = "K. Oshima",
  year = "2019",   
  title = "The use of vertical instability of L1 and L2 planar Lyapunov orbits for transfers from near rectilinear halo orbits to planar distant retrograde orbits in the {Earth-Moon} system",
  journal = "Celestial Mechanics and Dynamical Astronomy",
  volume = "131",
  pages = "53 (28 pp.)"
  }

@ ARTICLE{ kO19b,
  author = "K. Oshima",
  year = "2019",   
  title = "Linking low- to high-energy dynamics of invariant manifolds, transit orbits, and singular collision orbits in the planar circular restricted three-body problem",
  journal = "Celestial Mechanics and Dynamical Astronomy",
  volume = "131",
  pages = "14 (15 pp.)"
  }

@ ARTICLE{ kO21,
  author = "K. Oshima",
  year = "2021",   
  title = "Retrograde co-orbital orbits in the {Earth--Moon} system: planar stability region under solar gravitational perturbation",
  journal = "Astrophysics and Space Science",
  volume = "366",
  pages = "88 (11 pp.)"
  }

@ ARTICLE{ kOetal17,
  author = "K. Oshima and F. Topputo and S. Campagnola and T. Yanao",
  year = "2017",   
  title = "Analysis of medium-energy transfers to the {Moon}",
  journal = "Celestial Mechanics and Dynamical Astronomy",
  volume = "127",
  pages = "285--300"
  }

@ ARTICLE{ kOetal19,
  author = "K. Oshima and F. Topputo and T. Yanao",
  year = "2019",   
  title = "Low-energy transfers to the {Moon} with long transfer time",
  journal = "Celestial Mechanics and Dynamical Astronomy",
  volume = "131",
  pages = "4 (19 pp.)"
  }

@ ARTICLE{ sP26,
  author = "S. Pan and T. Urashi and M. Bando and Y. Yoshimura and H. Chen and T. Hanada",
  year = "2026",   
  title = "Data-driven prediction of chaotic transition in periapsis {Poincar{\'e}} maps",
  journal = "Nonlinear Dynamics",
  volume = "114",
  pages = "575"
  }

@ ARTICLE{ bPjM25,
  author = "B. Pang and J. J. Masdemont and D. Qiao",
  year = "2025",   
  title = "Temporary capture about the {Moon} involving {Sun--Earth} libration point dynamics",
  journal = "Communications in Nonlinear Science and Numerical Simulation",
  volume = "147",
  pages = "108792 (27 pp.)"
  }

@ ARTICLE{ bPkH24,
  author = "B. Park and K. C. Howell",
  year = "2024",   
  title = "Assessment of dynamical models for transitioning from the circular restricted three-body problem to an ephemeris model with applications",
  journal = "Celestial Mechanics and Dynamical Astronomy",
  volume = "136",
  pages = "6 (39 pp.)"
  }

@ ARTICLE{ cP24,
  author = "C. Peng and Y. Zhang and S. He",
  year = "2024",   
  title = "3:1/3:2 resonant orbits touring {$L_3$--$L_5$} in cislunar space",
  journal = "Advances in Space Research",
  volume = "73",
  pages = "2499--2514"
  }

@ ARTICLE{ lP25,
  author = "L. Peng and Y. Liang and X. He",
  year = "2025",   
  title = "Transfers to {Earth--Moon} triangular libration points by {Sun}-perturbed dynamics",
  journal = "Advances in Space Research",
  volume = "75",
  pages = "2837--2855"
  }

@ ARTICLE{ lPetal24,
  author = "L. T. Peterson and G. Brown and \'A. Jorba and D. J. Scheeres",
  year = "2024",
  title = "Dynamics around the {Earth--Moon} triangular points in the {Hill} restricted 4-body problem",
  journal = "Celestial Mechanics and Dynamical Astronomy",
  volume = "136",
  pages = "31 (41 pp.)"
  }

@ ARTICLE{ lPetal23,
  author = "L. T. Peterson and J. J. Rosales and D. J. Scheeres",
  year = "2023",
  title = "The vicinity of {Earth--Moon} $L_1$ and $L_2$ in the {Hill} restricted 4-body problem",
  journal = "Physica D",
  volume = "455",
  pages = "133889 (15 pp.)"
  }

@ ARTICLE{ jP62,
  author = "J. H. Piddington",
  year = "1962",   
  title = "The cis-lunar magnetic field",
  journal = "Planetary and Space Science",
  volume = "9",
  pages = "305--318"
  }

@ ARTICLE{ aPrB94,
  author = "A. F. B. A. Prado and R. Broucke",
  year = "1994",
  title = "Study of {H\'{e}non's} orbit transfer problem using the {Lambert} algorithm",
  journal = "Journal of Guidance Control and Dynamics",
  volume = "17",
  pages = "1075--1081"
  }

@ ARTICLE{ aPrB96,
  author = "A. F. B. A. Prado and R. Broucke",
  year = "1996",
  title = "Transfer obits in the {Earth}--{Moon} system using a regularized model",
  journal = "Journal of Guidance Control and Dynamics",
  volume = "19",
  pages = "929--933"
  }

@ ARTICLE{ vP14,
  author = "V. I. Prokhorenko",
  year = "2014",   
  title = "On the features of long-term evolution of a high-apogee orbit of the {Spektr-R} spacecraft",
  journal = "Kosmicheskie Issledovaniya", 
  volume = "52",
  pages = "132--152"
  }

@ ARTICLE{ yQ14,
  author = "Y. Qi and S. Xu and R. Qi",
  year = "2014",   
  title = "Gravitational lunar capture based on bicircular model in restricted four body problem",
  journal = "Celestial Mechanics and Dynamical Astronomy",
  volume = "120",
  pages = "1--17"
  }

@ ARTICLE{ gRetal02,
  author = "G. D. Racca and A. Marini and L. Stagnaro and others",
  year = "2002",   
  title = "{SMART-1} mission description and development status",
  journal = "Planetary and Space Science",
  volume = "50",
  pages = "1323--1337"
  }

@ ARTICLE{ aRbK26,
  author = "A. Rawat and B. Kumar and A. J. Rosengren and S. D. Ross",
  year = "2026",
  title = "Cislunar mean-motion resonances: {Definitions}, widths, and comparisons with resonant satellites",
  journal = "Journal of Guidance Control and Dynamics",
  volume = "49", 
  pages = "4 (15 pp.)"
  }

@ ARTICLE{ eRaS62,
  author = "E. Rabe and A. Schanzle",
  year = "1962",   
  title = "Periodic librations about the triangular solutions of the restricted {Earth--Moon} problem and their orbital stabilities",
  journal = "The Astronomical Journal",
  volume = "67",
  pages = "732--739"
  }

@ ARTICLE{ hR12,
  author = "H. Rein and S. -F. Liu",
  year = "2012",   
  title = "{REBOUND}: an open-source multi-purpose {N-body} code for collisional dynamics",
  journal = "Astronomy and Astrophysics",
  volume = "537",
  pages = "A128 (10 pp.)"
  }

@ ARTICLE{ hR15,
  author = "H. Rein and D. S. Spiegel",
  year = "2015",   
  title = "{IAS15}: a fast, adaptive, high-order integrator for gravitational dynamics, accurate to machine precision over a billion orbits",
  journal = "Monthly Notices of the Royal Astronomical Society",
  volume = "2",
  pages = "1424--1437"
  }

@ ARTICLE{ gR15,
  author = "G. R. Ricker and J. N. Winn and R. Vanderspek and others",
  year = "2015",   
  title = "{Transiting Exoplanet Survey Satellite}",
  journal = "Journal of Astronomical Telescopes, Instruments, and Systems",
  volume = "1",
  pages = "014003 (10 pp.)"
  }

@ ARTICLE{ jR21,
  author = "J. J. Rosales and {\'A}. Jorba and M. Jorba-Cusc{\'o}",
  year = "2021",   
  title = "Families of halo-like invariant tori around $L_2$ in the {Earth--Moon} bicircular problem",
  journal = "Celestial Mechanics and Dynamical Astronomy",
  volume = "113",
  pages = "16 (30 pp.)"
  }

@ ARTICLE{ aRdS14_CMDA,
  author = "A. J. Rosengren and D. J. Scheeres",
  year = "2014",
  title = "On the {Milankovitch} orbital elements for perturbed Keplerian motion",
  journal = "Celestial Mechanics and Dynamical Astronomy",
  volume = "118", 
  pages = "197--220"
  }

@ ARTICLE{ aRdS14_ApJ,
  author = "A. J. Rosengren and D. J. Scheeres",
  year = "2014",
  title = "Laplace plane modifications arising from solar radiation pressure",
  journal = "The Astrophysical Journal",
  volume = "786", 
  pages = "45 (13 pp.)"
  }

@ ARTICLE{ aRdS14_ASR,
  author = "A. J. Rosengren and D. J. Scheeres and J. W. McMahon",
  year = "2014",
  title = "The classical {Laplace} plane as a stable disposal orbit for geostationary satellites",
  journal = "Advances in Space Research",
  volume = "53", 
  pages = "1219--1228"
  }

@ ARTICLE{ aR19,
  author = "A. J. Rosengren and D. K. Skoulidou and K. Tsiganis and G. Voyatzis",
  year = "2019",
  title = "Dynamical cartography of {Earth} satellite orbits",
  journal = "Advances in Space Research",
  volume = "63", 
  pages = "443--460"
  }

@ ARTICLE{ aR20,
  author = "A. J. Rosengren and H. Namazyfard and G.E.O. Giacaglia",
  year = "2020",   
  title = "Effects of higher-order multipoles of the lunar disturbing potential on elongated orbits in cislunar space",
  journal = "European Physics Journal Special Topics",
  volume = "229",
  pages = "1545--1555"
  }

@ ARTICLE{ sR07,
  author = "S. D. Ross and D. J. Scheeres",
  year = "2007",   
  title = "Multiple gravity assists, capture, and escape in the restricted three-body problem",
  journal = "SIAM Journal on Applied Dynamical Systems",
  volume = "6",
  pages = "576--596"
  }

@ ARTICLE{ aR68a,
  author = "A. E. Roy",
  year = "1968",   
  title = "The theory of the motion of an artificial lunar satellite {I. Development} of the disturbing function",
  journal = "Icarus",
  volume = "9",
  pages = "82--132"
  }

@ ARTICLE{ aR68b,
  author = "A. E. Roy",
  year = "1968",   
  title = "The theory of the motion of an artificial lunar satellite {I. The} first-order and second-order theories",
  journal = "Icarus",
  volume = "9",
  pages = "133--161"
  }

@ ARTICLE{ aR69,
  author = "A. E. Roy",
  year = "1969",   
  title = "Luni-solar perturbations of an {Earth} satellite",
  journal = "Astrophysics and Space Science",
  volume = "4",
  pages = "375--386"
  }

@ ARTICLE{ rS26,
  author = "R. R. Sanaga and B. Park and K. C. Howell",
  year = "2026",   
  title = "Unified transition scheme for invariant tori across various models in the cislunar domain",
  journal = "Communications in Nonlinear Science and Numerical Simulation",
  volume = "152",
  pages = "109323 (28 pp.)"
  }

@ ARTICLE{ dS98,
  author = "D. J. Scheeres",
  year = "1998",   
  title = "The restricted {Hill} four-body problem with applications to the {Earth--Moon--Sun} system",
  journal = "Celestial Mechanics and Dynamical Astronomy",
  volume = "70",
  pages = "75--98"
  }

@ ARTICLE{ hS68,
  author = "H. B. Schechter",
  year = "1968",   
  title = "Three-dimensional nonlinear stability analysis of the {Sun}-perturbed {Earth--Moon} equilateral points",
  journal = "AIAA Journal",
  volume = "6",
  pages = "1223--1228"
  }

@ ARTICLE{ sS25,
  author = "S. T. {Scheuerle, Jr.} and K. C. Howell and D. C. Davis",
  year = "2025",   
  title = "Energy-informed pathways: {A} fundamental approach to designing ballistic lunar transfers",
  journal = "Advances in Space Research",
  volume = "75",
  pages = "1096--1117"
  }

@ ARTICLE{ cSeO97,
  author = "C. G. Schroer and E. Ott",
  year = "1997",   
  title = "Targeting in {Hamiltonian} systems that have mixed regular/chaotic phase spaces",
  journal = "Chaos",
  volume = "7",
  pages = "512--519"
  }

@ ARTICLE{ lS60,
  author = "L. I. Sedov",
  year = "1960",   
  title = "Orbits of cosmic rockets toward the {Moon}",
  journal = "American Rocket Society Journal",
  volume = "30",
  pages = "14--21"
  }

@ ARTICLE{ sS60,
  author = "L. Sehnal",
  year = "1960",   
  title = "The stability of the libration points $L_4$ and $L_5$ in the system {Earth--Moon}", 
  journal = "Bulletin of the Astronomical Institutes of Czechoslovakia",
  volume = "11",
  pages = "130--131"
  }

@ ARTICLE{ ySmE67,
  author = "Y. -Y. Shi and M. C. Eckstein",
  year = "1967",   
  title = "Uniformly valid asymptotic solution of nonplanar {Earth-to-Moon} trajectories in the restricted four-body problem",
  journal = "The Astronomical Journal",
  volume = "72",
  pages = "685--701"
  }

@ ARTICLE{ iS20,
  author = "I. I. Shevchenko",
  year = "2020",   
  title = "Lyapunov and clearing timescales in planetary chaotic zones",
  journal = "The Astronomical Journal",
  volume = "160",
  pages = "212 (12pp)"
  }

@ ARTICLE{ bS66,
  author = "B. E. Shute",
  year = "1966",   
  title = "Geocentric initial conditions of trajectories originating at the {Moon's} surface",
  journal = "The Astronomical Journal",
  volume = "71",
  pages = "602--609"
  }

@ ARTICLE{ bSjC66,
  author = "B. E. Shute and J. Chiville",
  year = "1966",   
  title = "The lunar-solar effect on the orbital lifetimes of artificial satellites with highly eccentric orbits",
  journal = "Planetary and Space Science",
  volume = "14",
  pages = "361--369"
  }

@ ARTICLE{ dSrE24,
  author = "D. Schwab and R. Eapen and P. Singla",
  year = "2024",   
  title = "Characterizing accuracy of normal forms to study trajectories in cislunar space",
  journal = "The Journal of the Astronautical Sciences",
  volume = "71",
  pages = "16 (38 pp.)"
  }

@ ARTICLE{ vS18,
  author = "V. V. Sidorenko",
  year = "2018",   
  title = "The eccentric {Kozai--Lidov} effect as a resonance phenomenon",
  journal = "Celestial Mechanics and Dynamical Astronomy",
  volume = "130",
  pages = "4 (23 pp.)"
  }

@ ARTICLE{ jS94,
  author = "J. L. Simon and P. Bretagnon and J. Chapront and others",
  year = "1994",   
  title = "Numerical expressions for precession formulae and mean elements for the {Moon} and the planets",
  journal = "Astronomy and Astrophysics",
  volume = "282",
  pages = "663--683"
  }

@ ARTICLE{ dSetal20,
  author = "D. Souami and J. Cresson and C. Biernacki and F. Pierret",
  year = "2020",   
  title = "On the local and global properties of gravitational spheres of influence",
  journal = "Monthly Notices of the Royal Astronomical Society",
  volume = "496",
  pages = "4287--4297"
  }

@ ARTICLE{ pSS12,
  author = "P. A. {Sousa Silva} and M. O. Terra",
  year = "2012",   
  title = "Applicability and dynamical characterization of the associated sets of the algorithmic weak stability boundary in the lunar sphere of influence",
  journal = "Celestial Mechanics and Dynamical Astronomy",
  volume = "113",
  pages = "141--168"
  }

@ ARTICLE{ lSdV66,
  author = "L. Steg and J. P. {de Vries}",
  year = "1966",   
  title = "{Earth--Moon} libration points: {Theory}, existence and applications",
  journal = "Space Science Reviews",
  volume = "5",
  pages = "210--233"
  }

@ ARTICLE{ dSt98,
  author = "D. Steichen",
  year = "1998",   
  title = "An averaging method to study the motion of lunar artificial satellites: {II: Averaging} and applications",
  journal = "Celestial Mechanics and Dynamical Astronomy",
  volume = "68",
  pages = "225--247"
  }

@ ARTICLE{ bS26,
  author = "B. J. Stivi and V. Mallik and L. Dell'Elce and A. J. Rosengren",
  year = "2026",   
  title = "A dynamical coherency gate for state recovery: {Statistical} requiem for the long arc of cislunar orbital mis-prediction",
  journal = "Advances in Space Research",
  volume = "77",
  pages = "12167--12183"
  }

@ ARTICLE{ hS58,
  author = "H. Strughold",
  year = "1958",   
  title = "Spatiography: {Geography} for space",
  journal = "Missiles and Rockets",
  volume = "3", 
  pages = "106--108"
  }

@ ARTICLE{ kTtK25,
  author = "K. Takeda and T. Kuwahara",
  year = "2025",   
  title = "Divergence evaluation criteria for lunar departure trajectories under bi-circular restricted four-body problem",
  journal = "Aerospace",
  volume = "12",
  pages = "918 (25 pp.)"
  }

@ ARTICLE{ bTbS68,
  author = "B. D. Tapley and B. E. Schutz",
  year = "1968",   
  title = "Persistent solar influenced libration point motion",
  journal = "AIAA Journal",
  volume = "6",
  pages = "1405--1406"
  }

@ ARTICLE{ bT05,
  author = "B. D. Tapley and J. C. Ries and S. Bettadpur and others",
  year = "2005",   
  title = "{GGM02} -- An improved {Earth} gravity field model from {GRACE}", 
  journal = "Journal of Geodesy",
  volume = "79",
  pages = "467--478"
  }

@ ARTICLE{ bT59,
  author = "Z. Th{\"u}ring",
  year = "1959",   
  title = "Zwei spezielle mondeinfang-bahnen in der raumfahrt um {Erde} und {Mond}",
  journal = "Astronautica Acta",
  volume = "5",
  pages = "241--250"
  }

@ ARTICLE{ fT08,
  author = "F. Topputo and E. Belbruno and M. Gidea",
  year = "2008",   
  title = "Resonant motion, ballistic escape, and their applications in astrodynamics",
  journal = "Advances in Space Research",
  volume = "42",
  pages = "1318--1329"
  }

@ ARTICLE{ fT13,
  author = "F. Topputo",
  year = "2013",   
  title = "On optimal two-impulse {Earth--Moon} transfers in a four-body model",
  journal = "Celestial Mechanics and Dynamical Astronomy",
  volume = "117",
  pages = "279--313"
  }

@ ARTICLE{ sTtY14,
  author = "S. Tremaine and T. D. Yavetz",
  year = "2014",   
  title = "Why do {Earth} Satellites Stay Up?",
  journal = "American Journal of Physics",
  volume = "82",
  pages = "769--777"
  }

@ ARTICLE{ sT09,
  author = "S. Tremaine and J. Touma and F. Namouni",
  year = "2009",   
  title = "Satellite Dynamics on the {Laplace} Surface",
  journal = "The Astronomical Journal",
  volume = "137",
  pages = "3706--3717"
  }

@ ARTICLE{ sTetal14,
  author = "S. Tzirti and A. Noullez and K. Tsiganis",
  year = "2014",   
  title = "Secular dynamics of a lunar orbiter: a global exploration using {Prony's} frequency analysis",
  journal = "Celestial Mechanics and Dynamical Astronomy",
  volume = "118",
  pages = "379--397"
  }

@ ARTICLE{ aU15,
  author = "A. Utku and L. Hagen and P. Palmer",
  year = "2015",   
  title = "Initial condition maps of subsets of the circular restricted three-body problem phase space",
  journal = "Celestial Mechanics and Dynamical Astronomy",
  volume = "123",
  pages = "387--410"
  }

@ ARTICLE{ mVkH14a,
  author = "M. Vaquero and K. C. Howell",
  year = "2014a",   
  title = "Design of transfer trajectories between resonant orbits in the {Earth--Moon} restricted problem", 
  journal = "Acta Astronautica",
  volume  = "94",
  pages = "302--317"
  }

@ ARTICLE{ mVkH14b,
  author = "M. Vaquero and K. C. Howell",
  year = "2014b",   
  title = "Leveraging resonant-orbit manifolds to design transfers between libration-point orbits", 
  journal = "Journal of Guidance, Control, and Dynamics",
  volume  = "37",
  pages = "1143--1157"
  }

@ ARTICLE{ dVP23,
  author = "D. Villegas-Pinto and N. Baresi and S. Locoche and D. Hestroffer",
  year = "2023",   
  title = "Resonant quasi-periodic near-rectilinear halo orbits in the elliptic-circular {Earth--Moon--Sun} problem",
  journal = "Advances in Space Research",
  volume = "73",
  pages = "64--84"
  }

@ ARTICLE{ bVeC18,
  author = "B. R. Vinson and E. Chiang",
  year = "2018",   
  title = "Secular dynamics of an exterior test particle: the inverse {Kozai} and other eccentricity--inclination resonances",
  journal = "Monthly Notices of the Royal Astronomical Society",
  volume = "474",
  pages = "4855--4869"
  }

@ ARTICLE{ gV05,
  author = "G. Voyatzis and T. Kotoulas and J. D. Hadjidemetriou",
  year = "2005",   
  title = "Symmetric and nonsymmetric periodic orbits in the exterior mean motion resonances with {Neptune}",
  journal = "Celestial Mechanics and Dynamical Astronomy",
  volume = "91",
  pages = "191--202"
  }

@ ARTICLE{ yWtF23,
  author = "Y. Wang and T. Fu",
  year = "2023",   
  title = "An orbit-flip mechanism by eccentric {Lidov--Kozai} effect with stellar oblateness",
  journal = "The Astronomical Journal",
  volume = "165",
  pages = "201 (14 pp.)"
  }

@ ARTICLE{ cW17,
  author = "C. M. Will",
  year = "2017",   
  title = "Orbital flips in hierarchical triple systems: {Relativistic} effects and third-body effects to hexadecapole order",
  journal = "Physical Review D",
  volume = "96",
  pages = "023017 (15 pp.)"
  }

@ ARTICLE{ aW25,
  author = "A. Wilmer and R. A. Bettinger and L. M. Shockley and M. J. Holzinger",
  year = "2025",   
  title = "Preliminary investigation and proposal of periodic orbits and their utilization for logistics in the cislunar regime", 
  journal = "Space Policy",
  pages = "101635 (17 pp.)"
  }

@ ARTICLE{ oWcM97,
  author = "O. C. Winter and C. D. Murray",
  year = "1997",   
  title = "Resonance and chaos: {II.} {Exterior} resonances and asymmetric libration",
  journal = "Astronomy and Astrophysics",
  volume = "328",
  pages = "399--408"
  }

@ ARTICLE{ jW80,
  author = "J. Wisdom",
  year = "1980",   
  title = "The resonance overlap criterion and the onset of stochastic behavior in the restricted three-body problem",
  journal = "The Astronomical Journal",
  volume = "85",
  pages = "1122--1133"
  }

@ ARTICLE{ kY04,
  author = "K. Yagasaki",
  year = "2004",   
  title = "Sun-perturbed {Earth-to-Moon} transfers with low energy and moderate flight time",
  journal = "Celestial Mechanics and Dynamical Astronomy",
  volume = "90",
  pages = "197--212"
  }

@ ARTICLE{ dGY13,
  author = "D. Garc{\'i}a Y{\'a}rnoz and J. P. Sanchez and C. R. McInnes",
  year = "2013",   
  title = "Easily retrievable objects among the {NEO} population",
  journal = "Celestial Mechanics and Dynamical Astronomy",
  volume = "116",
  pages = "367--388"
  }

@ ARTICLE{ kYeM04,
  author = "K. Yazdi and E. Messerschmid",
  year = "2004",   
  title = "Analysis of parking orbits and transfer trajectories for mission design of cis-lunar space stations",
  journal = "Acta Astronautica",
  volume = "55",
  pages = "759--771"
  }

@ ARTICLE{ gZ14,
  author = "G. S. Zaslavskiy and V. A. Stepan'yants and A. G. Tuchin and others",
  year = "2014",   
  title = "Trajectory correction of the {Spektr-R} spacecraft motion",
  journal = "Kosmicheskie Issledovaniya", 
  volume = "52",
  pages = "387--398"
  }

@ ARTICLE{ gZ17,
  author = "G. S. Zaslavskii and M. V. Zakhvatkin and N. S. Kardashev and others",
  year = "2017",   
  title = "Designing corrections for the trajectory of the {Spektr-R} spacecraft in the event of immersions into the {Moon's} sphere of influence",
  journal = "Kosmicheskie Issledovaniya", 
  volume = "55",
  pages = "305--320"
  }

@ ARTICLE{ eZ20,
  author = "E. M. Zimovan-Spreen and K. C. Howell and D. C. Davis",
  year = "2020",   
  title = "Near rectilinear halo orbits and nearby higher-period dynamical structures: orbital stability and resonance properties",
  journal = "Celestial Mechanics and Dynamical Astronomy",
  volume = "132",
  pages = "28 (25 pp.)"
  }

@ ARTICLE{ mZ13,
  author = "M. T. Zuber and D. E. Smith and M. M. Watkins and others",
  year = "2013",   
  title = "Gravity field of the {Moon} from the {Gravity Recovery and Interior Laboratory (GRAIL)} mission",
  journal = "Science",
  volume = "339",
  pages = "668--671"
  }

@ TECHREPORT{ rB68,
  author = "R. A. Broucke",
  year = "1968",
  title = "{Periodic Orbits in the Restricted Three-Body Problem with {Earth--Moon} Masses}",
  institution = "NASA",
  number = "JPL Technical Report 32--1168"
  }

@ TECHREPORT{ rB69,
  author = "R. A. Broucke",
  year = "1969",
  title = "{Periodic Orbits in the Elliptic Restricted Three-Body Problem}",
  institution = "NASA",
  number = "JPL Technical Report 32--1360"
  }

@ TECHREPORT{ rB56a,
  author = "R. W. Buchheim",
  year = "1956",   
  title = "{Motion of a Small Body in Earth--Moon Space}",
  institution = "RAND",
  number = "RM--1726"
  }

@ TECHREPORT{ rB56b,
  author = "R. W. Buchheim",
  year = "1956",   
  title = "{Artificial Satellites of the Moon}",
  institution = "RAND",
  number = "RM--1941"
  }

@ TECHREPORT{ pB80,
  author = "P. Butler",
  year = "1980",   
  title = "{Interplanetary Monitoring Platform}",
  institution = "NASA",
  number = "TM--80758"
  }

@ TECHREPORT{ gH64,
  author = "G. P. Herring",
  year = "1964",   
  title = "{A Comprehensive Astrodynamic Exposition and Classification of {Earth--Moon} Transits}",
  institution = "NASA",
  number = "TM X--53151"
  }

@ TECHREPORT{ rHbW68,
  author = "R. F. Hoelker and B. P. Winston",
  year = "1968",   
  title = "{A Comparison of a Class of {Earth--Moon} Orbits with a Class of Rotating Kepler Orbits}",
  institution = "NASA",
  number = "TN D--4903"
  }

@ TECHREPORT{ mH21,
  author = "M. J. Holzinger and C. C. Chow and P. Garretson",
  year = "2021",
  title = "{A Primer on Cislunar Space}",
  institution = "AFRL",
  number = "2021--1271"
  }

@ TECHREPORT{ fH21,
  author = "F. R. Hoots",
  year = "2021",
  title = "{SGP4-XP A New TLE Orbit Prediction Model}",
  institution = "Aerospace",
  number = "TOR--2021--00780"
  }

@ TECHREPORT{ sH60,
  author = "S. -S. Huang",
  year = "1960",   
  title = "{Very Restricted Four-Body Problem}",
  institution = "NASA",
  number = "TN D--501"
  }

@ TECHREPORT{ WEC59,
  author = "J. F. {Miller, Jr.}",
  year = "1959",   
  title = "{Trajectory Problems in Cislunar Space}",
  institution = "Westinghouse Electric Corporation, Air Arm Division", 
  address = "Baltimore, MD",
  number = "AFOSR--TN--59--1284"
  }

@ TECHREPORT{ jMrB62,
  author = "J. S. Miller and R. H. Battin",
  year = "1962",
  title = "{Preliminary Summary of Data for a Variety of Circumlunar Trajectories}",
  institution = "MIT Instrumentation Laboratory",
  number = "Engineering Report E--1131"
  }

@ TECHREPORT{ tP26,
  author = "T. W. Pennington",
  year = "2026",
  title = "{Understanding Space Frontier Areas: Strategy in Cislunar Space and Beyond}",
  institution = "Institute for National Strategic Studies, National Defense University",
  number = "45"
  }

@ TECHREPORT{ pP61,
  author = "P. A. Penzo",
  year = "1961",
  title = "{An Analysis of Moon-to-Earth Trajectories}",
  institution = "NASA",
  number = "CR--132100"  
  }

@ TECHREPORT{ aP61,
  author = "A. Petty and I. Jurkevich and M. Fabrize and T. Coffin",
  year = "1961",
  title = "{Lunar Trajectory Studies}",
  institution = "USAF",
  number = "AFCRL 507"  
  }

@ TECHREPORT{ hSwH64,
  author = "H. B. Schechter and W. C. Hollis",
  year = "1964",   
  title = "{Stability of the Trojan Points in the Four-Body Problem}",
  institution = "RAND",
  number = "RM-3992-PR"
  }

\end{document}